\documentclass[twocolumn]{aastex701}

\usepackage{multirow}
\usepackage{booktabs}
\usepackage{makecell}
\usepackage{array}
\usepackage{longtable}
\usepackage{bbding}
\usepackage{comment}

\usepackage{stackengine}

\usepackage{tcolorbox}
\tcbuselibrary{skins}

\usepackage[utf8]{inputenc}
\usepackage{natbib}
\usepackage{graphicx}
\usepackage{longtable}
\usepackage{enumerate}
\usepackage{amsmath}
\usepackage{hyperref}
\usepackage{CJK}
\usepackage[title]{appendix}
\usepackage{rotating}
\usepackage{longtable}

\usepackage{amsmath}
\def\hratio{\ifmmode r_{\mathrm {NW}} \else $r_{\mathrm {NW}}$\fi}
\def\wnratio{\ifmmode r_{\mathrm {CWN}} \else $r_{\mathrm {CWN}}$\fi}
\def\msun{\ifmmode {\rm M}_{\mathord\odot}\else $M_{\mathord\odot}$\fi}
\def\rsun{\ifmmode {\rm R}_{\mathord\odot}\else $R_{\mathord\odot}$\fi}
\def\lsun{\ifmmode {\rm L}_{\mathord\odot}\else $L_{\mathord\odot}$\fi}

\def\c18o{C$^{18}$O}
\def\h2{H$_{2}$}
\def\13co{$^{13}$CO}
\def\n2hp{$_{2}$H$^{+}$}

\def\cm2{cm$^{-2}$}

\newcommand{\kms}{km~s$^{-1}$}

\def\deg{$^{\circ}$}

\usepackage{xcolor}

\newcommand{\xd}[1]{\textcolor{black}{{#1}}}
\newcommand{\xdtwo}[1]{\textcolor{black}{{#1}}}
\newcommand{\pgm}[1]{%
  {\color{red} {\textbf{PGM: #1}}} %
}

\shorttitle{Assessing Zeeman Measurements of Magnetic Fields in Synthetic HI Observations}
\shortauthors{Xu et al.}

\begin{document}
\begin{CJK*}{UTF8}{gbsn}

\title{Assessing Zeeman Measurements of Magnetic Fields in Synthetic HI Observations}

\author[0000-0001-6216-8931]{Duo Xu}
\affiliation{Canadian Institute for Theoretical Astrophysics, University of Toronto, 60 St. George Street, Toronto, ON M5S 3H8, Canada}
\email[show]{xuduo@cita.utoronto.ca}

\author[0000-0002-5236-3896]{Peter G. Martin}
\affiliation{Canadian Institute for Theoretical Astrophysics, University of Toronto, 60 St. George Street, Toronto, ON M5S 3H8, Canada}
\email{pgmartin@cita.utoronto.ca}

\author[0000-0003-1252-9916]{Stella S. R. Offner}
\affiliation{Department of Astronomy, The University of Texas at Austin, Austin, TX 78712, USA}
\email{soffner@utexas.edu}

\author[0000-0002-6447-899X]{Robert Gutermuth}
\affiliation{Department of Astronomy, University of Massachusetts, Amherst, MA 01003, USA }
\email{rgutermu@astro.umass.edu}

\author[0000-0002-1655-5604]{Michael Y. Grudi\'{c}}
\affiliation{Center for Computational Astrophysics, Flatiron Institute, New York, NY 10010, USA}
\email{mike.grudich@gmail.com}

\author[0000-0003-2573-9832]{Joshua S. Speagle \begin{CJK*}{UTF8}{gbsn}(沈佳士)\end{CJK*}}
\affiliation{Department of Statistical Sciences, University of Toronto, 9th Floor, Ontario Power Building, 700 University Ave, Toronto, ON M5G 1Z5, Canada}
\affiliation{David A. Dunlap Department of Astronomy \& Astrophysics, University of Toronto, 50 St. George Street, Toronto, ON M5S 3H4, Canada}
\affiliation{Dunlap Institute for Astronomy \& Astrophysics, University of Toronto, 50 St. George Street, Toronto, ON M5S 3H4, Canada}
\affiliation{Data Sciences Institute, University of Toronto, 17th Floor, Ontario Power Building, 700 University Ave, Toronto, ON M5G 1Z5, Canada}
\email{j.speagle@utoronto.ca}


\begin{abstract}

Zeeman observations provide the only direct probe of line-of-sight (LOS) magnetic fields in the interstellar medium. To evaluate their accuracy and limitations, we generate synthetic HI Zeeman spectra from magnetohydrodynamic simulations and idealized cloud models, and analyze the resulting Stokes $I$ and $V$ profiles using two complementary methods. Approach I uses the classical relation between Stokes $V$ and $dI/d\nu$ to estimate LOS-averaged magnetic fields, achieving an upper-limit relative error of $\sim$16\% (half-width of 68.27\% confidence interval) for a representative noise level of 0.014~K. Approach II applies Gaussian decomposition to Stokes $I$ and $V$ to estimate component-level magnetic fields, \xd{yielding a $\sim$13\% relative error quantifying the same confidence range, reflecting the intrinsic uncertainty of such Zeeman estimates.} Both approaches recover the original fields under uniform-field  conditions and remain robust in turbulent environments. Approach I provides a simple and reliable LOS-averaged field estimate, while Approach II, although more complex, offers statistical insight into magnetic field variations along the LOS. We further show that joint fitting of Stokes $I$ and $V$ generally outperforms sequential fitting, particularly in the presence of attenuation. Increasing noise eight-fold produces a more modest rise in uncertainty, doubling to a $\sim$26\% relative error, while substantial optical depth introduces only a minor additional contribution to the overall uncertainty. Applying these methods to FAST observations of the L1544 star-forming region, we confirm the previously reported LOS magnetic field strength, demonstrating the validity of Zeeman analysis in this benchmark core.

\end{abstract}



\section{Introduction}
\label{Introduction}

Magnetic fields are a crucial component of the interstellar medium (ISM), influencing a broad range of astrophysical processes, from the dynamics of molecular clouds to star formation and stellar feedback \citep[e.g.,][]{2017ARA&A..55..111H, 2012ARA&A..50...29C}. These fields interact with interstellar gas and dust, providing support against gravitational collapse and shaping the large-scale structure of the ISM, including its filamentary and turbulent features \citep{2015MNRAS.450.4035F, 2024A&A...686L..11W}. Despite their fundamental role, accurately measuring the magnetic field strength remains a persistent challenge due to the inherently three-dimensional nature of magnetic fields and the limitations of current observational techniques.

Various methods have been developed to probe magnetic fields in the ISM. Polarized thermal dust emission \citep{1998ApJ...502L..75R,2016A&A...586A.138P}, starlight polarization \citep{1951ApJ...114..206D,2002ApJ...564..762F}, and synchrotron emission \citep{1982A&A...105..192B,2012ApJ...761L..11J} provide valuable constraints on the plane-of-sky (POS) magnetic field component. The line-of-sight (LOS) component can be inferred (estimated with some uncertainty) using Faraday rotation \citep{1966MNRAS.133...67B,2022A&A...657A..43H} and the Zeeman effect \citep{1986ApJ...301..339T,2010ApJ...725..466C}, but each method has limitations. Faraday rotation depends on assumptions about the electron density along the LOS, while the Davis-Chandrasekhar-Fermi (DCF) method estimates POS magnetic field strength from polarized dust emission by assuming equipartition between magnetic field energy and turbulent kinetic energy \citep{PhysRev.81.890.2,1953ApJ...118..113C,2015A&ARv..24....4B}. \xd{In contrast, the Zeeman effect offers the only direct estimate of the LOS magnetic field strength, although  depending on the observed species, it provides differently-density-weighted averages along the line of sight \citep{2008A&A...487..247F,2008ApJ...680..457T,2009ApJ...705L.176S,2022MNRAS.516L..48K}. While Zeeman analysis is an essential diagnostic for probing magnetic fields in the ISM,} its accuracy in real astrophysical conditions remains uncertain.

The Zeeman effect in the hyperfine 21 cm transition of neutral hydrogen (HI) is particularly valuable for probing magnetic fields in the diffuse and cold neutral medium (CNM), which serves as an intermediary phase between ionized and molecular regions of the ISM \citep{1968Natur.220.1207D,1999ApJ...520..706C}. Despite its importance, detecting the Zeeman effect in HI is extremely challenging because its magnitude of the line shifts, approximately 1.4 Hz ($\mu$G)$^{-1}$, is much smaller than the typical thermal and turbulent broadening of HI spectral lines. Additionally, HI emission is often composed of multiple velocity components along the LOS, leading to spectral blending that complicates the isolation of Zeeman signatures. Further challenges include the low signal-to-noise ratio (SNR) in circular polarization (Stokes $V$), instrumental systematics, and uncertainties in HI optical depth, all of which introduce significant uncertainties in magnetic field inference from Zeeman observations. These difficulties raise a fundamental question: how reliably can Zeeman HI observations constrain the magnetic field strength in the ISM?

In this study, we assess the impact of these observational and physical challenges on Zeeman HI measurements using synthetic observations generated from magnetohydrodynamic (MHD) simulations. %
Section~\ref{Data and Method} presents the numerical simulations, properties of the HI gas, radiative transfer to generate synthetic observations of Zeeman HI spectra (both Stokes $I$ and Stokes $V$), and the importance of optical depth.
Supported by Appendix \ref{app:smallfields}, Section \ref{TwoApproaches} lays out two primary approaches for analyzing Zeeman spectra: Approach I fits the $V$ spectrum as a whole to estimate $B_{zm}$ averaged along the line of sight, while Approach II focuses on fitting individual spectral component contributions to $I$ and $V$ to determine the $B_z$ associated with each component.  A quick alternative way to compute the radiative transfer
is described in Section \ref{gpb}.
In Section \ref{Simple Geometry Validation} we explore a simple five-layer validation model, ground truth contributions to Stokes $I$ and $V$ spectra with depth, and ground truth estimators for the modeled LOS magnetic field (see also Appendix \ref{Ground Truth II: Magnetic Field Weighted by Spectral Similarity}).
Further exploration appears in Appendices \ref{Smooth Magnetic Field Gradient Experiment} and \ref{Magnetic Field Structure}, all supplementing and complementing the validation. 

Section \ref{Analysis of Synthetic HI Spectra from MHD Simulations Using Approach I}
presents Approach I results of our Zeeman analysis of the synthetic spectra from MHD simulations, concentrating first on four selected lines of sight through the simulation cube and including the effect of noise.
Appendix \ref{Zeeman Fitting Gallery: Approach I Across Noise and Regions} reviews Approach I results across noise levels for two of the selected lines of sight. In Section~\ref{Assessing Uncertainties in Approach I: Derivative-Based Stokes $I$ Fitting}, we assess statistically the accuracy and limitations of Zeeman-based magnetic field measurements under a range of astrophysical conditions, scanning the entire cube.
To explore another key factor, optical depth,
in Appendix \ref{app:approachIstatistics0001} we carried out experiments lowering the densities so that the cube is optically thin. 

Section \ref{Analysis of Synthetic HI Spectra from MHD Simulations Using Approach II} presents results of our Zeeman analysis of the synthetic spectra using Approach II.
In Appendix \ref{app:approachII0001} we carried out experiments lowering the densities so that the cube is optically thin.
Appendix~\ref{app:flip} examines the physical reality of the fitted spectral components and their associated magnetic field properties by the device of analysing synthetic spectra as viewed from the other side of the cube. 
See also Appendix \ref{Zeeman Fitting Gallery: Approach II Across Noise and Regions} for Approach II results for spectra with different added noise levels. 
In Section~\ref{Assessing Uncertainties in Approach II: Gaussian Component Fitting}, we assess the accuracy and limitations of Zeeman-based magnetic field measurements under a range of astrophysical conditions. 

In Section~\ref{Re-examining HI Data from FAST}, we revisit analysis of recent HI Zeeman observations from FAST for a cold dense molecular core, to estimate the magnetic field strength using our own analysis framework. Finally, Section~\ref{Conclusions} summarizes our approaches to inferring LOS magnetic fields, enumerates key findings, and discusses implications for future HI Zeeman studies.

\section{MHD Simulations and Synthetic Observations of Zeeman HI Spectra}
\label{Data and Method}

\begin{figure*}[hbt!]
\centering
\includegraphics[width=0.75\linewidth]{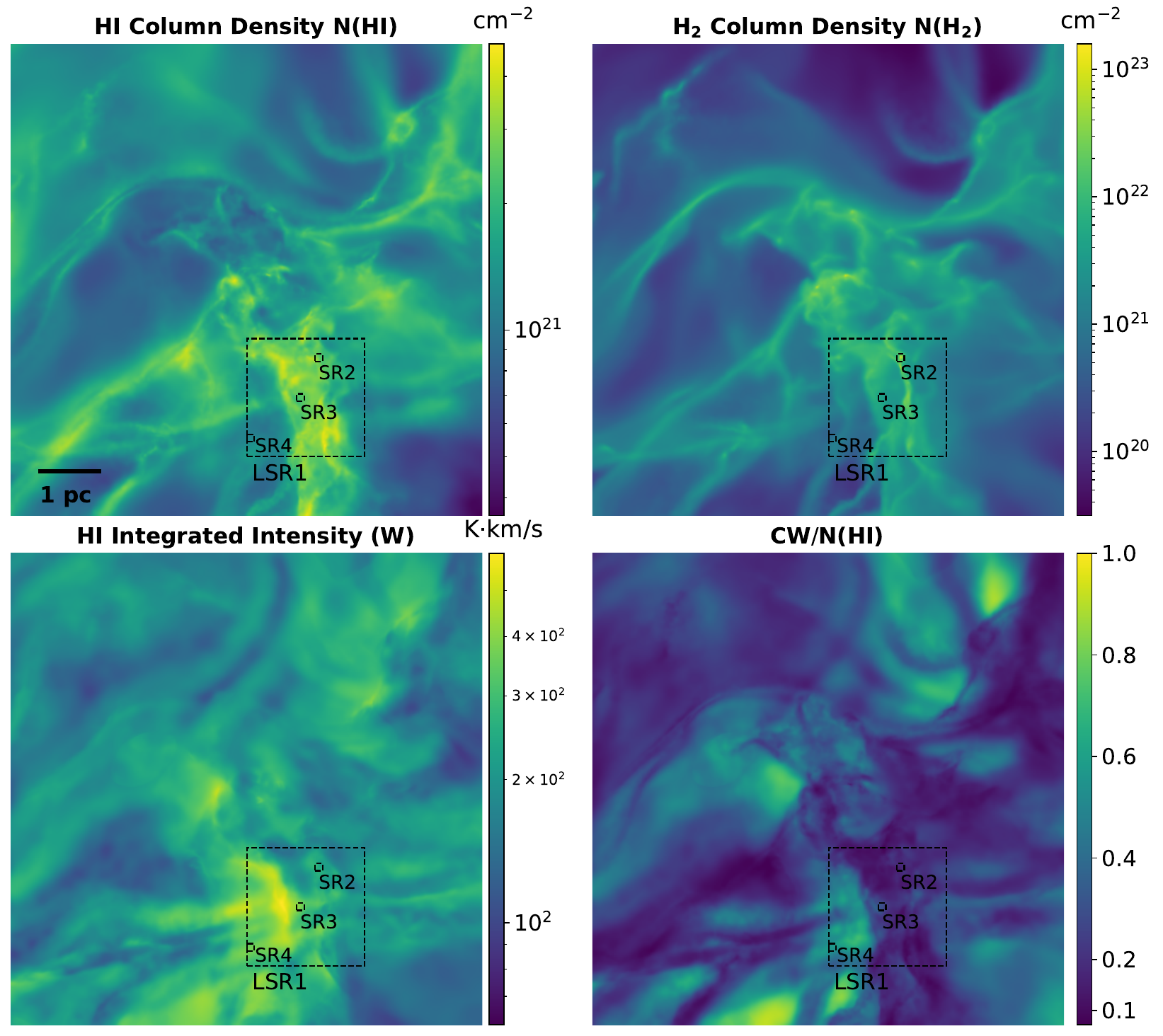}
\caption{HI and H$_2$ column density maps (upper left and right), HI integrated intensity map $W$ (lower left), and 
\xd{the ratio of $C\,W$ to HI column density (\wnratio\ -- see text) 
(lower right).} 
Subregions SR2, SR3, and SR4 have a box size of 0.125 pc, which at a distance of 140 pc subtends 3\arcmin.  LSR1 is 16 times larger in size.
}
\label{fig.column_density_full_4panel}
\end{figure*}

\subsection{Magnetohydrodynamics Simulations}
\label{Magnetohydrodynamics Simulations}

We use a simulation from the STARFORGE suite \citep{2021MNRAS.502.3646G,2021MNRAS.506.2199G,2022MNRAS.512..216G,2022MNRAS.515..167G,2022MNRAS.515.4929G,2023MNRAS.518.4693G,2024MNRAS.529.4128G}, which models star formation in giant molecular clouds with a comprehensive treatment of physical processes. These simulations track the formation, accretion, evolution, and dynamics of individual stars within massive clouds, incorporating gravity, magnetic fields, and cosmic ray ionization, as well as stellar feedback mechanisms such as jets, radiative heating and pressure, stellar winds, and supernovae. The simulations are conducted using the GIZMO code and employ the Lagrangian meshless finite-mass (MFM) method for MHD under the ideal MHD approximation \citep{2015MNRAS.450...53H,2016MNRAS.455...51H}.

The simulation (\texttt{M2e5\_\allowbreak R30\_\allowbreak Z1\_\allowbreak S0\_\allowbreak A2\_\allowbreak B0.1\_\allowbreak I1\_\allowbreak Res271\_\allowbreak n2\_\allowbreak sol0.5\_\allowbreak 42}) initializes with a uniform gas sphere containing 200,000 M$_\odot$ of material within a 30 pc radius. It begins as a relatively diffuse, predominantly atomic cloud with an initial temperature of 10 K. This cloud is surrounded by a warm ($T = 10^{4}$ K), diffuse medium that is 1000 times less dense, maintaining initial thermal pressure equilibrium with its surroundings. Turbulence is introduced as a Gaussian random velocity field with a power spectrum $E_k \propto k^{-2}$, scaled to achieve a turbulent virial parameter of 2. The gravitational and magnetic energies follow the relation $E_{\rm mag} = 0.1 E_{\rm grav}$. The simulation adopts local (solar neighborhood) interstellar radiation field (ISRF) conditions and solar metallicity. 

In this study, we focus on an early evolutionary stage at 2.9 Myr within an 8 pc region (3D cube) interpolated to a uniform spatial resolution of 0.015625 pc. 

\subsection{The HI Gas}
\label{HI gas}

\begin{figure*}[hbt!]
\centering
\includegraphics[width=0.48\linewidth]{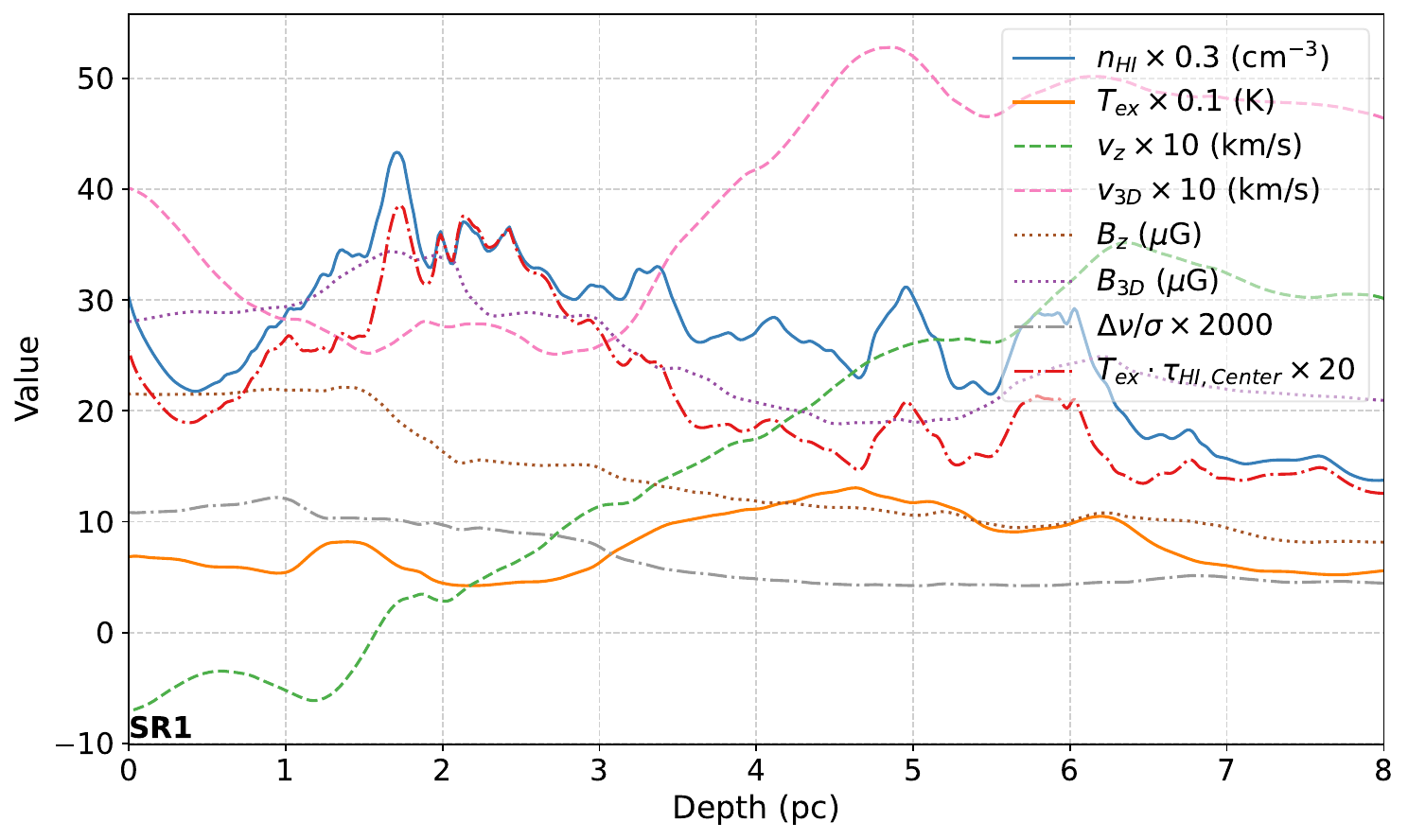}
\includegraphics[width=0.48\linewidth]{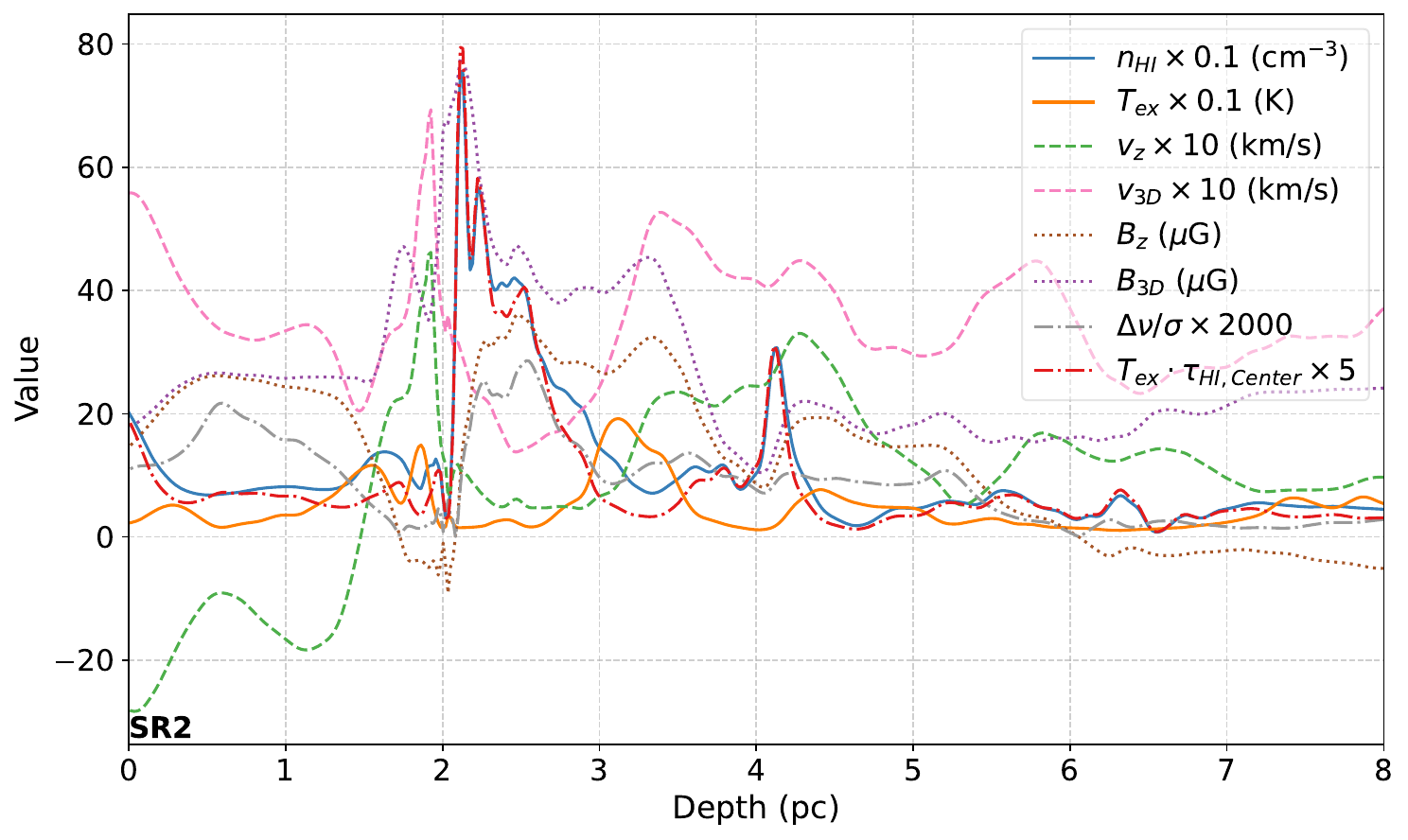}
\includegraphics[width=0.48\linewidth]{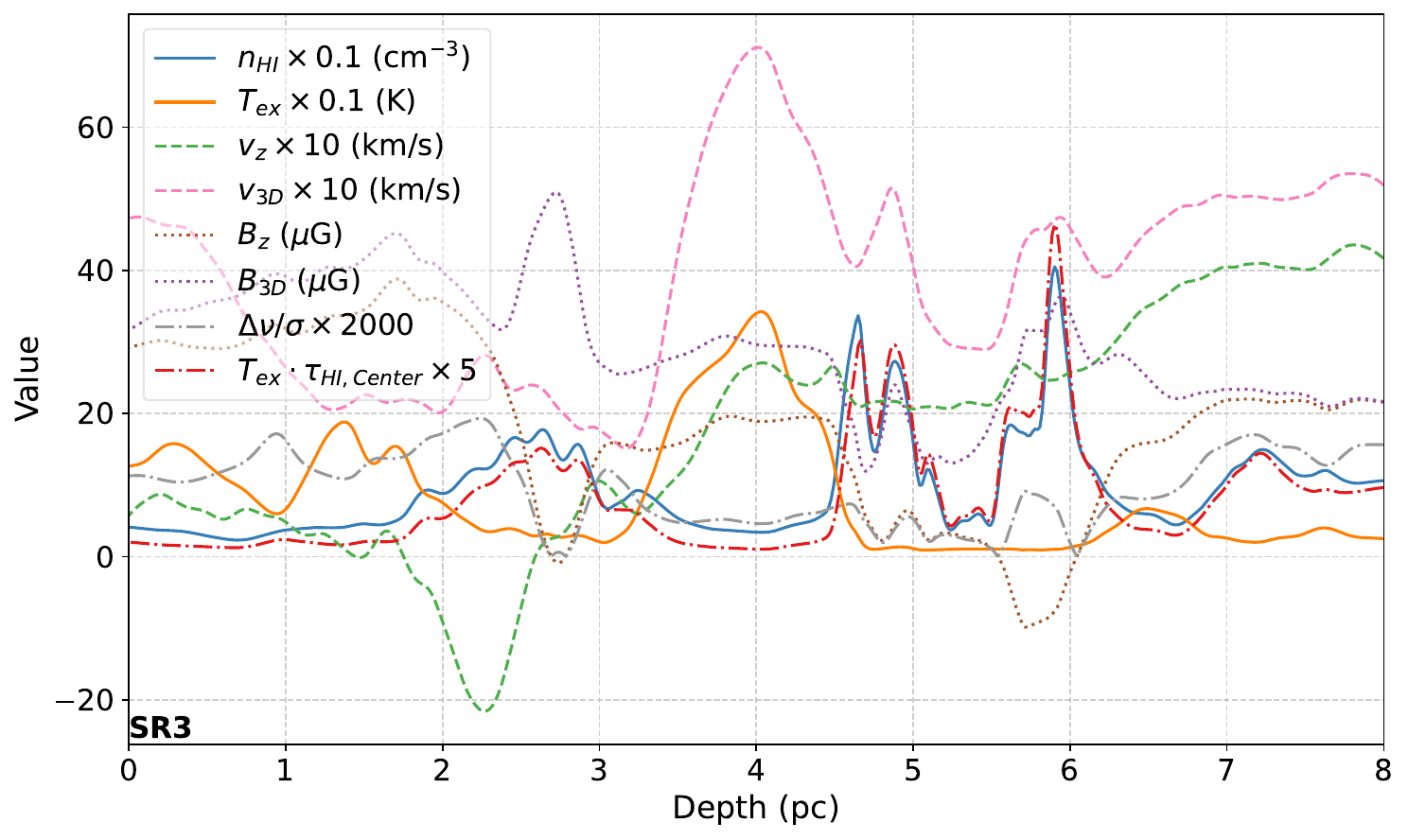}
\includegraphics[width=0.48\linewidth]{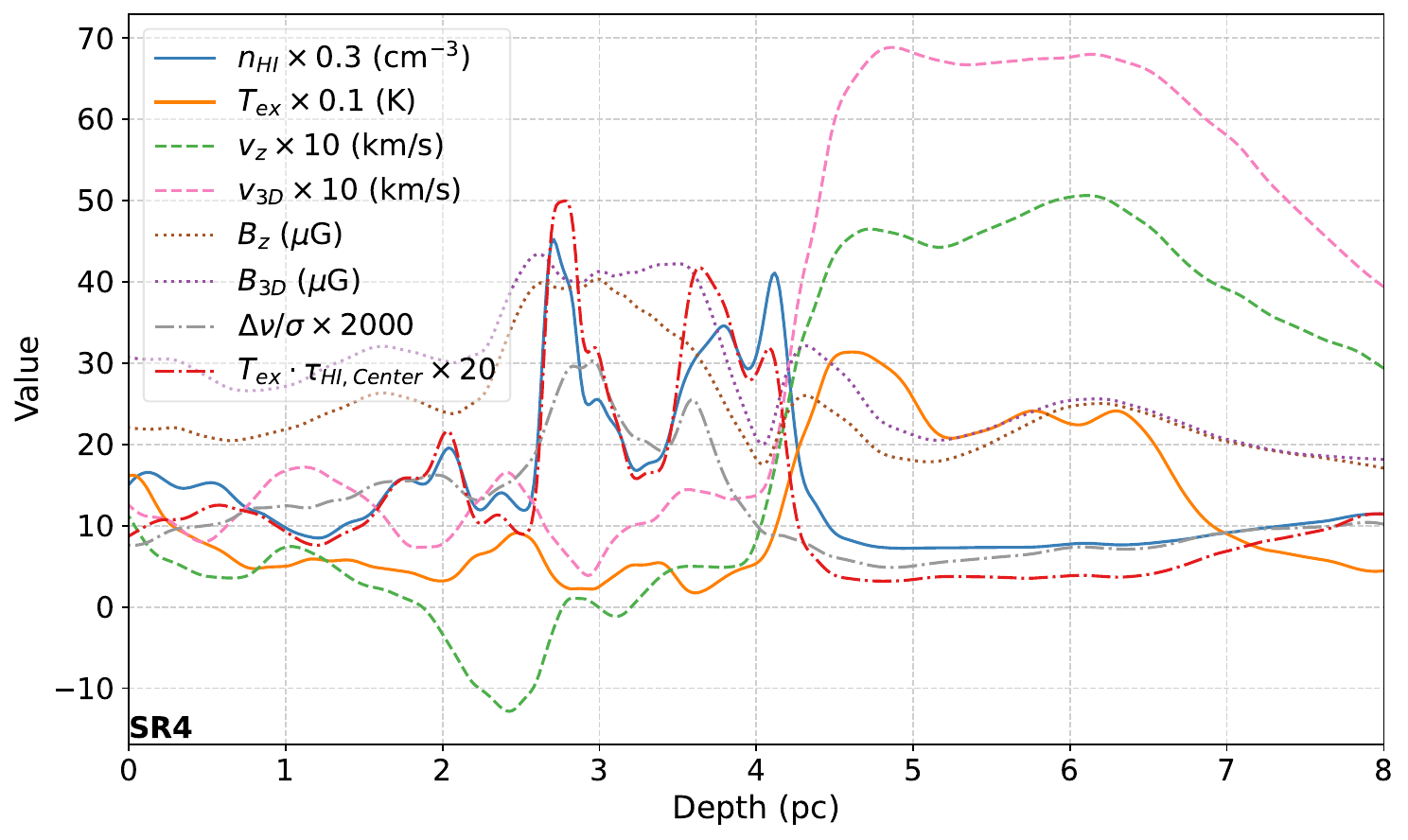}
\caption{LOS profiles of physical properties directly relevant to the Stokes $I$ and $V$ spectra for the four selected subregions: HI number density $n_{\mathrm{HI}}$, excitation temperature $T_{\mathrm{ex}}$, LOS velocity $v_z$, and
LOS magnetic field strength $B_z$.
Also shown for context, though not observable,
the 3D velocity magnitude $v_{\mathrm{3D}}$ and 3D magnetic field magnitude $B_{\mathrm{3D}}$.   To assess whether the small-shift approximation remains valid at the voxel level we plot the ratio of the Zeeman-induced frequency shift $\Delta \nu$ to the local velocity dispersion $\sigma$ (valid: see text). We also plot $T_{\mathrm{ex}} \cdot \tau_{\mathrm{HI,\, Center}}$ (see text). Each quantity has been suitably scaled to use the common value scale on the y axis.}
\label{fig.physical_Region_All}
\end{figure*}

The upper two panels of Figure~\ref{fig.column_density_full_4panel} present the column density maps of HI ($N({\mathrm {HI}})$) and H$_2$ ($N({\mathrm {H_2}})$)
for the selected 8 pc region from the simulation. 
Comparison of the color bar scales of the upper panels shows that the region is largely molecular with trace, but still detectable, HI.  


\begin{deluxetable}{rccccc}
\label{tab:srproperties}
\tablecaption{Basic densities in LSR1 and subregions}
\tablehead{Name& $N$(HI) & $N$(H$_2$) & HI mass & $n$(HI)$^b$ & $n$(H$_2$)$^b$\\
& \multicolumn{2}{c}{$10^{21}$ cm$^{-2}$} & fraction $^a$ & \multicolumn{2}{c}{cm$^{-3}$} } 
\startdata
LSR1 & 2.2 & 4.2  & 0.20 & 213 & 8797  \\ 
SR2  & 2.6 & 23 & 0.05 & 277 & 138589  \\ 
SR3  & 2.4 & 4.1  & 0.23 & 155 & 674 \\ 
SR4  & 1.2 & 0.77  & 0.43 & 68 & 164 \\ 
\enddata
\tablenotetext{}{$^a$ Derived from column densities.
}
\tablenotetext{}{$^b$ Mass-weighted along the line of sight.}
\end{deluxetable}

\xd{Within the selected 8 pc simulation region, we identify several subregions (SRs) that reflect different physical environments. As summarized in Table~\ref{tab:srproperties}, SR2 contains dense, high column density \h2\ molecular gas. Although its HI mass fraction is only 0.05, it still exhibits a moderate HI column density. In contrast, SR3 and SR4 are progressively more diffuse and increasingly atomic rather than molecular, although the molecular gas still accounts for more than half of the total mass.
}
\xd{Large subregion LSR1 spans 2 pc, while SR2, SR3, and SR4 each cover 0.125 pc. At a distance of 140 pc (e.g., for the Taurus molecular cloud), the latter SR coverage corresponds to an angular scale of approximately 3\arcmin\, comparable to the Five-hundred-meter Aperture Spherical Telescope (FAST) beam.} 

In later sections and appendices, we analyse synthetic spectra from these subregions (and the full LSR1).
\xd{These are important for exploring whether the profile of the LOS magnetic field along the line of sight might be recovered; as discussed below this is somewhat successful using the joint fitting strategy of Approach II (Section \ref{sec:jfs}).}
\xd{While the Zeeman analysis behaves qualitatively similarly across all regions, the detailed performance varies depending on additional local physical conditions such as the temperature and velocity profiles (Fig.~\ref{fig.physical_Region_All}) and the resulting attenuation (Section \ref{sec:shapespectra}), as well as the level of noise in the synthetic spectra being modeled (Section \ref{sec:addnoise}).}

We also carried out a full 64 by 64 scan (4096 spectra) at the SR box size across the entire 8 pc area to assess systematically the uncertainties in the Zeeman analysis, 
\xd{again as a function of noise level.}

\subsubsection{Physical conditions along the LOS shaping spectra}
\label{sec:shapespectra}

Figure~\ref{fig.physical_Region_All} shows the LOS distributions of key physical properties within each subregion. \xd{The kinetic temperature $T_{\mathrm{k}}$ is known from the simulation and the excitation temperature  $T_{\mathrm{ex}} = T_{\mathrm{k}}$ under assumed LTE.} This along with HI number density $n_{\mathrm{HI}}$ and LOS velocity $v_z$ determine the Stokes $I$ spectrum, typically expressed in terms of brightness temperature $T_b$ (also referred to as $T_{\mathrm{HI}}$ hereafter). The LOS magnetic field strength $B_z$, which governs the Stokes $V$ spectrum, is also shown for each region. While the full 3D velocity and magnetic field vectors, and their magnitudes $v_{\mathrm{3D}}$ and $B_{\mathrm{3D}}$, are available in the simulations and are included in the figure for context, they are not directly observable from the HI spectra.

The ratio of the Zeeman-induced frequency shift $\Delta \nu$ to the local velocity dispersion $\sigma$ in each subregion allows us to evaluate the validity of the small-shift approximation at the voxel level.  The largest value of the ratio is about 0.015, which is in the range of acceptability (Appendix \ref{app:smallfields}). 

We also plot $T_{\mathrm{ex}} \cdot \tau_{\mathrm{HI,Center}}$, where $T_{\mathrm{ex}}$ is the HI excitation temperature and $\tau_{\mathrm{HI,Center}}$ is the optical depth at the line center, defined as
\begin{equation}
\label{eq:tau_center}
\tau_{\mathrm{HI,Center}} = \left[ \frac{c^3 h}{32\pi k_B \nu_0^2 \sqrt{2\pi}} \frac{g_1}{g_0} A_{10} \right] \frac{N_{\mathrm{HI}}}{T_{\mathrm{ex}}\, s_v},
\end{equation}
with $c$ the speed of light, $h$ the Planck constant, $k_B$ the Boltzmann constant, $\nu_0$ the rest frequency of the HI hyperfine transition, $g_1$ and $g_0$ the degeneracies of the upper and lower states, $A_{10}$ the Einstein coefficient, $N_{\mathrm{HI}}$ the HI column density, $T_{\mathrm{ex}}$ the excitation temperature, and $s_v$ the velocity dispersion (not to be confused with $\sigma$, which denotes frequency dispersion in this work). The $T_{\mathrm{ex}} \cdot \tau_{\mathrm{HI,Center}}$ profile closely tracks $n_{\mathrm{HI}}$ in the optically thin limit, but exhibits clear inverse modulation with velocity dispersion, itself set by $T_{\mathrm{ex}}$. If the line were optically thin, as in the test cases in Appendix~\ref{app:approachIstatistics0001} and \ref{app:approachII0001} where $n_{\mathrm{HI}}$ is reduced by a factor 1000, lowering this product by the same factor, this is simply the maximum intrinsic brightness temperature of emission from each voxel. When optical depth effects are important across a voxel, as is often the case in this simulated cube, this product overestimates the brightness temperature, which cannot exceed $T_{\mathrm{ex}}$. Cold regions with high $n_{\mathrm{HI}}$ are more likely to be optically thick.

Some features to note along these lines of sight  are considerable changes in systemic velocity, which broaden the $I$ and $V$ profiles, cold regions with high $n_{\mathrm{HI}}$ particularly capable of producing absorption, and occasional field reversals.

\subsection{Radiative Transfer: POLARIS}
\label{Radiative Transfer}

We use the POLArized RadIation Simulator  \citep[POLARIS, ][]{2016A&A...593A..87R,2017A&A...601A..90B} to perform radiative transfer and generate synthetic Zeeman HI observations. The STARFORGE simulations track various hydrogen phases, including ionized, neutral, and molecular components \citep{2023MNRAS.519.3154H}. For our synthetic Zeeman HI spectra, we use the neutral hydrogen density as an input to compute the Stokes $I$ (expressed as the observable brightness temperature $T_{\mathrm {HI}}$) and Stokes $V$ spectra. Additionally, temperature, 3D velocity, and 3D magnetic field data from STARFORGE are incorporated into the radiative transfer calculations in POLARIS. 

To fully specify the model, we assume that the velocity dispersion $s$\footnote{When equations are written in terms of frequency, rather than velocity, the dispersion $\sigma$ is $\nu_0/c \times s$, where $\nu_0$ is the rest frequency.} includes a non-thermal component from microturbulence (for simplicity everywhere $s_{\rm {nt}} \simeq 0.5/\sqrt2$ km s$^{-1}$; e.g., \citealt{2023PASA...40...46K}), added in quadrature to the thermal dispersion which is
\begin{equation}
    s_{\rm th} = \sqrt{T_{\rm k}\,k_{\rm B}\,/\,{m_{\rm H}}}\, \simeq \sqrt{T_{\rm k}\,/\,{121}} \, \textrm{km s}^{-1} \, ,
    \label{eq:sig_th}
\end{equation}
where $T_{\rm k}$ is the kinetic temperature of the gas, $m_{\rm H}$ is the mass of hydrogen, $k_{B}$ is the Boltzmann constant, and the numerical factor is in units of K\,km$^{-2}$\,s$^{2}$. For $T_{\rm k}$ as small as 10\, K in the dense interior of the cloud $s_{\rm th} \simeq 0.29$, and so these two contributions are about equal and $s$ is as small as 0.456 km s$^{-1}$, or in terms of frequency,
$\sigma$ as small as 2.2 kHz.

For the cold dense cloud conditions in the simulation, it is likely that collisional excitation alone produces $T_{\rm {ex}} \simeq T_{\rm k}$ \citep{murray2015}.  In any case, we assume LTE conditions.

\xd{We used POLARIS to compute monochromatic spectral data at 0.04 km s$^{-1}$ intervals across the range $-20$ to 20 km s$^{-1}$. This can then be resampled to coarser spectral channels, for example 0.1 km s$^{-1}$ corresponding to the output of the L-band spectrometer of FAST.}

\subsubsection{Adding Noise to the Model Spectra}
\label{sec:addnoise}

\begin{figure*}[hbt!]
\centering
\includegraphics[width=0.95\linewidth]{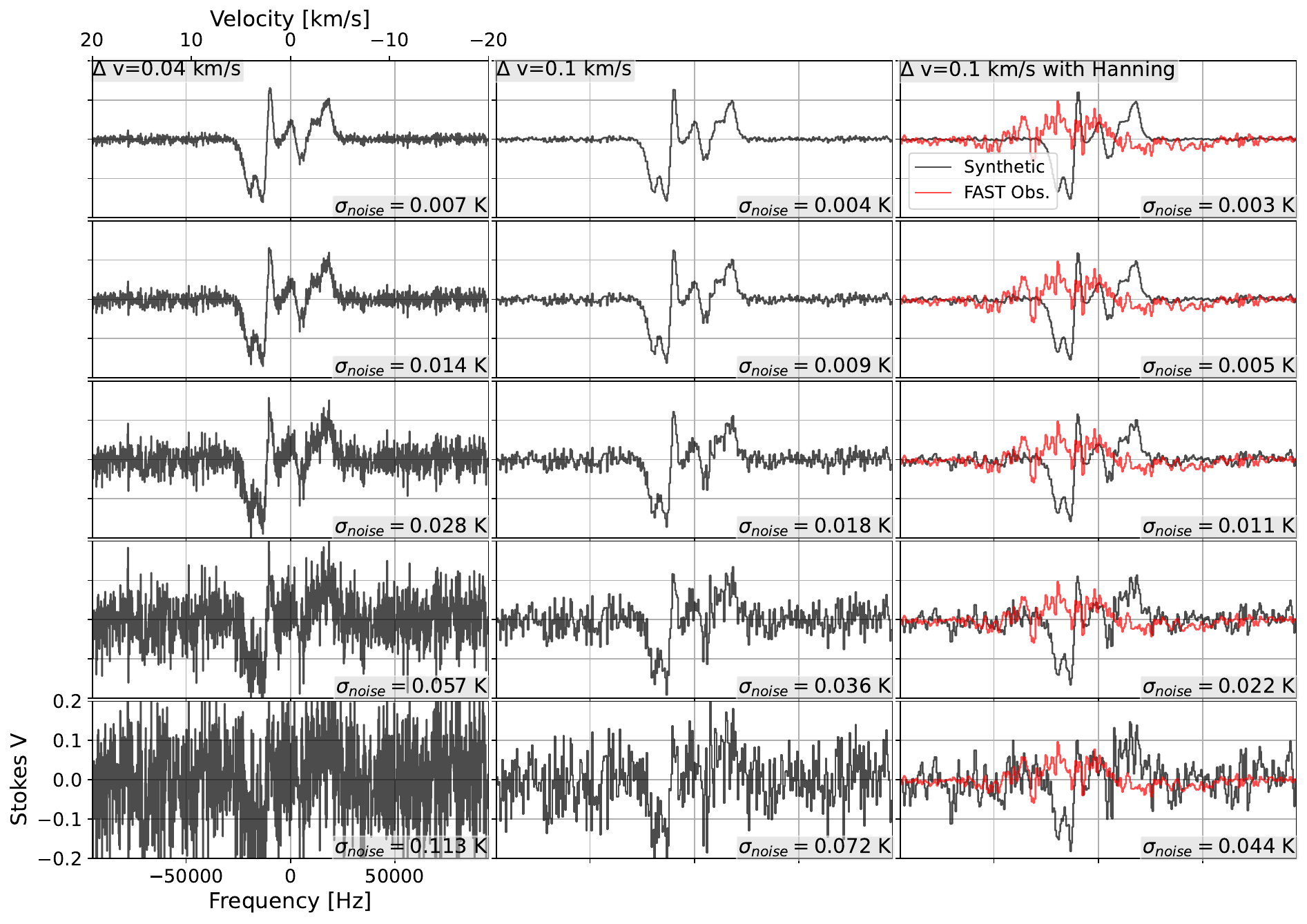}
\caption{Synthetic Stokes $V$ spectrum of the SR2 subregion with different noise levels added. The first column shows spectra at the native velocity intervals of 0.04 \kms. The second column shows the same spectra resampled to 0.1 \kms\ channels. The third column presents the Hanning-smoothed versions of the resampled spectra. For comparison, the red line shows the FAST spectrum from \citet{2022Natur.601...49C}, which has a noise level of 0.006 K after Hanning smoothing for 0.1 \kms\ channels, close to an injected noise level of 0.014 K in the left column.}
\label{fig.SR2_noiseALL}
\end{figure*} 

\xd{We assume that the right- and left-handed circularly polarized spectra have independent noise, drawn channel by channel ($ch$) in units of brightness temperature as ${\cal N}(ch)_r$ and ${\cal N}(ch)_l$ from normal distributions with the same dispersion $\sigma_{rl}$. This changes the model spectra $T(ch)_{b,I}$ and $T(ch)_{b,V}$ by $1/2\, [{\cal N}(ch)_r \pm {\cal N}(ch)_l]$, respectively.  The typical noise estimated as the dispersion in the signal-free end-channels of these $I$ and $V$ spectra is verified to be the same, $\sigma_{noise} = \sigma_{rl}/\sqrt 2$.}  

Stokes $I$ remains visually stable because of its strong signal, whereas Stokes $V$ appears noticeably noisier due to its weaker signal. \xd{The left column of Figure \ref{fig.SR2_noiseALL} presents an example of the synthetic POLARIS Stokes $V$ spectrum for SR2 with different levels of noise added at the native velocity intervals of 0.04 \kms: $\sigma_{noise} = [0.007$ K, 0.014 K, 0.028 K, 0.057 K, and 0.113 K].}

\xd{FAST spectra typically have a velocity resolution of 0.1 \kms\ and are Hanning smoothed.  Accordingly, we also show resampled (middle column) and Hanning-smoothed (right column) versions of our synthetic spectra with their reduced noise. The FAST spectrum overlaid in red from \citet{2022Natur.601...49C} has a post-Hanning-smoothed noise level of 0.006 K, which lies between our second and third noise levels but is much closer to the second (0.005 K).}\footnote{\xd{In terms of signal to noise, the third noise level would provide a better match.}}

\xd{LSR1 is a separate consideration, being 16 by 16 times the size of an SR region. With the same amount of integration time covering this larger region, averaging the signals, the noise would be the same as for the SR spectra. This is what we adopt. For a multi-beam system the noise would be reduced by about the square root of the number of beams (e.g., 19 beams for the FAST L-band receiver).}

\subsection{The importance of optical depth}
\label{Optical depth}

Both the radiative transfer calculations and the modeling of the spectra account for self-absorption by the neutral hydrogen gas.
From \citet{1978ppim.book.....S}, and assuming a Gaussian line profile centered at frequency 
$\nu_i$ reflecting bulk motion of the gas,\footnote{Both $\nu$ and $\nu_i$ can be considered to be relative to the rest frequency
$\nu_0 = 1.420406$\,GHz.}
the frequency dependent optical depth is
\begin{align}
\label{eq:tau}
\tau_{\nu,i} = &\bigg[
\bigg(\frac{N(HI)_i}{1.823\, 10^{13}}\bigg) 
\bigg(\frac{1}{T_{ex,i}}\bigg) 
\bigg(\frac{\nu_0}{c}\bigg)
\bigg(\frac{1}{\sqrt{2 \pi \sigma_i^2}} \bigg) \bigg] \, \times \\ \nonumber
&\exp{\bigg(-\frac{(\nu - \nu_i)^2}
{2\sigma_i^2}\bigg)}\, ,
\end{align}
where the quantity in square brackets is denoted below as $\tau_i$ to compactify the equations in Section \ref{Approach II: Gaussian decomposition}.
It contains factors involving distinct physical properties of component $i$ of the gas: neutral hydrogen column density $N(HI)_i$, thermal (excitation) temperature $T_{ex, i}$, and dispersion $\sigma_i$.
As defined, $\tau_i$ is the optical depth at the line center. It is the same value when expressed in terms of velocity $v$, because the velocity dispersion $s_i = \sigma_i c/\nu_0$. 
Using typical physical properties from Table \ref{tab:srproperties} and Fig.~\ref{fig.physical_Region_All}, significant effects of optical depth can be anticipated.

The lower left panel of Figure~\ref{fig.column_density_full_4panel} shows a map of $W = \int T_{\mathrm {HI}}\, dv$ from integrating the simulated HI spectra generated over velocity.
\xd{For optically thin gas under the conditions considered by \citet{1978ppim.book.....S},  $N({\mathrm {HI})} \simeq C \,W$, with $C = 1.823 \times 10^{18}$ cm$^{-2}$ (K km s$^{-1}$)$^{-1}$. In the lower right panel is a map of $\wnratio = C\,W/N({\mathrm {HI})}$.}
\xd{Optical depth effects are significant in this simulated cube, so that the observed integrated intensity $C\, W$ underestimates the true $N(\mathrm{HI})$ ($\wnratio < 1$, lower right panel).}\footnote{
\xd{There are no lines of sight where the HI becomes optically thin, for which we would expect \wnratio\ to approach unity.  To explore this, we lowered the density in the cube by a factor 1000, finding the asymptote to be 1.18.
We understand that the cause of this discrepancy is that what is called ``the LTE method'' 
in POLARIS uses LTE as an initial guess and then iteratively adjusts the level populations slightly following \citet{2003ARep...47..176P} to achieve consistency, which slightly affects the final HI intensity.\label{footnoteLTE}}}



\begin{figure*}[hbt!]
\centering
\includegraphics[width=0.75\linewidth]{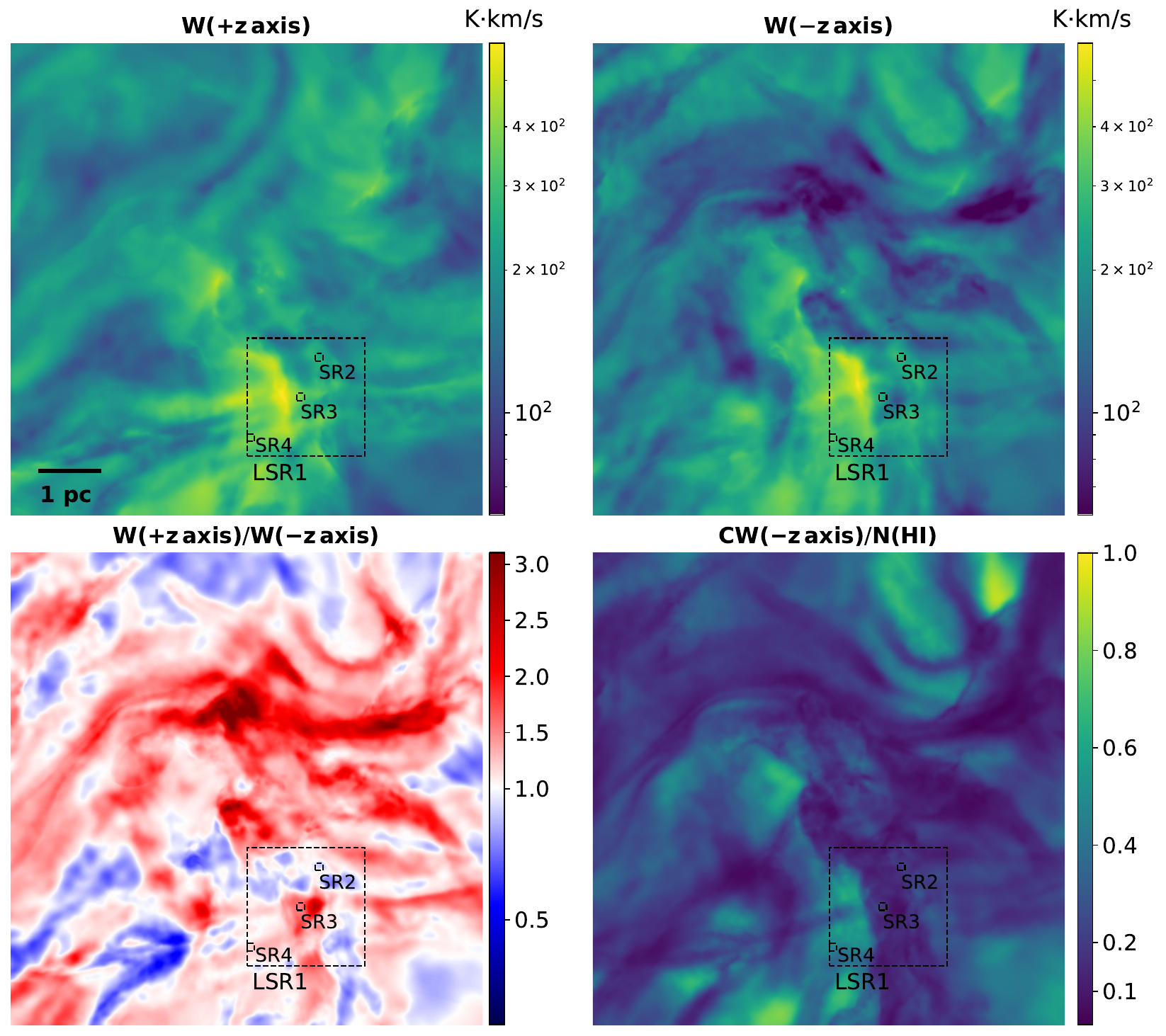}
\caption{Maps of HI integrated intensity $W$ viewing from the $+z$ direction (upper left -- same as Fig.~\ref{fig.column_density_full_4panel} lower left) and $-z$ direction (upper right); the ratio of the $+z$ to $-z$ $W$ maps (lower left); and \wnratio\ viewed from the $-z$ direction (lower right -- c.f., Fig.~\ref{fig.column_density_full_4panel} lower right, which shares the same denominator).}
\label{fig.column_density_flip_compare}
\end{figure*}

Viewing the same simulation cube from the opposite side (along the $-z$ axis) changes the LOS attenuation without altering intrinsic emission, allowing us to test the robustness of inferred $B_z$ values under varying optical depth conditions.  Figure~\ref{fig.column_density_flip_compare} presents the computed $W$ maps from both viewing directions, along with their ratio. The comparison clearly demonstrates that foreground attenuation significantly influences the observed HI integrated intensity, with variations reaching a factor of three or more.
\xd{Another way of appreciating this is to compare the lower right panels in Figures \ref{fig.column_density_full_4panel} and \ref{fig.column_density_flip_compare} in which the denominator $N({\mathrm {HI})}$ is in common.}
Optical depth has the benefit of distinguishing which components of emitting gas are affected by attenuation along the line of sight, because of their sorting in depth, but generally makes it more confusing to assess the magnetic field of obscured gas.

\section{Two Approaches to Analyzing Zeeman HI Spectra}
\label{TwoApproaches}

There are two primary approaches for analyzing Zeeman spectra: one fits the entire spectrum as a whole to estimate $B_{zm}$ averaged along the line of sight, while the other focuses on fitting individual spectral components to determine the $B_z$ associated with each.

\subsection{Zeeman Analysis Approach I: Fitting Stokes $V$ with $dI/d\nu$.}
\label{Approach I: Fitting Stokes V with dI/dnu}

Approach I relies on the weak-field approximation, which assumes that Zeeman shifts of polarized components are significantly smaller than the spectral line width, $\sigma$. Under this assumption (see Eq. \ref{eq:vderiv} in Appendix \ref{app:smallfields}), the model Stokes $V$ spectrum is given by
\begin{equation} \label{dI_dnu_eq1}
V \approx \Delta \nu \, \, dI/d\nu\, ,
\end{equation}
where shift $\Delta \nu= (\mu_{B}/h) B_{zm} = 1.4 B_{zm}$ Hz, with $\mu_{B}$ denoting the Bohr magneton, $h$ the Planck constant, and the LOS field $B_{zm} = \cos(\theta) B$ in $\mu$G. 

\xd{If the spectrum were} a simple single Gaussian line profile for an optically thin spectral line, $\exp\big(-(\nu - \nu_1 )^2/(2 \sigma_1^2)\big)$, this yields the characteristic ``S-shaped'' Stokes $V$ spectrum, with extrema located at 
$\nu_1 \pm \sigma_1$. The amplitude of the extrema is approximately 
$|\exp(-1/2)\, \, \Delta\nu_1/\sigma_1|$, so that
the $V$ signal is suppressed relative to $I$ by this factor, which makes $V$ much more susceptible to noise.

\xd{Although actual spectra are much more complex,} this method simply involves numerically differentiating the observed (simulated) Stokes $I$ spectrum and \xd{treating that as a model to fit} the observed Stokes $V$ spectrum \xd{using only one free scaling parameter, $B_{zm}$, an averaged projected global magnetic field strength.}\footnote{Observationally, the
amplitude of $B_{\mathrm{3D}}$ and the projection of the field on the line of sight by factor $\cos(\theta)$ are inseparable in their product, though they are accessible in the simulations. 
}

\subsection{Zeeman Analysis Approach II: Gaussian Decomposition}
\label{Approach II: Gaussian decomposition}

\xd{Approach II involves fitting observed (synthetic) $I$ and $V$ spectra using models of $I$ and $V$ calculated for multiple Gaussian-profile-based
components (GPB-C model spectra).  Each component quantifies emission from HI gas with distinct physical properties, allowing for optical depth effects.}  

In particular, the goal is to estimate $B_{zm}$ for each component. Because the components are ordered, the approach has the potential of probing the profile of the magnetic field along the LOS. 

\xd{A less ambitious but still useful goal is to estimate $B_{zm}$ in components with different velocity.  This might for example be possible toward the wings of lines broadened by bulk motions in the gas.}





\subsubsection{Models for $I$ and $V$}
\label{sec:miv}

For small shifts, the Stokes $I$ profile 
is to leading order unchanged by the Zeeman effect (see Appendix \ref{app:smallfields}) so that allowing for optical depth using Eq. \ref{eq:tau} and expressing $I$ in terms of brightness temperature $T_{b,I}$ we have
%
\begin{align} \label{TB_gauss_tauabs}
T_{b,I}=&\sum_{\ell} T_{ex,\ell} \left\{1-\exp\left[-\tau_{\ell} \exp\left(-\frac{(\nu-\nu_{\ell})^{2}}{2\sigma_{\ell}^{2}}\right)\right]\right\}\times \nonumber \\
&\exp\left[-\sum_{m<\ell}\tau_{m}\exp\left(-\frac{\left(\nu-\nu_{m}\right)^{2}}{2\sigma_{m}^{2}}\right)\right]\, ,
\end{align}
where $T_{ex,\ell}$ is the excitation temperature of the $\ell^{\text{th}}$ component, $\tau_{\ell}$ represents the optical depth, and $\sigma_{\ell}$ denotes the velocity dispersion.  The first line of each term is simply the intrinsic emission of the $\ell^{\text{th}}$ component. The second factor accounts for attenuation by all components in its foreground; \xd{the component indexing increases with distance from the observer, whence $m < \ell$.}

The Stokes $V$ model is based on a generalization of the small-shift approximate $I_\pm$ for opposite circular polarizations in Eq. \ref{eq:approxIpm}, then allowing for both the optical depth of each component and the attenuation from foreground components as in Eq.\ref{TB_gauss_tauabs}.
Here, the intrinsic polarized emission \xd{in terms of brightness temperature} for each component is
\begin{align}\label{stokesV_fit_tauabs_I}
I_{\ell,\pm}=T_{ex,\ell}\bigg\{&1-\exp\bigg[-\tau_{\ell}\exp \bigg( \\ \nonumber
&-\frac{(\nu-(\nu_{\ell}\pm\Delta\nu_{\ell}))^{2}}{2\sigma_{\ell}^{2}}\bigg)\bigg]\bigg\},
\end{align}
and the cumulative optical depth due to foreground absorption is:
\begin{equation} \label{stokesV_fit_tauabs_tau}
\tau_{\ell,atten,\pm}=\sum_{m<\ell}\tau_{m}\exp\left\{-\frac{\left[\nu-(\nu_{m}\pm \Delta\nu_{m})\right]^{2}}{2\sigma_{m}^{2}}\right\} \, . 
\end{equation}
$\Delta \nu_{i}$ of each component is implicitly linked to the projected field $B_{z,i}$.

Thus
\begin{align} 
\label{stokesV_model}
T_{b,V}=\sum_{\ell}& \frac{1}{2}\bigg[I_{\ell,+}\, \, \exp(-\tau_{\ell,atten,+}) \, \, - \nonumber\\
&I_{\ell,-} \, \, \exp(-\tau_{\ell,atten,-})\bigg]\, ,
\end{align}
where $\tau_{\ell,\mathrm{atten},+}$ and $\tau_{\ell,\mathrm{atten},-}$ denote the optical depths of the right- and left-handed circularly polarized spectra, respectively.

We fit up to 10 components, selecting the optimal number using BIC (Section \ref{sec:ngc}). The spatial order of the components is also optimized. The parameters fit for each component are $T_{ex}$, central $\tau$, velocity dispersion expressed in frequency $\sigma$, central velocity, and $B_{z}$.  Only for $T_{ex}$ is the range limited, from 10 K to 10,000 K. A lower limit to $\sigma$ comes from the thermal broadening through Equation \ref{eq:sig_th} and so there can be some co-variance.  However, $\sigma$ can be larger because of bulk motions in the gas. 

\xd{The central optical depth $\tau$ depends inversely on $T_{\mathrm{ex}}$ and $\sigma$ through the prefactor in Equation~\ref{eq:tau}, leading to strong covariance between $\tau$, $T_{\mathrm{ex}}$, and $\sigma$. Thus, $\tau$ is not an independent variable. The truly independent quantity is the column density $N(\mathrm{HI})$ of the gas parcel. However, we found that directly fitting for $\tau$ yielded more stable and reliable results than fitting for $N(\mathrm{HI})$ using either \texttt{curve\_fit} or \texttt{lmfit}.}

\subsubsection{Number of Gaussian Components}
\label{sec:ngc}

It is worth noting that we do not manually select the number of Gaussian components, but we do carry out fits to models with successively more components. Subsequently, for model comparison and selection we favor a low value of the Bayesian Information Criterion (BIC), defined as:
\begin{align}
\text{BIC} &= - 2 \ln(\hat{L}) + k \ln(p) \, ; \\
\hat{L}(\theta) &= \prod_{i=1}^p \frac{1}{\sqrt{2\pi\sigma_{noise}^2}} \exp\left( -\frac{(y_i - f(x_i;\theta))^2}{2\sigma_{noise}^2} \right)\, .
\end{align}

In the first term $\hat{L}$ is the maximum likelihood, ensuring the goodness of fit; $y_i$ are the observations, $f(x_i; \theta)$ is the model, and $\sigma_{noise}$ is the noise level. (For simulated spectra with no noise, we assume a signal-to-noise ratio of 1000, corresponding to $\sigma_{noise}$ = 0.1\% of the Stokes $V$ amplitude.) The second term penalizes a more complex model; $k$ is the number of free parameters and
$p$ is the number of data points. Long stretches of continuum in which the spectra are barely varying can be prejudicial to both terms.  In any case, differences in BIC values are the main discriminant. To ensure consistency and fairness (avoiding  arbitrary decisions) across all tests, we adopt this uniform fitting strategy throughout this work, without manual fine-tuning of component numbers or fit parameters. In practice, we explore the fitting with 1 to 10 components. For each case, we generate 30 different initial guesses for \texttt{curve\_fit}, allowing up to 50 million iterations per run, and select the result with the minimum BIC to reduce bias from the choice of initial guess. Because the number of fitting parameters is large and the initial guesses can vary, some degeneracy exists in the fitting, meaning that repeated runs may yield slightly different results for certain spectra. To ensure consistency and avoid fine-tuning biases, we focus on the statistical trends of the fitting results rather than individual variations.

\subsubsection{Sequential Fitting Strategy} 
\label{sec:sfs}

In standard fitting procedures \citep[e.g.,][]{2022Natur.601...49C}, the Stokes $I$ spectrum is typically fitted first, determining the spatial ordering of the components and their derived parameters-central frequency, velocity dispersion, excitation temperature, and optical depth. The BIC for this fit of Stokes $I$ is used to assess the optimal number of Gaussian components.  The properties of these Gaussian components are then fixed and the Stokes $V$ spectrum is subsequently fitted using Equation~\ref{stokesV_model}, with $B_{zm}$ the sole free parameter for each component, entering through the new term $\Delta \nu$ in Eqs. \ref{stokesV_fit_tauabs_I} and \ref{stokesV_fit_tauabs_tau}. There is no BIC assessment for the $V$ fit that would alter the number of Gaussian components.

However, this ``sequential fitting strategy'' has a critical limitation that can render it sub-optimal: it assumes that the Stokes $V$ profile offers no additional constraints on the number of components or the line shape. In practice, if the Stokes $V$ profile contains distinct features not well represented in Stokes $I$, this assumption can lead to poor fits.

\subsubsection{Joint Fitting Strategy} 
\label{sec:jfs}

To address this, we introduce a ``joint fitting strategy'' in which Stokes $I$ and Stokes $V$ spectra are fitted simultaneously. This approach allows the fitting process to utilize information from both profiles fully, potentially improving the robustness and accuracy of magnetic field strength estimates. \xd{During the fitting process, 
the Stokes $V$ weight is increased by the ratio of the maximum amplitude of Stokes $I$ to that of Stokes $V$, ensuring that both spectra have comparable importance constraining the fit.} The BIC is modified to include the goodness of both fits in the maximum likelihood and the number of data points in the penalty term is doubled. The free parameters for the Gaussians are shared by the fits and additional parameters are added for the $B_{zm}$ of each component needed for the $V$ fit.


\subsection{Alternative Radiative Transfer: GPB-S Spectra}
\label{gpb}

The ground truth simulated observations of Stokes $I$ and $V$ (their synthetic spectra) are obtained by performing radiative transfer along the LOS through the simulated data cube using POLARIS.
Alternatively to POLARIS, we computed synthetic spectra assuming LTE and Gaussian-profile-based emission from each slice crossed along the LOS (GPB-S synthetic spectra),
allowing for polarization and optical depth using the same Eqs.~\ref{TB_gauss_tauabs} through \ref{stokesV_model} that we use below to calculate the multiple Gaussian-profile-based component (GPB-C) models to fit to the synthetic spectra.
In this case, the index $\ell$ is for a slice of the cube along $z$ (i.e., voxel by voxel, not by Gaussian component) and the evaluation of $I_{\ell,\pm}$ is for the known physical properties of the voxel in the simulated cube. Likewise, the line center optical depth $\tau_i$ is evaluated using Eq. \ref{eq:tau} for the known physical properties in the relevant voxel $\ell$ or in the foreground voxels $m$ causing attenuation.  All values of the dispersion $\sigma$ include thermal and non-thermal contributions added in quadrature as above.
\xd{These alternative GPB-S synthetic spectra agree with the POLARIS results to about 15\% for both Stokes $I$ and $V$
and the overall spectral shapes remain highly consistent. See also footnote \ref{footnoteLTE}.}

Because we are working from a simulation, the GPB-S approach enables a quick way to compute  the radiative transfer, including for example to evaluate the contributions to the $I$ and $V$ profiles from different depth ranges (layers) along the LOS.  This source of ``ground truth'' to aid in the understanding of the Zeeman analysis is available in the simulation environment, but not of course in real observations.

In this light, Approach II, which is also Gaussian component-based (GPB-C), can be appreciated as an attempt to model the synthetic spectra using a substantially reduced set of components arranged in an optimized spatial order and appropriately attenuated.

\section{Validation and Ground Truth Estimators of the LOS Field Using a Simple Five-Layer Cloud}
\label{Simple Geometry Validation}

\begin{figure}[hbt!]
\centering
\includegraphics[width=0.98\linewidth]{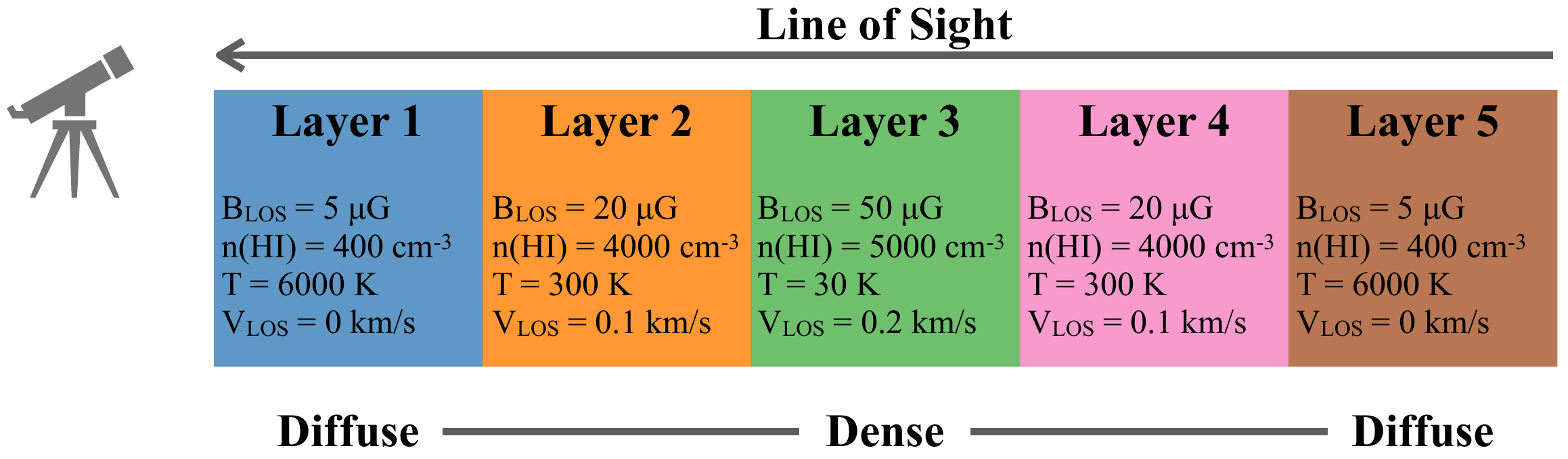}
\caption{Schematic illustration of the five-layer cloud structure.  Note that layer 5 (aka L5) and L1 have the same properties, as do L4 and L2, but they produce quite different contributions to the $I$ and $V$ spectra because of attenuation (see Figure \ref{fig.5_Layer_GPB}).  Not encoded here is that L2 and L4 are half as thick as the other three, so that individual layers span 12.5\% (orange, pink) or 25\% (blue, green, brown) of the total spatial extent (see Figure \ref{fig.f_sim_5_layers}).}
\label{fig.5_layer}
\end{figure}

To assess the performance of Zeeman analysis techniques, we conduct a synthetic observation experiment using a simple symmetric five-layer cloud model, as summarized in Figure~\ref{fig.5_layer}. 
At the center of the 128-slice cube the gas is most dense, coolest, and has the highest $B_z$. The LOS velocities of the layers are only slightly offset, at 0, 0.1, 0.2, 0.1, and 0~\kms,
respectively. This geometric/physical setup does not purport to correspond to a known astrophysical scenario but is intentionally designed to test whether Zeeman analysis can effectively disentangle contributions from different layers within the cloud structure, in the correct order,  and to provide insight into what Zeeman measurements actually represent.

\begin{figure}[hbt!]
\centering
\includegraphics[width=0.98\linewidth]{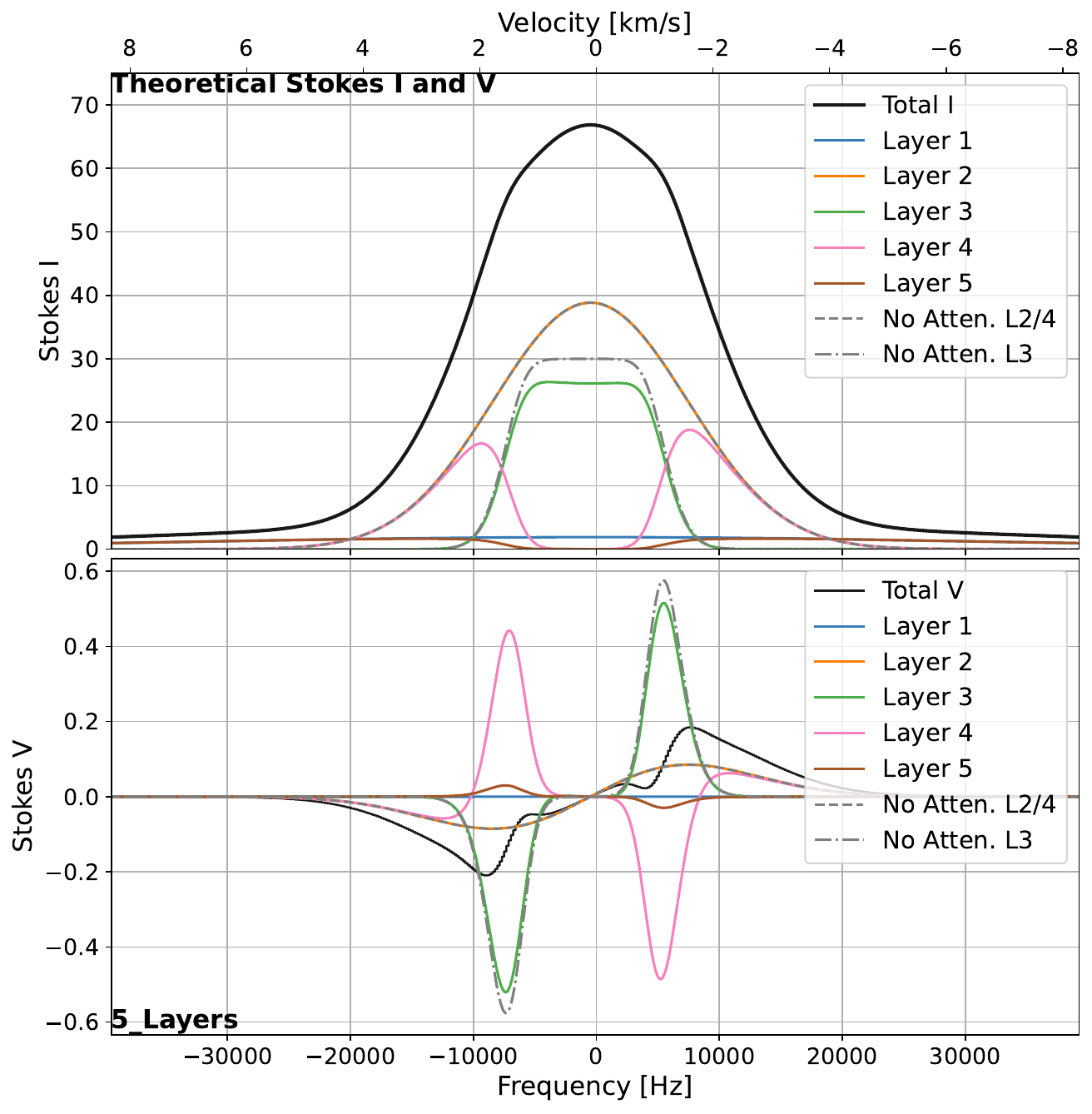}
\caption{Stokes $I$ (upper panel) and Stokes $V$ (lower panel) spectra for the five-layer cloud experiment calculated with the GPB-S method. Total emission for Stokes $I$ and $V$ GPB-S spectra is in black. The individual contributions from the five layers accounting for self (internal) and foreground attenuation are shown by colored lines coded as in Figure \ref{fig.5_layer}. Removal of foreground attenuation results in the broken lines (see legend and text). See text for explanation of the unusual L4 (pink) $V$ spectrum.
}
\label{fig.5_Layer_GPB}
\end{figure}

\subsection{Ground Truth Layer Contributions to $I$ and $V$ with Depth}
\label{sec:gtlc}

Using the GPB-S approach, we compute the $I$ and $V$ spectra and the individual ground truth contributions from each of the five layers.  These are shown in Figure \ref{fig.5_Layer_GPB}.  In both $I$ and $V$ spectra, the effects of attenuation by the foreground layers is apparent, providing clues as to the depth at which the emission arises.

Note that L1 (blue) produces very weak and broad $I$ emission peaking at 0 \kms\ and the low LOS field accentuates this in the imperceptible contribution to $V$. 

L2 (orange) is not much affected by self-absorption or attenuation by the warmer L1 and produces a tell-tale signature ``S'' contribution to $V$ with a negative excursion on the left and cross-over at 0.1 \kms.  The rising trend in $V$ through the cross-over is entirely due to L2.

L3 (green) has a lot of self-absorption (attenuation across the slices in the layer), making the $I$ profile flat-topped.  Normally the saturation level would be 30 K (dash-dot line), but it is lower because of polarized attenuation by L2. The $I$ profile is symmetrical and centered at about 0.2 \kms. The $V$ profile has distinctive relatively sharp features with mirror symmetry that arise because of polarized self-absorption, not a normal Zeeman emission effect.  If the foreground absorption were removed, the extrema increase a bit (dash-dot lines).

The contribution of L4 (pink) is very different than its companion layer L2 (dash line for no foreground attenuation),
because of attenuation by the cold gas in L3. The Stokes $I$ profile shows a central dip, while \xdtwo{$V$ exhibits complex features arising from both Zeeman emission (see ``No Atten. L3'') and polarized attenuation by cold foreground gas, sharp reversed excursions whose prominence depends on the LOS optical-depth structure.
}


L5 (brown) is similarly affected and very different than L1. There is again a reversed profile in $V$ from polarized attenuation, to be distinguished from an intrinsic normal ``S'' of negligible amplitude from the normal Zeeman emission effect.

\begin{figure}[hbt!]
\centering
\includegraphics[width=0.98\linewidth]{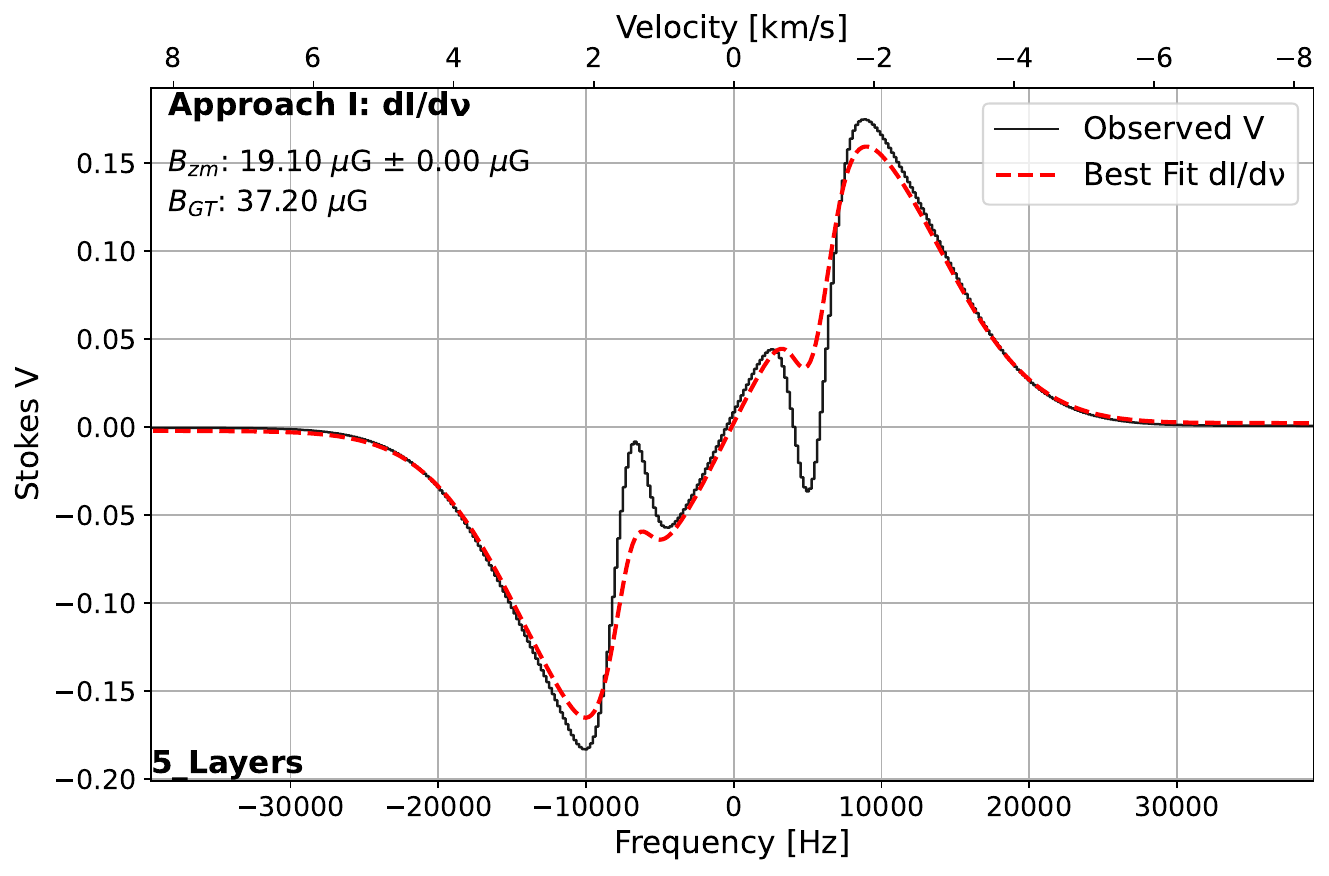}
\caption{Result of the Approach I fitting of the Stokes $V$ spectrum (black line) with the scaled differentiated Stokes $I$ spectrum (red dashed line) for the five-layer cloud experiment using Equation~\ref{dI_dnu_eq1}. 
The top-left corner legend lists the fitted magnetic field ($B_{zm}$) and the corresponding ground truth value ($B_{GT}$).
}
\label{fig.ZeemanFitting_dI_dnu_5_layers}
\end{figure}

\subsection{Approach I and $B_{GT}$ Ground Truth Estimator}
\label{sec:approachI5layer}

A simple estimate of an average LOS magnetic field strength is obtained by fitting the Stokes $V$ spectrum using the derivative of the Stokes $I$ spectrum (Eq. \ref{dI_dnu_eq1}). 
The fit to $V$ is shown in Figure~\ref{fig.ZeemanFitting_dI_dnu_5_layers}, yielding $B_z=19.1~\mu$G. The fine structure in $V$ is missed because this simple model cannot resolve the internal structure of the cloud.

As a ground truth estimator, a mass-weighted LOS magnetic field strength can be defined as
\begin{align}\label{eq.BGT_approch2}
B_{GT}(x,y) = \frac{\int B_{x,y}(z) \times n_{x,y}(z)\, dz}{\int n_{x,y}(z)\, dz},
\end{align}
where $n_{x,y}(z)$ is the number density of HI 
and $B_{x,y}(z)$ is the magnetic field at each voxel. For this configuration, this yields a significantly higher value of 37.20~$\mu$G. This discrepancy is affected by the optical thickness of the denser regions, which tend to host stronger magnetic fields, as well as foreground attenuation by the more diffuse gas, both of which affect the $V$  spectrum but neither of which is accounted for in the simple estimator $B_{GT}(x,y)$.

\begin{figure*}[hbt!]
\centering
\includegraphics[width=0.98\linewidth]{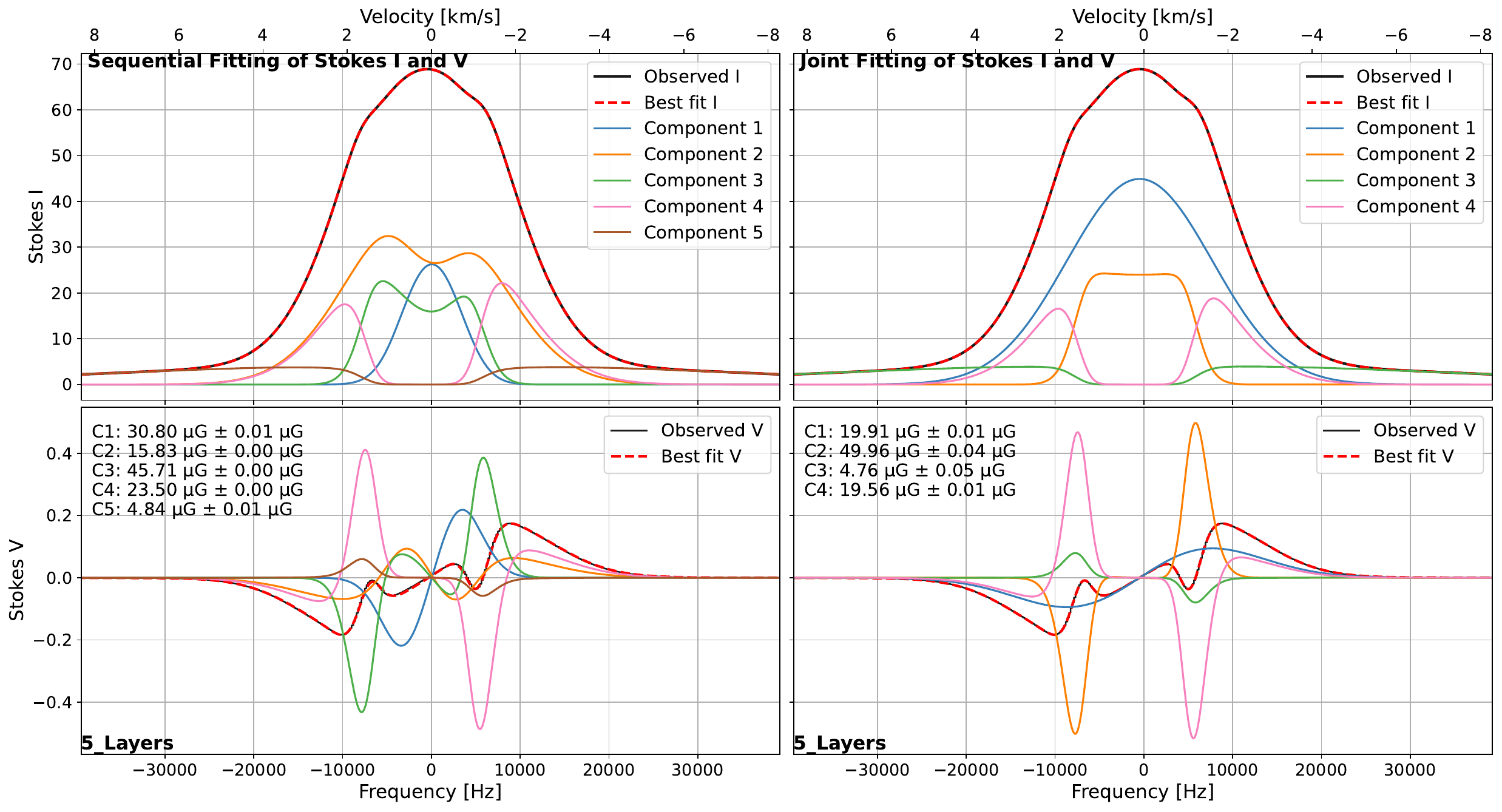}
\caption{Approach II fitting results for the Stokes $I$ and Stokes $V$ spectra in the five-layer cloud experiment, sequential strategy on the left and joint on the right. The upper panels show the synthetic (``observed'') Stokes $I$ spectrum from POLARIS (black solid line), the best-fit model based on Equation~\ref{TB_gauss_tauabs} (red dashed line), and the individual component contributions accounting for foreground attenuation (colored lines). Components are in the order determined by the fit, with C1 being the most foreground. The lower panels display information for the Stokes $V$ spectrum, color-coded in the same way.
The fitted magnetic field strengths for each component are listed in the top-left corners of the lower panels.}
\label{fig.ZeemanFitting_5_Layer_IV_noise0}
\end{figure*}

\begin{figure*}[hbt!]
\centering
\includegraphics[width=0.98\linewidth]{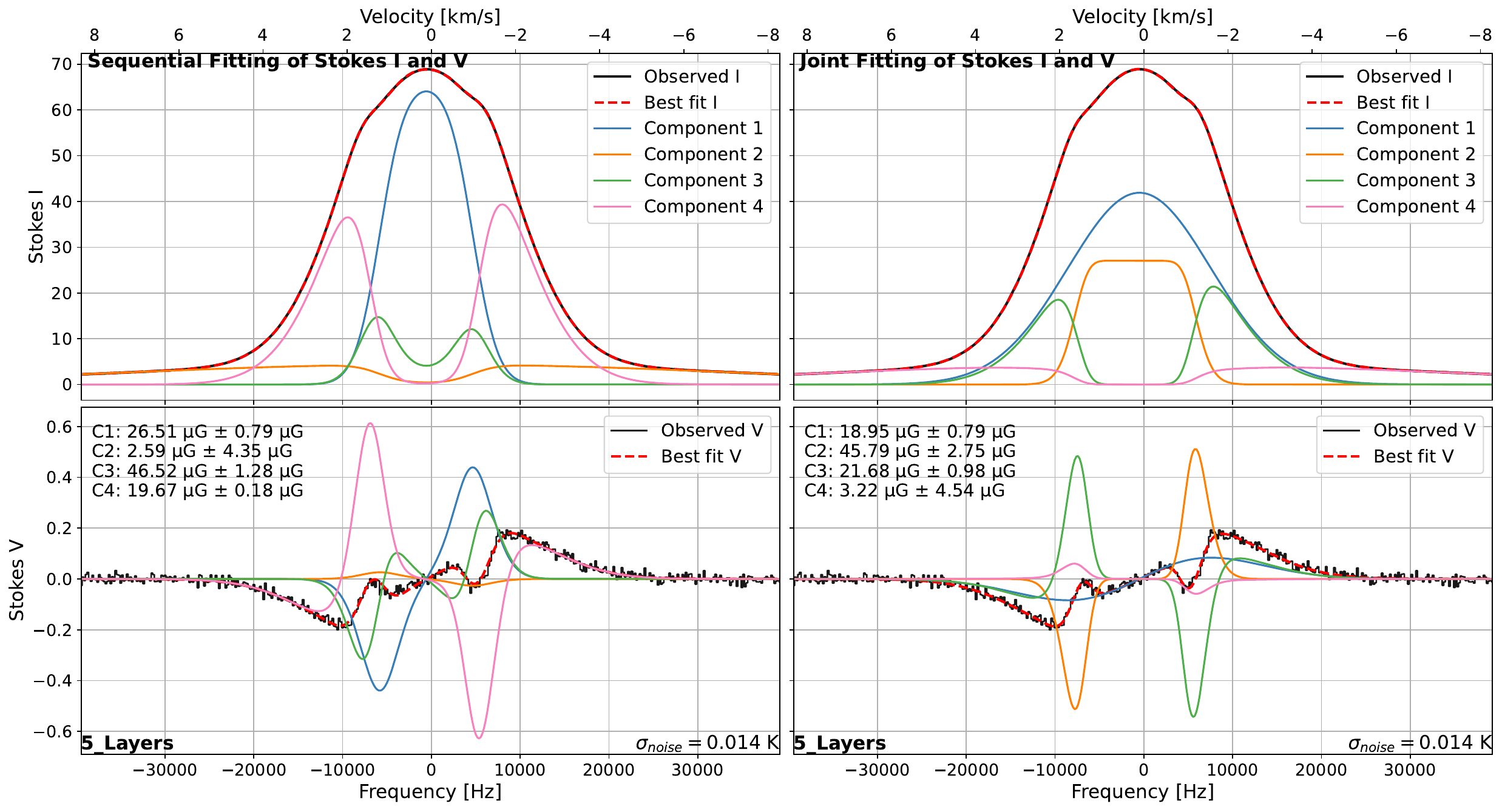}
\caption{Same as Figure~\ref{fig.ZeemanFitting_5_Layer_IV_noise0}, but showing the Approach II fitting results for the five-layer cloud experiment with a noise level of 0.014 K.}
\label{fig.5_Layers_I_V_fitting_noise01}
\end{figure*} 

\begin{figure}[hbt!]
\centering
\includegraphics[width=0.98\linewidth]{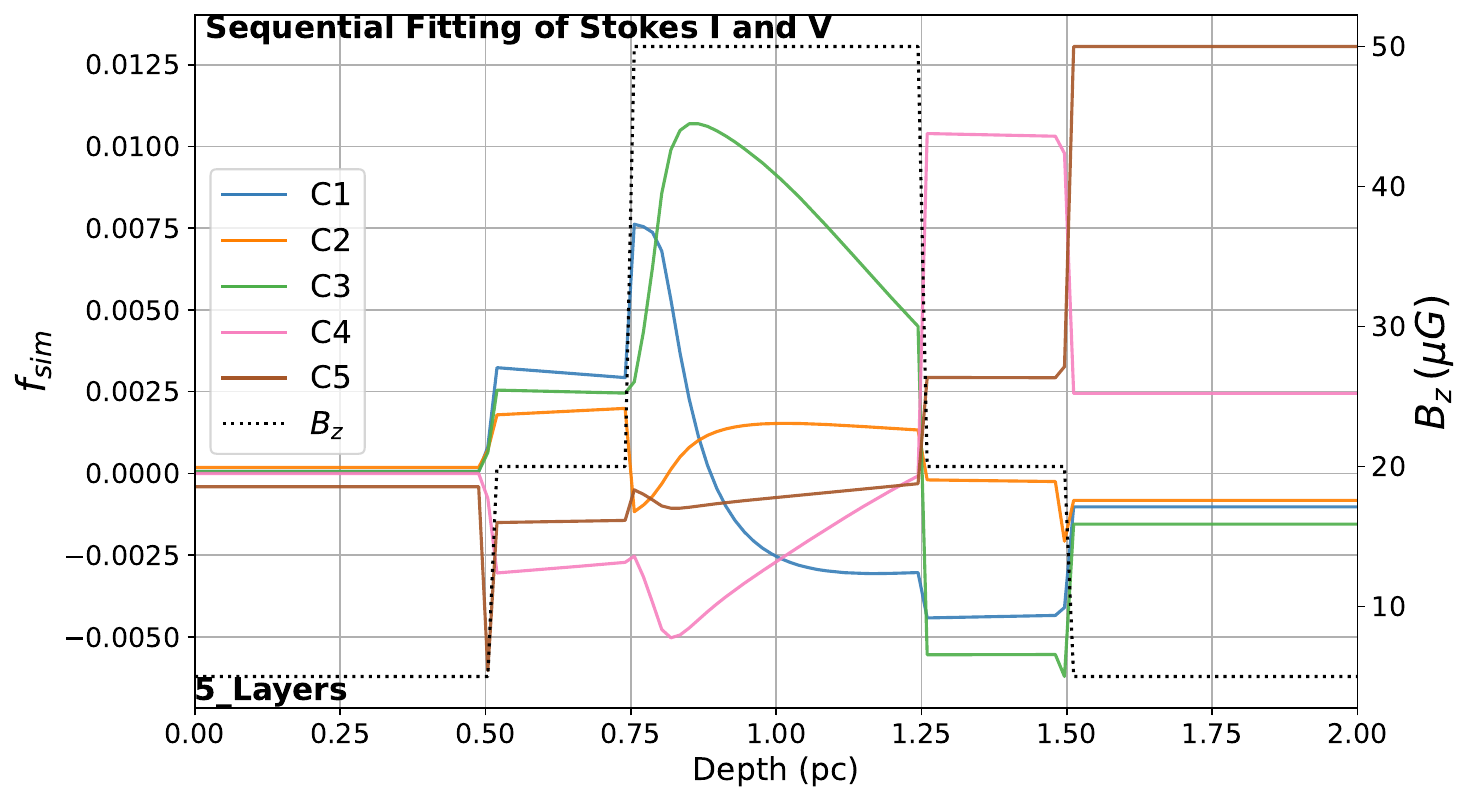}
\includegraphics[width=0.98\linewidth]{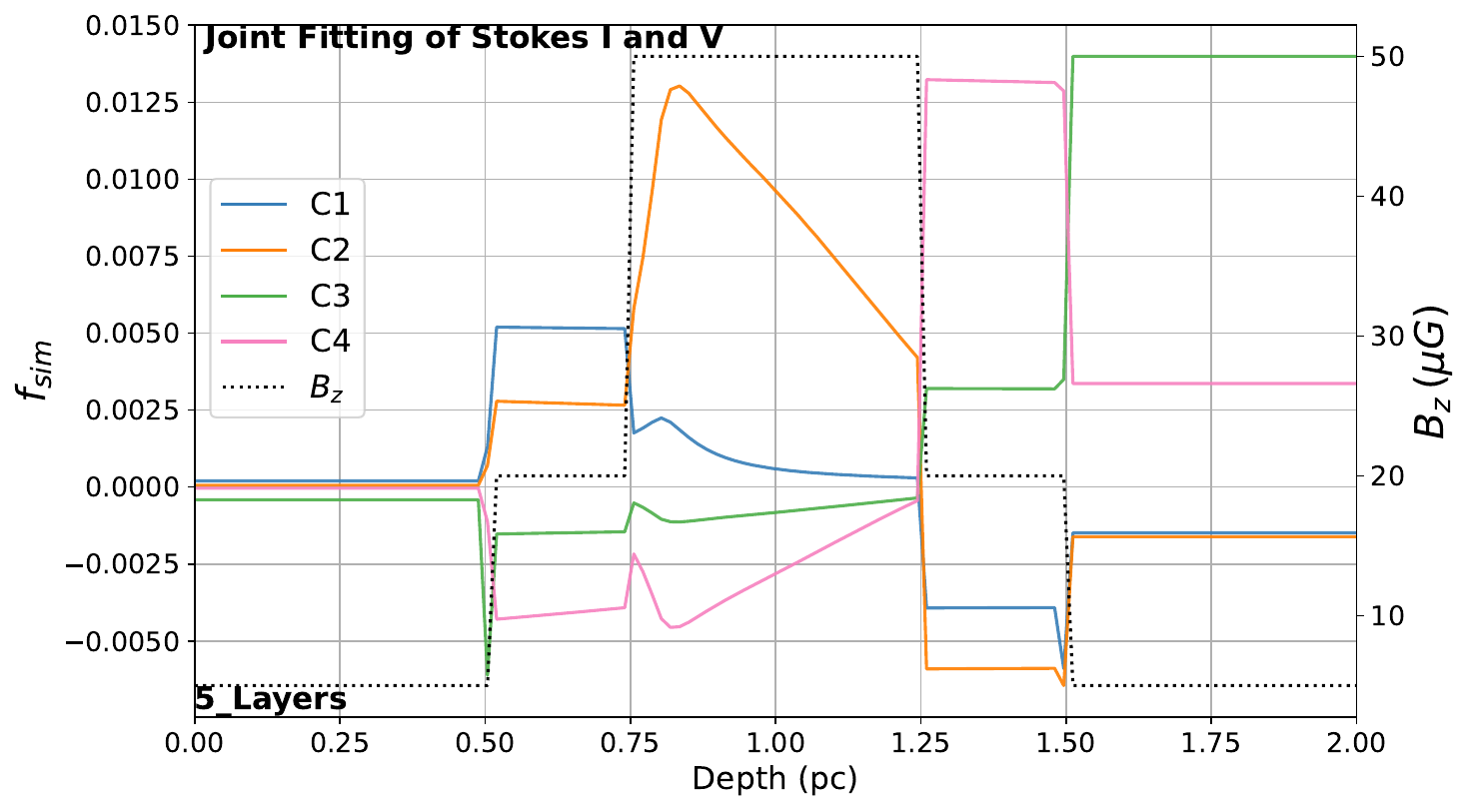}
\caption{Profile of the similarity function $f_{\text{sim}}$ (Equation~\ref{eq.fsim}) along the LOS in the five-layer cloud experiment for each fitted Gaussian component obtained using the two fitting strategies under Approach II. The actual magnetic field structure is also illustrated by the dotted line (right y-axis scale), showing the different spatial extents of the layers.}
\label{fig.f_sim_5_layers}
\end{figure}

\subsection{Approach II}
\label{sec:approachII5layer}

To address this, we next apply Approach II to decompose the spectrum into individual Gaussian components and analyze the magnetic field associated with each component separately.  The spatial order of the components is optimized during the fit.

\xd{Figure~\ref{fig.ZeemanFitting_5_Layer_IV_noise0} presents the Zeeman results using both  sequential and joint fitting strategies. The model results for $B_{zm}$ are tabulated in Table \ref{tab.Bz_BZeeman_5_layer} for comparison to the actual configuration and two more sophisticated estimators described below. We also present the Zeeman results for a noise level of 0.014 K in Figure~\ref{fig.5_Layers_I_V_fitting_noise01}, with the modeled $B_{\mathrm{zm}}$ values given in Table~\ref{tab.Bz_BZeeman_5_layer_noise01}.}


\xd{For the noisy case, we tested fitting both with and without rescaling the weighting of Stokes $V$ (see Section~\ref{sec:jfs}). Without rescaling, the model (not shown) fails to capture the two small peaks and valleys in Stokes $V$ between $-2$ and $+2$ \kms, whereas with rescaling the model successfully reproduces these features (Figure \ref{fig.5_Layers_I_V_fitting_noise01}). This highlights the importance of appropriately adjusting the Stokes $V$ weighting in the presence of noise.}

We adopt the noise-free case in the following analysis to study and interpret the results, including the definition of the ground truth. \xd{This is without prejudice, because the solution with noise is quite similar, with as favorable an ordering of components in the joint strategy.}

\subsubsection{Sequential strategy}
\label{sec:sequentialstrategy5layer}

In the left panel, despite a good fit to the $I$ and $V$ spectra, the components arising from the sequential strategy appear to be averaged blends of emission from multiple layers, compromising the ability to discern the actual stratified structure of the cloud.  The central region is particularly messy, with the linear rise in $V$ being the result of cancellation of many components.  Although there are five components, the foreground L1 is missing, but the general features of the rearmost two layer contributions appear to be recovered. \xd{The goal of finding the correct spatial ordering of components is not achieved.}

\subsubsection{Joint strategy}
\label{sec:jointstrategy5layer}

On the other hand the joint strategy is able to recover the overall cloud structure reasonably well, despite using one fewer component (see right panel). The component spectra look remarkably like the layer contributions in Figure \ref{fig.5_Layer_GPB}. 
The fitted components correctly capture the relative ordering (C1 is L2, C2 is L3, and C4 is L4), with fitted fields near the true values 20, 50, and 20~$\mu$G, respectively. The weak component C3 subsumes emission from L1 and L5 and is out of order, though correctly behind C2 (L3).
The faint foremost 5~$\mu$G layer is missing as a separate component, but recall that L1 has no perceptible signature in $V$. The central region shares the simplicity of Figure \ref{fig.5_Layer_GPB}.  

Despite some limitations, the overall structure recovered by the joint strategy much more closely resembles the true cloud configuration.  \xd{This correct spatial ordering of components holds up with the addition of noise.}
This ability of the joint strategy to recover the correct internal structure signals a huge advance relative to the normal sequential strategy that we will see repeated in our Zeeman analysis of the synthetic observations from the simulated MHD cube below. 

\subsection{Ground truth estimators for components in Approach II}

To further understand what is recovered by the Zeeman fitting, we define different estimators for the ``ground truth'' magnetic field strength associated with each fitted component. These allow for a more objective and quantitative evaluation of the performance of Approach II in estimating magnetic field strengths. 

\begin{deluxetable*}{cc|cccccc|cccccc}
\label{tab.Bz_BZeeman_5_layer}
\tablecaption{Summary of Zeeman component fitting results and the ground truth LOS magnetic field estimates for the five-layer cloud experiment$^a$ }
\tablehead{\multicolumn{2}{c|}{Cloud Structure} & \multicolumn{6}{c|}{Sequential Strategy} & \multicolumn{6}{c}{Joint Strategy} \\ \hline
\multirow{2}{*}{Layer} & $B_{real}$ & \multirow{2}{*}{Comp.}  & $B_{zm}$ & $B_{GT1}$ & $B_{GT2}$ & $\delta_{GT1}$$^b$ & $\delta_{GT2}$$^b$  & \multirow{2}{*}{Comp.}  & $B_{zm}$ & $B_{GT1}$ & $B_{GT2}$ &  $\delta_{GT1}$$^b$ & $\delta_{GT2}$$^b$ \\
 & $(\mu G)$ &   & $(\mu G)$ & $(\mu G)$ & $(\mu G)$ &  &  &  & $(\mu G)$ & $(\mu G)$ & $(\mu G)$ &  & } 
\startdata
L1 & 5.0 & C1 & 30.8 & 39.6 & 39.0$^c$ & -28.4\% & -26.6\% & C1 & 19.9 & 33.5 & 20.0 & -68.1\% & -0.4\% \\ 
        L2 & 20.0 & C2 & 15.8 & 33.7 & 20.0 & -113.0\% & -26.3\% & C2 & 50.0 & 40.1 & 50.0 & 24.5\% & -0.1\% \\ 
        L3 & 50.0 & C3 & 45.7 & 40.0 & 50.0 & 14.3\% & -9.4\% & C3 & 4.8 & 28.4 & 5.0 & -495.4\% & -5.0\% \\ 
        L4 & 20.0 & C4 & 23.5 & 33.6 & 20.0 & -43.0\% & 17.5\% & C4 & 19.6 & 34.0 & 20.0 & -73.8\% & -2.3\% \\ 
        L5 & 5.0 & C5 & 4.8 & 28.4 & 5.0 & -486.1\% & -3.3\% &  & ~ & ~ & ~ & ~ & ~ \\ 
 \enddata
 \tablenotetext{}{Note:}
 \tablenotetext{}{$^a$ This table summarizes the cloud layer structure (Layer) and associated magnetic field strengths ($B_{real}$). It includes fitted components and magnetic field estimates from both the sequential and joint fitting strategies ($B_{zm}$), along with their corresponding ground truth values as defined by Equations~\ref{eq.BGT1} ($B_{GT1}$) and~\ref{eq.BGT2} ($B_{GT2}$).}
 \tablenotetext{}{$^b$ This relative error is given by $\delta_{B}=\frac{B_{zm}-B_{GT}}{\min(|B_{zm}|, |B_{GT}|)}$, as in Equation~\ref{eqn_rela_error}.}
 \tablenotetext{}{$^c$ This result is derived from Equation~\ref{eq.BGT2}. Unlike other components, where the value closely matches the actual layer magnetic field strength, this component is primarily contributed by Layer 2 and Layer 3, as indicated by the similarity function in the upper panel of Figure~\ref{fig.f_sim_5_layers}. The corresponding ground truth is 39.0 $\mu$G, which does not match the magnetic field strength of any individual layer.}
 \end{deluxetable*}

\begin{deluxetable*}{cc|cccccc|cccccc}
\label{tab.Bz_BZeeman_5_layer_noise01}
\tablecaption{\xd{Same as Table~\ref{tab.Bz_BZeeman_5_layer}, but summarizing the Zeeman component fitting results for the five-layer cloud experiment with a noise level of 0.014 K}}
\tablehead{\multicolumn{2}{c|}{Cloud Structure} & \multicolumn{6}{c|}{Sequential Strategy} & \multicolumn{6}{c}{Joint Strategy} \\ \hline
\multirow{2}{*}{Layer} & $B_{real}$ & \multirow{2}{*}{Comp.}  & $B_{zm}$ & $B_{GT1}$ & $B_{GT2}$ & $\delta_{GT1}$ & $\delta_{GT2}$  & \multirow{2}{*}{Comp.}  & $B_{zm}$ & $B_{GT1}$ & $B_{GT2}$ &  $\delta_{GT1}$ & $\delta_{GT2}$ \\
 & $(\mu G)$ &   & $(\mu G)$ & $(\mu G)$ & $(\mu G)$ &  &  &  & $(\mu G)$ & $(\mu G)$ & $(\mu G)$ &  & } 
\startdata
L1 & 5.0 & C1 & 26.5 & 39.3 & 36.3 & -48.4\% & -37.1\% & C1 & 19.0 & 33.4 & 20.0 & -76.4\% & -5.5\% \\ 
        L2 & 20.0 & C2 & 2.6 & 28.6 & 5.0 & -1004.3\% & -93.1\% & C2 & 45.8 & 40.0 & 50.0 & 14.5\% & -9.2\% \\ 
        L3 & 50.0 & C3 & 46.5 & 39.4 & 50.0 & 18.1\% & -7.5\% & C3 & 21.7 & 33.9 & 20.0 & -56.2\% & 8.4\% \\ 
        L4 & 20.0 & C4 & 19.7 & 33.8 & 20.0 & -71.7\% & -1.7\% & C4 & 3.2 & 28.6 & 5.0 & -788.3\% & -55.3\% \\ 
        L5 & 5.0 &  &   &   &   &  &   &  & ~ & ~ & ~ & ~ & ~ \\ 
 \enddata
 \end{deluxetable*}

\subsubsection{$B_{GT1}$: Incorporating velocity dispersion}

The most direct way to define the ground truth magnetic field strength for each spectral component is to compute a mass-weighted average along the LOS, incorporating velocity dispersion to account for the contribution of different components. In this case, the ground truth magnetic field for \xd{the $\ell^\text{th}$ component at position $(x, y)$ is given by:
\begin{align}
\label{eq.BGT1}
B_{GT1}(x,y,\ell) = \frac{\int B_{x,y}(v) \times \Phi_{x,y,\ell}(v)\, dv}{\int \Phi_{x,y,\ell}(v)\, dv} \,  ,
\end{align}
where the factor $\Phi_{x,y,\ell}(v)$ corresponds to the fitted line profile of the $\ell^\text{th}$ component including attenuation, which is the $\ell^\text{th}$} term in the Stokes $I$ model spectrum (Equation \ref{TB_gauss_tauabs}). By excluding attenuation, we recover the true ground-truth magnetic field weighted by the intrinsic emission (i.e., mass in optically thin limit).
The factor $B_{x,y}(v)$ represents the mass-weighted LOS magnetic field strength as a function of velocity, given by:

\begin{align}
B_{x,y}(v) = \frac{\int B(x,y,z) \times n(x,y,z) \times \exp\left(-\frac{(v-v_{z})^{2}}{2 s_{z}^{2}}\right)\, dz}{\int n(x,y,z) \times \exp\left(-\frac{(v-v_{z})^{2}}{2 s_{z}^{2}}\right) \, dz}\, ,
\label{weightvoxel}
\end{align}
where velocity dispersion $s_{z}$ in the cube includes both thermal and turbulent broadening effects (Section \ref{Radiative Transfer}). Equation \ref{weightvoxel} favors fields for emission close to the peak of the emitted spectrum in a voxel (in common to all components), whereas Equation \ref{eq.BGT1} focuses on fields where emission contributes strongly to the $\ell^\text{th}$ fitted component profile. Note how this estimator is ``conditional'' on the component for which it is being estimated.  It is to be compared to the inferred magnetic field from the fit.

Table~\ref{tab.Bz_BZeeman_5_layer} presents this basic estimator of the ground truth LOS magnetic field strength for each fitted component, calculated using Equation~\ref{eq.BGT1}. However, this definition does not adequately capture the true structure of the cloud. As seen in the Stokes $I$ spectrum in Figure~\ref{fig.ZeemanFitting_5_Layer_IV_noise0}, the gas along the line of sight shares very similar central velocities, resulting in all layers contributing—via mass weighting—to each component's profile. This leads to an overly averaged magnetic field strength, which fails to resolve and distinguish between the distinct layers within the cloud.

\subsubsection{$B_{\text{GT2}}$: Exploiting the $V$ spectrum} 

Importantly, the Zeeman fitting strategies also incorporate information from the Stokes $V$ spectrum, which can partially disentangle different magnetic field strengths and structures along the line of sight. For example, even if two gas layers have identical density, velocity, and temperature—and thus produce identical Stokes $I$ spectra—differences in their magnetic fields can generate distinguishable features in the Stokes $V$ spectrum, allowing the Zeeman analysis to separate them.

To better reflect this aspect, we introduce an alternative estimator of the ground truth magnetic field strength for each fitted component. This new estimator incorporates contributions to the Stokes $V$ spectrum as part of the weighting function, enabling a more accurate comparison between the fitted and true magnetic field structures.

We begin by computing $T^\nu_{b,V,\text{atten}}(x,y,z)$, the voxel by voxel contributions to the GPB-S $V$ spectrum including polarized foreground attenuation, using the known physical properties of the simulated cube (Section \ref{gpb}).

The $\ell^{\rm th}$ fitted Gaussian component contributes to Stokes $V$ according to a
profile denoted $\Psi^\nu_{x,y,\ell}$ that is simply the 
$\ell^{\rm th}$ term in the $V$ model in Eq. \ref{stokesV_model}, which also includes foreground attenuation.

The similarity between a voxel's attenuated $V$ signal and the fitted component profile is quantified using a normalized dot product (cosine similarity), defined as:

\begin{align}
\label{eq.fsim}
f_{sim}(x,y,z,\ell) = \frac{T^\nu_{b,V,atten}(x,y,z)\cdot\Psi^{\nu}_{x,y,\ell}}{\lVert{T^\nu_{b,V,atten}(x,y,z)}\lVert\cdot\lVert{\Psi^{\nu}_{x,y,\ell}}\lVert }.
\end{align}
This similarity function $f_{\text{sim}}$ reflects how strongly each voxel contributes to a given fitted Zeeman Gaussian component. Because the spectral shapes vary across voxels due to differences in central velocities and dispersions, perfect matches are rare, and $f_{\text{sim}}$ values tend to be small.
In Figure~\ref{fig.f_sim_5_layers}, we present the distribution of $f_{\text{sim}}$ along the line of sight for the five-layer cloud experiment. 

To extract a meaningful estimate of the true LOS magnetic field strength associated with each fitted Gaussian component, we select the top 10\% of voxels (``TOP'') with the highest similarity scores and compute a similarity-weighted LOS magnetic field:
\begin{align}\label{eq.BGT2}
B_{\text{GT2}}(x,y,\ell) = \frac{\int_{\text {TOP}}  B(x,y,z)\, f_{\text{sim}}(x,y,z,\ell)\, dz}{\int_{\text{TOP}} f_{\text{sim}}(x,y,z,\ell)\, dz}\, ,
\end{align}
where $B(x, y, z)$ is the known LOS field $B_z$ in the simulated cube.
The 10\% threshold is an empirical choice; a more detailed exploration of its impact is presented in Appendix~\ref{Ground Truth II: Magnetic Field Weighted by Spectral Similarity}. Importantly, using the full distribution of $f_{\text{sim}}$ across all voxels would produce an overly averaged magnetic field value, which fails to capture the localized magnetic field structure traced by each Zeeman component.  Again it is important to note that this estimator too is conditional on the component.

Table~\ref{tab.Bz_BZeeman_5_layer} also reports the ground truth magnetic field strengths derived using this new estimator from Equation~\ref{eq.BGT2}. Compared to the earlier definition, these values not only agree well with the inferred fitted field (as a conditional estimator hopefully would) but also align closely with the original setup parameters and thus more accurately reflect the true magnetic field structure of the cloud, validating the improved physical interpretability of this estimator.

To examine how this arises, we return to the behavior of component
$f_{\text{sim}}$ distributions with $z$ in Figure~\ref{fig.f_sim_5_layers} and relate that to the contributions of the layers in Figure \ref{fig.5_Layer_GPB} (discussed in Section \ref{sec:gtlc}) and the spectra of individual components in Figure \ref{fig.ZeemanFitting_5_Layer_IV_noise0} (discussed in Section \ref{sec:approachII5layer}). In the joint fitting strategy (lower panel), maxima in each component $f_{\text{sim}}$ clearly align with distinct magnetic field structures, providing the basis for  how this method effectively disentangles the magnetic field contributions from different layers along the LOS in the five-layer cloud experiment. Walking this through in detail for a individual layer contribution, say L2 which corresponds to component C1 (blue), the TOP 10\% of $f_{\text{sim}}$ values for this component lie entirely within L2. Likewise for the L3 central component (C2, orange) and the background L4 (C4, pink). Thus, $B_{GT2}$ for the joint strategy is able to quantify magnetic fields successfully in the correctly ordered central and two main foreground and background layers on the basis of the distinct respective TOP $f_{\text{sim}}$ values therein.  

In contrast, the sequential strategy performs less effectively. The fitted components tend to blend contributions from multiple layers, resulting in averaged magnetic field strengths that do not clarify the true underlying structure. For instance, in Figure~\ref{fig.f_sim_5_layers}, the first component from the sequential fit (C1, blue) yields $B_{\text{GT2}} = 39\, \mu$G, which reflects a combination of contributions from both the foreground 20~$\mu$G and central 50~$\mu$G layers (layers 2 and 3), rather than isolating them individually.  This mixing over-complicates the $I$ and $V$ spectra in the region near zero velocity, and makes interpretation unreliable.

Further validation and exploration is important because discerning the LOS field structure is so challenging.
In that spirit, Appendix \ref{Smooth Magnetic Field Gradient Experiment} analyses a model cloud with smoothly varying physical properties along the line of sight.

Analysis of the STARFORGE simulation data is more complex still.  To examine further how magnetic field structure could affect magnetic field estimates from Zeeman measurements, Appendix \ref{Magnetic Field Structure} modifies the magnetic field structure in the simulation cube in a simple controlled way.


\section{Analysis of Synthetic HI Spectra from MHD Simulations Using Approach I}
\label{Analysis of Synthetic HI Spectra from MHD Simulations Using Approach I}

\begin{figure*}[hbt!]
\centering
\includegraphics[width=0.48\linewidth]{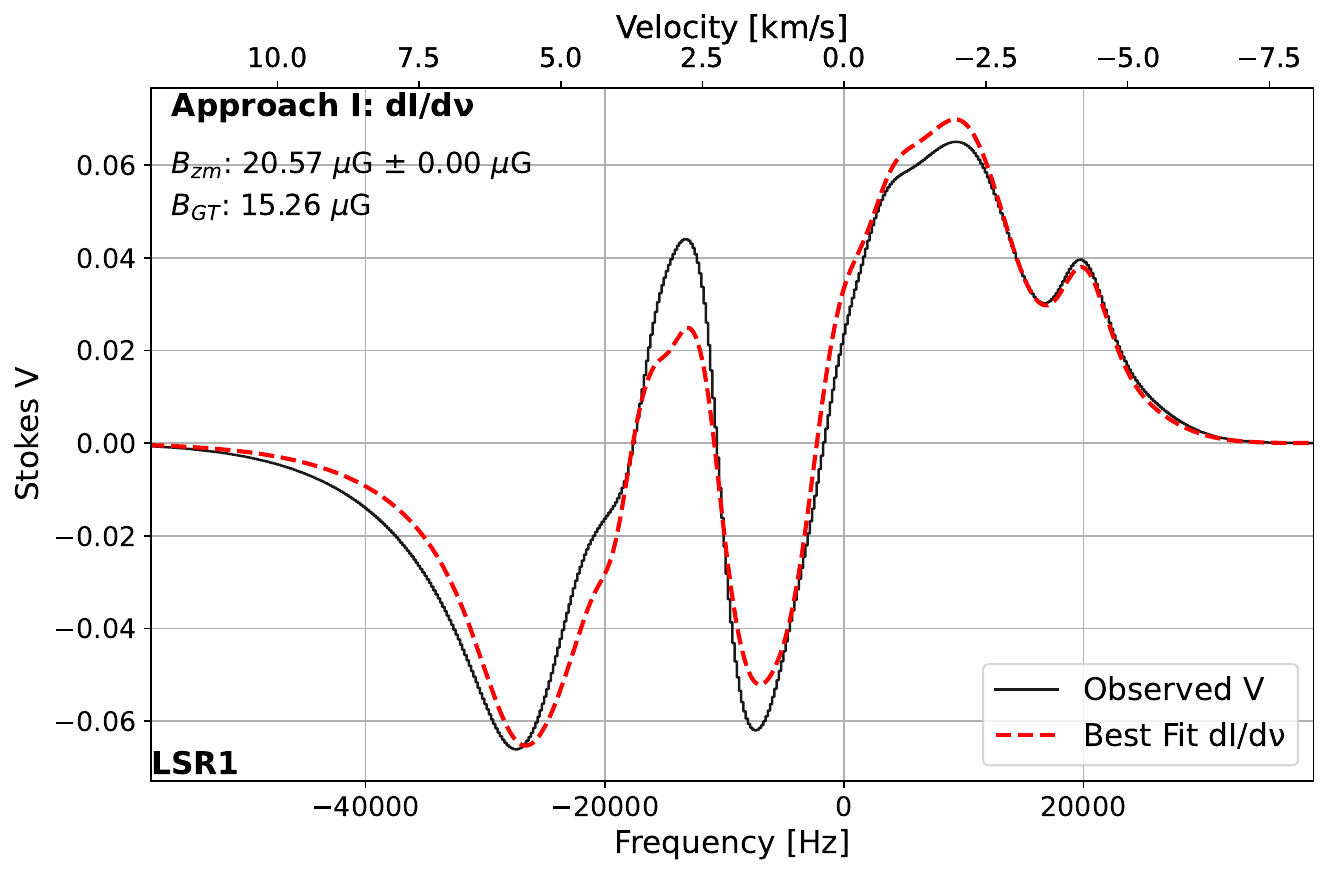}
\includegraphics[width=0.48\linewidth]{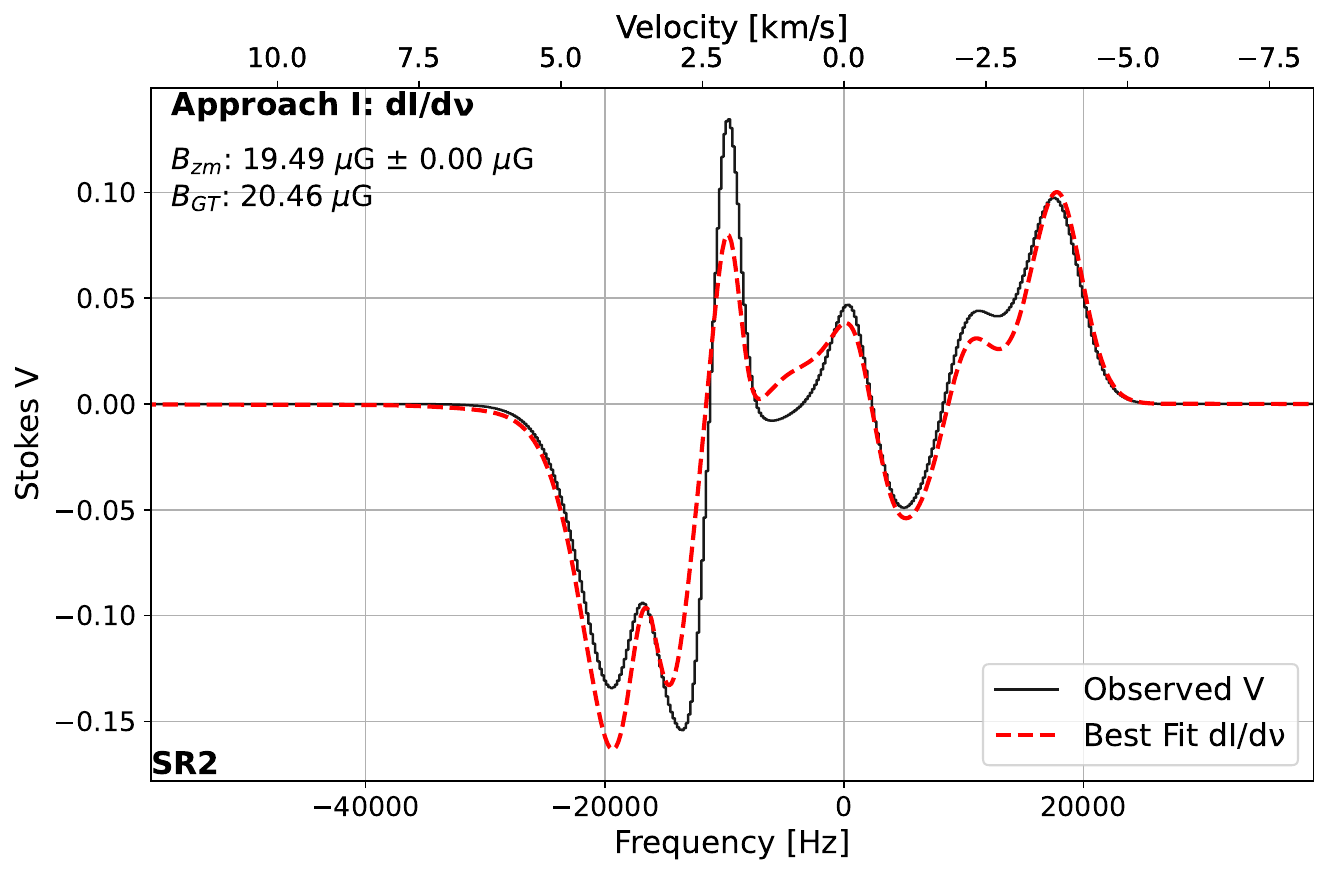}
\includegraphics[width=0.48\linewidth]{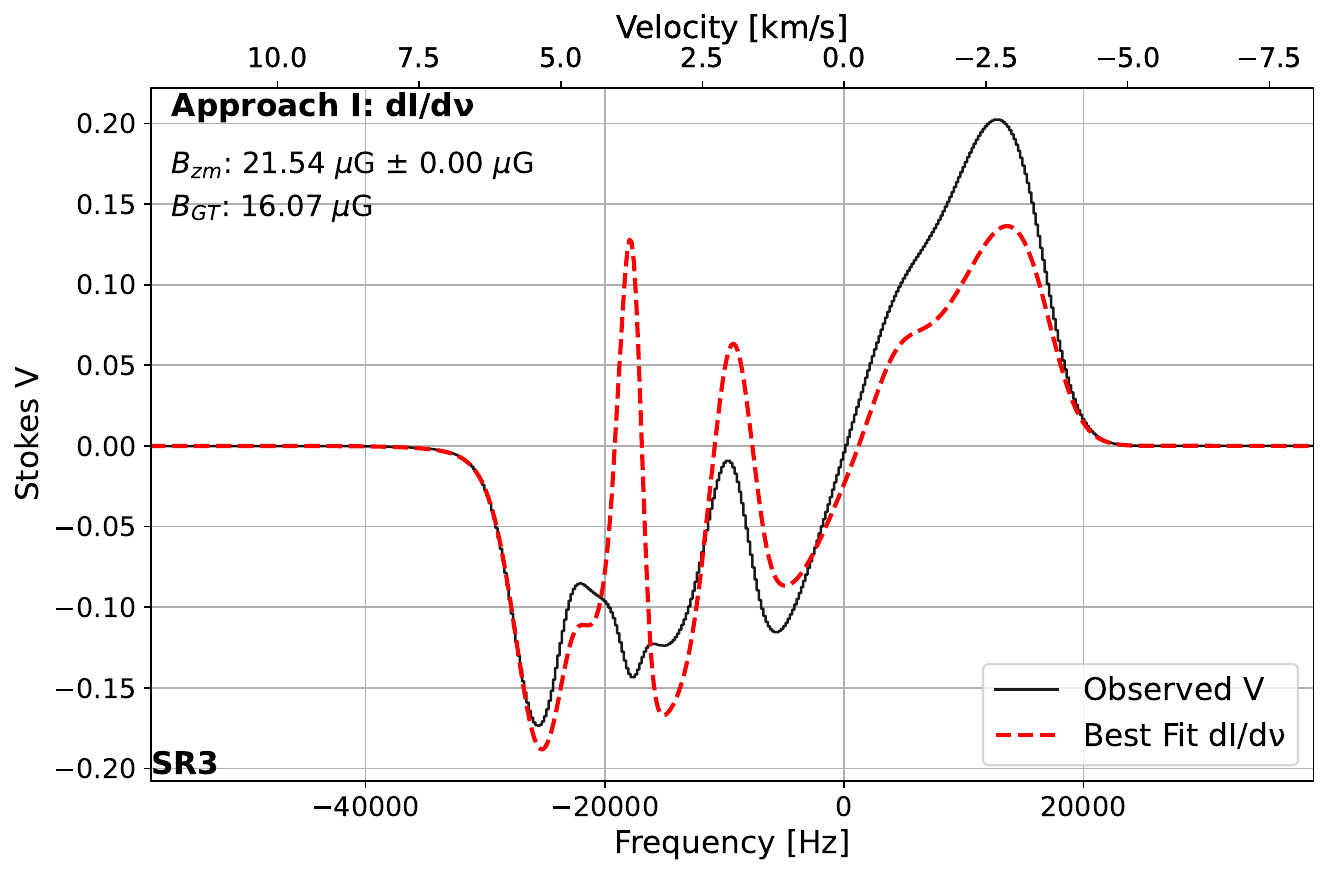}
\includegraphics[width=0.48\linewidth]{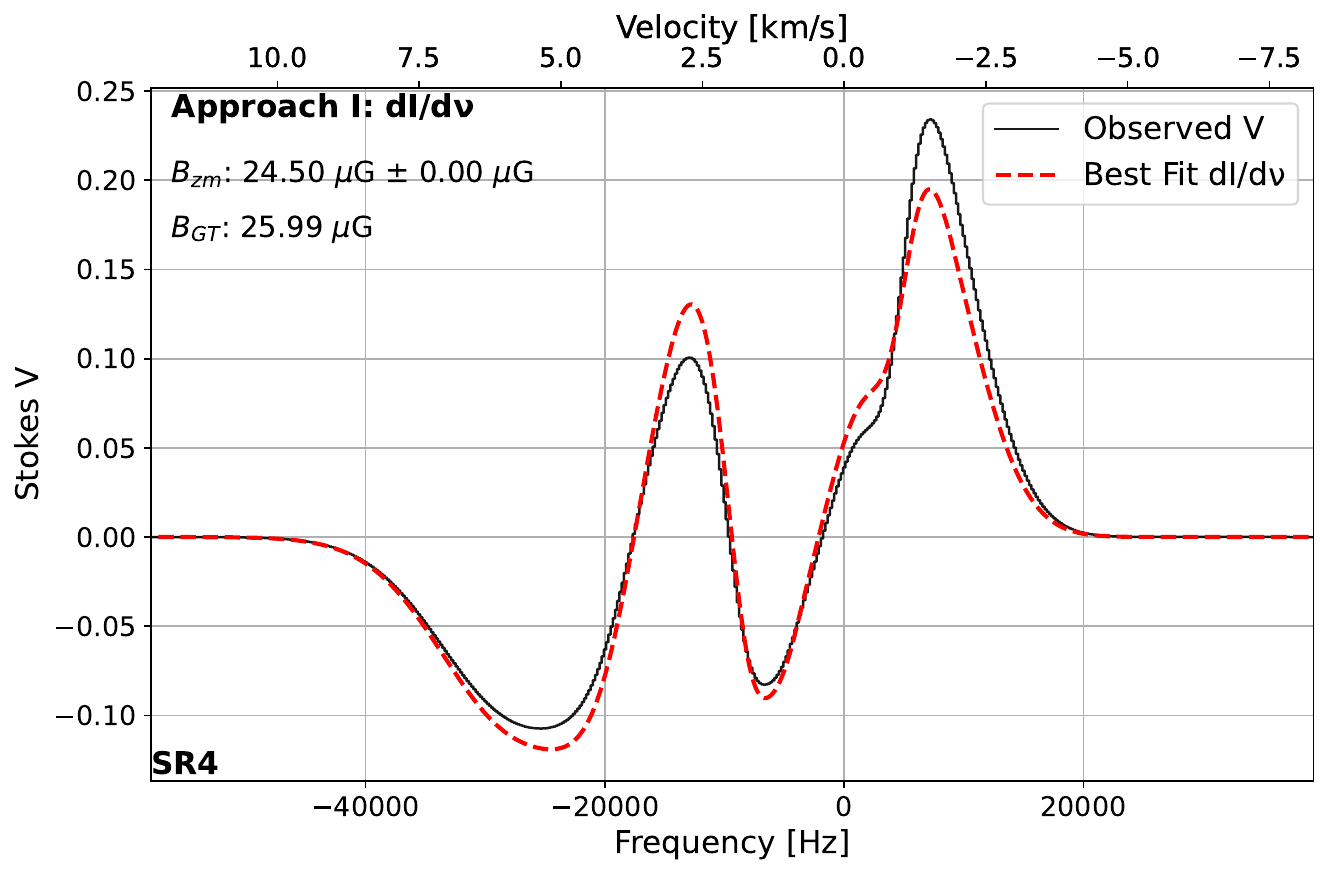}
\caption{
\xd{Result of the Approach I fitting of the Stokes $V$ spectrum (black line) with the scaled differentiated Stokes $I$ spectrum (red dashed line) using Equation~\ref{dI_dnu_eq1}, as in Figure~\ref{fig.ZeemanFitting_dI_dnu_5_layers}, but here for spectra of all four subregions of the simulated cube.}
The overall fit roughly captures the amplitude and profile, but the optimized scaled model
(red dash) does not precisely match the Stokes $V$ line shape (black), presenting challenges for accurate fitting and interpretation.}
\label{fig.ZeemanFitting_dI__SR_Breal_noise0}
\end{figure*}

\begin{figure*}[hbt!]
\centering
\includegraphics[width=0.48\linewidth]{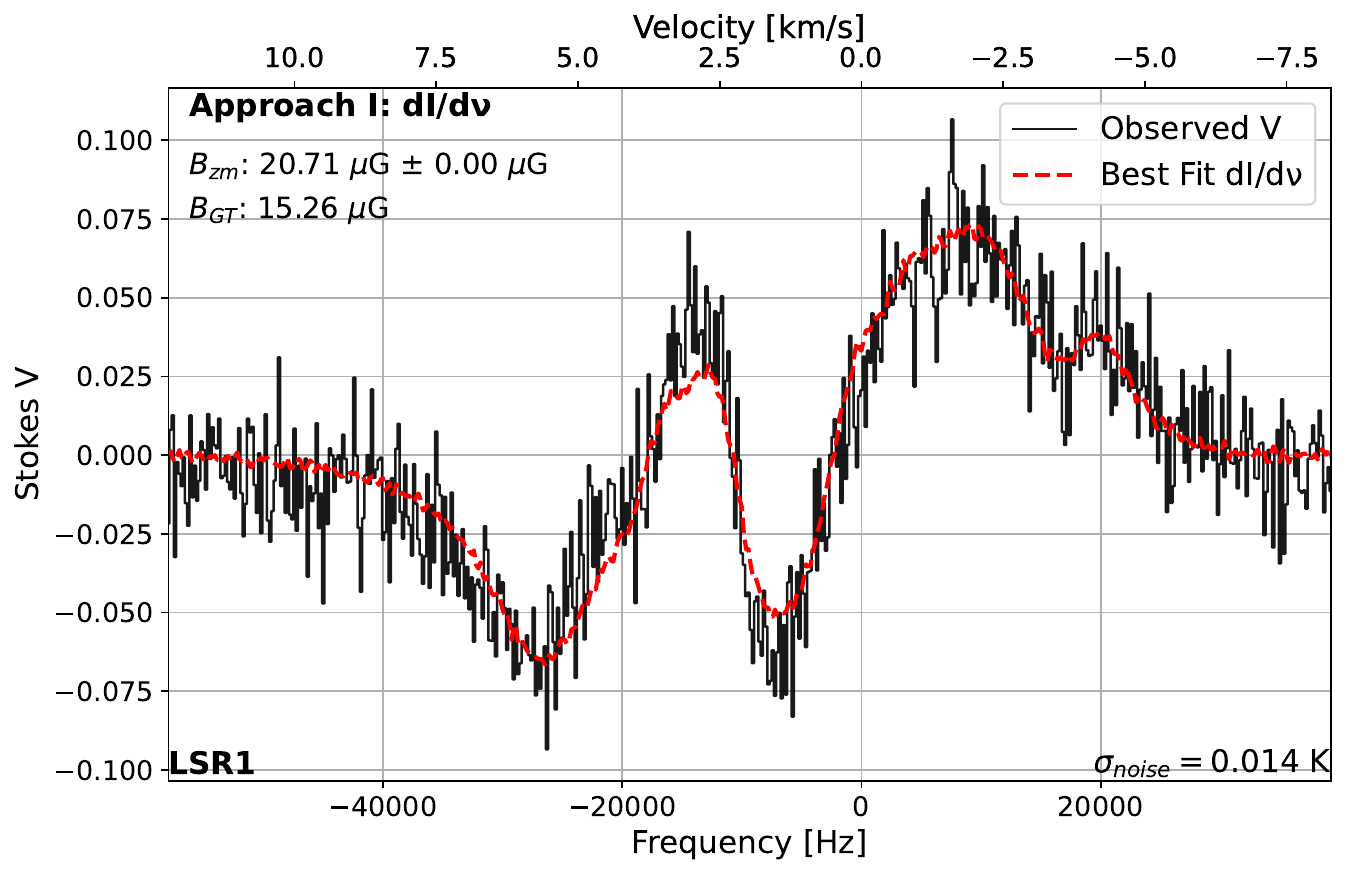}
\includegraphics[width=0.48\linewidth]{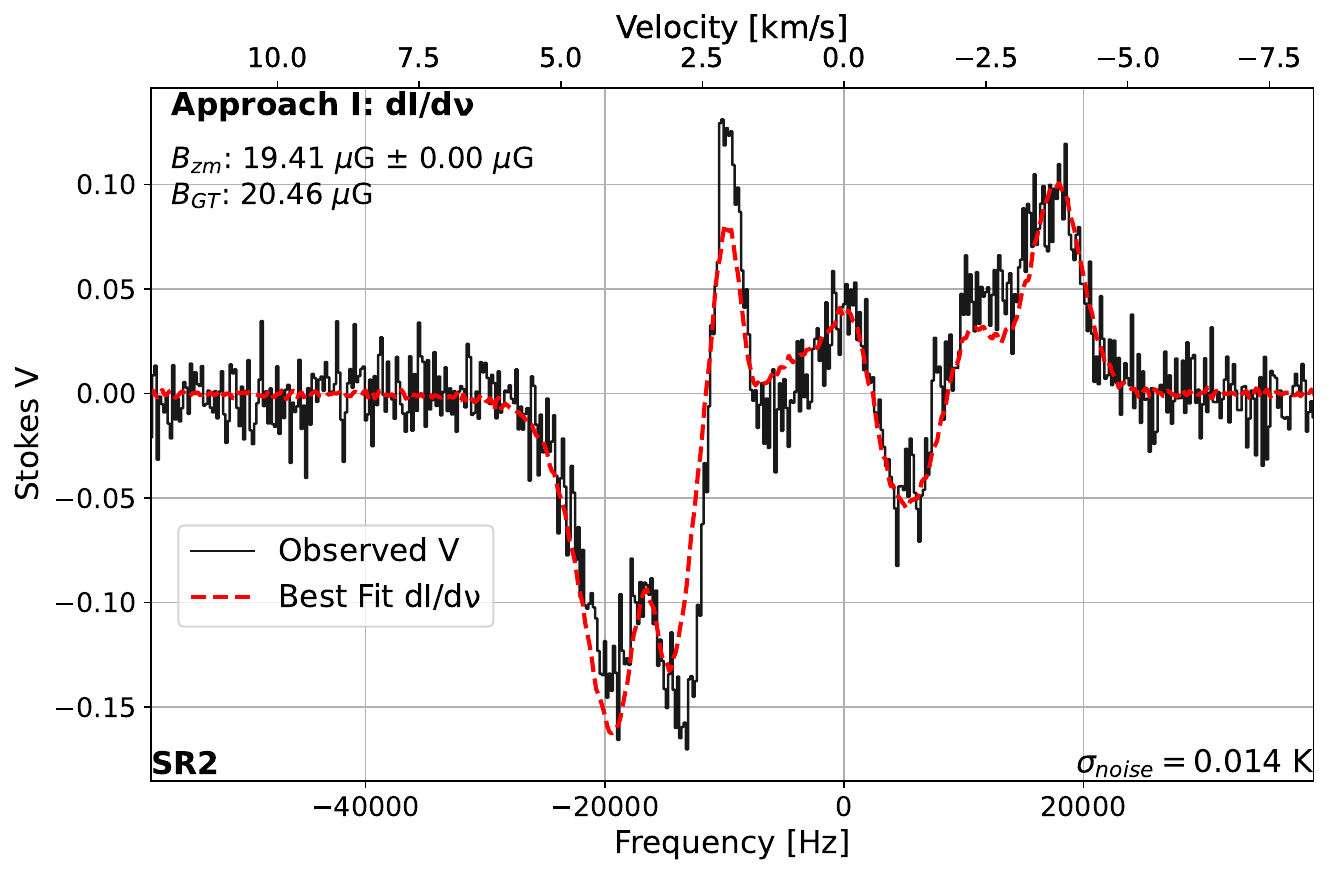}
\includegraphics[width=0.48\linewidth]{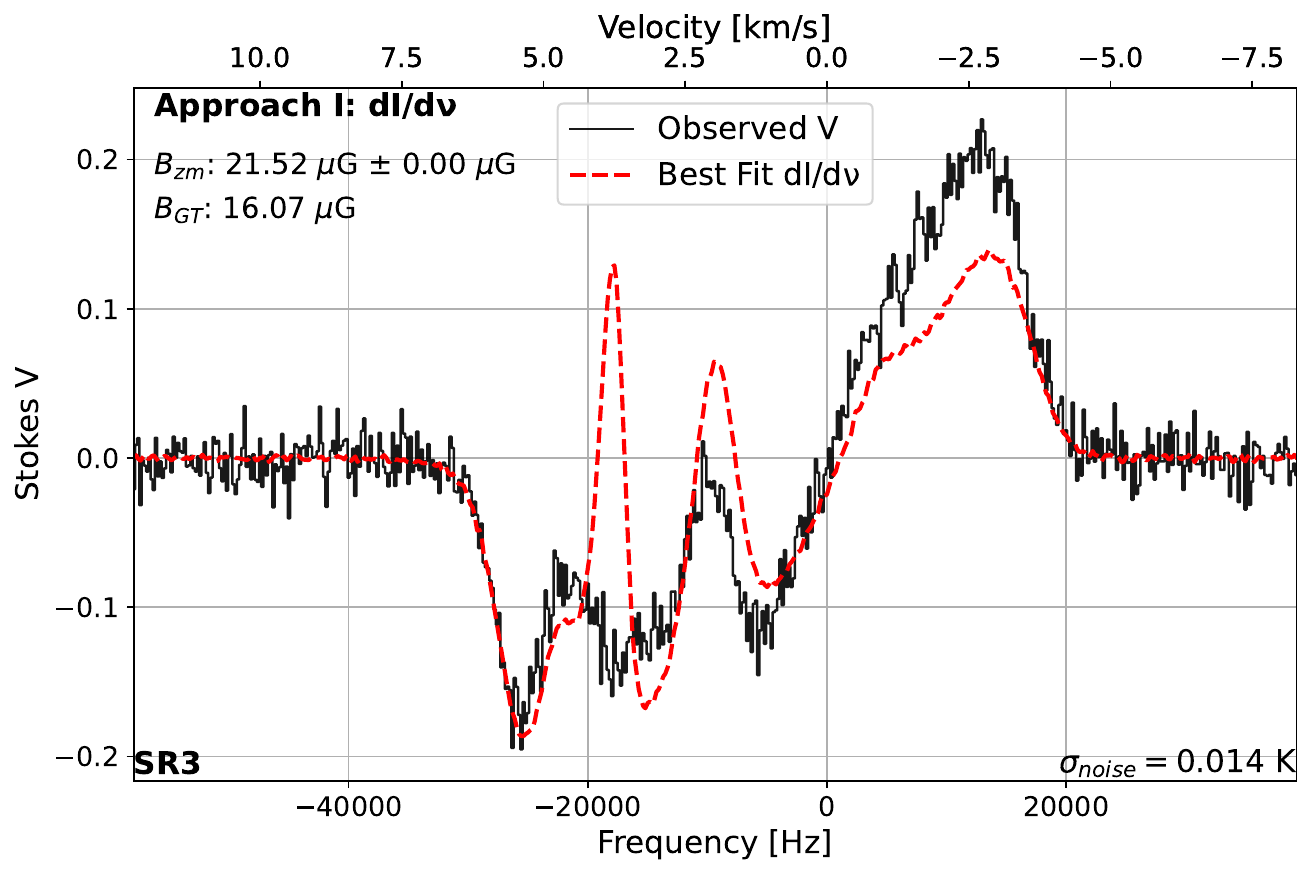}
\includegraphics[width=0.48\linewidth]{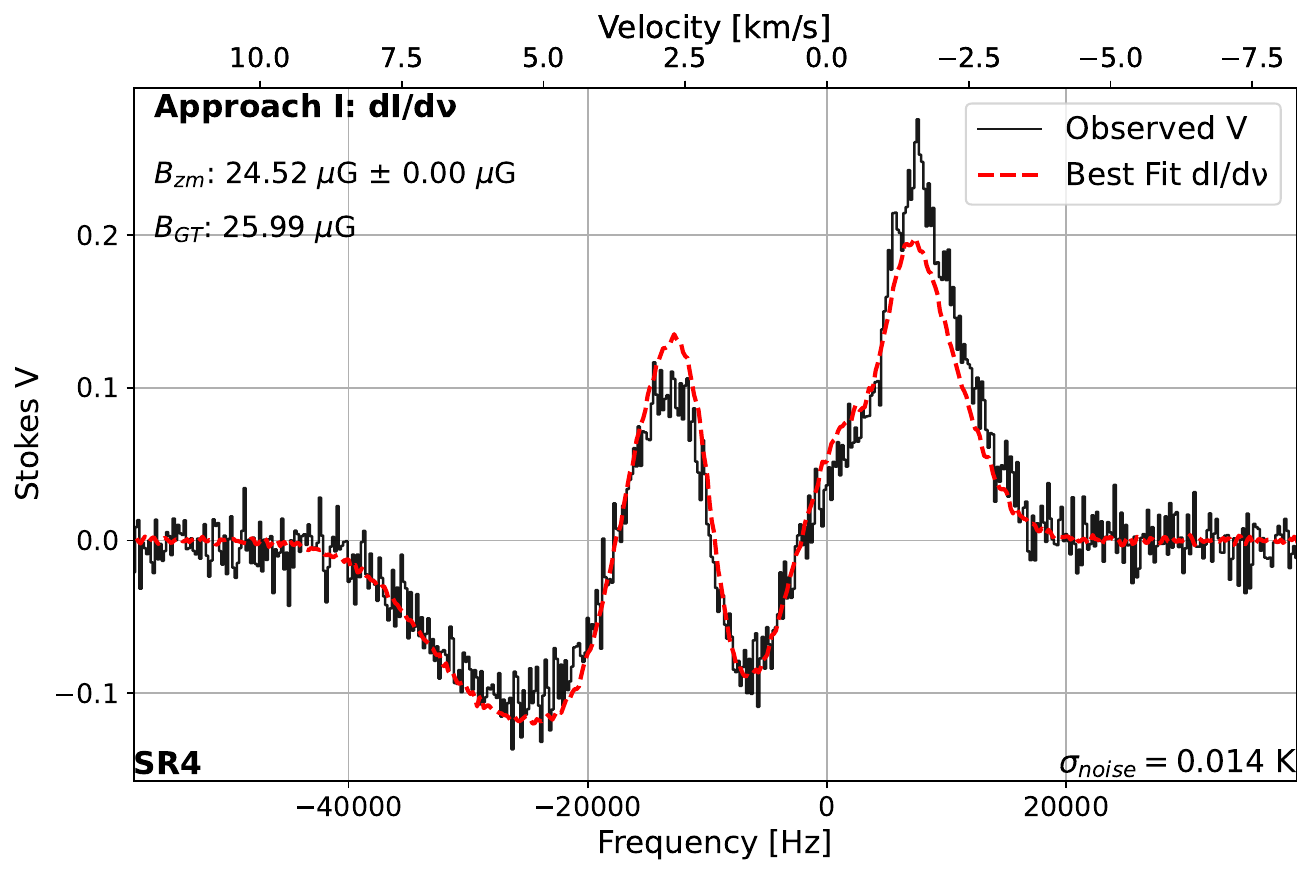}
\caption{\xd{Same as Figure~\ref{fig.ZeemanFitting_dI__SR_Breal_noise0}, but showing the Approach I fitting results for the Stokes $V$ spectra and the differentiated Stokes $I$ spectra in all four subregions at a noise level of 0.014 K (see Section \ref{sec:addnoise}).}}
\label{fig.ZeemanFitting_dI__SR_Breal_noise01}
\end{figure*}

\xd{In this section, we apply Approach I to synthetic HI spectra from four subregions of the MHD simulation and then extend the analysis systematically to a full set of synthetic spectra obtained from a scan across the entire 8 pc map.}

\subsection{Individual Region Analysis of the Four Subregions}
\label{Individual Region Analysis of the Four Subregions}

Figure~\ref{fig.ZeemanFitting_dI__SR_Breal_noise0} shows the results from Approach I for all four subregions, a weighted average of $B_{zm}$ along the line of sight across each subregion. For reference, the corresponding values of $B_{GT}$, the HI mass-weighted LOS magnetic field from the simulation, are also shown in the figure legend.

When noise is added, the $V$ spectrum to be fit is noticeably noisy as seen in Figure~\ref{fig.ZeemanFitting_dI__SR_Breal_noise01} that presents the Approach I fitting results for the Stokes $V$ spectra in all four subregions at a noise level of 0.014 K.
Fits of spectra for subregions SR2 and SR3 at a range of noise levels are shown in 
Appendix~\ref{Zeeman Fitting Gallery: Approach I Across Noise and Regions}. Despite the added noise, the inferred magnetic field strength $B_{zm}$, and thus its agreement with the constant ground-truth value $B_{GT}$, are only mildly affected, \xdtwo{changing by less than 10\% even for an eight-fold increase in noise.}  This is because the differentiated $I$ model to be scaled for the fit (red curves) is virtually the same.

While the fitted values are in very good agreement with the ground true values for some lines of sight, discrepancies are apparent in others. These differences arise from the complex magnetic field structures along the LOS that are not fully captured by a single averaged value. We investigate the typical size of these discrepancies in the following section.

\begin{figure}[hbt!]
\centering
\includegraphics[width=0.98\linewidth]{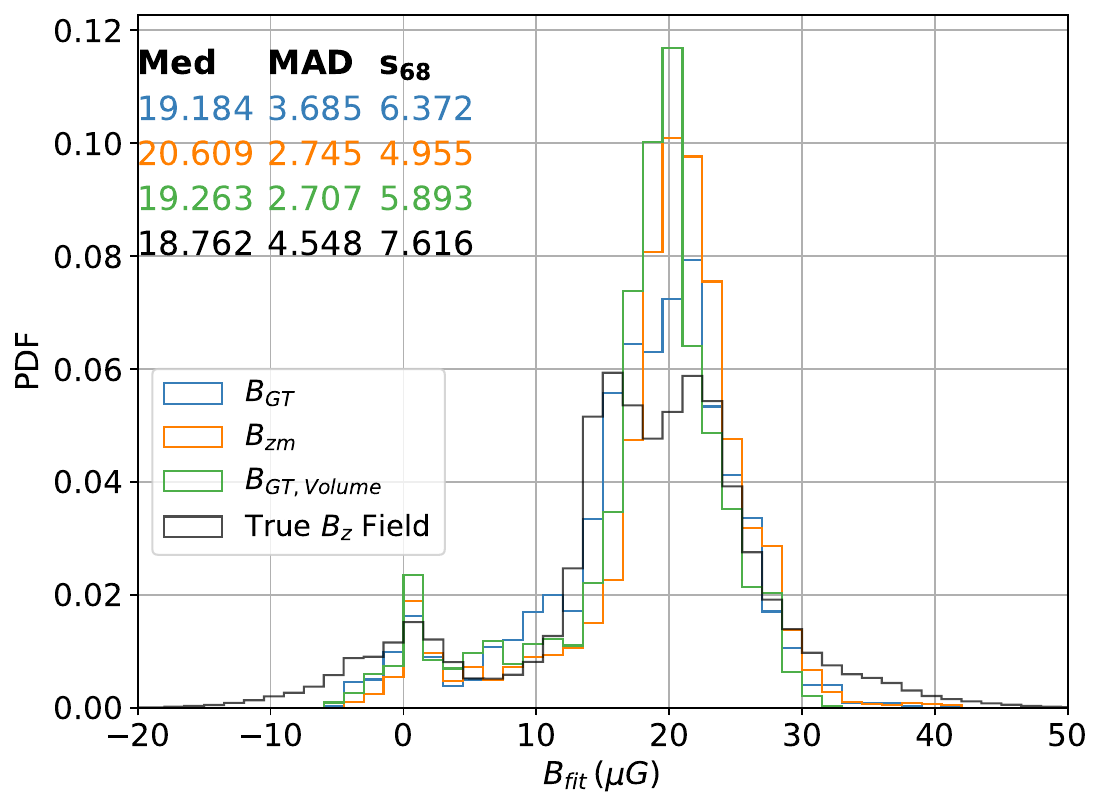}
\caption{\xd{Probability distribution function (PDF) of the inferred magnetic field strengths $B_{zm}$ from Zeeman analysis using Approach I for 4096 independent spectra across the face of the cube (orange).  This is compared with PDFs for estimators of ground-truth fields: $B_{\mathrm{GT}}$ that is mass weighted as defined in Equation~\ref{eq.BGT_approch2} (blue) and $B_{\mathrm{GT,volume}}$ that is volume-weighted (green). 
The PDF of the true LOS magnetic fields in all analyzed voxels (black) is identical to the black curve in Figure~\ref{fig.hist_plot_Bfit_all_noise0}.}
\xdtwo{In the legend are reported important robust statistics for comparing the distributions: the median, MAD, and s68 (see text).}
}
\label{fig.hist_plot_Bfit_all_App1}
\end{figure}

\begin{figure}[hbt!]
\centering
\includegraphics[width=0.98\linewidth]{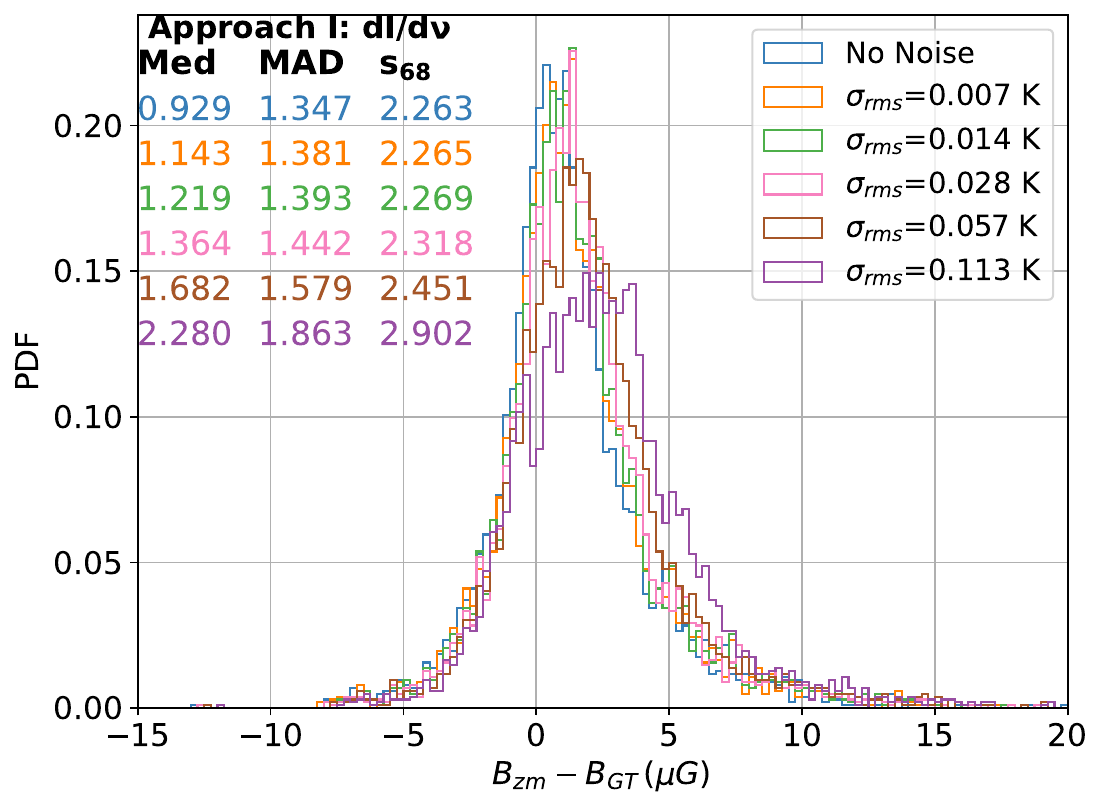}
\caption{The effect of different noise levels on the PDF of the difference between the magnetic field strength inferred using Approach I and the ground truth estimator $B_{GT}$ from Equation \ref{eq.BGT_approch2} (mass-weighted LOS magnetic field) for 4096 independent spectra across the face of the cube.
}
\label{fig.hist_plot_Bmag_error_dI_dnu}
\end{figure}

\begin{figure}[hbt!]
\centering
\includegraphics[width=0.98\linewidth]{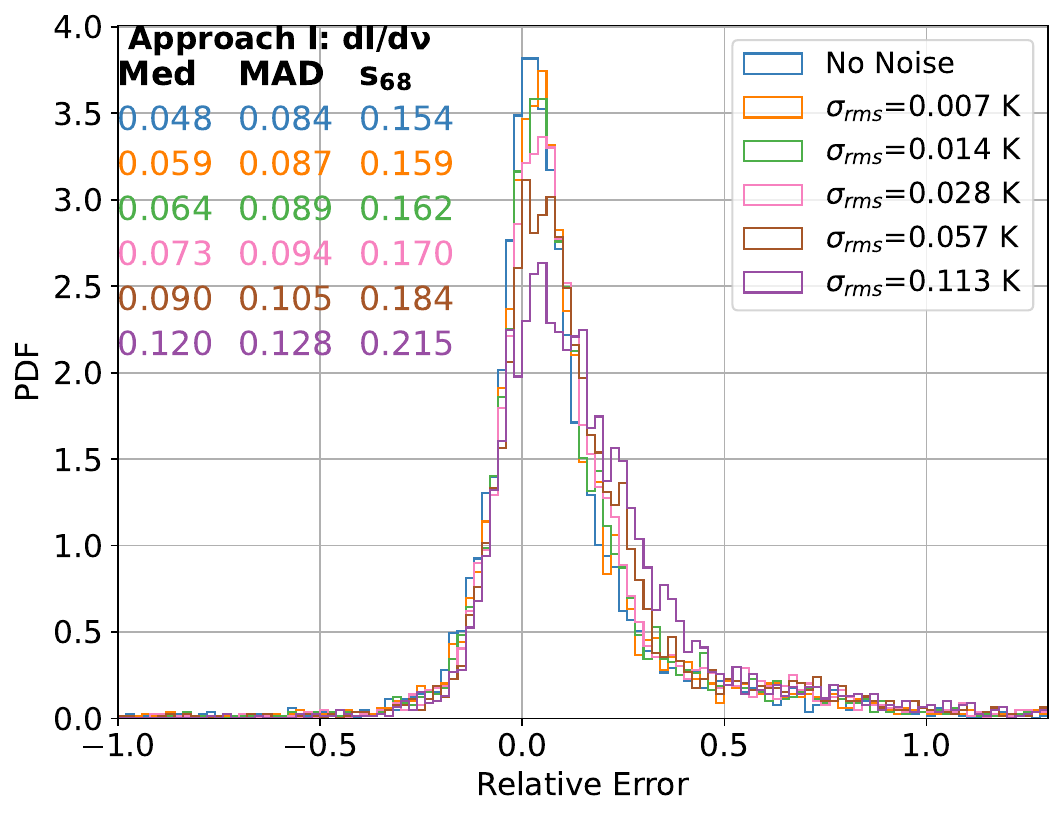}
\caption{PDF of the relative error between the magnetic field strength inferred using Approach I and the ground truth estimator $B_{GT}$ from Equation \ref{eq.BGT_approch2} (mass-weighted LOS magnetic field) for 4096 independent spectra across the face of the cube.}
\label{fig.hist_plot_Bmag_relative_error_dI_dnu_noise}
\end{figure}

\subsection{Assessing Uncertainties in Approach I}
\label{Assessing Uncertainties in Approach I: Derivative-Based Stokes $I$ Fitting}

A full set of synthetic spectra was created from a scan across the entire 8 pc map. We adopted a consistent box size of 0.125 pc, as used for SR2, SR3, and SR4, which corresponds to a 3\arcmin\ beam size. Each box was treated as an independent measurement with no overlap between adjacent regions, ensuring that every Zeeman analysis was performed independently on 4096 spectra without spatial redundancy. We applied Approach I to this set.

Figure~\ref{fig.hist_plot_Bfit_all_App1} shows the probability distribution functions (PDFs) of magnetic field strengths $B_{zm}$ estimated by Zeeman analysis using Approach I, compared with PDFs for the ground-truth estimators of LOS magnetic fields, shown in both mass-weighted ($B_{GT}$ from Equation \ref{eq.BGT_approch2}) and volume-weighted (straight average) forms. The PDF of the actual (true) LOS magnetic fields across all analyzed voxels is somewhat broader because it does not involve averaging along individual lines of sight.
The overall appearance of the PDFs 
indicate that despite uncertainties (errors) in all estimators, including $B_{zm}$, these average estimators are all tracking the same underlying distribution faithfully.

More quantitatively, \xdtwo{for histograms like in Figure~\ref{fig.hist_plot_Bfit_all_App1} and especially like in} Figure~\ref{fig.hist_plot_Bmag_error_dI_dnu}, which presents the PDF of differences between $B_{zm}$ and $B_{GT}$, we computed from the raw data three statistics that characterize the distributions \xdtwo{and tabulated them in the legends.}
The first is simply the median, a location statistic. The other two are scale parameters: \xdtwo{the median absolute deviation from the median (MAD, aka MADFM) and $s_\textrm{68}$, which} is half the width between the lower and upper (roughly 16th and 84th) percentiles defining the 68.27\% confidence (coverage) interval, in the same spirit as 1-$\sigma$ is for a Gaussian. 
These statistics are robust despite these distributions having “heavy (fat)” tails compared to a Gaussian and some asymmetry, and so are appropriate choices.


We investigated how varying noise levels (Section~\ref{sec:addnoise}) affect $B_{zm}$ 
and the resulting difference histograms, noting that $B_{GT}$ remains constant. 
These results show that Approach I is fairly robust in recovering the estimator of the averaged LOS magnetic field strength, \xd{with $s_\textrm{68}$ about 2.3 $\mu$G and a small bias, about 1.2 $\mu$G at 0.014 K noise level, both increasing only slightly with the noise level (see legend in figure).}



To explore another key factor, optical depth,
in Appendix \ref{app:approachIstatistics0001} we carried out experiments lowering the densities so that the cube is optically thin. For Approach I, Figure \ref{fig.hist_plot_Bmag_error_dI_dnu_rho0001} in Appendix \ref{app:approachIstatistics0001} shows that the statistics for the no-noise case are similar, \xd{Median, MAD, and $s_\textrm{68}$ being $-0.842$, 1.216, and 2.042 $\mu$G, respectively.}

\xd{Figure~\ref{fig.hist_plot_Bmag_relative_error_dI_dnu_noise} shows the PDF of the relative error (Equation \ref{eqn_rela_error}) between the magnetic field strength estimated with Approach I analysis, $B_{zm}$, and the ground truth estimator, $B_{GT}$. The relative error is approximately 16\% in the noise-free and low to moderate noise cases, and remains modest even at the highest tested noise level of 0.113~K, increasing to only about 22\%.
}

\xdtwo{In this analysis, we cannot separate how much the uncertainty of the individual estimators $B_{zm}$ and $B_{GT}$ contributes to $s_\textrm{68}$ of the difference distributions; therefore, whether expressed in absolute or relative terms (Figure~\ref{fig.hist_plot_Bmag_error_dI_dnu} or Figure~\ref{fig.hist_plot_Bmag_relative_error_dI_dnu_noise}), the value of $s_\textrm{68}$ should be regarded as an upper limit on the uncertainty of $B_{zm}$.
}

In observational studies, Approach I offers a reliable estimate of an averaged LOS magnetic field.  However, it lacks the ability to resolve multiple velocity components along the line of sight, which \xd{precludes probing detailed magnetic field structure in the cloud.}

\section{Analysis of Synthetic HI Spectra from MHD Simulations Using Approach II}
\label{Analysis of Synthetic HI Spectra from MHD Simulations Using Approach II}

\xd{Approach II involves component-based model Stokes $I$ and $V$ spectra fitted to synthetic HI spectra from the MHD simulation. We examine the fitting results for the SR3 LOS in detail, and then extend the analysis systematically to the full set of synthetic spectra obtained from a scan across the entire 8 pc map.}

\subsection{Individual Region Analysis: Spotlight on SR3}
\label{Individual Region Analysis Spotlight on SR3}





\begin{figure}[hbt!]
\centering
\includegraphics[width=0.98\linewidth]{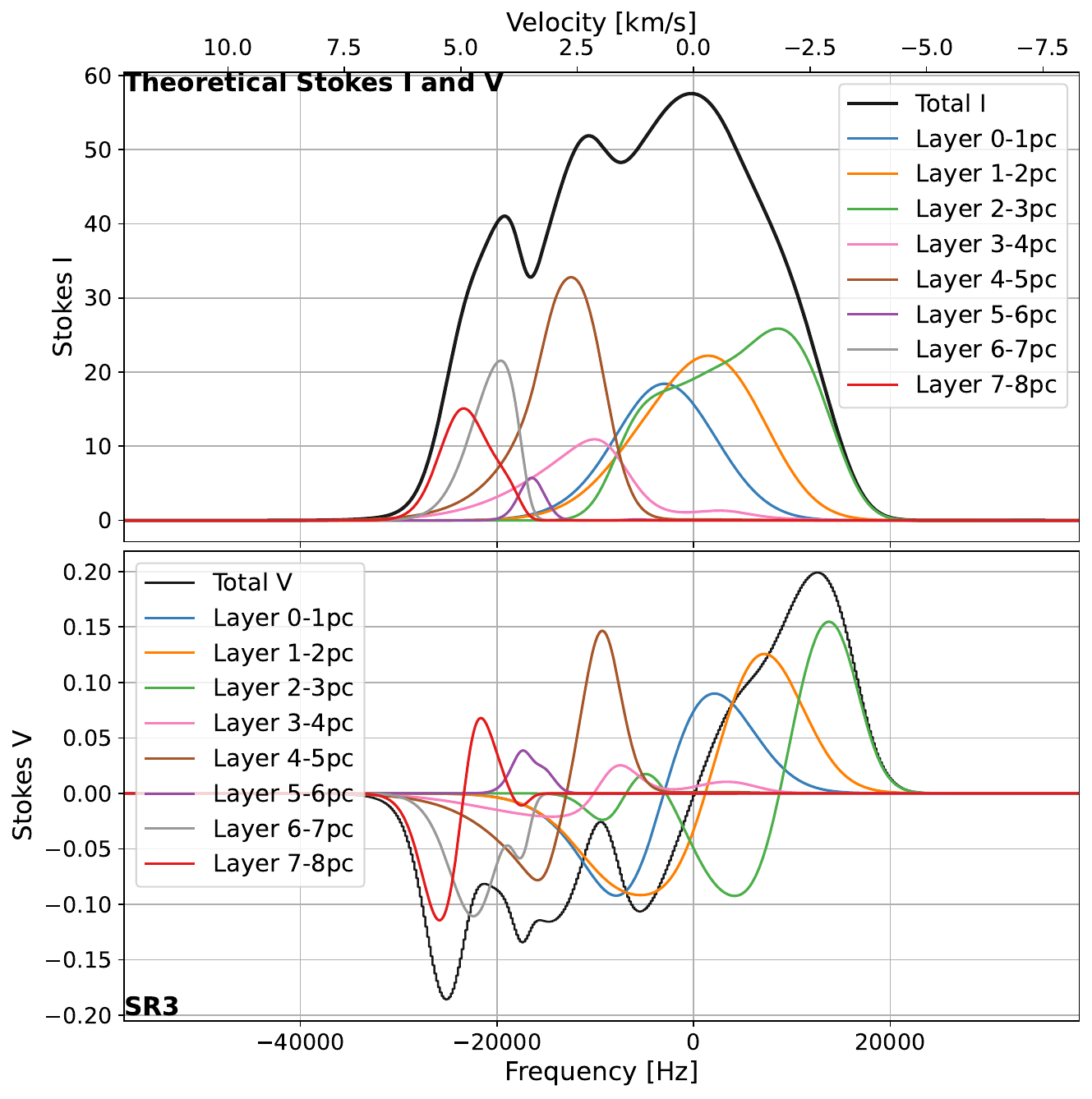}
\caption{Layer contributions as in Figure~\ref{fig.5_Layer_GPB}, but for the SR3 subregion. Each layer has a depth of 1 pc, ranging from the foreground (Layer 0-1 pc, aka L01) to the background (L78).  (See Figure \ref{fig.SR3_rho0001_GPB} for a reduced density (optically thin) experiment and Figure \ref{fig.SR3_flip_GPB} for the $-z$ perspective.)
}
\label{fig.SR3_GPB}
\end{figure} 

\subsubsection{Ground Truth Layer Contributions for SR3}
\label{sec:gtlsr3}

Figure \ref{fig.SR3_GPB} shows the ground truth layer contributions to $I$ and $V$ spectra from GPB-S calculations for eight layers along the SR3 line of sight.  L01 is for emission from the first layer in the foreground, from 0 to 1 pc.  The last is L78.
The properties of these contributions can be understood by reference to the gas properties in Figure \ref{fig.physical_Region_All} lower left.

For example, L01 and L12 (blue, orange) have a velocity only slightly positive and a field of about 30 to 35 $\mu$G. L23 (green) ought to include the most negative velocity emission but on average a lower field below 20 $\mu$G. All three contribute to the $I$ spectrum in the expected way, with L23 broad and showing signs of attenuation by the foreground layers. All three also contribute to the broad peak in $V$ at negative velocity.  The gas producing the most negative velocity wing (at about 2.2 pc) has a field near 30 $\mu$G. 

L45 (brown) should have strong intrinsic emission at 2 \kms\ according to the 
$n_{\mathrm{HI}}$ and 
$T_{\mathrm{ex}} \cdot \tau_{\mathrm{HI,\, Center}}$ depth profiles, but it suffers from foreground extinction on its low-velocity tail.  The field is low which would make the amplitude of a normal Zeeman profile small in the $V$ spectrum contribution; however, as we have seen with the five-layer cloud, the polarized attenuation can have a significant effect too.

L56 (purple) has intrinsically strong emission too at about the same velocity.  There is a field reversal that would be hard to detect in the $V$ profile because of the attenuation 
\xd{(but see Figure \ref{fig.SR3_flip_GPB} in which the cube is viewed from the back side and this layer is then closer to being in the foreground).}

L78 has $v_z \simeq 40$ \kms\ and field about 20 $\mu$G and so it contributes to the positive velocity wing of the theoretical spectra in the expected way.  Toward zero velocity, it is of course affected by attenuation.

\begin{figure*}[hbt!]
\centering
\includegraphics[width=0.98\linewidth]{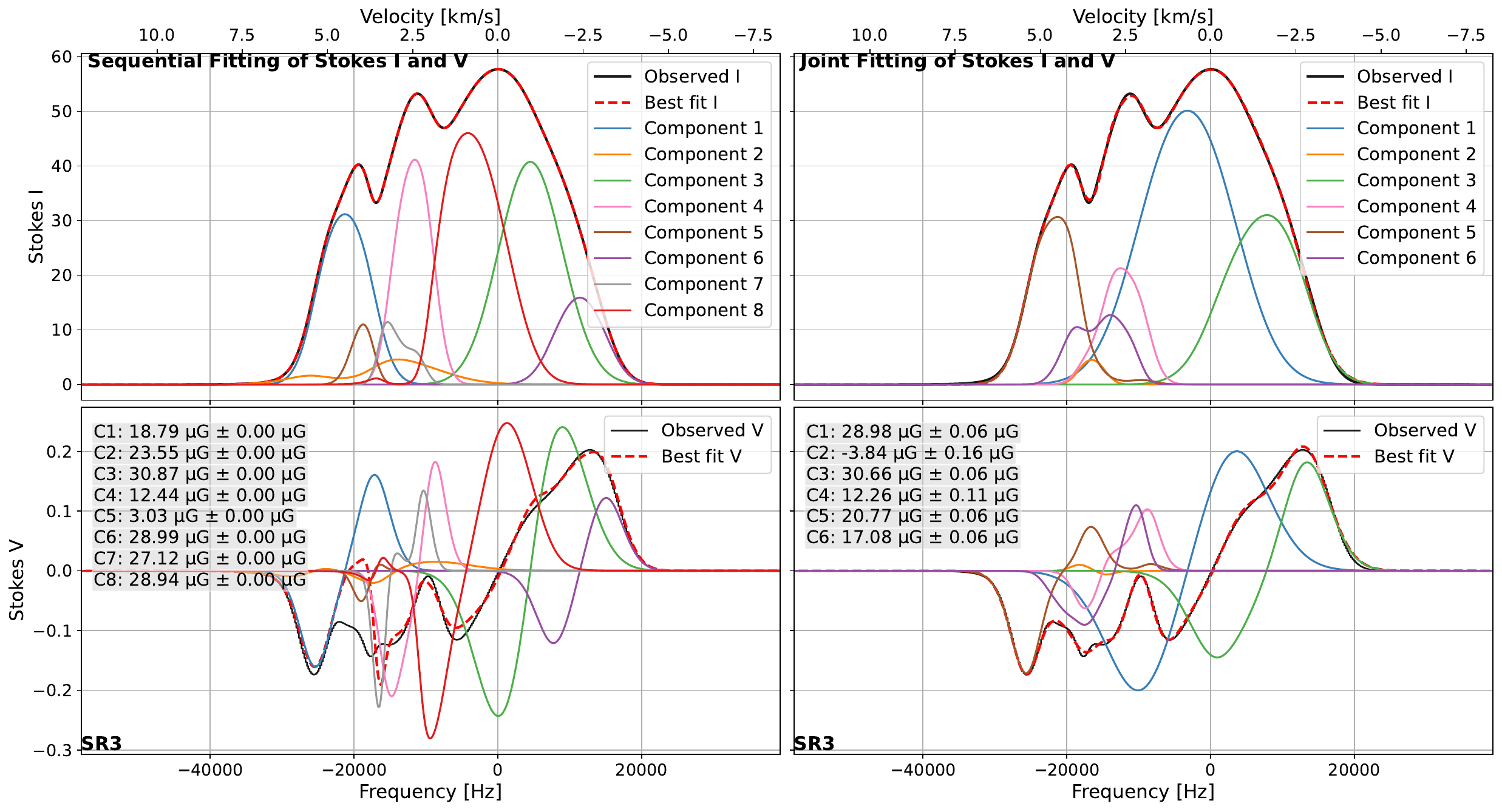}
\caption{\xd{Approach II fitting results for the Stokes $I$ and Stokes $V$ spectra for the SR3 subregion}, 
displayed as in Figure~\ref{fig.ZeemanFitting_5_Layer_IV_noise0}. 
\xd{The joint strategy effectively captures both the Stokes $I$ and $V$ spectra, whereas the sequential strategy fits Stokes $I$ well but performs poorly for Stokes $V$.} (See Figure \ref{fig.SR3_I_V_fitting_rho0001_} for the spectral fits in the low optical depth case and Figure \ref{fig.ZeemanFitting_SR3_I_V_fitting_flip_} from the $-z$ axis perspective.)}
\label{fig.ZeemanFitting_SR3_Breal_noise0}
\end{figure*} 

\begin{figure}[hbt!]
\centering
\includegraphics[width=0.98\linewidth]{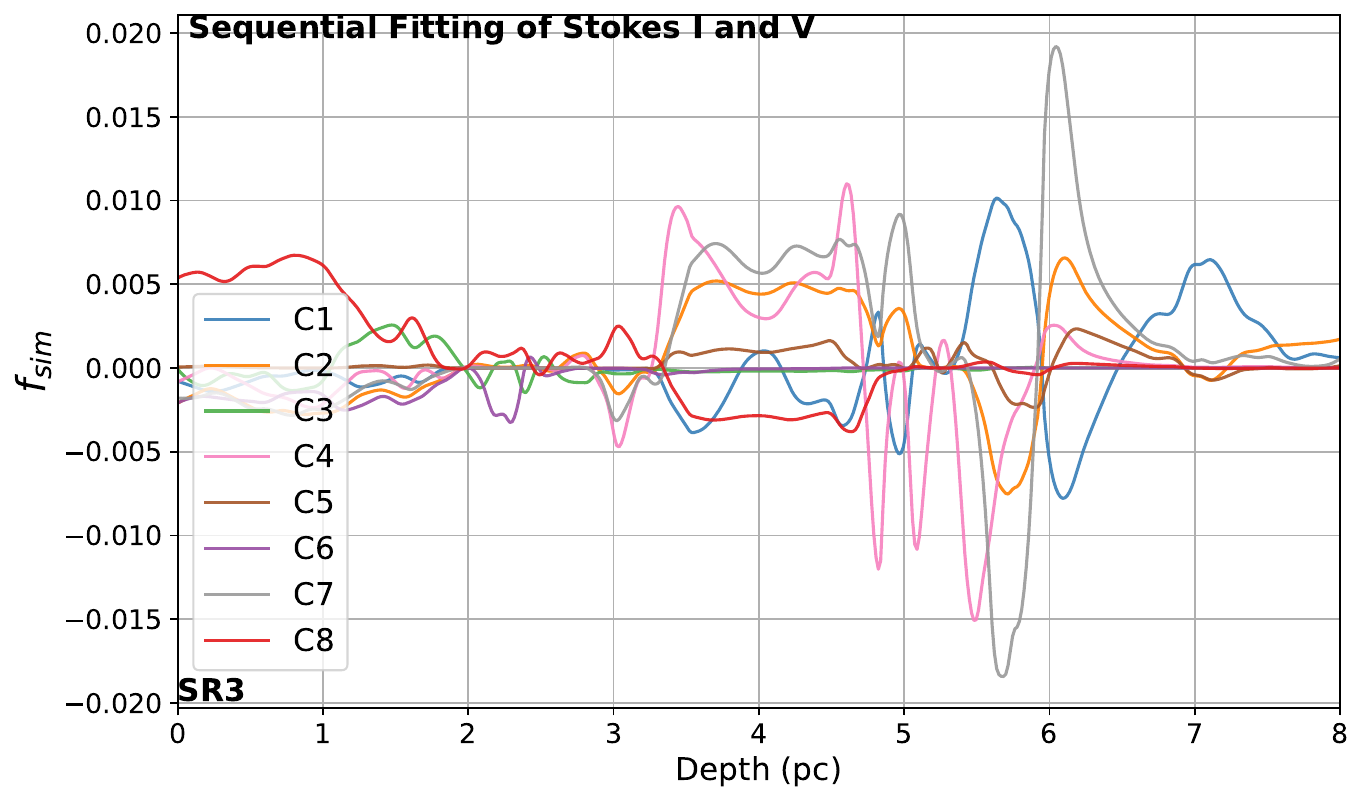}
\includegraphics[width=0.98\linewidth]{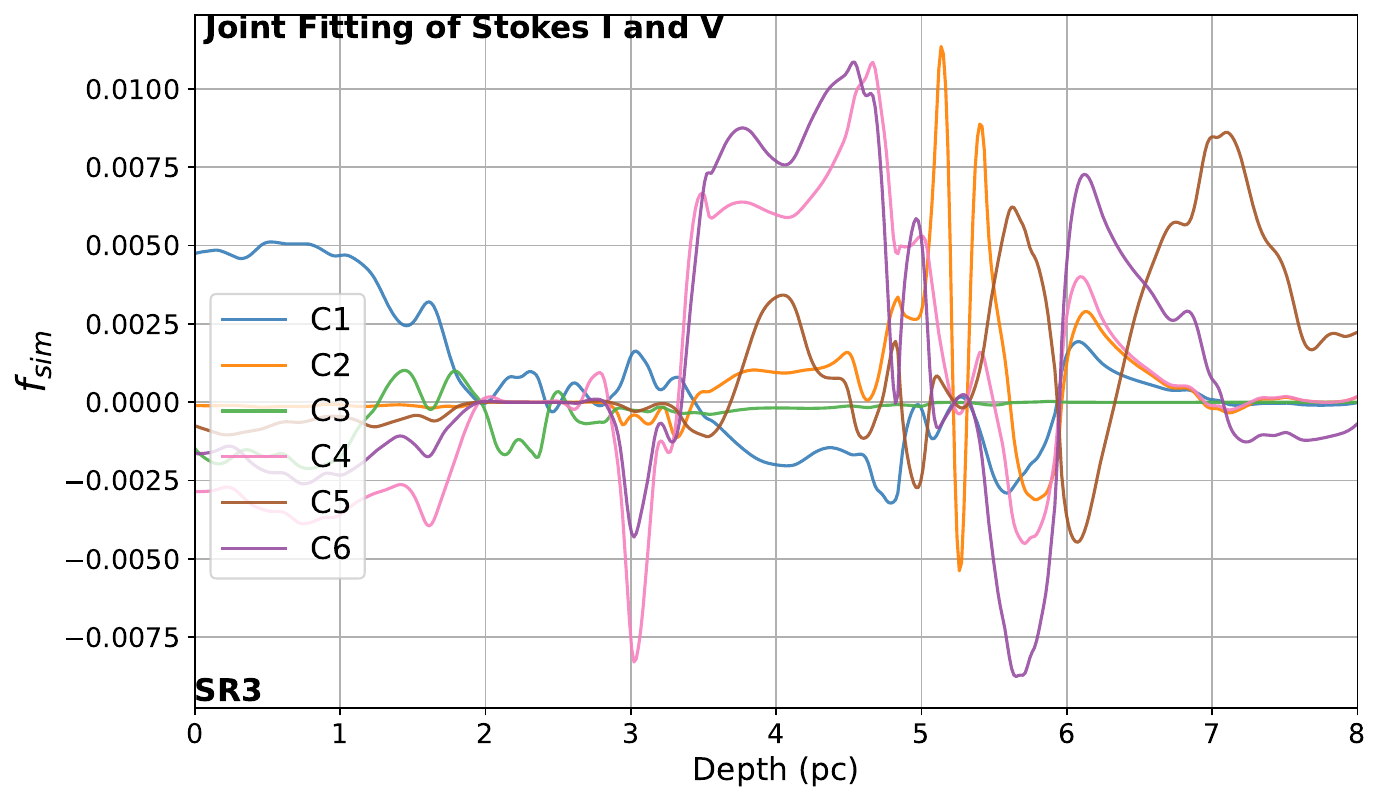}
\caption{Profile of the similarity function $f_{\text{sim}}$ (Equation~\ref{eq.fsim}) for each fitted component obtained using the two fitting strategies under Approach II along the LOS in the SR3 subregion.
(See Figure \ref{fig.f_sim_SR3_flip} for $f_{\text{sim}}$ for the spectral fits from the $-z$ axis perspective.)}
\label{fig.f_sim_SR3}
\end{figure}

\begin{figure*}[hbt!]
\centering
\includegraphics[width=0.98\linewidth]{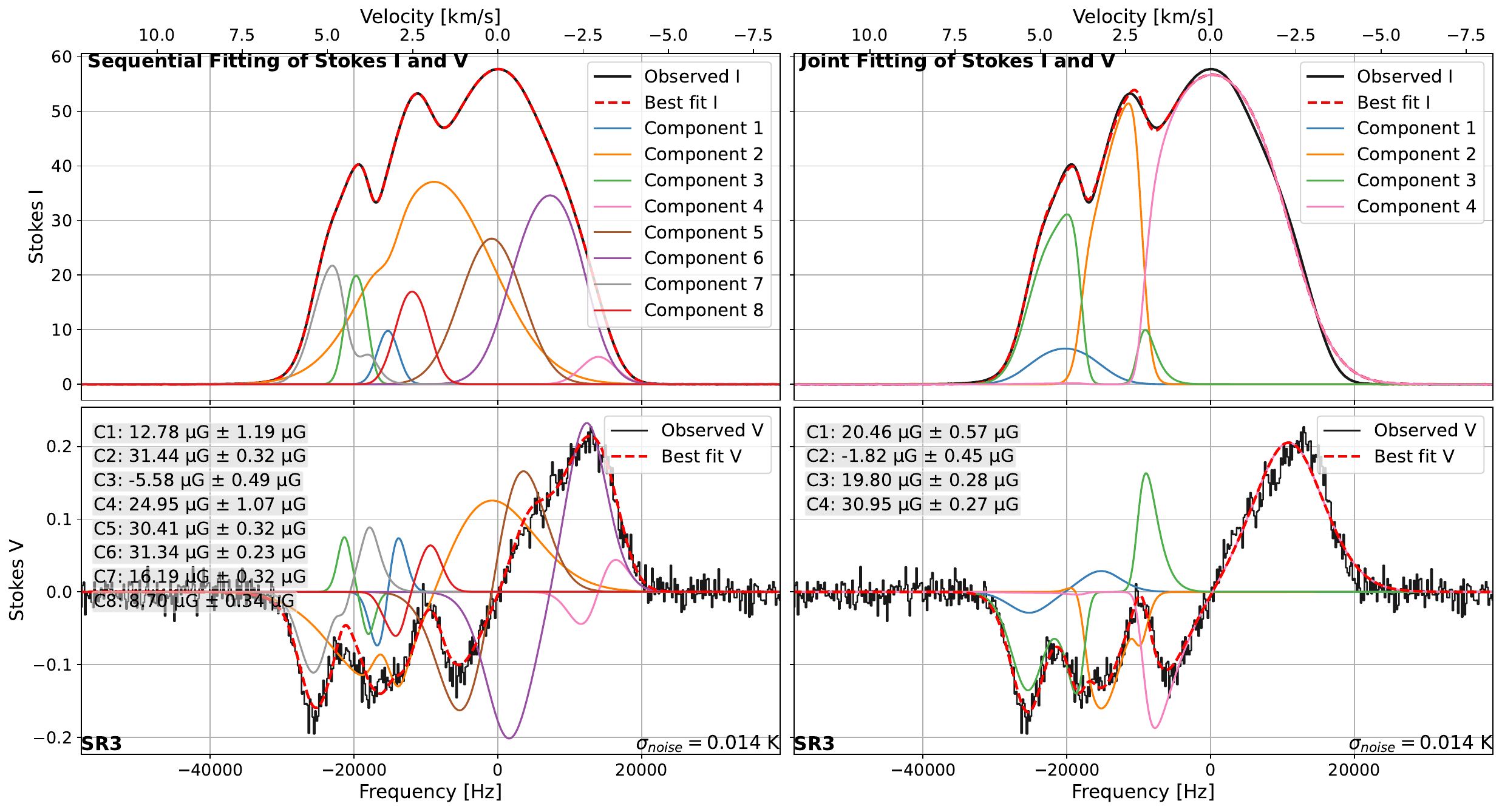}
\caption{Same as Figure~\ref{fig.ZeemanFitting_SR3_Breal_noise0}, \xd{but showing Approach II fitting results for low-noise ($\sigma_{\mathrm{noise}} = 0.014$ K) Stokes $I$ and Stokes $V$ synthetic spectra for} the SR3 subregion. 
The sequential strategy provides a satisfactory fit to both Stokes $I$ and $V$ using eight components, whereas the joint strategy with only four components is starting to reveal slight deficiencies in both.}
\label{fig.ZeemanFitting_SR3_Breal_noise01}
\end{figure*} 

\begin{figure}[hbt!]
\centering
\includegraphics[width=0.98\linewidth]{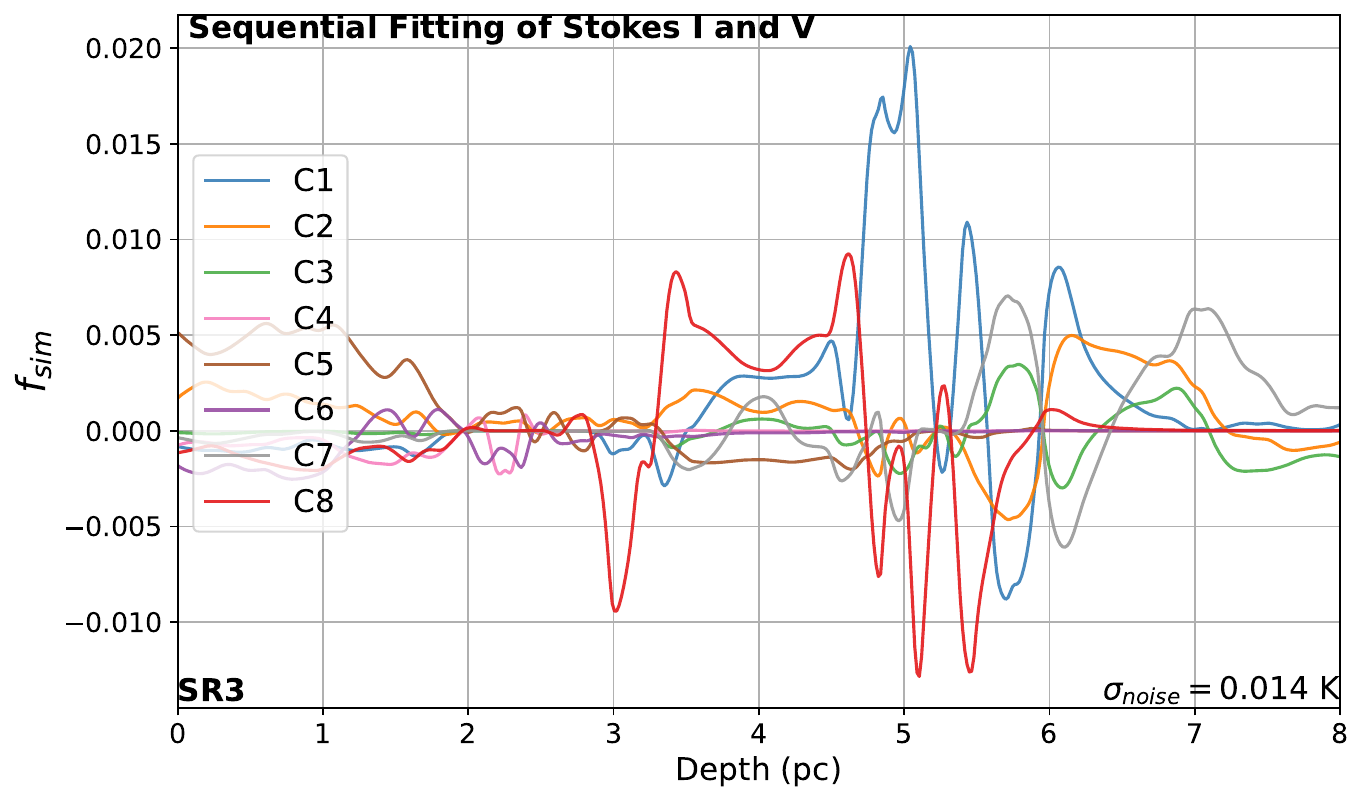}
\includegraphics[width=0.98\linewidth]{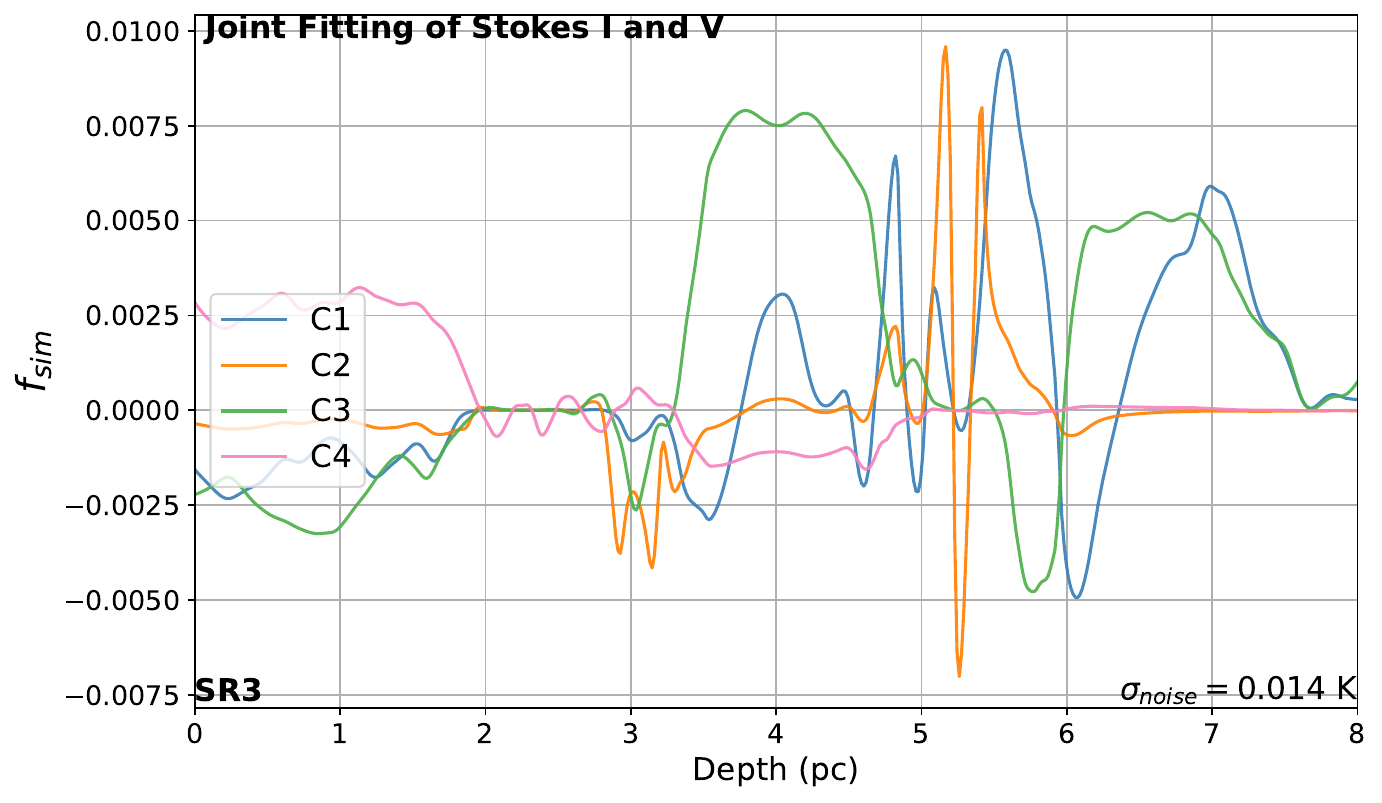}
\caption{\xd{Same as Figure~\ref{fig.f_sim_SR3}, but showing the similarity function $f_{\text{sim}}$ (Equation~\ref{eq.fsim}) for each fitted component obtained with the two fitting strategies under Approach II, for a LOS in the SR3 subregion with a noise level of 0.014 K.}}
\label{fig.f_sim_SR3_noise01}
\end{figure} 


\subsubsection{Approach II Fitting of SR3}
\label{sec:sr30}

Figure~\ref{fig.ZeemanFitting_SR3_Breal_noise0} compares results of sequential and joint fitting strategies for the SR3 subregion in the case of no noise.
As shown on the left, the sequential strategy struggles to reproduce the Stokes $V$ profile accurately, even with a larger number of fitted components (eight, as limited by the BIC). Also, the components are out of order relative to the expectations of the layer contributions.

In contrast, on the right the joint strategy simultaneously fits both Stokes $I$ and $V$ more effectively, despite requiring fewer components (six). Although the joint strategy may slightly compromise the precision of the Stokes $I$ fit, it generally yields more consistent and physically meaningful results for the Stokes $V$ profile.  See Table \ref{tab.Bz_B3D_sum_subregion} for the inferred fields and ground truth estimators by component. 


\xd{To assess how well the spatially-ordered fitted spectral components relate to the underlying cloud structure, we compared the fitting results with the LOS physical properties in Figure  \ref{fig.physical_Region_All} lower left, the layer contributions in Figure \ref{fig.SR3_GPB}, and the similarity function $f_{\text{sim}}$ in Figure \ref{fig.f_sim_SR3}. Effects of component ordering and attenuation are clearly evident and the joint fitting strategy generally provides a more physically plausible association between fitted components and cloud layers than the sequential strategy. Particularly near the line center, strong spectral blending along the LOS introduces significant degeneracy, limiting the ability of Zeeman analysis alone to recover a unique depth ordering and thus the profile of the field strength.  However, determining the field strength associated with gas responsible for emission in the line wings is more robust.}

\subsubsection{Low optical depth}
\label{sec:lowtausr3}

Appendix~\ref{app:approachII0001} examines the low optical depth case for Approach II, including the SR3 line of sight in Figure\ref{fig.SR3_rho0001_GPB} (layer contributions) and Figure~\ref{fig.SR3_I_V_fitting_rho0001_} (sequential and joint decompositions). 
\xd{The main result is that both decomposition strategies fail to recover the correct depth ordering of components along the line of sight, because the extremely low optical depth provides insufficient information from attenuation effects to constrain their relative ordering.}

\subsubsection{The SR3 LOS from the Reverse Perspective}
\label{sec:reversesr3}

As described in Appendix \ref{app:flip}, we further tested the physical reality of components and their magnetic fields by analysing the same LOS through the simulation cube from the opposite direction. 
%
%
The key question is whether the inferred magnetic field is consistent between physically corresponding components observed from the two viewing directions.
\xd{Agreement is found in the line wings, whereas near the line center the correspondence is weaker, because strong spectral overlap and attenuation prevent a unique depth ordering and a clear one-to-one matching of fitted components between the two perspectives.}

\subsubsection{The Impact of noise on Approach II results for SR3}
\label{sec:sr301}

\xd{When realistic noise is introduced, the Approach II results are considerably different (Figure~\ref{fig.ZeemanFitting_SR3_Breal_noise01} for low-noise, $\sigma_{\mathrm{noise}} = 0.014$ K) and correspondingly the profiles of $f_{\mathrm{sim}}$ as well (Figure \ref{fig.f_sim_SR3_noise01}). 
Noise ultimately suppresses the fine structure in Stokes $V$ that is essential for distinguishing individual components through attenuation effects, and when the number of components is reduced by the BIC the fitted components represent more LOS-averaged emission rather than distinct physical gas layers. Although acceptable spectral fits can still be obtained, the joint strategy loses much of its discriminatory power to determine the LOS profile of the magnetic field (the hoped-for inferred depth ordering of components becomes unreliable).}

\subsubsection{Magnetic Field Associated with the Gas Producing Emission in the Line Wings}
\label{sec:sr3velocity}

\xd{But the field associated with the gas producing emission in the lines wings is quite robust for the SR3 LOS. From the relevant components in Figures \ref{fig.ZeemanFitting_SR3_Breal_noise0} and \ref{fig.ZeemanFitting_SR3_Breal_noise01} (with 0.014 K noise), 
the field corresponding to the negative velocity wing is about $30\, \mu$G, quite distinct from $20\, \mu$G corresponding to the positive velocity wing.  This is the case for both sequential and joint strategies, though the component numbering differs.
This field difference is confirmed to be physical by the analysis of the cube viewed from the reverse direction in Section \ref{sec:reversesr3} Figure \ref{fig.ZeemanFitting_SR3_I_V_fitting_flip_}.  It persists in the presence of noise as examined for the joint strategy in detail in Appendix \ref{Zeeman Fitting Gallery: Approach II Across Noise and Regions}, Figure \ref{fig.ZeemanFitting_SR3_I_V_fitting_noise_All}.
It is also in accord with the velocity and magnetic field profiles in Figure \ref{fig.physical_Region_All} lower left and the discussion of layer contributions in Section \ref{sec:gtlsr3} (Figure \ref{fig.SR3_GPB}) and Section \ref{sec:reversesr3} (Figure \ref{fig.SR3_flip_GPB}}.

\xd{This motivated us to examine the statistical accuracy of the fields of the components accounting for the line wings (see Section \ref{Assessing Uncertainties in Approach II: Gaussian Component Fitting}). The ``Wings'' subset of components was isolated as those for which $\nu_\ell \pm \sigma_\ell$ (Equation \ref{TB_gauss_tauabs}) was most extreme (low and high) for each LOS scanned.}

\bigskip\noindent
Overall, the results in this subsection spotlighting SR3 
\xd{(and other results, some presented in a gallery using the Approach II joint strategy for various noise levels and subregions 
in Appendix~\ref{Zeeman Fitting Gallery: Approach II Across Noise and Regions})} 
indicate that while the joint fitting strategy can improve spectral fitting quality under idealized conditions, neither strategy robustly recovers the true three-dimensional cloud structure in the presence of noise. This underscores the intrinsic limitations of Zeeman analysis for detailed LOS decomposition and highlights the need for additional, complementary constraints beyond Stokes $I$ and $V$ alone.



\begin{figure}[hbt!]
\centering
\includegraphics[width=0.98\linewidth]{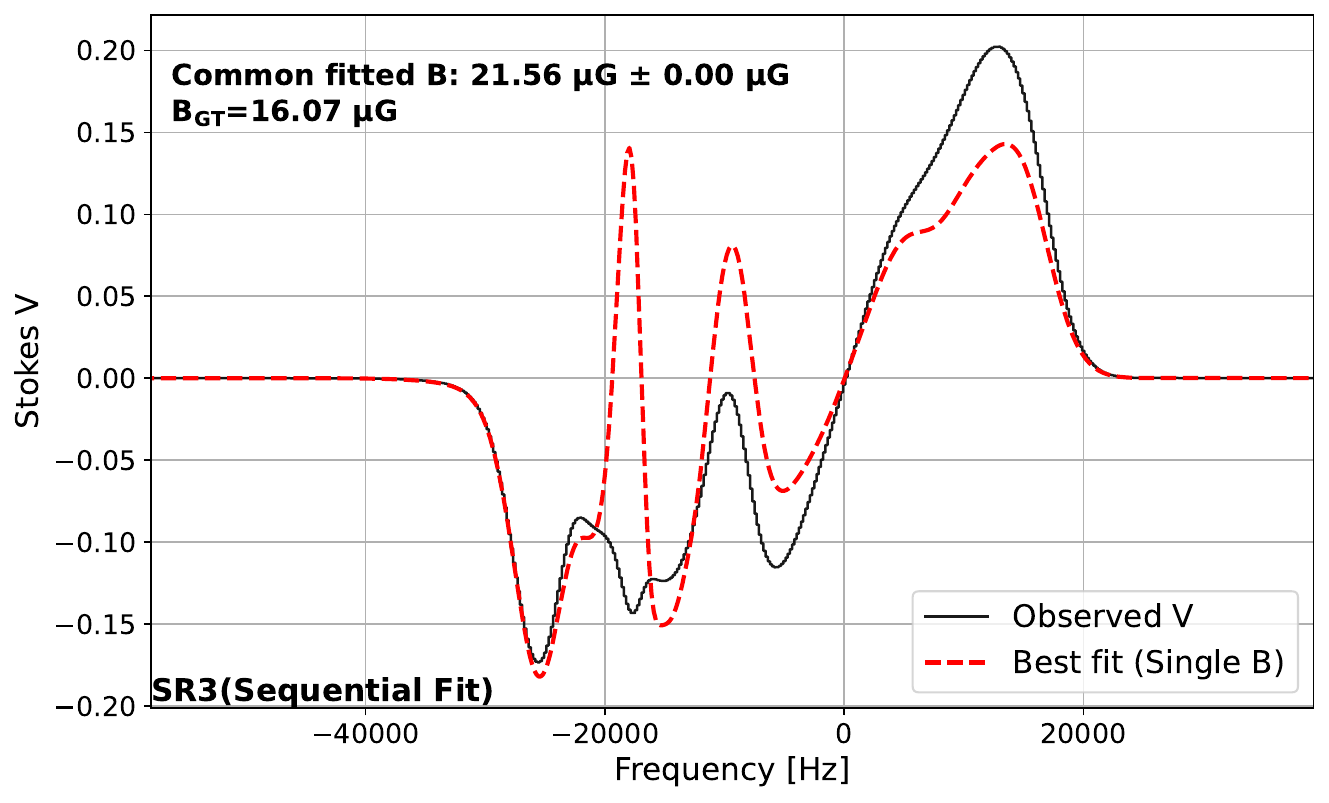}
\caption{\xd{Sequential fitting strategy (Approach II) applied to the SR3 subregion, but enforcing a single common magnetic field for all eight Gaussian components from Figure~\ref{fig.ZeemanFitting_SR3_Breal_noise0}. As expected from Equation \ref{eq:vderiv}, the shape of the model $V$ spectrum is like $dI/\nu$ (see also Figures \ref{fig.SR3_dI_dnu_B10_} and \ref{fig.SR3_I_V_fitting_B10_}). The fitted field strength is virtually identical to that obtained with Approach I in Figure~\ref{fig.ZeemanFitting_dI__SR_Breal_noise0}. }}
\label{fig.SR3_I_V_fitting_AppII}
\end{figure} 


\subsection{Conceptual Relationship Between Approach I and Approach II}
\label{sec:Conceptual Relationship Between Approach I and Approach II}

Both approaches operate in the weak-field limit, where the Zeeman-induced frequency shift is much smaller than the line width, but they differ fundamentally in whether the spectrum is decomposed.

As introduced in Section \ref{Approach I: Fitting Stokes V with dI/dnu}, in Approach I there is no reference to Gaussian components along the line of sight. The line profile of Stokes $I$ from the radiative transfer (observed or simulated) is considered as a whole and assuming that there is a uniform LOS magnetic field $B_z$ then Equation \ref{eq:vderiv} predicts the model Stokes $V$ profile to be that of $dI/d\nu$ scaled in proportion to $B_z$. The computational complexity fitting for $B_z$ is small whatever the spectral complexity of the observed/simulated Stokes $I$ and $V$.

The model in Approach II is an explicit decomposition of Stokes $I$ into multiple Gaussian components (Section \ref{Approach II: Gaussian decomposition}, Equation \ref{TB_gauss_tauabs}). The model of Stokes $V$ is based on the same components (Equation \ref{stokesV_model}), with the field $B_{z,i}$ estimated for individual components by fitting the observed/simulated Stokes $V$ spectrum.  This enables component-level estimates of $B_{z,i}$ along the LOS within the cloud, but increases computational cost as the number of components grows.

The two approaches become mathematically equivalent when Approach II is applied in sequential strategy and in addition a single common magnetic field is enforced across all Gaussian components in the Stokes $V$ model, abandoning the goal of gauging LOS field structure. In this case, the resulting single averaged field estimated is identical to that obtained with Approach I, as demonstrated in Figure~\ref{fig.SR3_I_V_fitting_AppII}.  When a common field is enforced, the model $V$ profile, despite its underlying multi-component complexity incorporating both internal and foreground attenuation, is indeed shaped like $dI/d\nu$ as predicted by Equation \ref{eq:vderiv}.  
A corollary is that when the simulated Stokes spectra to be fit arise from a cube with a uniform field, the correct field value is recovered by both approaches
as expected (Figures \ref{fig.SR3_dI_dnu_B10_} and \ref{fig.SR3_I_V_fitting_B10_}).



\begin{figure}[hbt!]
\centering
\includegraphics[width=0.98\linewidth]{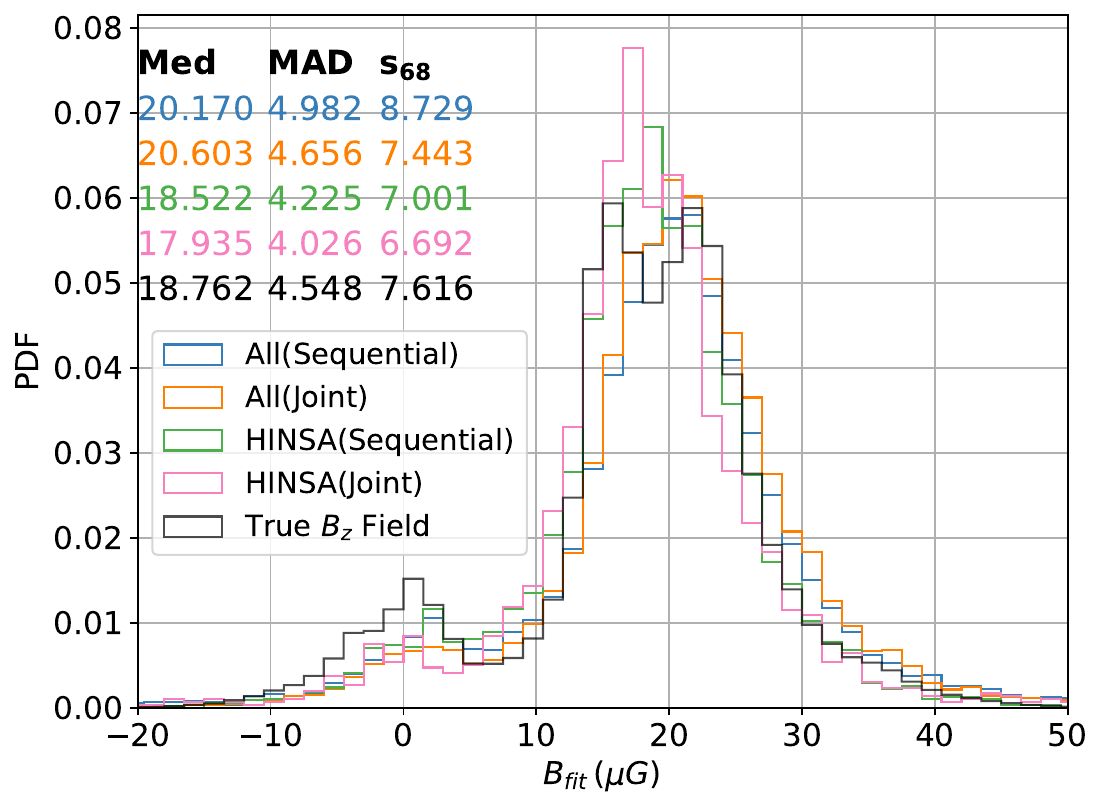}
\caption{PDF of the estimated component magnetic field strength via Zeeman fits using Approach II (both strategies). The statistics in the legend allow intercomparison of the results of different analyses and subsets. Components are from analysis of 4096 independent spectra across the face of the cube, with low noise as in Figure \ref{fig.ZeemanFitting_SR3_Breal_noise01} ($\sigma_{\mathrm{noise}} = 0.014$ K).   Members of the HINSA (HI Narrow Self-Absorption) subset are identified by their narrow velocity dispersion and foreground positioning in the sequence of fitted components (see text).
}
\label{fig.hist_plot_Bfit_all_noise0}
\end{figure}

\begin{figure*}[hbt!]
\centering
\includegraphics[width=0.48\linewidth]{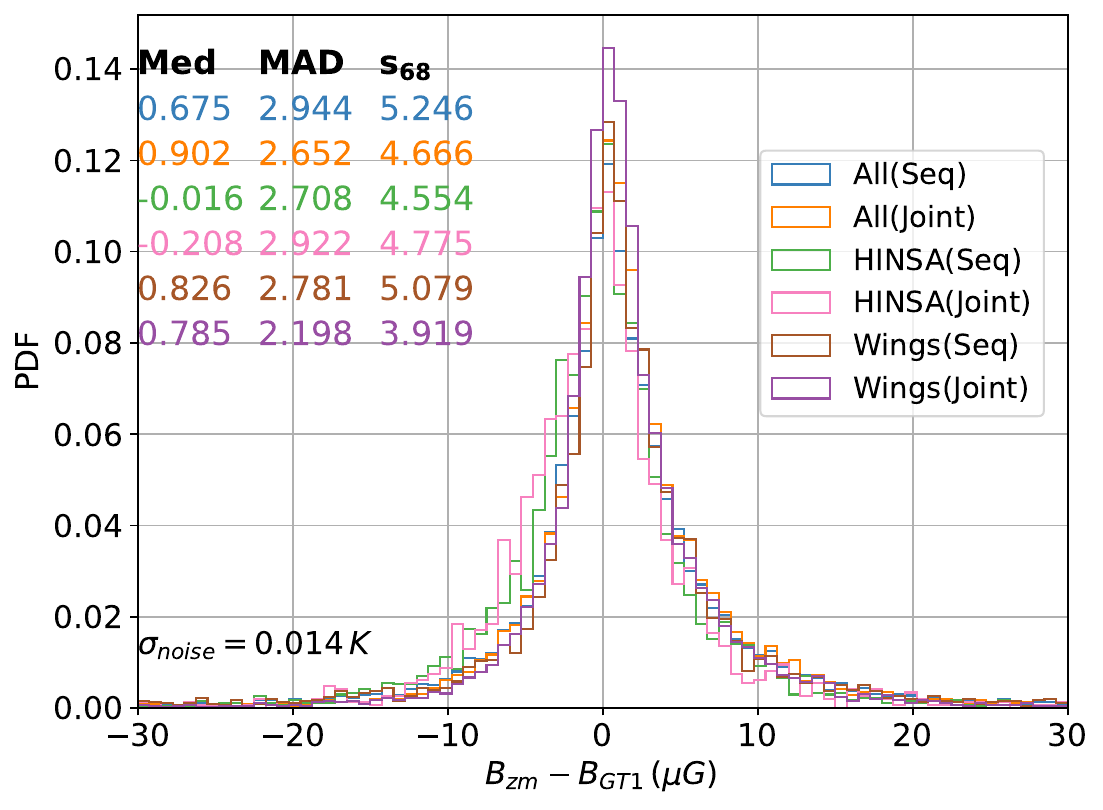}
\includegraphics[width=0.48\linewidth]{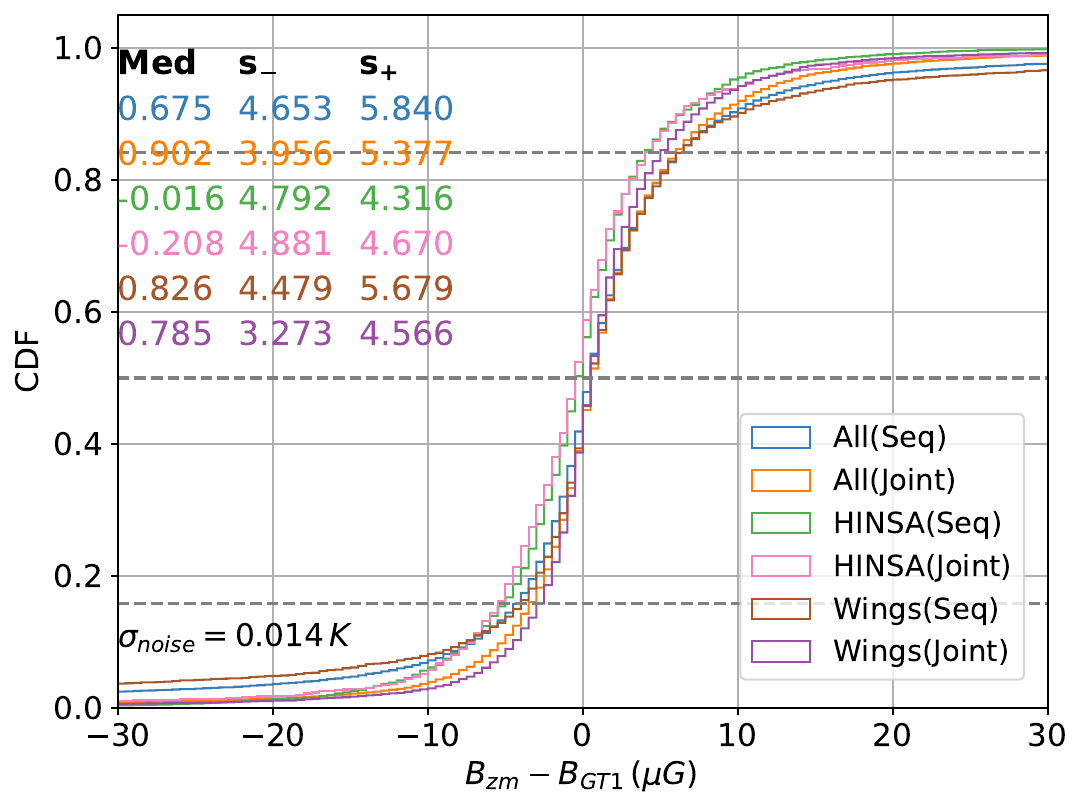}
\includegraphics[width=0.48\linewidth]{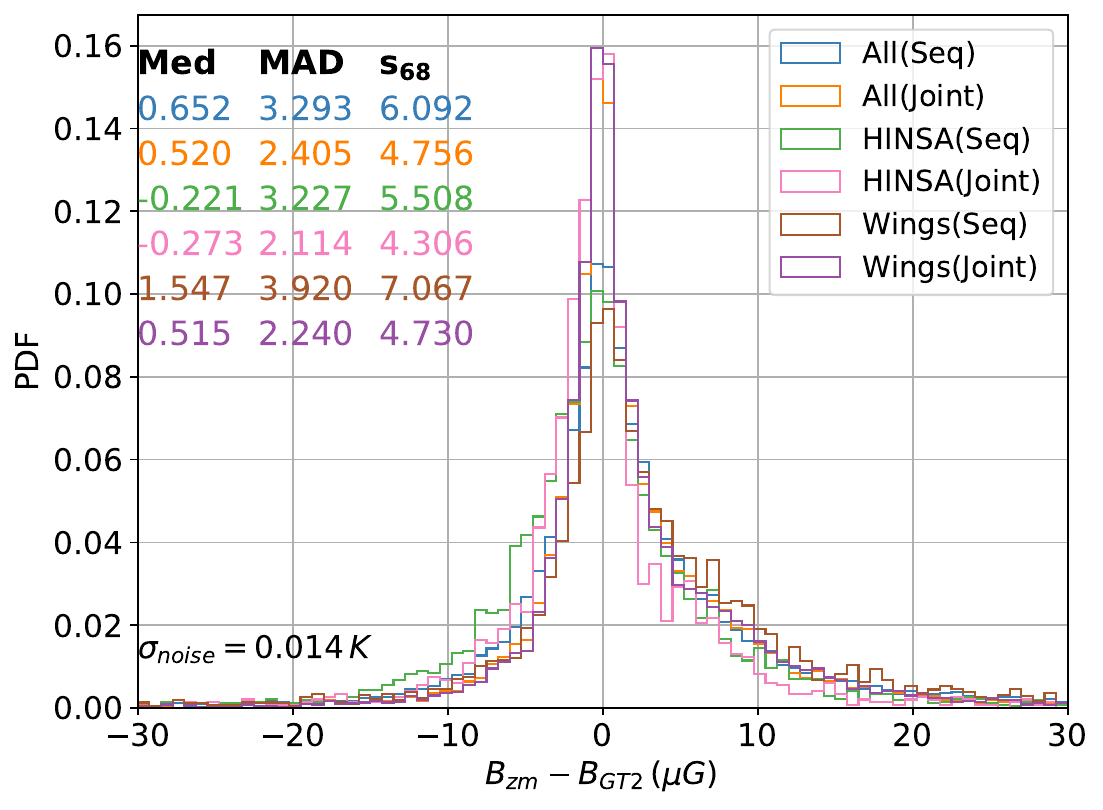}
\includegraphics[width=0.48\linewidth]{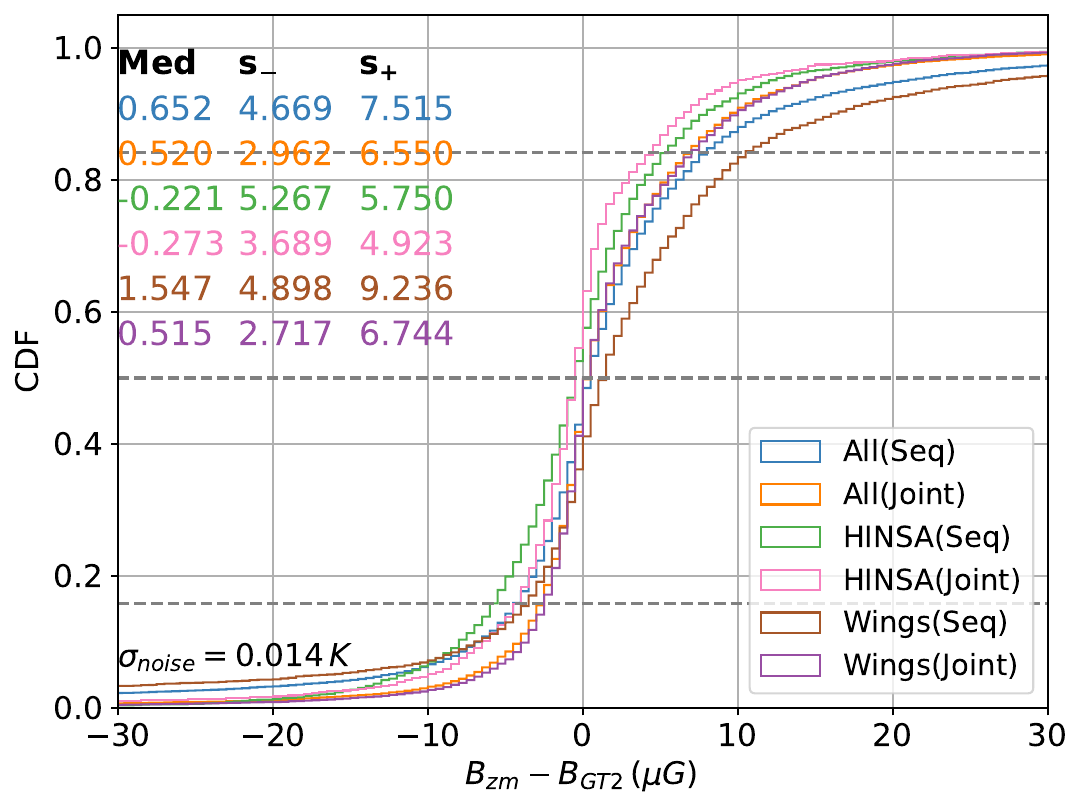}
\caption{PDFs (left column) and cumulative distribution functions (CDFs, right column) of the differences between the Zeeman-inferred magnetic field strengths \xd{for individual fitted components}  (Approach II, using both sequential and joint strategies) and the corresponding ground-truth estimators defined in Equations~\ref{eq.BGT1} (upper panels) and \ref{eq.BGT2} (lower panels). The statistics are computed from components fitted to 4096 independent synthetic spectra across the face of the cube, with low noise as in Figure \ref{fig.ZeemanFitting_SR3_Breal_noise01} ($\sigma_{\mathrm{noise}} = 0.014$ K). In the legend in the PDF panels \xdtwo{we tabulate the statistics.}
For the definitions of the subsets HINSA and Wings, see text.
In the CDF panels, the tabulated values denote the 
\xd{the 50th percentile (the median) 
and increments in x to reach the 16th ($s_-$ downward) and 84th percentiles ($s_+$ upward), the latter two defining the 68.27\% coverage interval. We note the slight asymmetry and that $s_\textrm{68} = (s_+ + s_-)/2.$}
}
\label{fig.hist_plot_Bmag_error_GT1_GT2}
\end{figure*}

\begin{figure}[hbt!]
\centering
\includegraphics[width=0.98\linewidth]{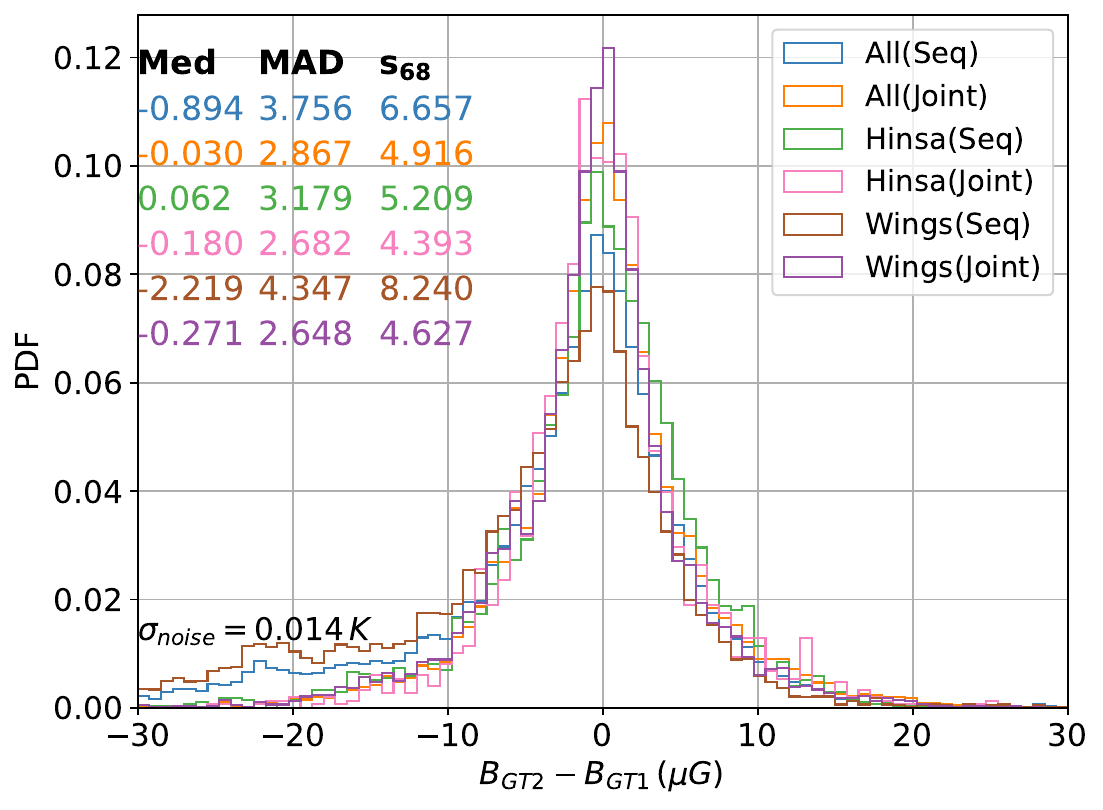}
\caption{PDF of the differences in magnetic field strength between the two ground truth estimators defined by Equations~\ref{eq.BGT2} and \ref{eq.BGT1} for components from fits of 4096 independent synthetic spectra across the face of the cube.}
\label{fig.hist_plot_Bmag_error_GT1_GT2_comp}
\end{figure}

\begin{figure*}
\centering
\includegraphics[width=0.48\linewidth]{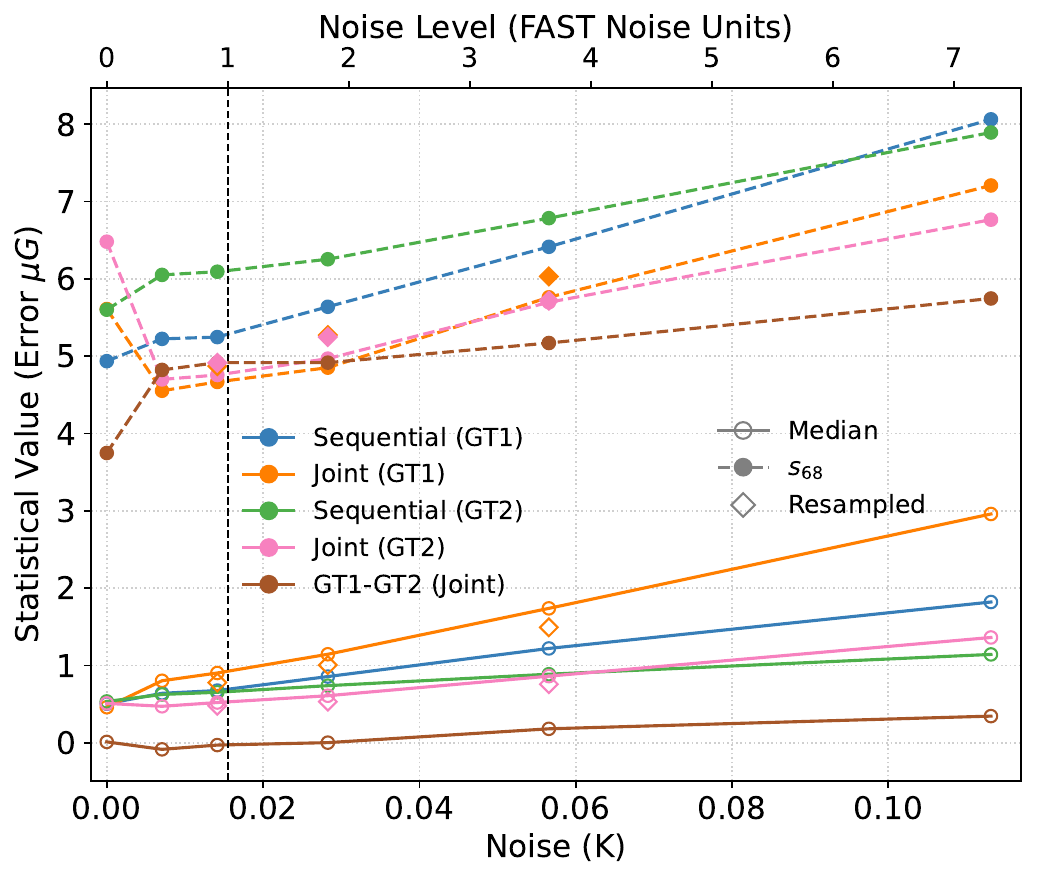}
\includegraphics[width=0.50\linewidth]{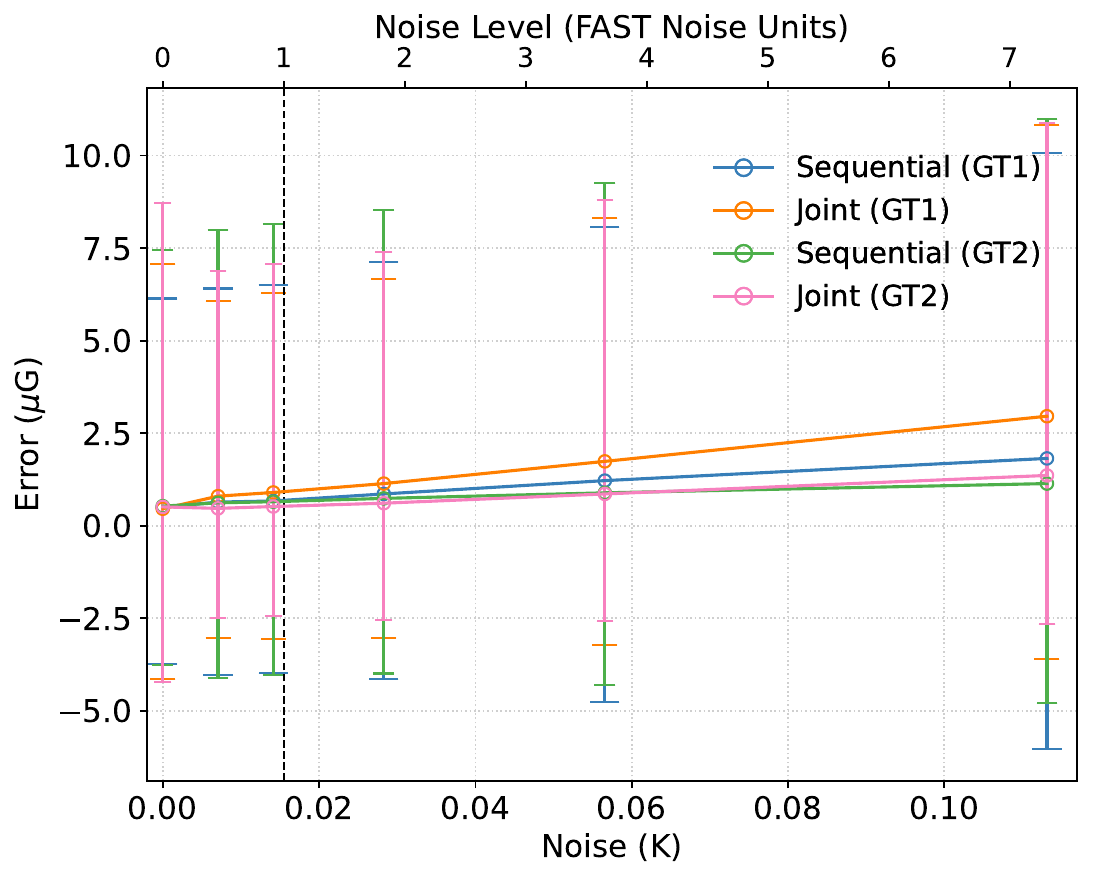}
\caption{Summary of the statistics of absolute differences between the Zeeman-estimated magnetic field strengths (Approach II, using both sequential and joint strategies) and the corresponding ground-truth estimators GT1 and GT2, as well as the difference of the latter, all as a function of noise level. Statistics are computed for individual components from 4096 independent spectra across the cube for all tested noise levels (see Section~\ref{sec:addnoise}). \xdtwo{The left panel shows the median (open symbols) and $s_\textrm{{68}}$ (half-width of the 68.27\% confidence interval, filled symbols).} Diamond markers denote results from resampled and Hanning-smoothed spectra (analogous to FAST data), with the indicated noise levels referring to the pre-resampling values. The right panel shows the median with increments $s_-$ and $s_+$ to reach the 16th and 84th percentiles defining that confidence interval.
}
\label{fig.hist_plot_Bmag_error_GT12_All}
\end{figure*}

\begin{figure}[hbt!]
\centering
\includegraphics[width=0.98\linewidth]{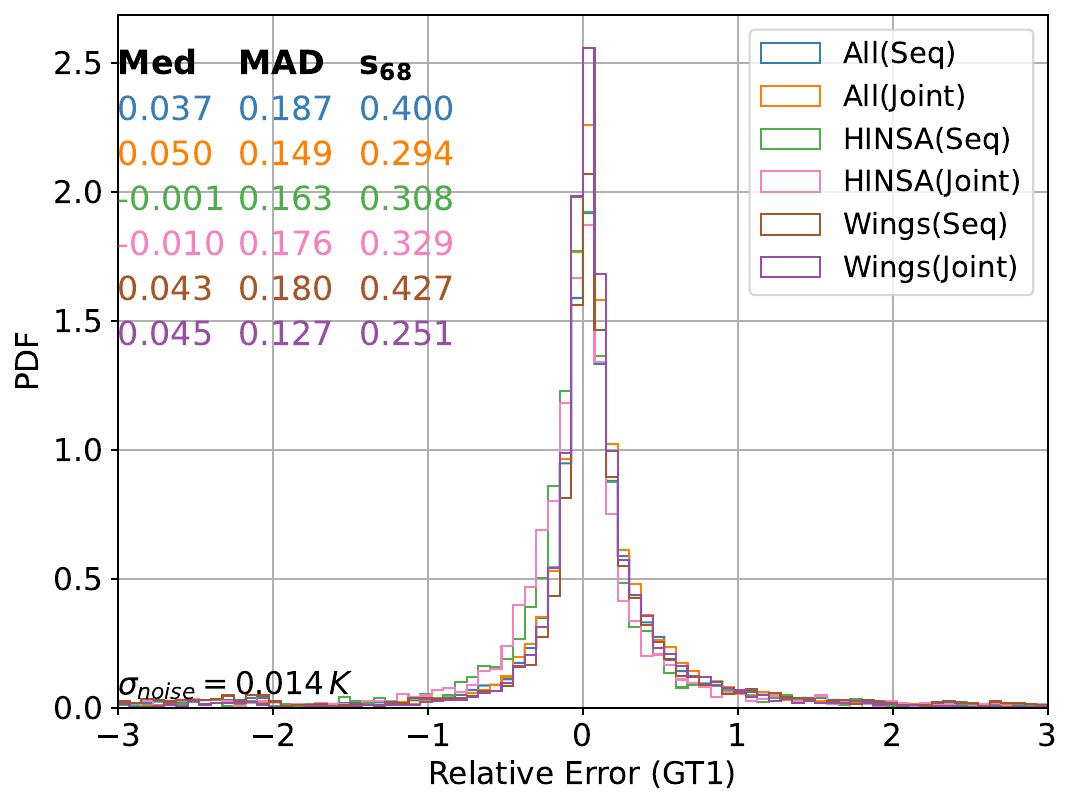}
\includegraphics[width=0.98\linewidth]{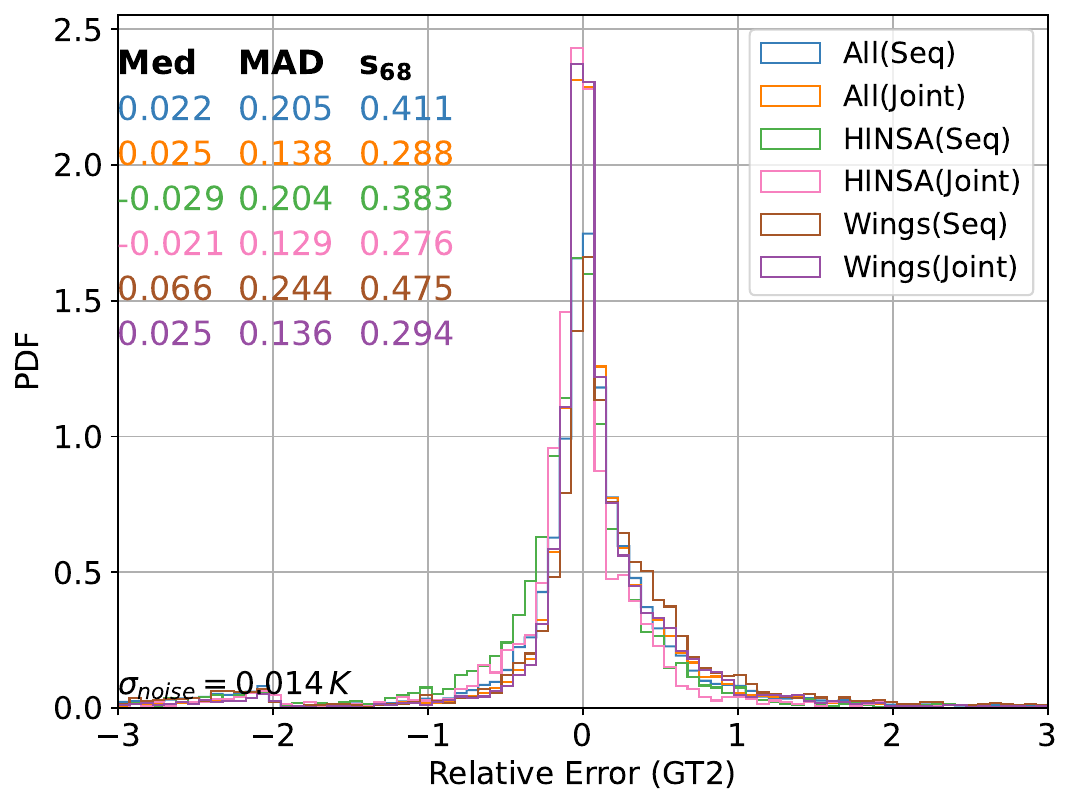}
\caption{PDF of the relative error of differences between the Zeeman-estimated magnetic field strength and the ground truth values defined by Equations~\ref{eq.BGT1} (upper panel) and \ref{eq.BGT2} (lower panel). Statistical measures of the distributions are tabulated for various subsets.
}
\label{fig.hist_plot_Bmag_error_GT1_GT2_relative}
\end{figure}

\begin{figure}[hbt!]
\centering
\includegraphics[width=0.98\linewidth]{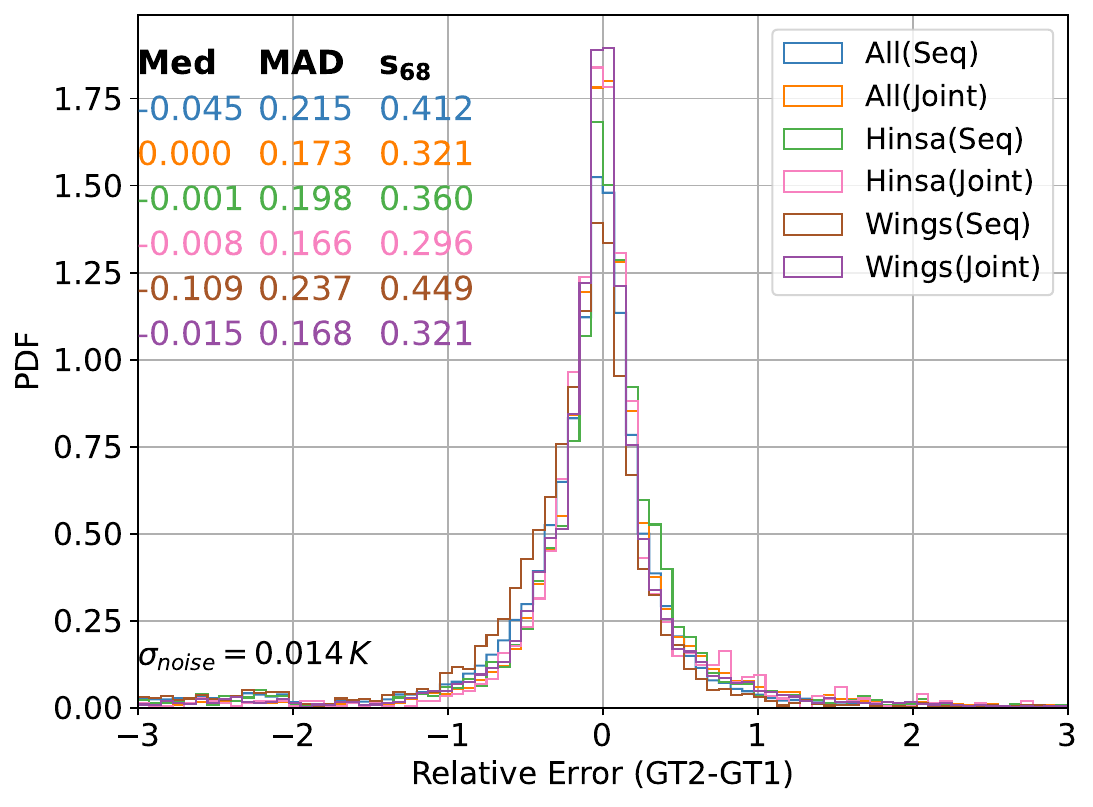}
\caption{Similar to Figure \ref{fig.hist_plot_Bmag_error_GT1_GT2_relative}, but for the PDF of the relative errors of the differences in magnetic field strength between the ground truth values given by Equations~\ref{eq.BGT2} and \ref{eq.BGT1}.}
\label{fig.hist_plot_Bmag_error_GT1_GT2_comp_relative}
\end{figure}

\begin{figure}
\centering
\includegraphics[width=0.99\linewidth]{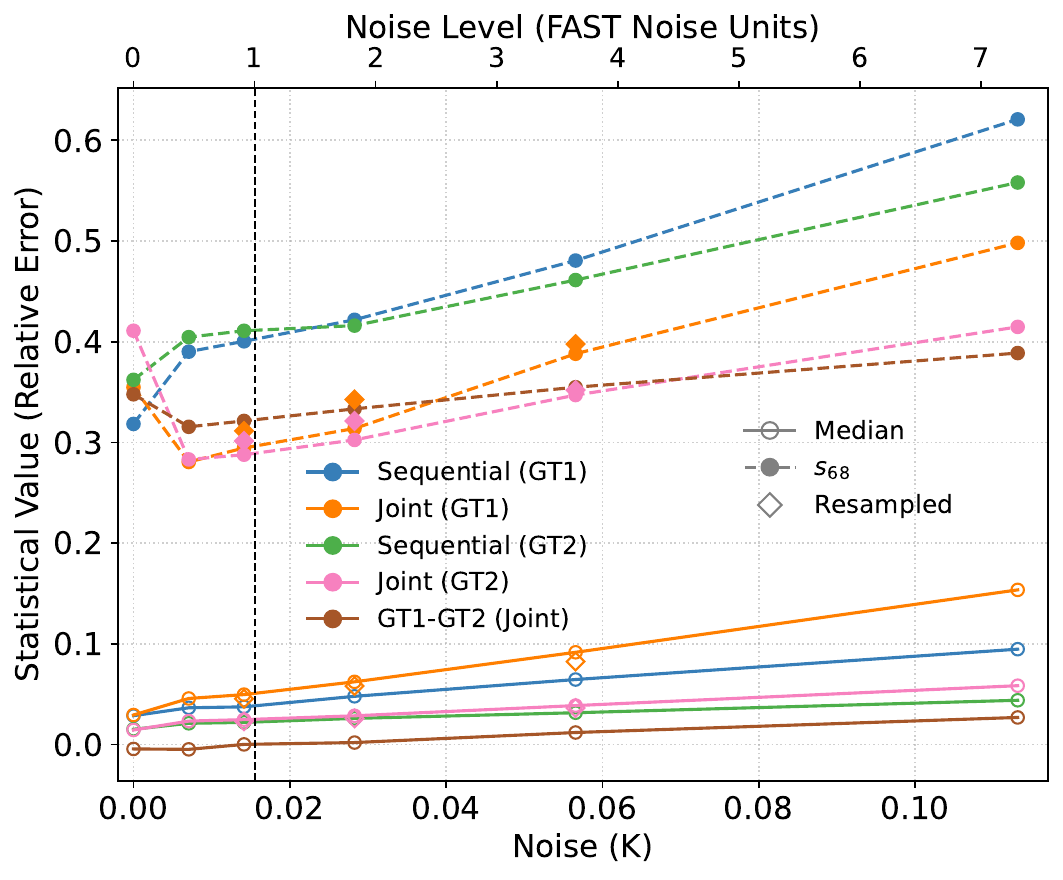}
\caption{Similar to Figure~\ref{fig.hist_plot_Bmag_error_GT12_All} left, but for the \xd{median and half of the 68\% confidence interval ($s_\textrm{{68}}$)} of the relative errors of the difference between the Zeeman-estimated  magnetic field strengths and ground truth estimators.}
\label{fig.hist_plot_Bmag_relative_error_GT12_All}
\end{figure}

\subsection{Assessing Uncertainties in Approach II: Gaussian Component Fitting}
\label{Assessing Uncertainties in Approach II: Gaussian Component Fitting}

We subsequently analyzed the complete set of 4096 synthetic HI spectra using Approach II. It is important to note that during the fitting we did not impose a constraint requiring the velocity dispersion to exceed the thermal broadening implied by the fitted excitation temperature. This is because the \texttt{curve\_fit} function only allows fixed hard bounds, rather than equation-based constraints. 

We tested the alternative fitting package \texttt{LMFIT}, which can enforce such equation-based constraints to ensure that fitted components obey the $\sigma - T_{ex}$ relation. However, when applying this constraint, the fitted components failed to reproduce the overall Stokes $I$ and $V$ profiles, even when using up to 15 Gaussian components. The reason is that the observed spectrum results from the accumulation of contributions from all 512 voxel-level Gaussian components with foreground attenuation. While these voxels inherently satisfy the $\sigma - T_{ex}$ relation, the limited number of components in the fitting makes it difficult for \texttt{LMFIT} to capture the full Stokes $I$ and $V$ profiles under this restriction. When the constraint is removed, the fitting successfully reproduces the spectra, but about 44\% of the fitted components violate the $\sigma - T_{ex}$ relation. This suggests that the fitted Gaussian components do not correspond directly to individual gas parcels in the simulation, but instead represent effective combinations of multiple parcels along the line of sight within certain velocity ranges.

Figure~\ref{fig.hist_plot_Bfit_all_noise0} presents the PDF of the estimated magnetic field strengths $B_{fit}$ by component.
Both the sequential and joint fitting strategies produce similar distributions.  

We also distinguish a subset of Gaussian components that exhibit characteristics similar to HINSA (HI Narrow Self-Absorption). In particular, we specify a narrow velocity dispersion ($< 1$ \kms; \citealt{2003ApJ...585..823L}), $\tau > 0.2$, and foreground positioning in the sequence of fitted components (in the first half). 
The PDF for HINSA fields is similar to that for the entire population.

For comparison, the PDF of actual (true) LOS fields in the voxels of all spectra analysed is also plotted Figure~\ref{fig.hist_plot_Bfit_all_noise0}  (black). 
This PDF agrees well with the PDF of the Zeeman-estimated magnetic field strengths obtained using Approach II under both strategies, including the statistical properties of the HINSA subset. \xdtwo{The overall appearance of the PDFs and the statistics in the legend indicate that despite uncertainties (errors) in individual estimators, $B_{fit}$, these distributions are tracking the same underlying field-strength distribution faithfully.}

\xd{Figure~\ref{fig.hist_plot_Bmag_error_GT1_GT2} shows the PDF of the differences between the Zeeman-fitted magnetic field strengths ($B_{zm} \equiv B_{fit}$) by Gaussian component and estimators $B_{GT1}$ (Equation~\ref{eq.BGT1}, upper left panel) and $B_{GT2}$ (Equation~\ref{eq.BGT2}, lower left panel), with low noise as in Figure \ref{fig.ZeemanFitting_SR3_Breal_noise01} ($\sigma_{\mathrm{noise}} = 0.014$ K). The statistics for the low optical depth case in Appendix \ref{app:appIIstats001} Figure \ref{fig.hist_plot_Bmag_error_GT1_GT2_rho0001} are quite similar.}

The right panels of Figure \ref{fig.hist_plot_Bmag_error_GT1_GT2} show the cumulative distribution functions (CDFs) for the PDFs on the left.  
\xd{This gives a visualization of $s_-$ and $s_+$ as x-intercepts of the CDFs with the two dashed horizontal lines defining the 68.27\% confidence interval relative to the intercept at y = 0.5. Averaging over the slight asyymmetry, we define $s_\textrm{68} = (s_+ + s_-)/2.$}  All calculations of the statistics were  performed on the raw data using percentiles.

As shown in Figure~\ref{fig.hist_plot_Bmag_error_GT1_GT2_comp}, the PDF of differences between the two ground truth estimators $B_{GT1}$ and $B_{GT2}$ reveals a measurable statistical difference with MAD and $s_\textrm{68}$ similar to those found in the other difference distributions in Figure \ref{fig.hist_plot_Bmag_error_GT1_GT2}.

\subsubsection{Effect of noise level}

To assess how varying spectral noise levels influence the uncertainty in Zeeman analysis,  Figure~\ref{fig.hist_plot_Bmag_error_GT12_All} summarizes the median and $s_\textrm{68}$ for the three difference PDFs in Figures \ref{fig.hist_plot_Bmag_error_GT1_GT2} and \ref{fig.hist_plot_Bmag_error_GT1_GT2_comp} for different noise levels other than the 0.014 K used above.
\xdtwo{Overall, despite an eight-fold increase in noise from 0.014 K, $s_\textrm{68}$ grows by less than a factor of 1.5.}


\xd{
The unexpectedly sharp changes in the no-noise joint strategy case likely results from the algorithm fitting additional weak components to match both Stokes $I$ and $V$ more perfectly, leading to poorer recovery of those magnetic fields, while the strong components remain accurate.}

\xd{To evaluate the effect of spectral resolution, we also resampled the spectra from 0.04 km s$^{-1}$ to 0.1 km s$^{-1}$ and applied Hanning smoothing,  matching FAST observations, for three noise levels (0.014, 0.028, 0.057 K). The diamond markers in Figure~\ref{fig.hist_plot_Bmag_error_GT12_All} show the resulting Zeeman analysis uncertainties, plotted at noise levels corresponding to the pre-resampling values. The results demonstrate that velocity resolution and Hanning smoothing have negligible influence on the uncertainty.}

In summary, 
the overall impact remains modest and the Zeeman analysis remains robust across a range of realistic observational noise conditions. 


\subsubsection{Apportioning the uncertainties}
\label{sec:apportion}

The observed scatter in the 
three difference PDFs in Figures \ref{fig.hist_plot_Bmag_error_GT1_GT2} and \ref{fig.hist_plot_Bmag_error_GT1_GT2_comp} arises because each difference inherits uncertainties from the calculation of the three quantities $B_{zm}$, $B_{GT1}$, and $B_{GT2}$ that are each tracking the same underlying truth.

The PDF distributions have non-Gaussian heavy (fat) tails, so that these uncertainties do not add in quadrature as they do for normally-distributed uncertainties. If these are treated as Stable distributions, the $\alpha$ of the characteristic function (the stability index that simultaneously controls the shape of the core and the heaviness of the tail of the distribution) can be estimated from the ratio $s_\textrm{68}$/MAD. This ratio is 1.486 for a Gaussian and 1.837 for a Lorentzian (Cauchy) distribution, corresponding to $\alpha = 2$ and 1, respectively. For typical measured ratios for our three distributions, $\alpha$ is about 1.1. Independent uncertainties for two primary quantities $s_i$ and $s_j$ that are differenced add approximately according to the generalized addition rule: $s_{i\, \textrm{and}\, j}^\alpha = s_i^\alpha + s_j^\alpha$, which is very close to linear for the typical $\alpha$.  Given that the heavy tails are not perfectly described and have some asymmetry, linear combination is adequate (perhaps a slight over estimate), certainly better than in quadrature which would lead to an underestimate.

For our three-node system of 
$B_{zm}$, $B_{GT1}$, and $B_{GT2}$, the uncertainties from the pairwise difference PDFs are known, say 
$s(B_{zm}- B_{GT1})$,
$s(B_{zm}- B_{GT2})$, and
$s(B_{GT1}- B_{GT2})$.
Assuming the linear addition rule, the portion attributed to each individual estimator, $s(B_{zm})$, 
$s(B_{GT1})$, and 
$s(B_{GT2})$
can be computed with simple linear algebra.  To the point, the uncertainty of $B_{zm}$, the estimator of interest, is $s(B_{zm}) = [s(B_{zm}- B_{GT1}) +
s(B_{zm}- B_{GT2}) -
s(B_{GT1}- B_{GT2})]/2.$\footnote{This could be generalized to any $\alpha$ by raising each uncertainty to that power, but is not warranted given our caveats about the PDFs.}

Using the values of $s_\textrm{68}$ subset by subset in Figures \ref{fig.hist_plot_Bmag_error_GT1_GT2} and \ref{fig.hist_plot_Bmag_error_GT1_GT2_comp} (for which the noise level is 0.014 K),
we find $s_\textrm{68}(B_{zm})$ is [2.43, 2.25, 2.43, 2.34, 1.95, 2.01] $\mu$G.
There is marginal evidence that the joint strategy is better than sequential and that the ``Wings'' are better than ``All''.
\xdtwo{As a summary value, we adopt $s_\textrm{68}(B_{zm}) = 2.25\, \mu$G from the All(joint) solution. For the highest noise level, 0.113 K, an eight-fold increase, this summary value is 4.11 $\mu$G, an increase by a more modest factor 1.8.} 

\subsection{Relative errors}
\label{Relative errors}

\xd{We found that if the field in the simulation cube is artificially increased, the absolute errors go up proportionately and so relative errors are more appropriate to report.  Because fields are of order 20 $\mu$G, we can anticipate relative errors of 10 or 15\%.}

More quantitatively, we computed a relative error
defined symmetrically for the $B_{zm}-B_{GT1}$ differences as:
\begin{equation} \label{eqn_rela_error}
\delta_{Bi}=\frac{B_{zm}-B_{GTi}}{\min(|B_{zm}|, |B_{GTi}|)}\, .
\end{equation}
This symmetric formulation offers a more balanced metric than the classical relative error that uses $B_{GTi}$ in the denominator. 

In Figure~\ref{fig.hist_plot_Bmag_error_GT1_GT2_relative} we show the PDFs
of relative errors $\delta_{B1}$ and $\delta_{B2}$, corresponding to the analysis for subsets in Figure \ref{fig.hist_plot_Bmag_error_GT1_GT2} for which the noise level is 0.014 K.
Figure \ref{fig.hist_plot_Bmag_error_GT1_GT2_comp_relative} shows the relative errors corresponding to 
Figure \ref{fig.hist_plot_Bmag_error_GT1_GT2_comp} for the same noise level.  
In the inset panels we report the median, MAD, and $s_\textrm{68}$ for each. 

In Figure \ref{fig.hist_plot_Bmag_relative_error_GT12_All} we show for the entire set (``All'') how $s_\textrm{68}$ of the relative differences increases modestly as the noise level in the synthetic spectra increases.
\xd{All three PDFs are non-Gaussian and heavy tailed and we again adopt the linear addition rule to apportion the errors.
For the 0.014 K noise level, the value of $s_\textrm{68}$ attributable to the distribution of relative errors for $B_{zm}$ for the six subsets is
[0.200, 0.131, 0.166, 0.155,    0.227, 0.112].
%
Because we favor the joint strategy analysis, the summary number reported is 0.13 (13\%) for the entire set, All(Joint).  HINSA(Joint) and Wings(Joint) are consistent with this (0.15, 0.12).
While we argue that this is a reasonable estimate, the caveats that the distributions are not precisely Lorentzian and are asymmetric do suggest some caution.}

\xdtwo{Even under an eight-fold rise in noise relative to 0.014 K, the summary value of $s_\textrm{68}$ grows to only $\sim$26\% relative error, up by a factor 2.}

\section{Re-examining Taurus HI Data from FAST}
\label{Re-examining HI Data from FAST}

\begin{figure}[hbt!]
\centering
\includegraphics[width=0.98\linewidth]{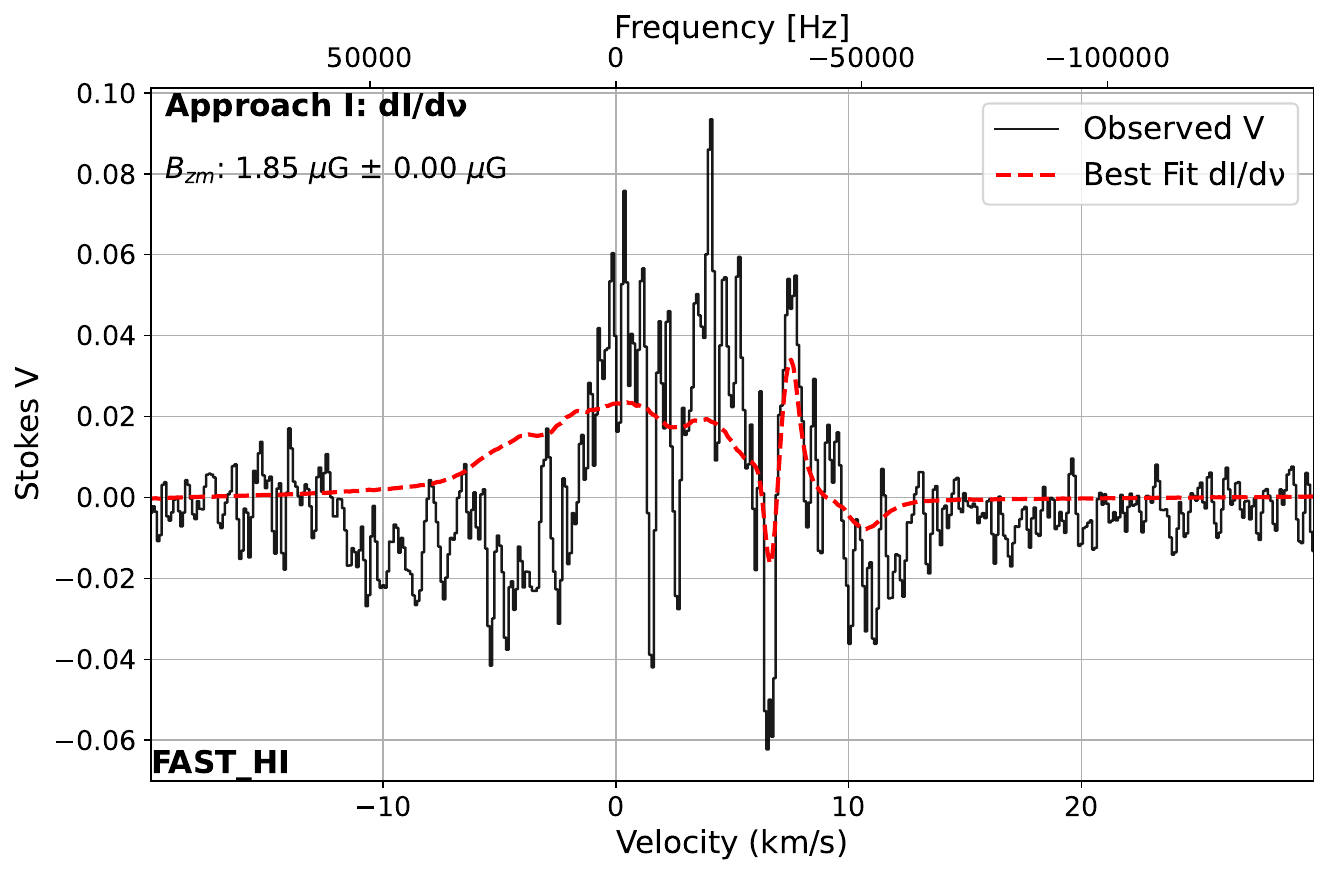}
\caption{Same as Figure~\ref{fig.ZeemanFitting_dI_dnu_5_layers}, \xd{but showing Approach I fitting result of the Stokes $V$ spectrum and the differentiated Stokes $I$ spectrum from} real FAST HI observations, as presented in \citet{2022Natur.601...49C}. }
\label{fig.FAST_HI_dI_dnu}
\end{figure}

\begin{figure*}[hbt!]
\centering
\includegraphics[width=0.98\linewidth]{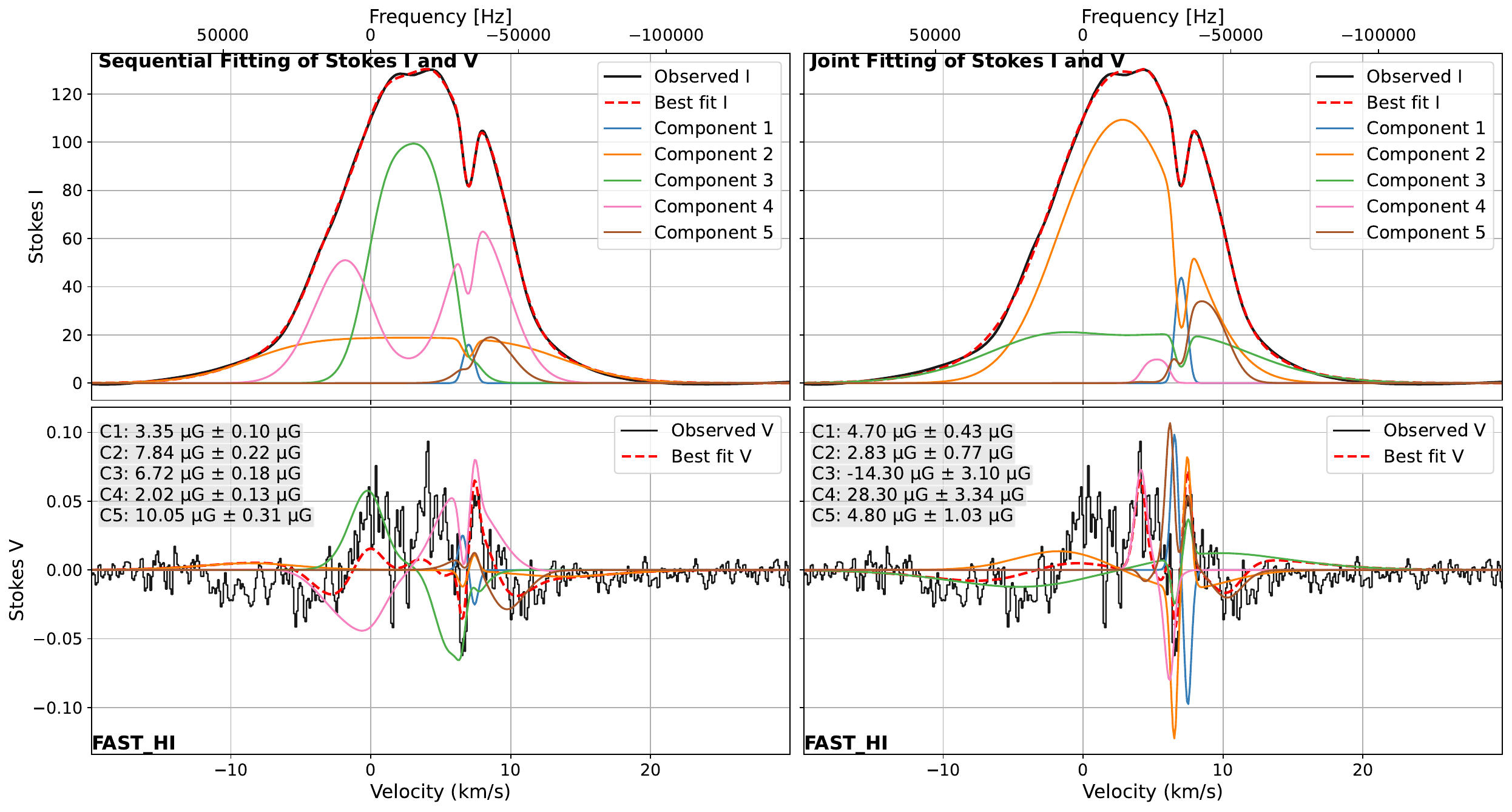}
\caption{\xd{Same as Figure~\ref{fig.ZeemanFitting_5_Layer_IV_noise0}, but showing Approach II fitting results for the Stokes $I$ and Stokes $V$ spectra from real FAST HI observations \citep{2022Natur.601...49C}. Sequential results are shown on the left and joint results on the right.}}
\label{fig.ZeemanFitting_FAST_HI_I_V_fitting}
\end{figure*}

Arecibo OH Zeeman observations of the L1544 (dark starless) dense core within the extended Taurus molecular cloud complex by \citet{Crutcher_2000} revealed a LOS field of $11\pm 2\, \mu$G. Using the FAST telescope, \citet{2022Natur.601...49C} targeted this core to investigate magnetic fields via the Zeeman effect in HI. They reported a magnetic field strength of $3.8\pm0.3\, \mu$G for the narrow HI absorption (HINSA) component. They modeled the HINSA as arising from cold dense HI gas mixed in trace amounts with predominantly molecular hydrogen, by association with narrow-line CO emission, foreground to HI emission from cold neutral medium (CNM) and warm neutral medium (WNM) components further along the LOS.  Barring an explanation in terms of projection, finding a low field associated with cold condensed gas is remarkable. Are fields not compressed?

We adopted their $I$ and $V$ spectra available online and applied our own Zeeman fitting approaches to this real data. As shown in Figure~\ref{fig.FAST_HI_dI_dnu}, Approach I indicates a weak LOS field of $1.85\, \mu$G. However, this averaging approach could result in cancellation effects due to field reversals along the line of sight.  The HINSA component is not well fit and is suggestive of a somewhat higher (but still low) field.

Figure~\ref{fig.ZeemanFitting_FAST_HI_I_V_fitting} presents results from our component-based Approach II, where the spatial ordering of components is not assumed but instead determined through fitting. We find a HINSA absorption component in the foreground and our fits yield values of 3.35 and $4.7\, \mu$G with the sequential and joint fitting strategies, respectively. These results support the presence of a weak LOS magnetic field in the HINSA.

\xd{As discussed in Section \ref{Relative errors}, component-based Zeeman fitting inherently carries a relative uncertainty of 
about 15\%, 
including for HINSA features. This is based on $s_\textrm{68}$ and so a larger error can occur in the heavy wings of the distribution. But it seems there is not a large likelihood that the LOS magnetic field strength is as large as that measured in OH.}

Furthermore, because the orientation with respect to the line of sight is unknown, a low LOS field might not necessarily demand a weak field in 3D. An extensive study by \citet{2022MNRAS.510.6085L} of the plane of sky (POS) magnetic field in a different filamentary Taurus subregion L1495/B211 finds values of 10's of $\mu$G.  However, that subregion is about 10\deg\ to the west of the FAST target L1544 core.
Analysis by \citet{2015MNRAS.452.2500L} finds Zeeman-estimated LOS fields in molecular clumps to be 10's of $\mu$G.
Perhaps the orientation of the field in this FAST target is just unfavorable.

A broader, systematic survey of dense cores with FAST would help to better constrain typical LOS magnetic field strengths revealed by HI. Structured line profiles, as with HINSA, would be beneficial and could be pre-selected without the investment in Zeeman measurements.
Note that strong fields might be detected by the Zeeman effect with much less integration time than invested in this particular observation by \citet{2022Natur.601...49C}, making a systematic survey a tractable proposition.

\section{Summary, Conclusions, and Discussion}
\label{Conclusions}

\begin{deluxetable*}{c|c|cc}
\label{tab.fitting_approach}       
\tablecaption{Comparison of Different Fitting Approaches }
\tablehead{  & \multicolumn{1}{c|}{\multirow{3}{3.8cm}{\textbf{Approach I (Derivative Method)}}}  & \multicolumn{2}{c}{\textbf{Approach II (Gaussian Decomposition)}} \\ 
 & & \multirow{3}{3.8cm}{\textbf{Sequential Strategy}} &\multirow{3}{3.8cm}{\textbf{Joint Strategy}}\\
&&&
  } 
\startdata
 \multirow{3}{3.3cm}{\textbf{Primary Goal}} &  \multirow{3}{3.3cm}{LOS-averaged field estimates} &  \multirow{3}{3.3cm}{Component-level field estimates} &  \multirow{3}{3.3cm}{Component-level field estimates and capture LOS structure }\\
 &&& \\
 &&& \\
 \multirow{3}{3.3cm}{\textbf{Bias \& Scatter}} & \multirow{3}{3.3cm}{Minimal Bias, Minimal Scatter }& \multirow{3}{3.3cm}{Larger Scatter, Larger Uncertainty }& \multirow{3}{3.3cm}{Larger Scatter, Larger Uncertainty}\\
 &&& \\
 \multirow{3}{3.3cm}{\textbf{Relative Error}} &  \multirow{3}{3.3cm}{$\sim$16\% {$\downarrow^{a}$}}& \multirow{3}{3.3cm}{$\sim$20\%} & \multirow{3}{3.3cm}{$\sim$13\%} \\
 &&& \\
 \multirow{3}{3.3cm}{\textbf{Sensitivity to Noise}} &  \multirow{3}{3.3cm}{Low} &  \multirow{3}{3.3cm}{Modest} &  \multirow{3}{3.3cm}{Low, but convergence is susceptible to initial conditions} \\
 &&& \\
 \multirow{3}{3.3cm}{\textbf{Performance in Uniform Fields (B10)}} &  \multirow{3}{3.3cm}{\textbf{$\checkmark$ }}&  \multirow{3}{3.3cm}{\textbf{$\checkmark$}} &  \multirow{3}{3.3cm}{\textbf{$\checkmark$}} \\
 &&& \\
 &&& \\
 \multirow{3}{3.3cm}{\textbf{Performance in Two Uniform Fields (B-10/10)}} &   \multirow{3}{3.3cm}{\textbf{$\times$}} &  \multirow{3}{3.3cm}{\textbf{$\checkmark$}} &  \multirow{3}{3.3cm}{\textbf{$\checkmark$}} \\
 &&& \\
 &&& \\
 \multirow{3}{3.3cm}{\textbf{Performance in Gradient Fields (Bgrad)}} &  \multirow{3}{3.3cm}{\textbf{$\checkmark$}} &  \multirow{3}{3.3cm}{\textbf{$\checkmark$}} &  \multirow{3}{3.3cm}{\textbf{$\checkmark$}} \\
 &&& \\
 &&& \\
 \multirow{3}{3.3cm}{\textbf{Performance in Low Density ($\mathbf{\rho_0=0.001}$)}} &  \multirow{3}{3.3cm}{\textbf{$\checkmark$}} &  \multirow{3}{3.3cm}{\textbf{$\checkmark$}} &  \multirow{3}{3.3cm}{\textbf{$\checkmark$}} \\
 &&& \\
 &&& \\
 \multirow{3}{3.3cm}{\textbf{Physical Interpretation}} &  \multirow{3}{3.3cm}{Provides a reliable mean-field estimate} &  \multirow{3}{3.3cm}{Provides statistical distributions of fields} &  \multirow{3}{3.3cm}{Clear, fit-dependent structural recovery} \\
 &&& \\
 &&& \\
 \multirow{3}{3.3cm}{\textbf{Strengths}} &  \multirow{3}{3.3cm}{Robust mean-field measurement; minimal bias} &  \multirow{3}{3.3cm}{Component-level insight; statistical distributions} &  \multirow{3}{3.3cm}{Better agreement in Stokes $V$; marginally lower uncertainty} \\
 &&& \\
 &&& \\
 \multirow{3}{3.3cm}{\textbf{Limitations}} &  \multirow{3}{3.3cm}{Cannot resolve LOS variations} &  \multirow{3}{3.3cm}{Larger uncertainty, larger scatter} &  \multirow{3}{3.3cm}{Susceptible to fitting inaccuracies; computationally intensive}\\ 
 &&& \\
 &&& \\ 
 &&& \\ 
\enddata
\tablenotetext{}{Note:}
 \tablenotetext{}{$^a$ This should be regarded as an upper limit, since we cannot disentangle or apportion the individual contributions to the relative error from the two estimators, $B_{zm}$ and $B_{GT}$.}
\end{deluxetable*}

We created synthetic HI Zeeman spectral observations using both MHD simulations and simplified cloud models and developed Zeeman analyses to assess the uncertainties in deriving the line of sight (LOS) magnetic field strength.

Our Approach I is an adaptation of the ``classical'' result that the observed Stokes $V$ should be proportional to the (numerical) derivative of the observed Stokes $I$ in idealized conditions, with the relative scale providing an estimate of the LOS field.

In Approach II, we apply Gaussian decomposition to Stokes $I$ (and $V$).
The traditional strategy is to decompose Stokes $I$ and then estimate fields for individual components by optimizing the fit to the Stokes $V$ spectrum. We found that often the result of this ``sequential'' strategy was a poor fit to $V$. It was obvious that $V$ spectra could provide valuable information on Gaussian component properties beyond their LOS field, and so we innovated a ``joint'' strategy fitting Stokes $I$ and $V$ simultaneously. 

Our key findings are summarized as follows:
\begin{enumerate}

\item Both Approach I and Approach II reliably recover the ground truth magnetic field strength in scenarios with a uniform magnetic field, even under realistic density, temperature, and velocity conditions, confirming a fundamental robustness of these methods.


\item \xdtwo{In the realistic MHD simulation with complex magnetic field structures and turbulence,} Approach I recovers the simple mass-weighted ground-truth estimator of the LOS magnetic field strength well: the difference has no significant systematic bias (about 1.2 $\mu$G) and a half-width of the 68\% confidence interval $s_\textrm{{68}}$ of about 2.3 $\mu$G. However, Approach I  provides no information on the magnetic field structure along the line of sight.

\item From Approach II, the probability distribution function (PDF) of estimated LOS magnetic fields by component is a good match to the PDF of actual LOS fields in the voxels of all spectra analysed across the face of the simulation cube.

\item Approach II can recover the ground-truth estimators of the LOS magnetic field for individual components without significant systematic bias (less than a $\mu$G). \xd{For the difference PDFs for the joint fitting strategy at noise levels comparable to FAST observations, $s_\textrm{{68}}$ is about 
4.7~$\mu$G and about 29\% as a relative error.
When uncertainties are apportioned between contributions from the two ground-truth estimators and the Zeeman $B_{zm}$ estimates for components, $s_\textrm{{68}}$ for the latter is 2.3~$\mu$G and 13\%.
The error distribution exhibits a heavy tail such that rare but more significant deviations can occur, with typical 
99.7\%-confidence-interval errors (akin to $3\sigma$ confidence) being substantially larger.}

\item \xdtwo{For the joint fitting strategy, an eight-fold increase in noise results in a more modest rise in the uncertainty apportioned to $B_{zm}$, to 4.11 $\mu$G and 26\% (factors 1.8 and 2, respectively).}
Thus the overall impact remains limited, and the Zeeman analysis stays robust across a broad range of realistic observational noise conditions. 

\item The optical depth has minimal impact on the above-mentioned statistical uncertainties of magnetic field estimates in both Approach I and Approach II.

\item \xd{Joint fitting can improve Zeeman analysis spectral fits under idealized, noise-free conditions, but this degrades 
once realistic noise and spectral blending are present, emphasizing the need for complementary observational or physical constraints beyond Stokes $I$ and $V$ to determine the true 3D magnetic field structure.}


\item We applied both Approach I and Approach II to the actual HI Zeeman observations from FAST, originally analyzed by \citet{2022Natur.601...49C}. Our fitting results are consistent with the magnetic field strength of the narrow HI absorption (HINSA) component reported in their study.

\end{enumerate}



We compare the different fitting approaches and summarize their performance, strengths, and limitations in Table~\ref{tab.fitting_approach}.




Looking ahead, the unprecedented sensitivity of FAST and the Square Kilometre Array (SKA) will make it possible to carry out large-scale, detailed magnetic field mapping using Zeeman HI observations, and the insights gained from this work will help guide the interpretation of those future results. 

\begin{acknowledgements}
We thank the referee for many questions and suggestions that have led to improvements in this paper.
D.X. acknowledges the support of the Natural Sciences and Engineering Research Council of Canada (NSERC), [funding reference number 568580]. D.X. also acknowledges support from the Eric and Wendy Schmidt AI in Science Postdoctoral Fellowship Program, a program of Schmidt Sciences. JSS is supported by NSERC Discovery Grant RGPIN-2023-04849 and a University of Toronto Connaught New Researcher Award. The Dunlap Institute is funded through an endowment established by the David Dunlap family and the University of Toronto. The author acknowledges the use of open-source software essential for this work, including NumPy and SciPy for data analysis and numerical operations, and the SciPy \texttt{curve\_fit} function and \texttt{LMFIT} library for model fitting and optimization.

\end{acknowledgements}

\clearpage
\appendix
\counterwithin{table}{section}
\counterwithin{figure}{section} 

\section{Approximate Equations for RCP and LCP for the Case of Small Fields}
\label{app:smallfields}

Stokes $I$ and $V$ are calculated as $I = (I_+ +
 I_-)/2$ and $V = (I_+ - I_-)/2$ from
the general fundamental equations for the intensities detected in right and left handed circular polarization ($\pm$ for RCP and LCP, respectively), which are \citep{1990ApJS...74..437S}
\begin{equation}
I_\pm = \frac{1}{4} \big[(\cos(\theta)\pm 1)^2 I_0(\nu +\delta \nu)\, +
(\cos(\theta)\mp 1)^2 I_0(\nu -\delta \nu)\, + 2 \sin^2(\theta)I_0(\nu)
\big]\, ,
\label{eq:sault2p1}  
\end{equation}
where $\cos(\theta)$ is explicitly for the projection of the magnetic field on the LOS, $\delta \nu = \mu_b B/h$ is the (unprojected) frequency shift in the presence of a magnetic field of amplitude $B$, and from the defintion $I_0(\nu)$ is the $I$ profile when $\delta \nu = 0$.  

For illustrative purposes, we take the line profile to be Gaussian:\footnote{Here $\nu$ can be thought of as the frequency relative to the line center.}
\begin{equation}
 I_0(\nu) = \exp\big(-(\nu^2/(2 \sigma^2)\big)\, ,   
\label{simpleGaussian}
\end{equation}
where $\sigma$ is the dispersion accounting for thermal motions and a non-thermal component from turbulence, in quadrature.

To evaluate Equation \ref{eq:sault2p1}, whence $I$ and $V$, $\cos(\theta)$ and $\delta \nu$ need to be known separately, as they are for simulations. However, for analysis of observations, further development is needed based on the small-shift (weak-field) approximation with ratio $\delta \nu/\sigma << 1$.\footnote{How small? Numerical evaluations of the equations below indicate that a ratio 0.02 is adequate and 0.01 remarkably good.  For conditions encountered in the simulation (e.g., Fig. \ref{HI gas}), $B$ is no more than 100 $\mu$G, making $\delta \nu$ smaller than 140 Hz. The (thermal plus non-thermal) dispersion of an HI component is likely to be larger than 0.5 km s$^{-1}$ (see discussion accompanying Eq. \ref{eq:sig_th})
, so that $\sigma$ is larger than 2.35\, kHz. From these rather extreme estimates, $\delta \nu/\sigma < 0.06$, which is only marginally adequate especially for Eq. \ref{eq:vderiv}.  However, the latter is used only in Approach I for the entire spectral line, which is much broader than the above extreme and also the ``averaged'' field for all of the gas would be lower too.}

In this approximation, the profiles $I_\pm$ are very close in shape to $I_0$ but shifted slightly by $\mp \cos(\theta) \delta\nu \equiv \mp \Delta \nu = \mp \mu_b \cos(\theta) B/h$,\footnote{The size of the shift is readily found analytically solving $dI_\pm/d\nu = 0$ in this small-shift approximation.} in which the product $\cos(\theta) B = B_z$ is the relevant quantity.
Thus, $I_\pm \simeq I_0(\nu \pm \Delta\nu)$ and $I(\nu) \simeq [I_0(\nu + \Delta\nu) + I_0(\nu-\Delta\nu)]/2 \simeq I_0(\nu)$.
For our illustrative simple Gaussian profile, in this small-shift approximation
\begin{equation}
I_\pm \simeq \exp\big(-(\nu \pm \Delta \nu)^2/(2 \sigma^2)\big)\, ,
\label{eq:approxIpm}
\end{equation}
and again Stokes $\I \simeq I_0$.

Generally Stokes $V = (I_+ - I_-)/2$.  There are many ways to compute $V$.  (i) V can be computed directly the full expressions of $I_\pm$ in Eq. \ref{eq:sault2p1}. (ii) The definition can also be reduced analytically, without the small-shift approximation, to \citep{1990ApJS...74..437S}
\begin{equation}
V = \frac{1}{2} \cos(\theta) \big[I_0(\nu +\delta \nu)\, - I_0(\nu -\delta \nu)\big]\,  ,
\label{eq:sault2p2}  
\end{equation}
again, as in Equation \ref{eq:sault2p1}, requiring $cos(\theta)$ and $\delta \nu$ independently.  

But when $\delta \nu/\sigma  << 1$, then (iii)
\begin{equation}
    V = \cos(\theta) \delta \nu\, dI_0/d\nu = \Delta \nu\, dI_0/d\nu\, ,
\label{eq:vderiv}
\end{equation}
wherein it is the product $B_z$ that is important to determining the amplitude. 

Alternatively, it can be convenient for modeling to (iv) calculate $V = (I_+ - I_-)/2$ directly from the illustrative simple Gaussians in Eq. \ref{eq:approxIpm}. (For simple Gaussians, Eq. \ref{eq:vderiv} is also readily derived in the small shift limit.)

Numerical computations of $V$ using all four of these approaches confirm that they are closely equivalent in the small-shift approximation. Note that Eq. \ref{eq:vderiv} is in accord with \citet{2017A&A...601A..90B} but apparently a factor 2 smaller than some previous formulations \citep[e.g.,][]{2019FrASS...6...66C}.  Our equation is as required for consistent definitions of Stokes $I$ and $V$ from $I_\pm$.

\section{Ground Truth 2: Magnetic Field Weighted by Spectral Similarity}
\label{Ground Truth II: Magnetic Field Weighted by Spectral Similarity}

\begin{figure}[hbt!]
\centering
\includegraphics[width=0.48\linewidth]{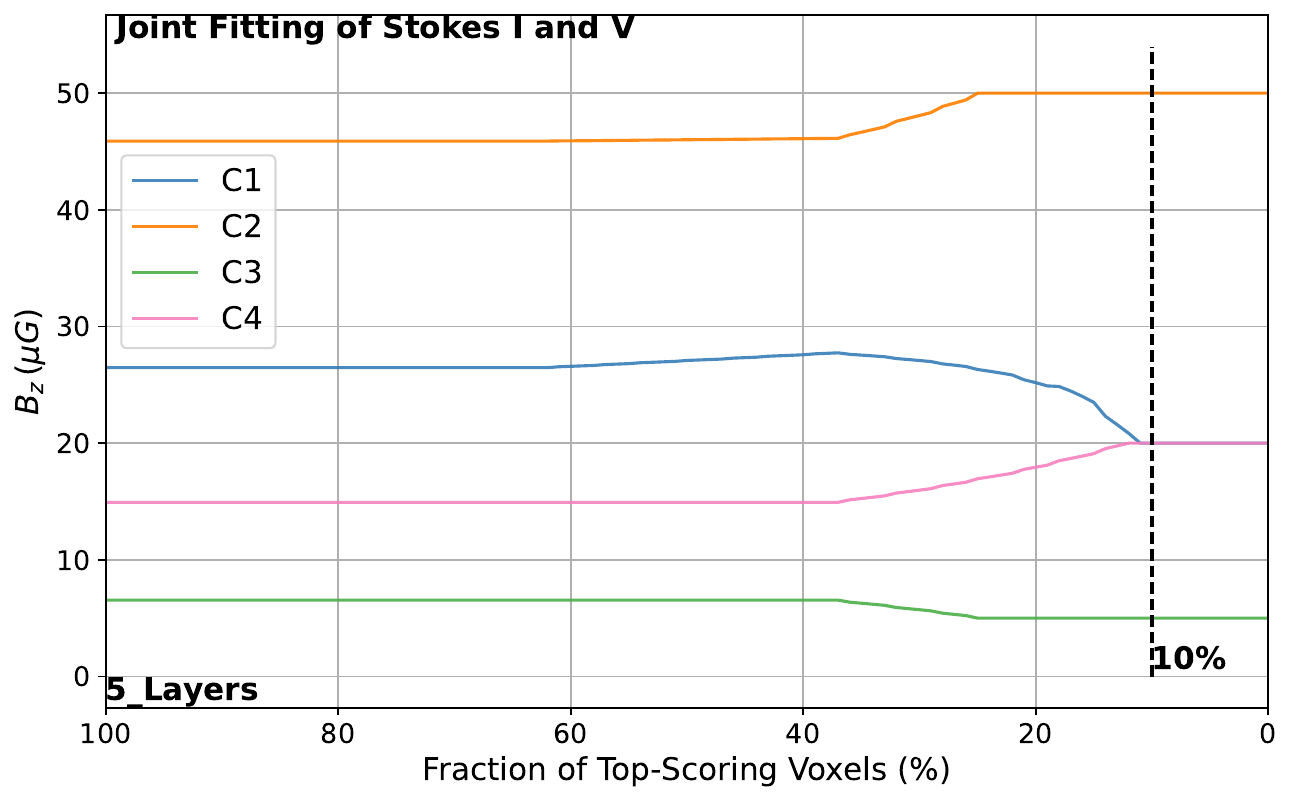}
\includegraphics[width=0.48\linewidth]{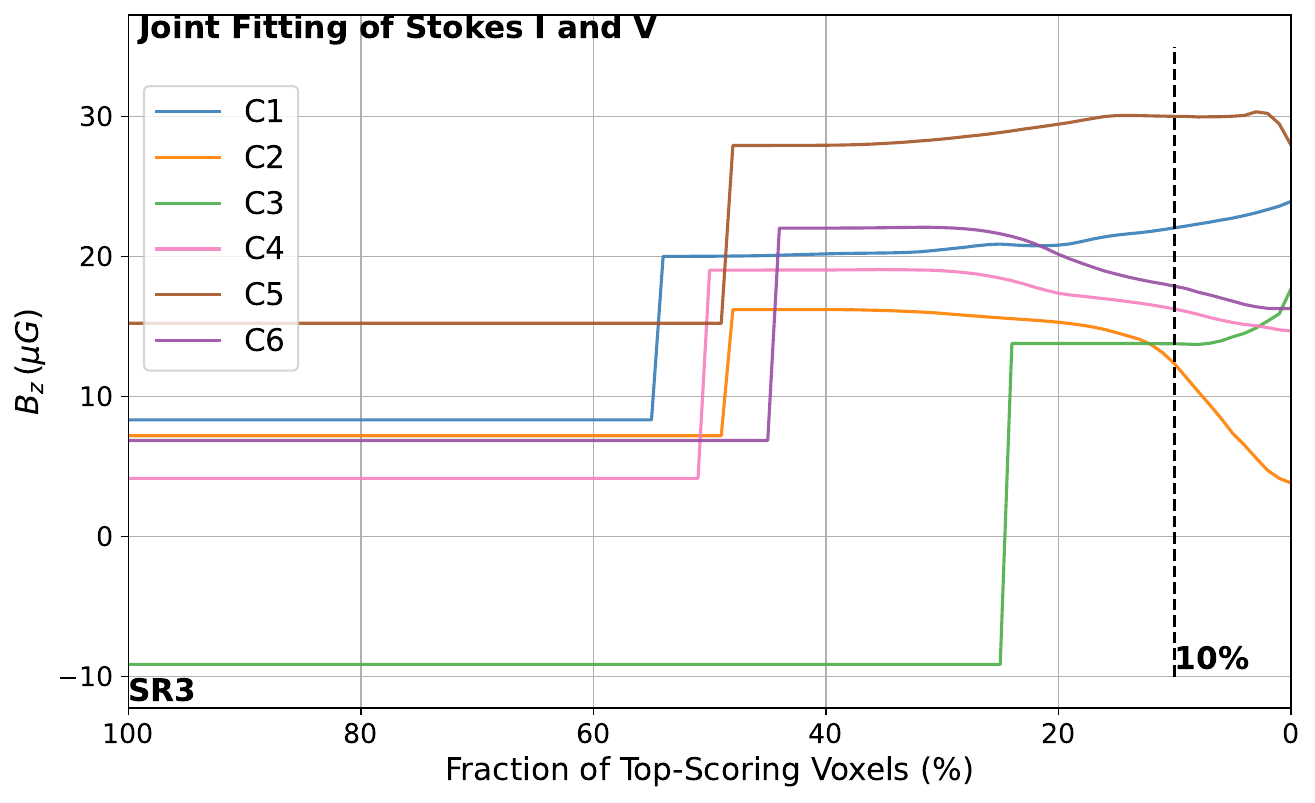}
\caption{Inferred similarity-weighted magnetic field as a function of the selected top-scoring voxel fraction in the five-layer cloud experiment (left panel) and the SR3 subregion (right panel), derived using the joint fitting strategy under Approach II. \xd{The ground-truth magnetic fields in the five-layer model are 5, 20, 50, 20, and 5 $\mu$G from the near to the far side. The fitted component numbering likewise increases with distance from the observer.} The vertical dashed line indicates the fiducial selection being the top 10\%.}
\label{fig.f_sim_fraction_5_layers}
\end{figure}

In this appendix, we explore how selecting voxels based on their highest similarity scores influences the value of the inferred similarity-weighted magnetic field $B_{GT2}$, as given by Equation~\ref{eq.BGT2}.

Figure~\ref{fig.f_sim_fraction_5_layers} presents the inferred similarity-weighted magnetic field as a function of the selected top-scoring voxel fraction in both the five-layer cloud experiment and the SR3 subregion, derived using the joint fitting strategy under Approach II. In the five-layer cloud experiment, where the ground truth is well-defined (each layer containing 20\% of the voxels), the results clearly indicate that selection of only the top-scoring voxels has the desired impact through the Stokes $V$ components. In contrast, for the SR3 subregion, identifying the true contribution of each component is more challenging. Ideally, any voxel with a positive similarity score ($f_{\text{sim}}$) should contribute to the observed Stokes $V$ signal. Conversely, relying on only the single top-scoring voxel would be unrealistic, as this would reduce the similarity-weighted magnetic field to the value at that single point. Therefore, the choice of using the top 10\% of voxels represents a compromise that captures the most influential contributors while averaging over a meaningful portion of the spatial structure.

Examples of the depth profile of $f_{\text{sim}}$ for component results for the SR3 LOS appear in Figures \ref{fig.f_sim_SR3} and \ref{fig.f_sim_SR3_flip}.

\section{Smooth Magnetic Field Gradient Experiment}
\label{Smooth Magnetic Field Gradient Experiment}

\begin{figure}[hbt!]
\centering
\includegraphics[width=0.48\linewidth]{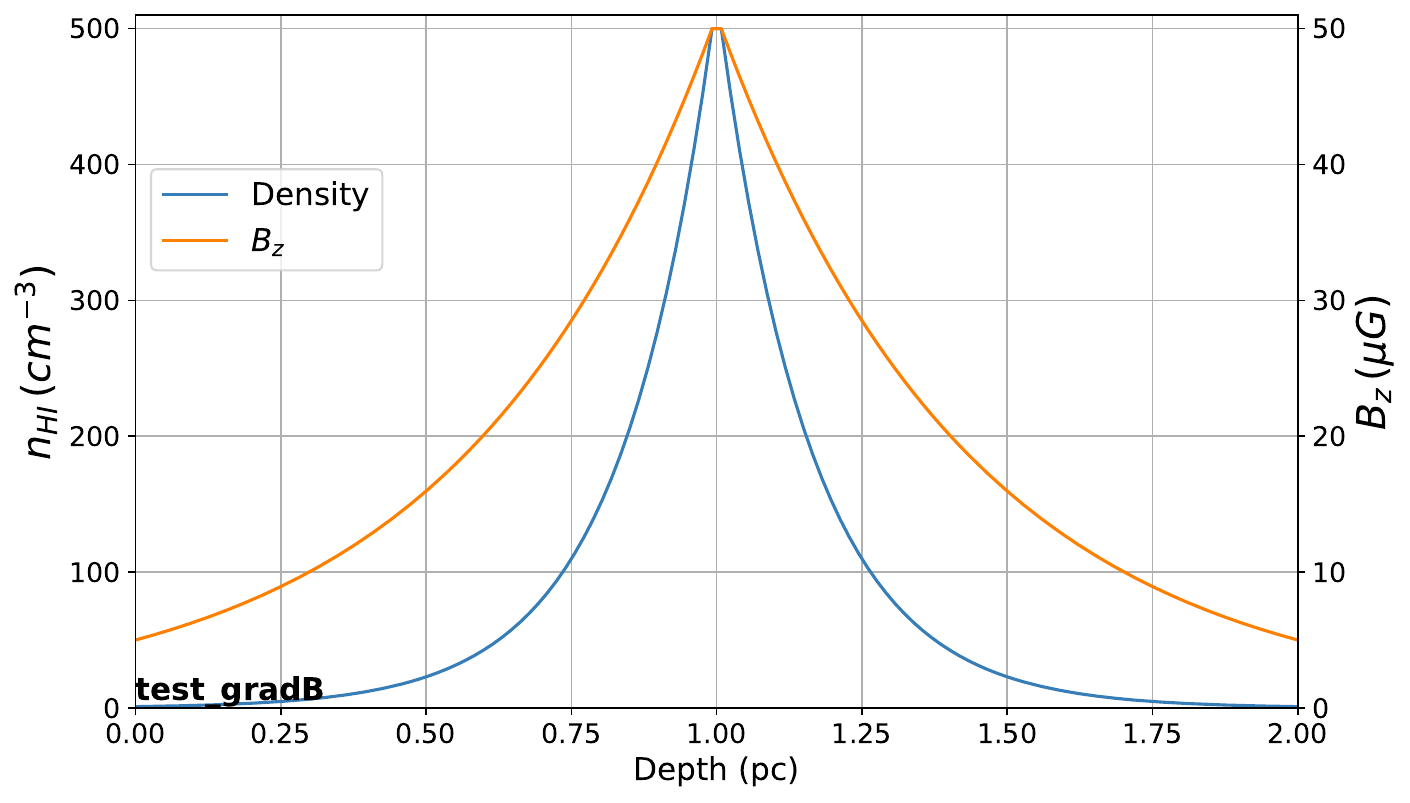}
\includegraphics[width=0.48\linewidth]{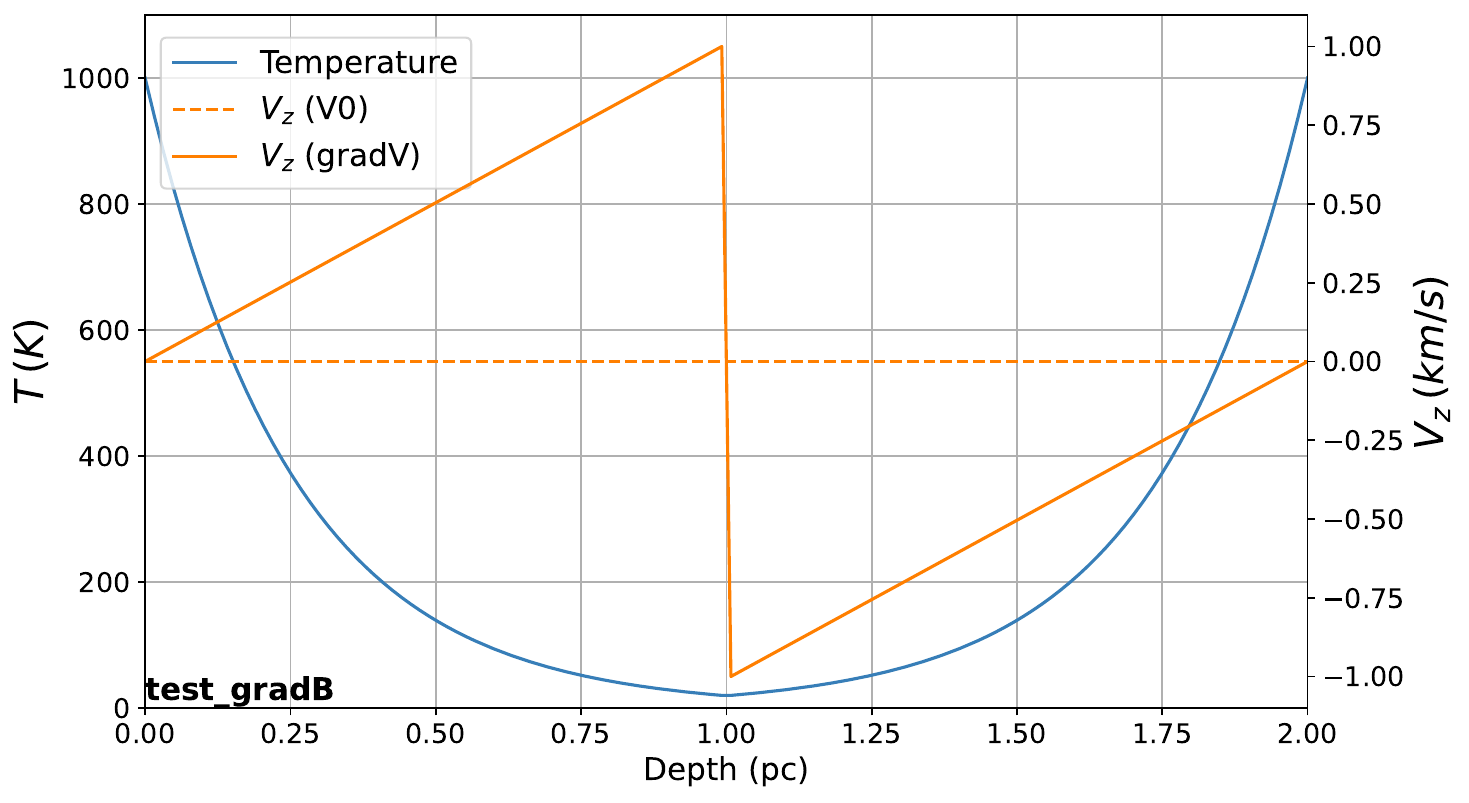}
\caption{LOS distributions of $n_{\mathrm {HI}}$ and magnetic field (left panel), and gas temperature and velocity (right panel) in the smooth magnetic field gradient test. In the right panel, the orange dashed line marks a uniform zero LOS velocity, while the solid orange line represents a velocity profile corresponding to a contracting spherical structure.  These gradients are notional, not representing an actual cloud model.}
\label{fig.Bz_pho_test_gradB}
\end{figure}

\begin{figure}[hbt!]
\centering
\includegraphics[width=0.48\linewidth]{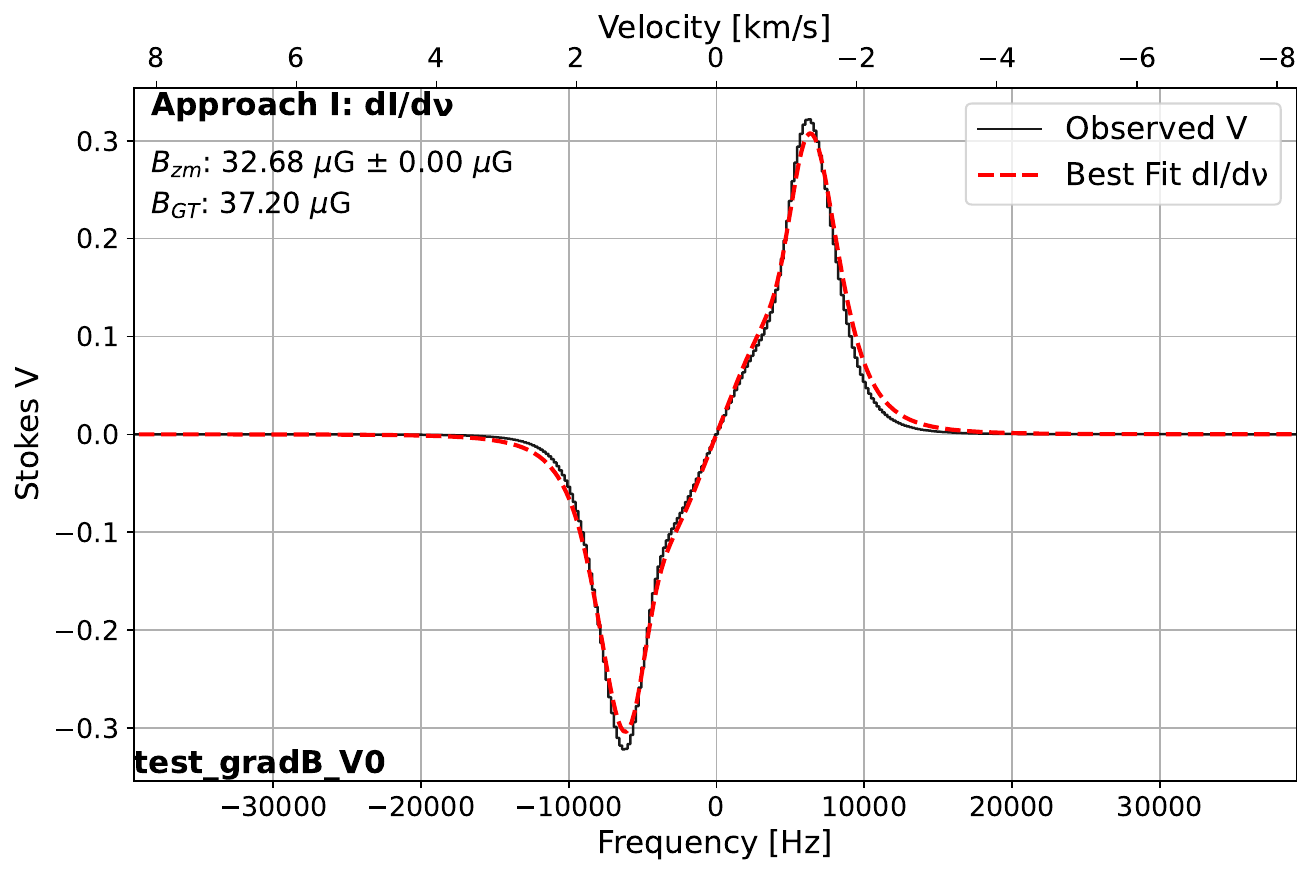}
\includegraphics[width=0.48\linewidth]{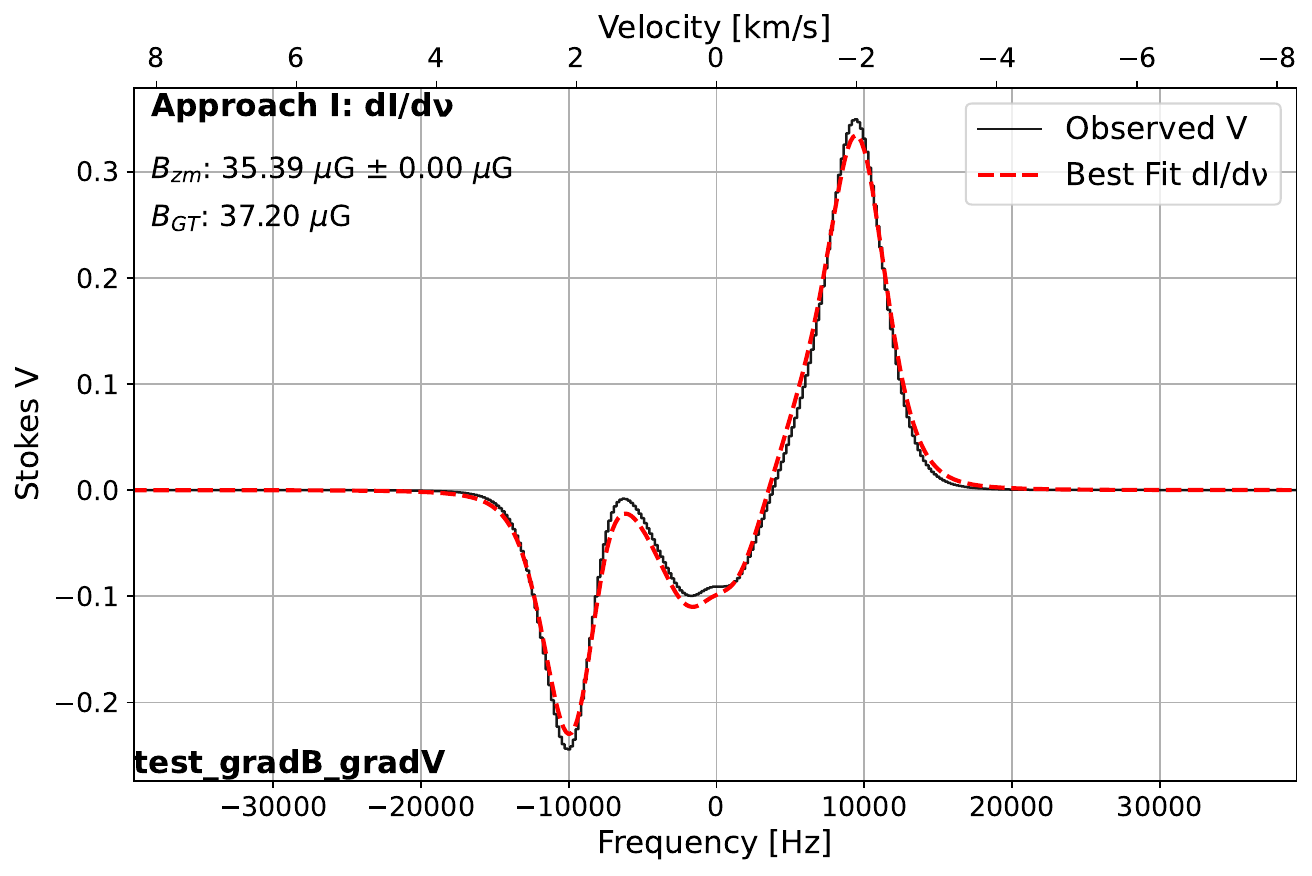}
\caption{\xd{Approach I result of fitting the Stokes $V$ spectrum with the scaled differentiated Stokes $I$ spectrum,} 
as in Figure~\ref{fig.ZeemanFitting_dI_dnu_5_layers}, 
\xd{but for the smooth magnetic field gradient test (Figure \ref{fig.Bz_pho_test_gradB}). Left: zero LOS velocities across all gas. Right: LOS velocities crudely reminiscent of a contracting spherical configuration.}
}
\label{fig.ZeemanFitting_dI_dnu_gradB_gradV}
\end{figure}

\begin{figure}[hbt!]
\centering
\includegraphics[width=0.98\linewidth]{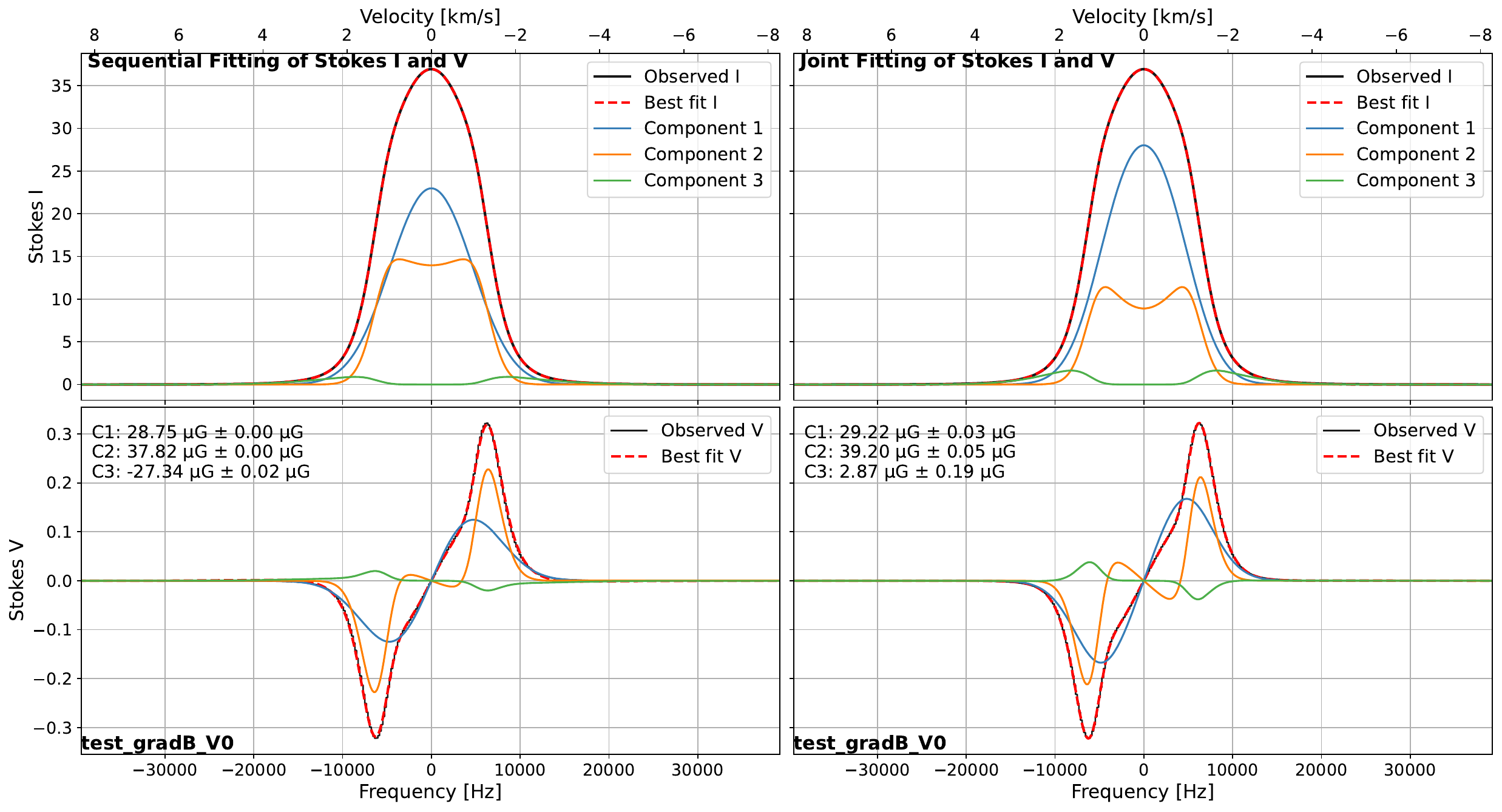}
\caption{\xd{Approach II fitting results for the Stokes $I$ and Stokes $V$ spectra,}
as in Figure~\ref{fig.ZeemanFitting_5_Layer_IV_noise0}, 
\xd{but for the test with a smooth magnetic field gradient and zero LOS velocities across all gas. Components are in the order determined by the fit, with C1 being the most foreground. Sequential strategy results are shown on the left and joint results on the right.} }
\label{fig.ZeemanFitting_test_gradB_V0_I_V_fitting}
\end{figure} 

\begin{figure}[hbt!]
\centering
\includegraphics[width=0.98\linewidth]{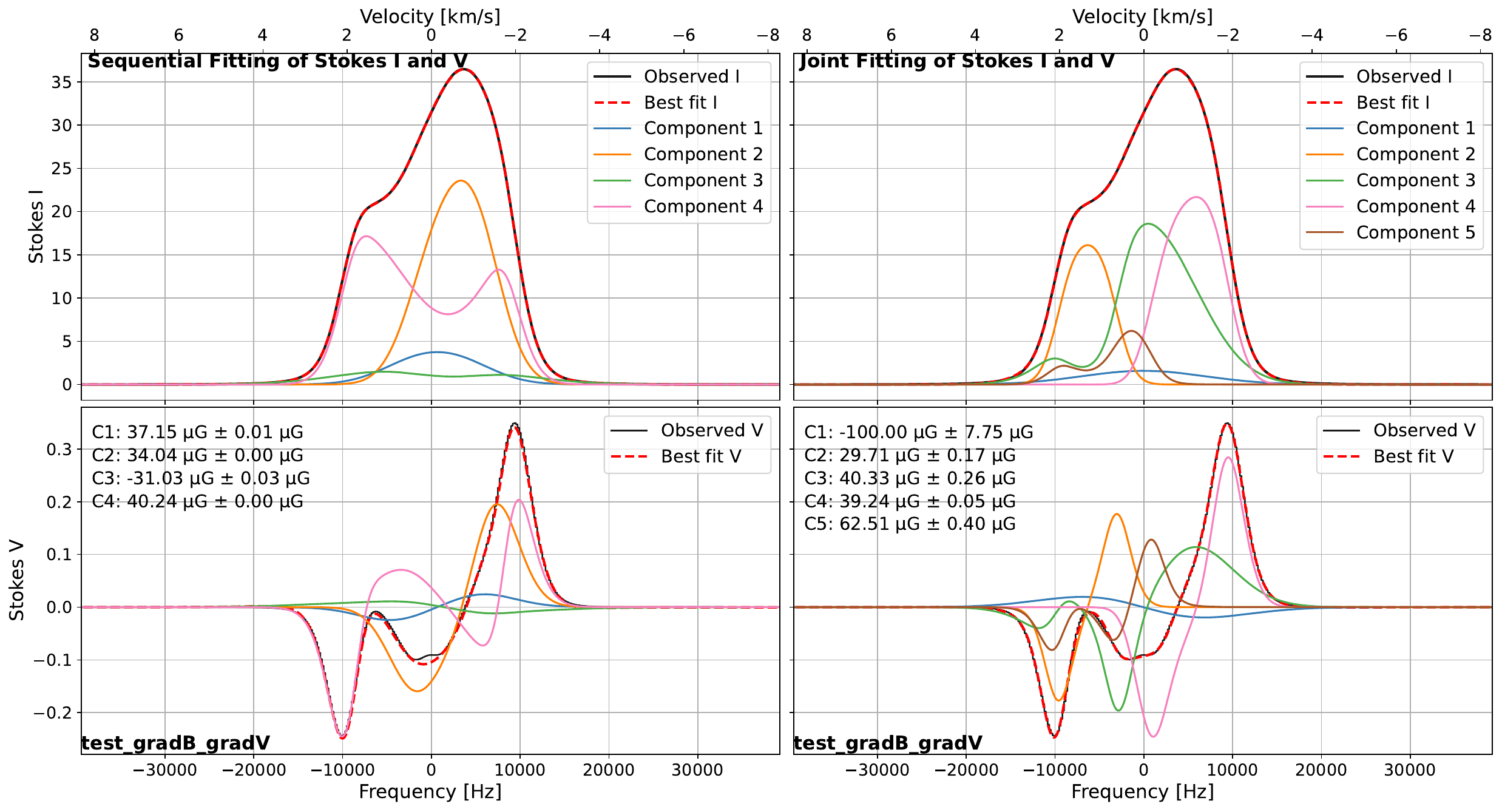}
\caption{\xd{Approach II fitting results for the Stokes $I$ and Stokes $V$ spectra for the test with a smooth magnetic field gradient,}
as in Figure~\ref{fig.ZeemanFitting_test_gradB_V0_I_V_fitting}, 
\xd{but for LOS velocities representing a contracting spherical configuration.}}
\label{fig.ZeemanFitting_test_gradB_gradV_I_V_fitting}
\end{figure}

This appendix describes Zeeman analysis of an idealized scenario with smoothly varying HI density, temperature, velocity, and magnetic field along the LOS, as illustrated in Figure~\ref{fig.Bz_pho_test_gradB}. \xd{These gradients are notional in analytic form for the purposes of this exercise, 
and do not purport to derive from any actual cloud structure.}  We evaluated two LOS velocity scenarios: one assumes a simplified dynamical configuration with zero LOS velocity for all gas (V0), while the other incorporates additional complexity by imposing LOS velocities notionally motivated by a contracting spherical cloud (gradV).
 
\subsection{Approach I}

First, we applied Approach I to estimate the LOS magnetic field strength. The fitting results for both the zero LOS velocity and contracting cloud velocity setups are shown in Figure~\ref{fig.ZeemanFitting_dI_dnu_gradB_gradV}.  In both cases, the fit is remarkably good using the single parameter $B_{zm}$ and the inferred $B_{zm}$ is reasonably close to the ground truth LOS mass-weighted estimator $B_{GT}$. Note that most emission arises in the central 10\% of the cloud.

Overall, Approach I, as an averaging method, appears to perform best in cases with smoothly varying physical conditions, particularly $B_z$. This aligns with what was seen in Figure \ref{fig.ZeemanFitting_dI__SR_Breal_noise0} for the actual cube, where the better fits were for the lines of sight with the smoother fields, ranked LSR1, SR4, SR2, SR3 in increasing complexity on the basis of Figure \ref{fig.physical_Region_All}.

\subsection{Approach II}

Next, we applied Approach II assuming the zero LOS velocity setup, V0. Figure~\ref{fig.ZeemanFitting_test_gradB_V0_I_V_fitting} presents the Stokes $I$ and Stokes $V$ fitting results. Both sequential and joint fitting strategies capture the spectral line profiles reasonably well.  Foreground component C1 produces significant attenuation of C2.  Combined, they account for most of the emission, with component C3 of lesser significance in either $I$ or $V$.  C2 has a higher $B_{zm}$, arising from deeper in the cloud core.
However, the weak component C3 has unrealistically low $B_{zm}$ not present in the setup, including a large negative value for the sequential strategy.

Finally, we applied Approach II assuming the contracting cloud velocity setup, grad V. The corresponding fitting results are shown in Figure~\ref{fig.ZeemanFitting_test_gradB_gradV_I_V_fitting}. 
The sequential fit is fairly good, but the component C2 produces emission at negative velocities and C4 at positive (also negative) velocities, the opposite of the setup.  Their fields are of plausible size. Component C3 is quite inconsequential in both $I$ and $V$ and has an unrealistically negative field.

The joint fit is better in two respects, the fit itself to $V$ and getting the depth order correct, C2 at positive velocity, C3 at intermediate velocity, and C4 at negative velocity.  The highly attenuated C5 is weak and has an unrealistically high field, whereas the other weaker component C1 has an unrealistically high negative field, though ultimately its contribution to $V$ is not substantial.

In contrast to Approach I, the problem being tackled in Approach II -- probing ISM structure along the entire line of sight -- is more demanding and these results highlight important limitations. The combined complexity of realistic gas dynamics and magnetic field structures can make accurate recovery of the ground truth challenging.
Caution should be taken not to overinterpret components that make only a small contribution to the spectra; while tweaking the fit with such components is mathematically possible, results are often unphysical.

\section{Zeeman Magnetic Field Estimates for Cubes with Modified Magnetic Field Structure}
\label{Magnetic Field Structure}

In this appendix, we examine one factor that could affect magnetic field estimates from Zeeman measurements, namely the structure of the magnetic field. We conducted a controlled experiment where the magnetic field in the simulation cube was first set to a uniform value, while all other physical properties, including density, velocity, and temperature, remain unchanged. Specifically, we assigned a constant value of 10~$\mu$G to the magnetic field along the line of sight and the two perpendicular directions (the latter having negligible effect). 
Using this modified setup, we generated synthetic HI observations. 

\begin{figure}[hbt!]
\centering
\includegraphics[width=0.68\linewidth]{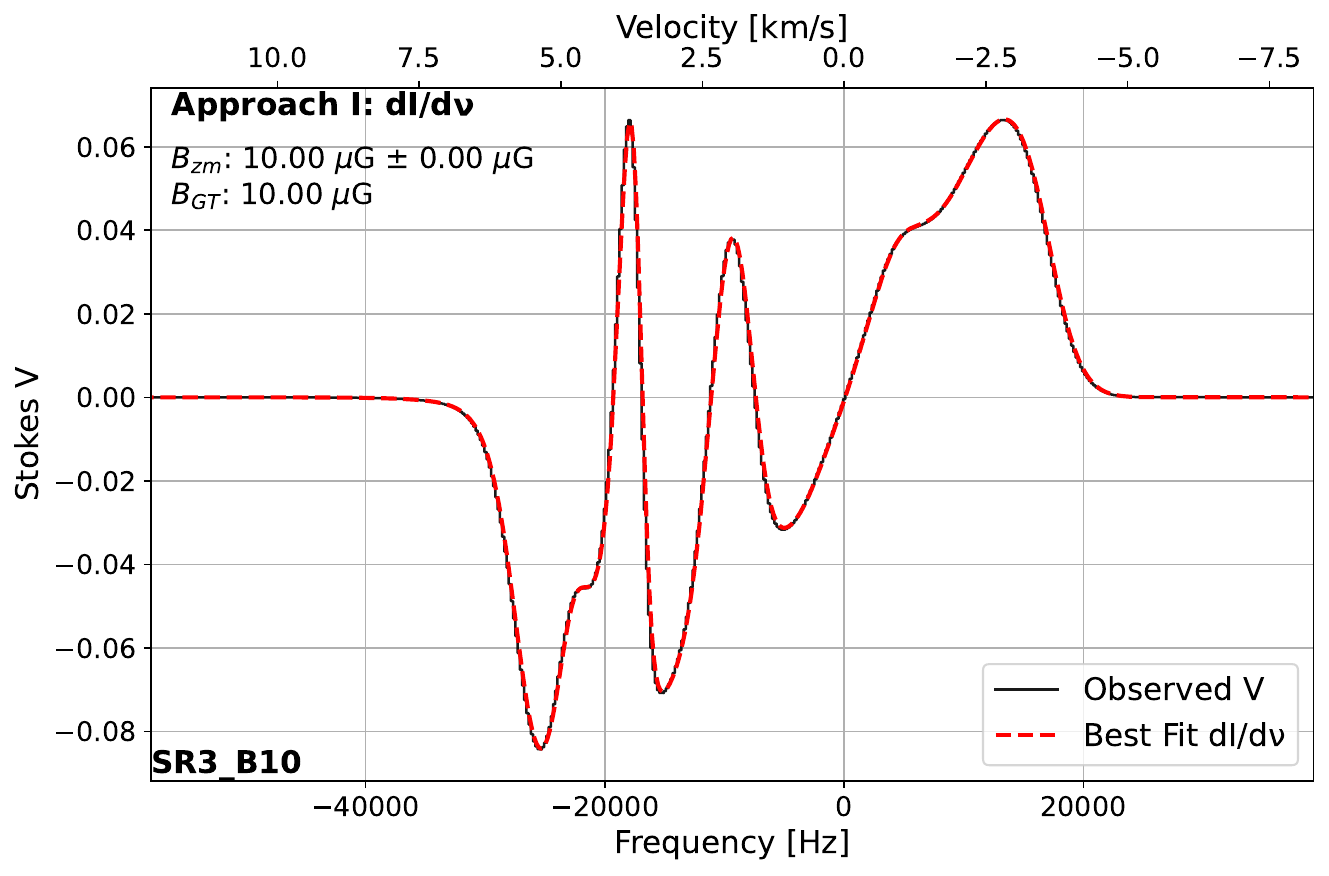}
\caption{\xd{Approach I result of fitting the synthetic Stokes $V$ spectrum (black) with the scaled the differentiated Stokes $I$ spectrum (red),} as in Figure~\ref{fig.ZeemanFitting_dI_dnu_5_layers}, \xd{but for the SR3 subregion with a uniform magnetic field of 10~$\mu$G. There is a perfect model match for 10~$\mu$G.}
}
\label{fig.SR3_dI_dnu_B10_}
\end{figure}

\begin{figure}[hbt!]
\centering
\includegraphics[width=0.68\linewidth]{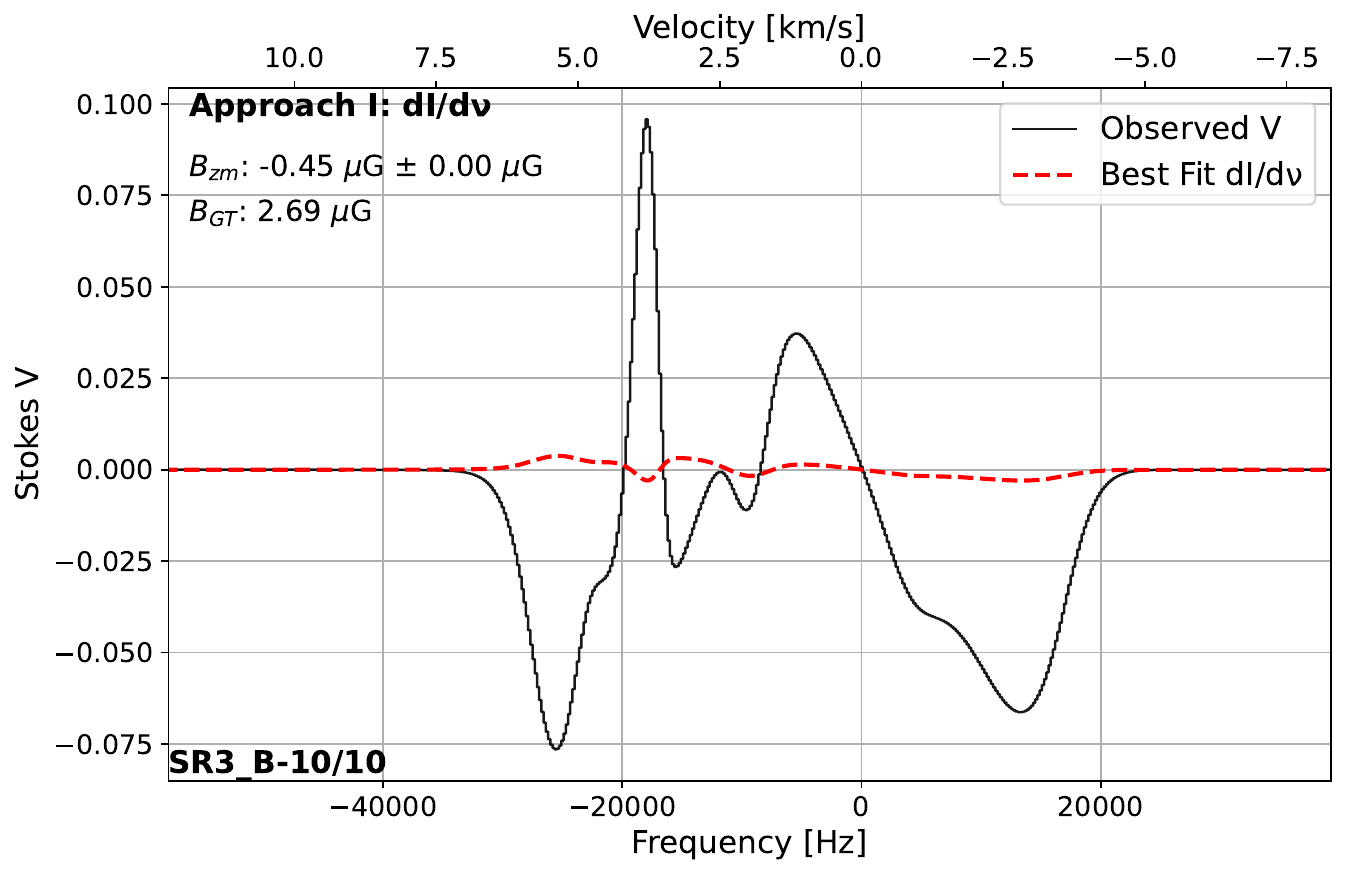}
\caption{Same as Figure~\ref{fig.SR3_dI_dnu_B10_}, but Approach I result for the LOS magnetic field set to $-10\, \mu$G in the front half and $+10\, \mu$G in the back half of the simulation box. The model spectrum to be scaled is the same (red).
}
\label{fig.SR3_dI_dnu_B10neg10_}
\end{figure}

\begin{figure*}[hbt!]
\centering
\includegraphics[width=0.98\linewidth]{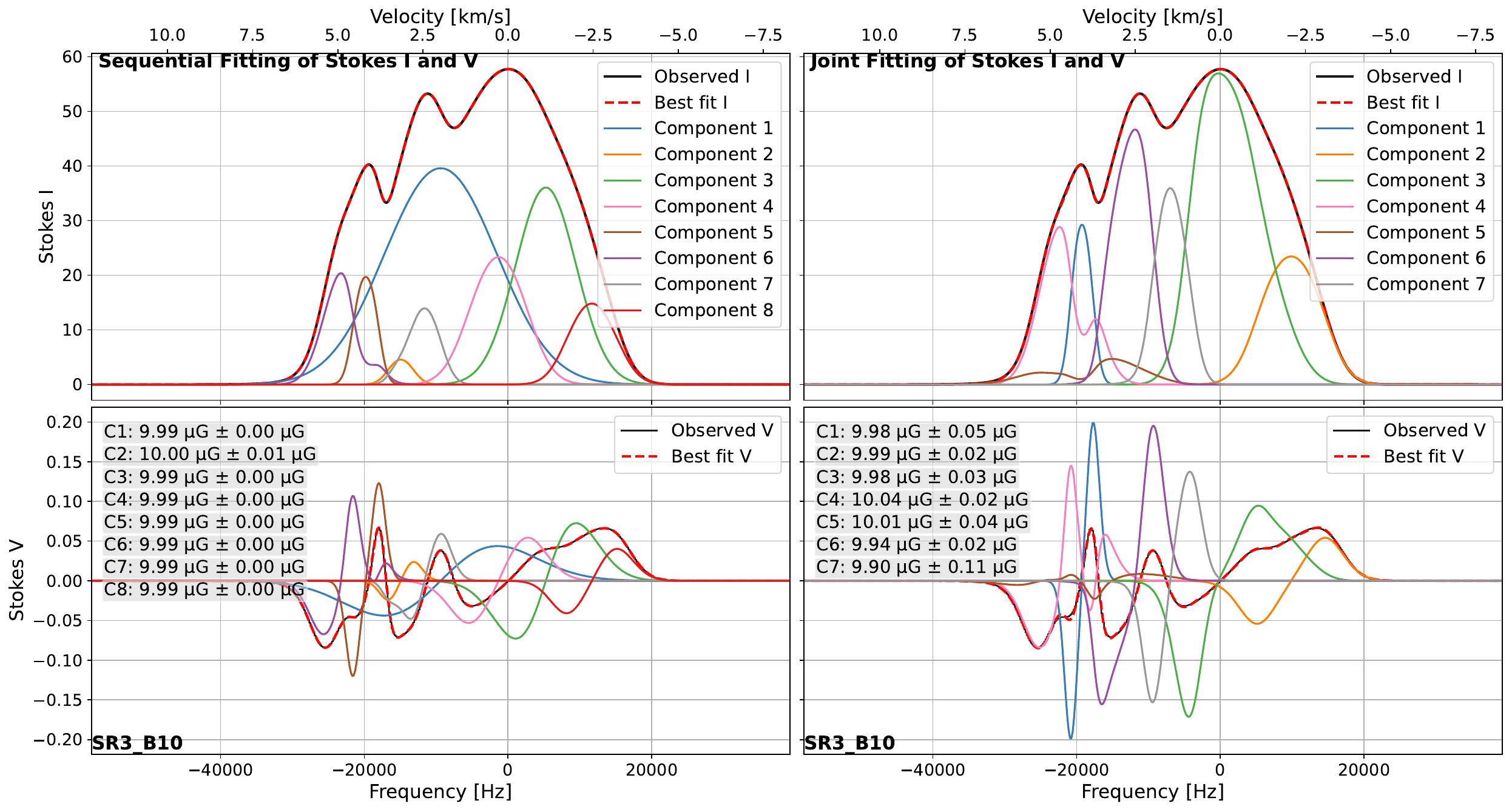}
\caption{\xd{Approach II results fitting Stokes $I$ and Stokes $V$ spectra (black) for the SR3 subregion modified to have a uniform magnetic field of 10~$\mu$G,} 
presented as in Figures~\ref{fig.ZeemanFitting_5_Layer_IV_noise0} and \ref{fig.ZeemanFitting_SR3_Breal_noise0}.  
Both sequential and joint strategies successfully recover the magnetic field.
}
\label{fig.SR3_I_V_fitting_B10_}
\end{figure*}

\begin{figure*}[hbt!]
\centering
\includegraphics[width=0.98\linewidth]{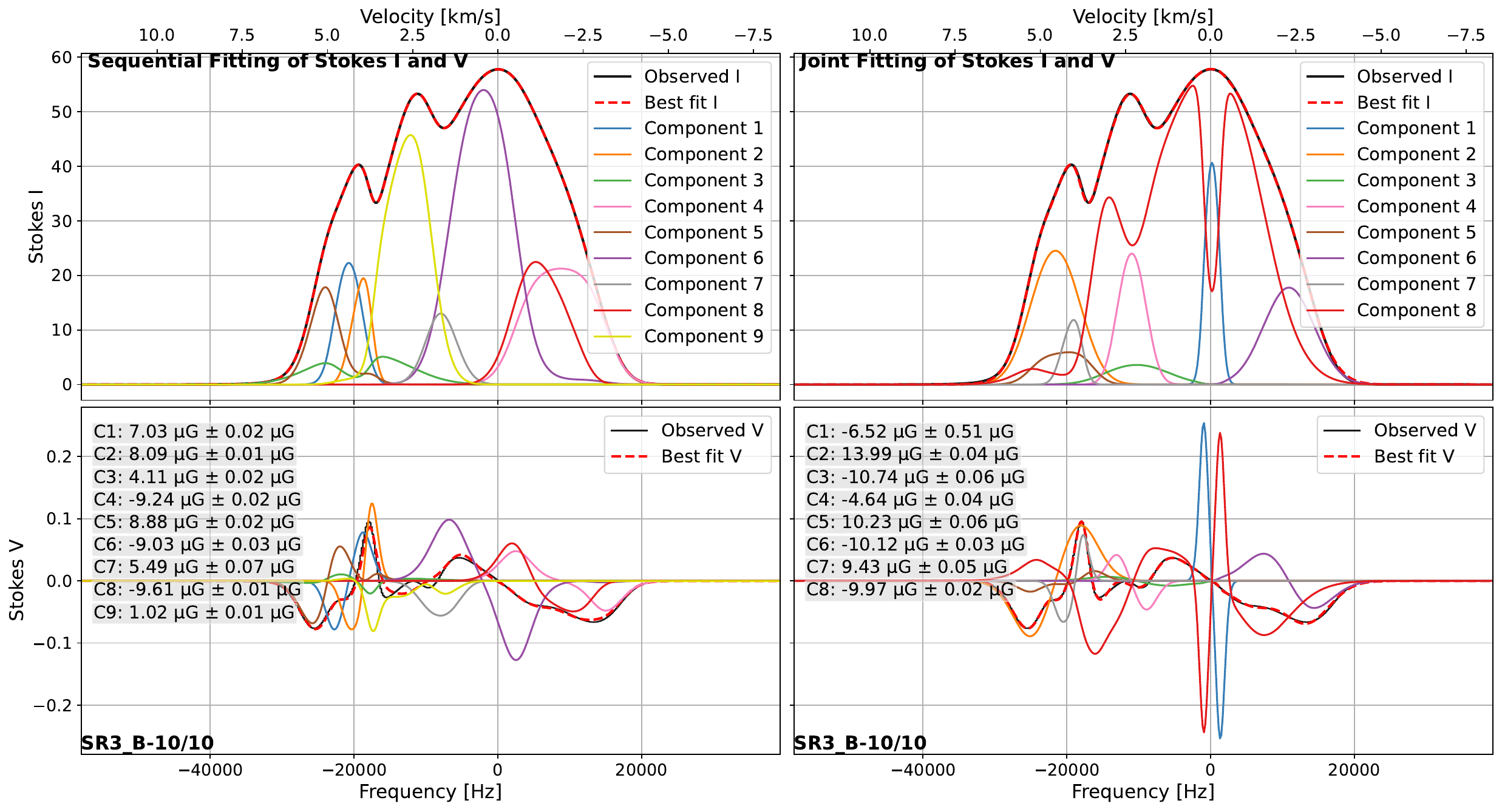}
\caption{Same as Figure~\ref{fig.SR3_I_V_fitting_B10_}, \xd{but Approach II fitting results for the LOS magnetic field set to $-10\, \mu$G in the front half and $+10\, \mu$G in the back half of the simulation box.} Compared to the sequential strategy, the joint strategy yields a better match to the Stokes $V$ spectrum, even using fewer components.
}
\label{fig.SR3_I_V_fitting_B10neg10_}
\end{figure*}

Next, we conducted an experiment in which, instead of being uniform, the LOS magnetic field was set to $-10\, \mu$G in the front half of the simulation box and $+10\, \mu$G in the back half (the $B-10/10$ case). This does not change the synthetic (``observed'') Stokes $I$ spectrum, or the shape of the differentiated Stokes $I$ spectrum, which remains the same as from the original cube (see red curve in Figure \ref{fig.ZeemanFitting_dI__SR_Breal_noise0}, lower left). However, it has a profound effect on the synthetic observations of the $V$ spectrum: compare black curves in Figures \ref{fig.SR3_dI_dnu_B10_} and \ref{fig.SR3_dI_dnu_B10neg10_} or 
Figures \ref{fig.SR3_I_V_fitting_B10_} and \ref{fig.SR3_I_V_fitting_B10neg10_}, where spectral features from gas in the front half of the cube change sign in $V$ (most obvious in SR3 for HI gas with low to negative $v_z$ in Figure \ref{fig.physical_Region_All} and layer contributions L01, L12, and L23 in Figure \ref{fig.SR3_GPB}).  

We then applied both Approach I and Approach II to test the ability of Zeeman analysis to recover the ground truth estimators of the field in these modified cubes. This can be appreciated as complementary to the validation of the analysis methods using the five-layer cloud (simpler geometry and physical conditions) in Section \ref{Simple Geometry Validation}.

\subsection{Approach I} 

Using the SR3 subregion as an example for Zeeman analysis, Figure~\ref{fig.SR3_dI_dnu_B10_} shows that Approach I, which fits the synthetic Stokes $V$ spectrum with the scaled differentiated Stokes $I$ spectrum via Equation~\ref{dI_dnu_eq1}, exactly retrieves the true LOS magnetic field strength. This is not at all surprising, because the model for $V$ is perfectly appropriate.  

In the $B-10/10$ case in Figure \ref{fig.SR3_dI_dnu_B10neg10_}, the $dI/d\nu$ spectrum (red curve) is unchanged from the uniform field case, but the synthetic $V$ spectrum (black) is dramatically different. There is no longer a single factor that would scale that red profile to match the profoundly changed synthetic spectrum. \xdtwo{Because it averages, Approach I completely fails to find a signature of either strong field in the foreground or background.} Instead, as Figure~\ref{fig.SR3_dI_dnu_B10neg10_} shows, fitting with Equation~\ref{dI_dnu_eq1} inevitably results in a very low average $B_{zm}$. \xd{The negative field fit inverts the red model profile seen in Figure \ref{fig.SR3_dI_dnu_B10_} and the small value compresses the profile amplitude.}

\subsection{Approach II} 

Figure~\ref{fig.SR3_I_V_fitting_B10_} presents Stokes $I$ and Stokes $V$ fitting results for the uniform field case, using sequential and joint strategies.
The sequential strategy uses eight components to fit $I$ (though quite different than those in Figure \ref{fig.ZeemanFitting_SR3_Breal_noise0}, upper left, from fitting the original cube, because of different starting conditions). Regardless of strategy, each of the multiple components recovers the ground truth magnetic field, because any parcel of gas has the same $\Delta \nu$ and so follows Equation \ref{dI_dnu_eq1} component by component.


Figure~\ref{fig.SR3_I_V_fitting_B10neg10_} presents the fitting results for the $B-10/10$ case.  The sequential strategy uses yet another set of components (now nine). Each component contribution to optimize the overall fit of $V$ yields a different value of $B_{zm}$. These are in the range $-10\, \mu$G to $+10\, \mu$G, but there is clearly some mixing occurring between purely $\pm10\, \mu$G components. One would not conclude from these values that the field had such a simple structure. 

The joint fitting strategy achieves visually good fits for both Stokes $I$ and $V$, using eight components, five of which could qualify as $\pm10\, \mu$G components. But there are still three with inferred values of $B_{zm}$ that deviate significantly from the input setup, including one, component C2 with 13.99~$\mu$G lying outside the range and obviously out of order (gas in the foreground should have a negative field).

In fact, the order is probably meaningless, with the fields in successive components alternating sign, whereas again the foreground components ought to have negative fields. These results underscore the limitations of Zeeman analysis in accurately recovering magnetic field strengths and finding the correct spatial ordering simultaneously.

\xd{It is interesting to note that for the gas producing the negative velocity line wing, both sequential and joint strategies agree on the field being about $-10\, \mu$G (C4 and C6, respectively).  The modeling of the positive velocity wing is more complex, but the consensus is a positive field (C5 and C2, respectively).}

The above test with the abrupt change in field might be artificially challenging. 
The results were more encouraging for the related and complementary experiment in which the magnetic field and other physical quantities have smoothly varying LOS profiles (Appendix~\ref{Smooth Magnetic Field Gradient Experiment}).

\section{Zeeman Fitting Gallery: Approach I for SR2 and SR3 across Noise Levels}
\label{Zeeman Fitting Gallery: Approach I Across Noise and Regions}

\begin{figure}[hbt!]
\centering
\includegraphics[width=0.48\linewidth]{{SR2_dI_dnu}.pdf}
\includegraphics[width=0.48\linewidth]{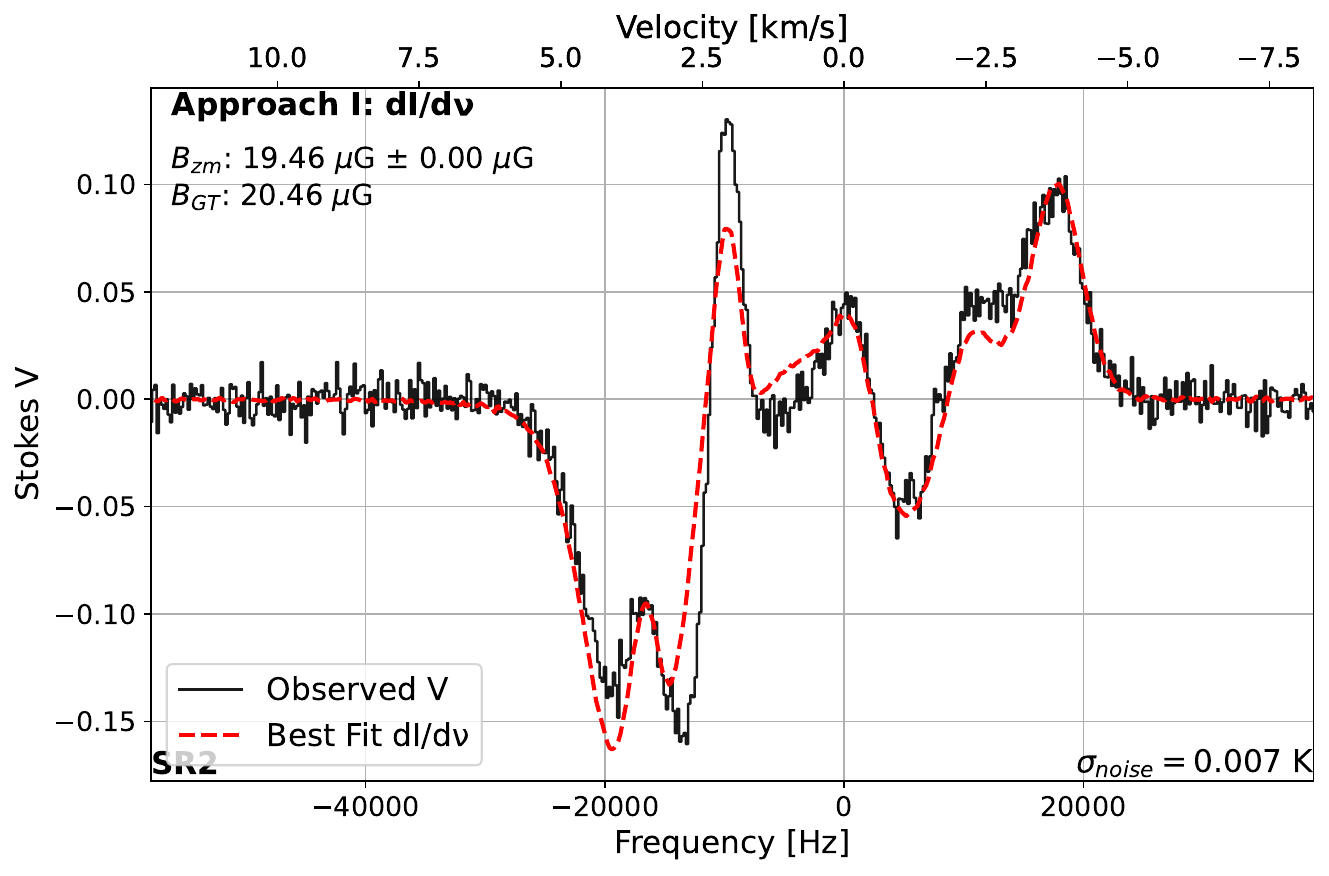}
\includegraphics[width=0.48\linewidth]{{SR2_dI_dnu01}.pdf}
\includegraphics[width=0.48\linewidth]{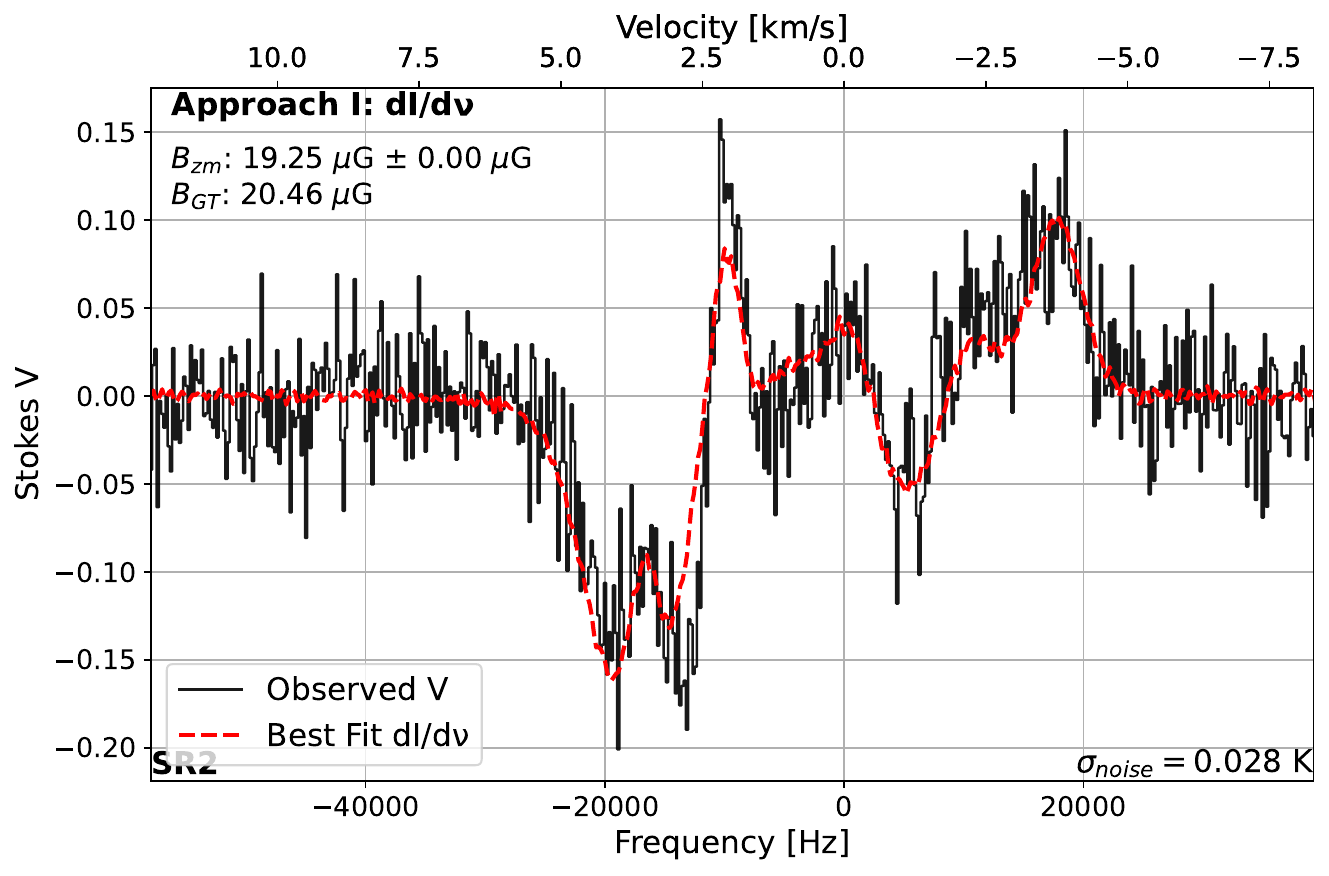}
\includegraphics[width=0.48\linewidth]{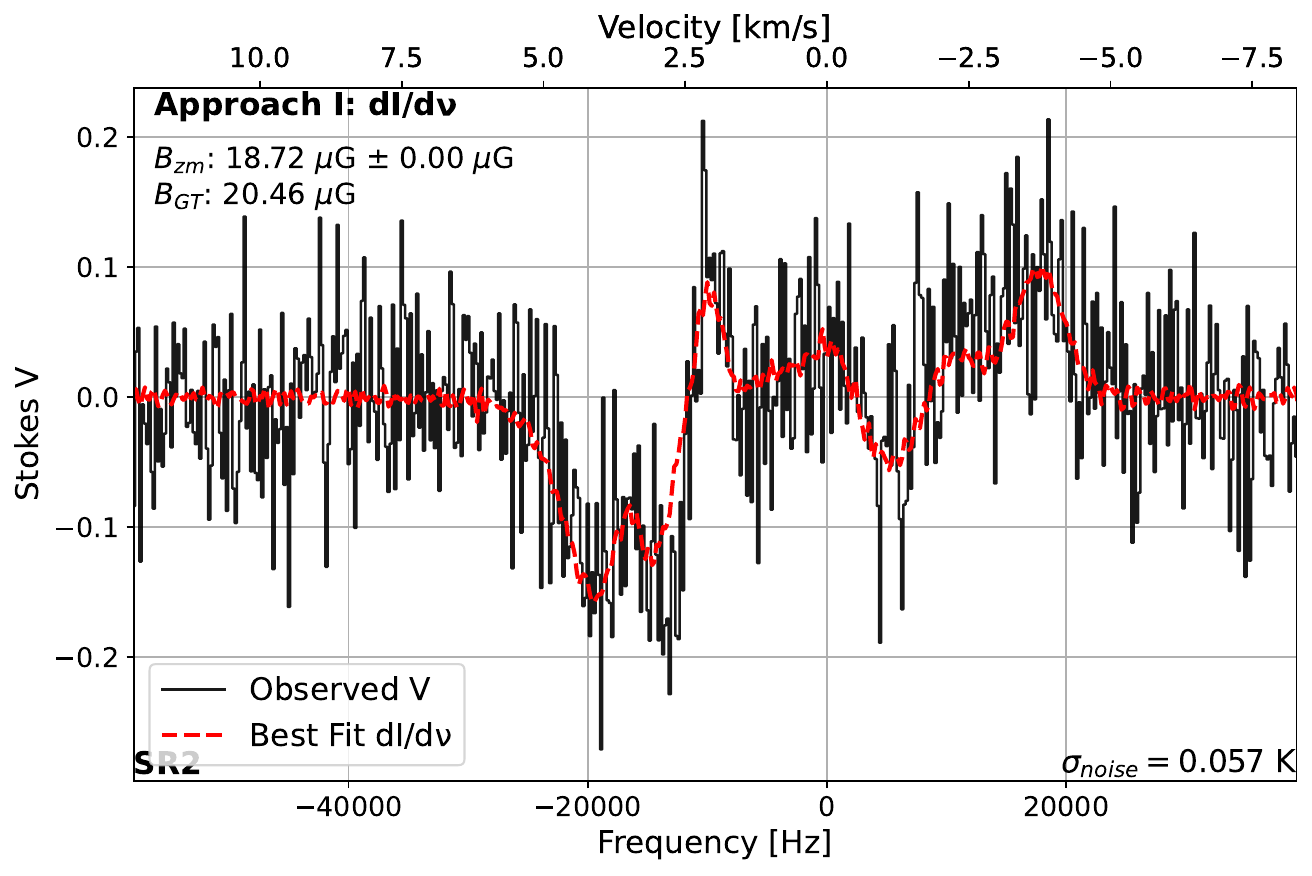}
\includegraphics[width=0.48\linewidth]{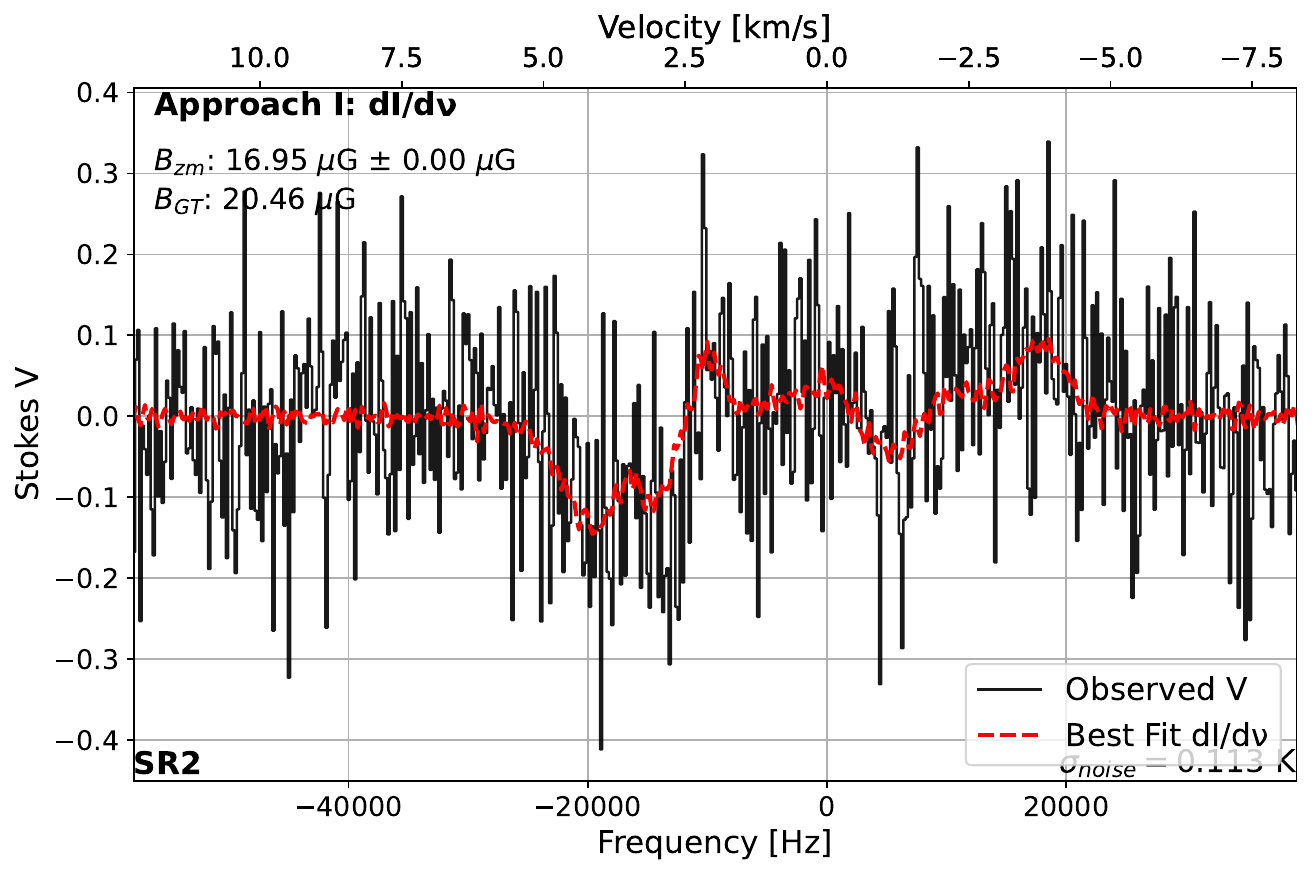}
\caption{Approach I fitting results for the Stokes $V$ spectrum (black line) with the scaled differentiated Stokes $I$ spectrum (red dashed line) for subregion SR2. The panels show the no-noise case (upper left) followed by spectra with increasing noise levels \xdtwo{as indicated in the legends (see Section \ref{sec:addnoise}).}}
\label{fig.ZeemanFitting_SR2_dI_dnu_fitting_noise_All}
\end{figure}

\begin{figure}[hbt!]
\centering
\includegraphics[width=0.48\linewidth]{{SR3_dI_dnu}.pdf}
\includegraphics[width=0.48\linewidth]{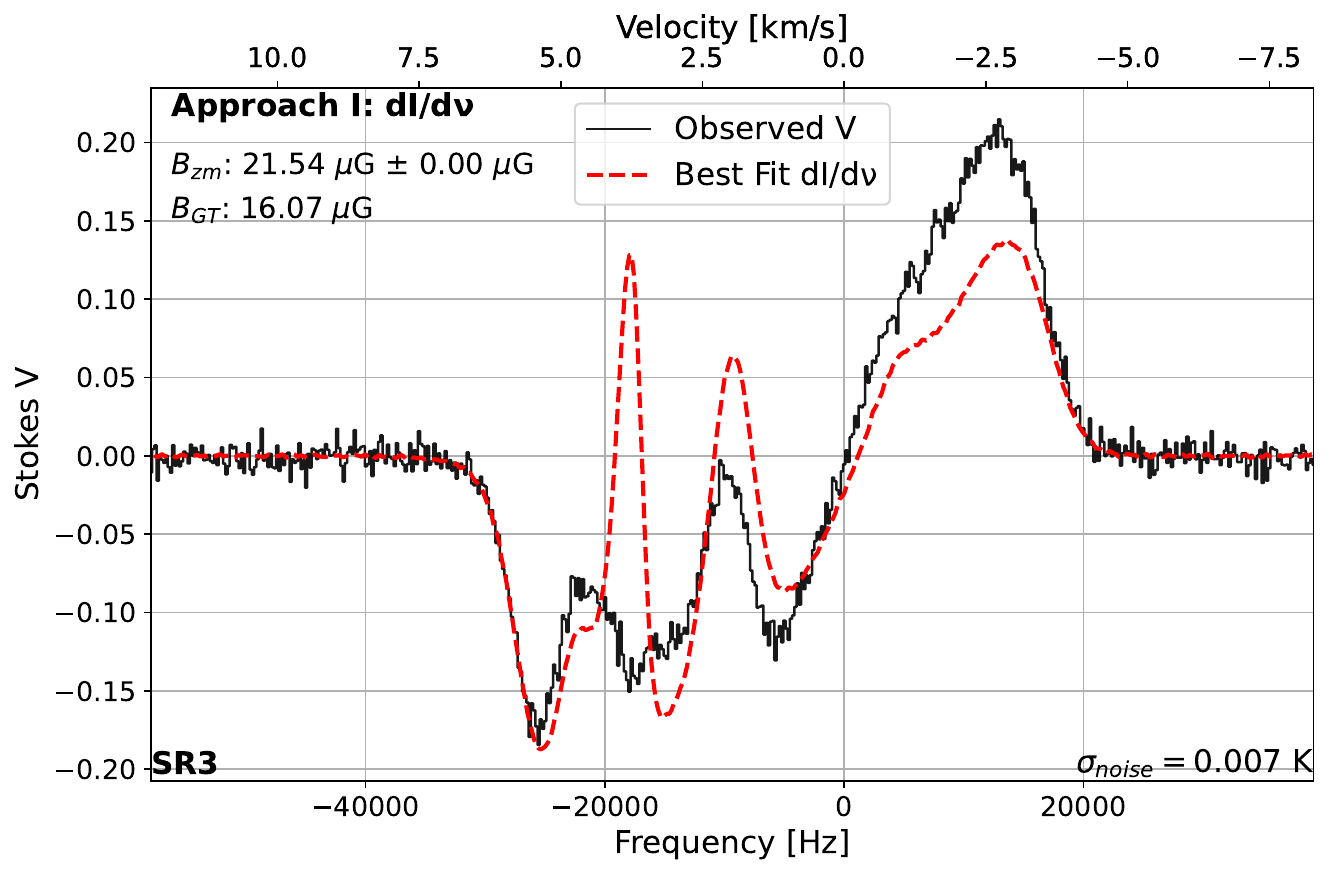}
\includegraphics[width=0.48\linewidth]{{SR3_dI_dnu01}.pdf}
\includegraphics[width=0.48\linewidth]{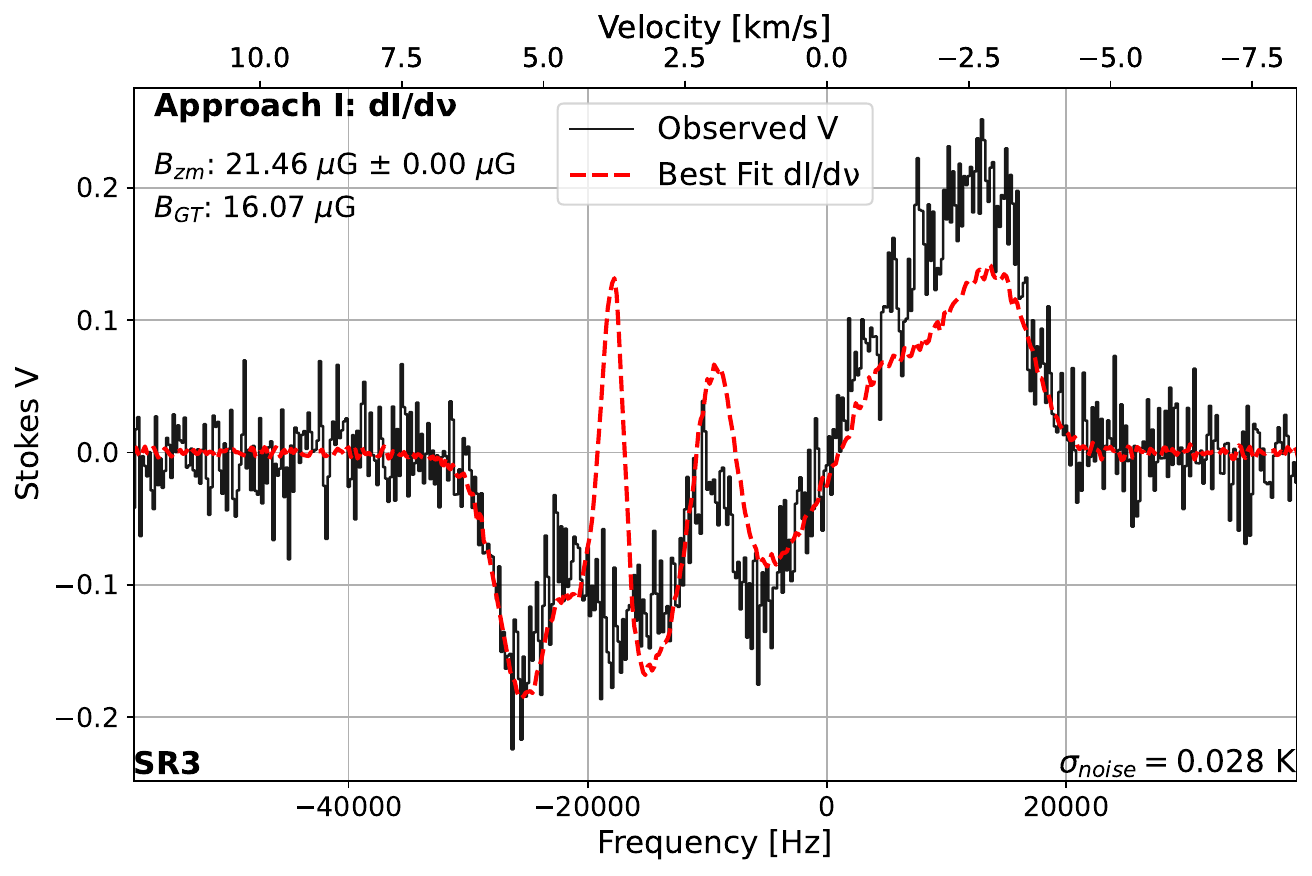}
\includegraphics[width=0.48\linewidth]{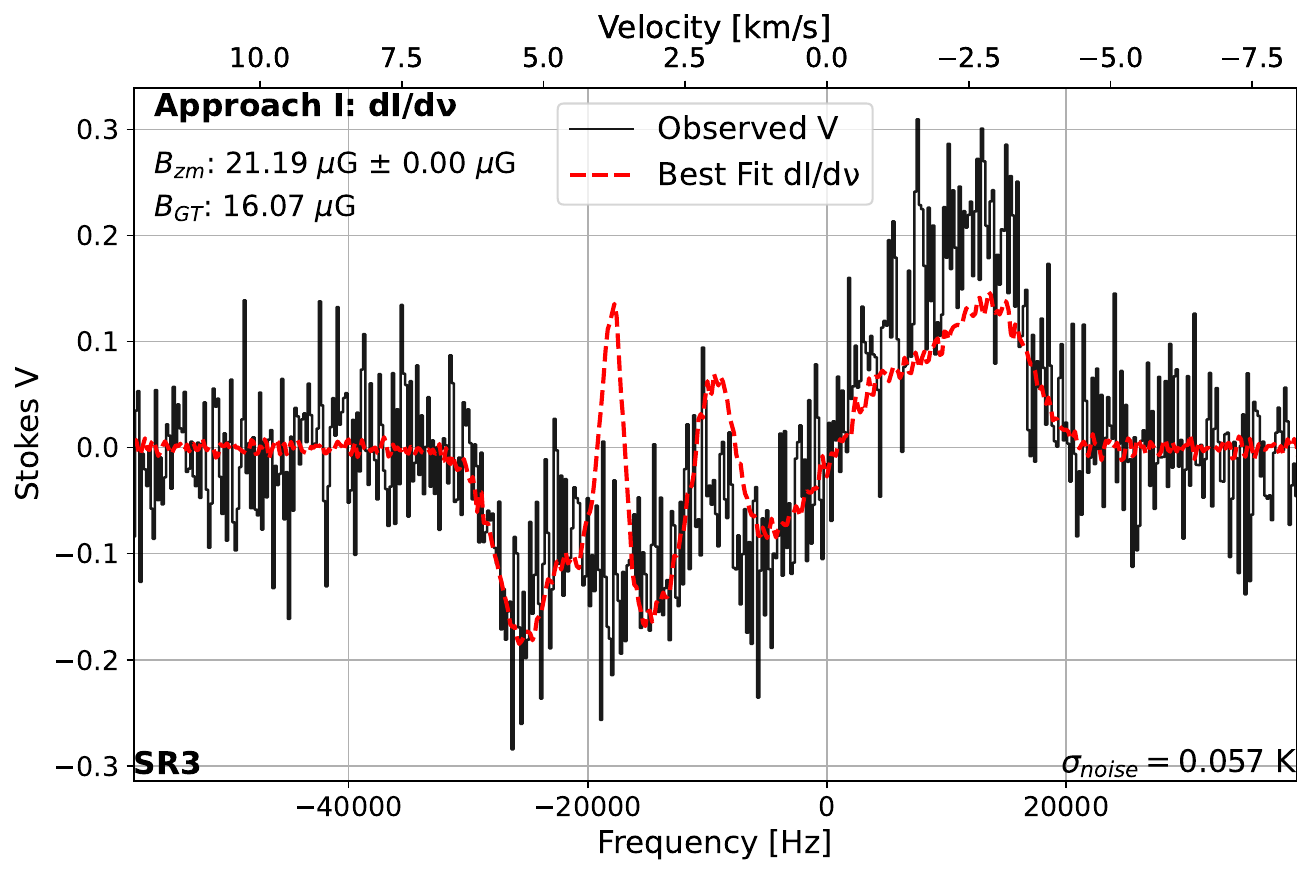}
\includegraphics[width=0.48\linewidth]{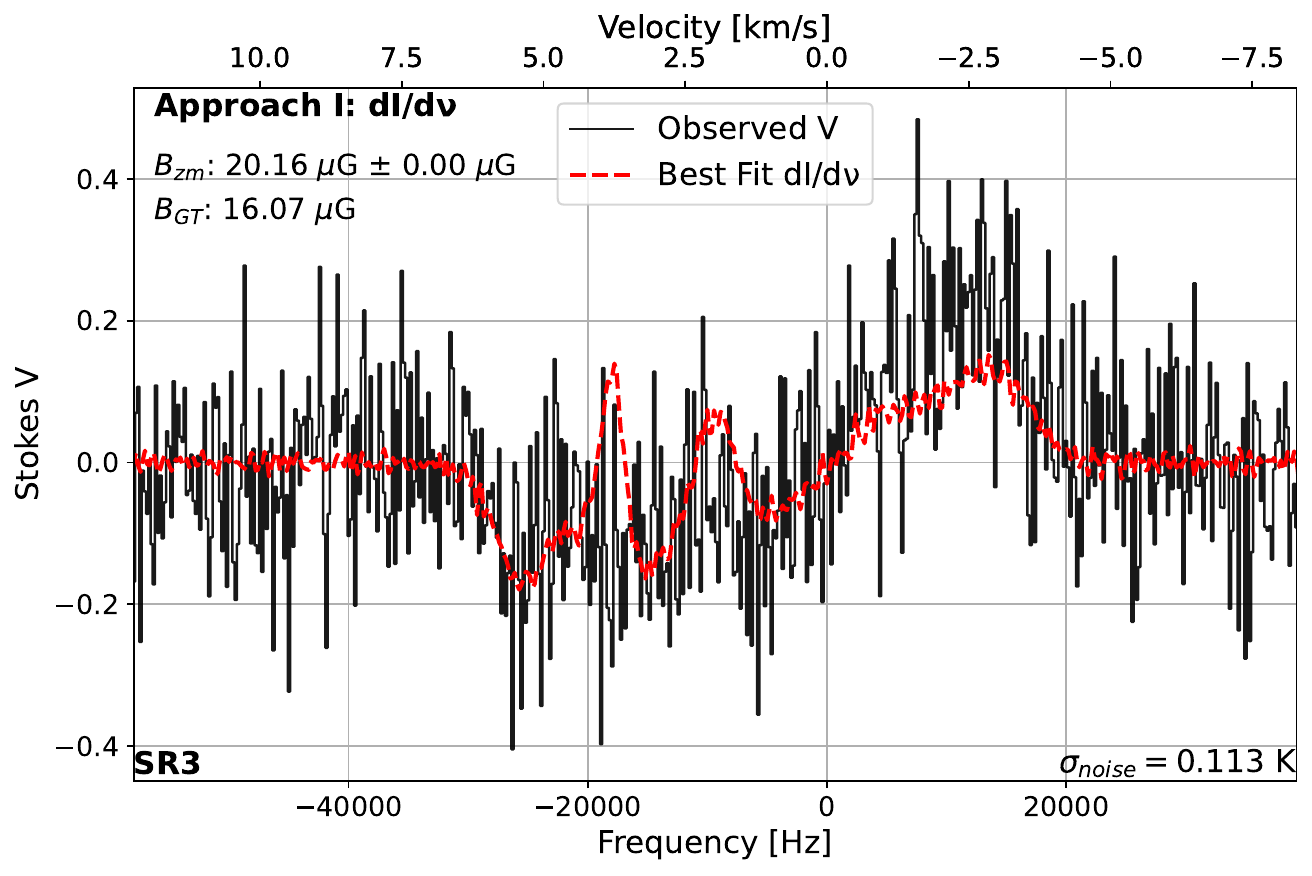}
\caption{Same as Figure~\ref{fig.ZeemanFitting_SR2_dI_dnu_fitting_noise_All}, but for the SR3 subregion.
}
\label{fig.ZeemanFitting_SR3_dI_dnu_fitting_noise_All}
\end{figure}


In this appendix, we present the Approach I fitting results for the Stokes $V$ spectra in two representative subregions, SR2 and SR3, under a range of noise conditions, including the noise-free case and noise levels of 0.007, 0.014, 0.028, 0.057, and 0.113 K (see Section \ref{sec:addnoise}).

\xdtwo{As shown in Figures~\ref{fig.ZeemanFitting_SR2_dI_dnu_fitting_noise_All} and \ref{fig.ZeemanFitting_SR3_dI_dnu_fitting_noise_All}, at all noise levels the overall fit roughly captures the amplitude and profile, even if the optimized scaled model does not precisely match the Stokes $V$ line shape. The value of $B_{zm}$ from the fit decreases only slightly as the noise increases,} while the ground truth estimator $B_{GT}$ is unchanged. This confirms that the inferred magnetic field strength from the fit remains robust across noise levels, consistent with the statistical analysis in Section~\ref{Assessing Uncertainties in Approach I: Derivative-Based Stokes $I$ Fitting} (Figure \ref{fig.hist_plot_Bmag_error_dI_dnu}).

\section{Zeeman Magnetic Field Estimates for Modified Cubes with Reduced Optical Depth: Approach I}
\label{app:approachIstatistics0001}

In this appendix, by controlled modification of the cube, we examined the effect of optical depth, another key factor that could affect magnetic field estimates from Zeeman measurements. 
Optical depth complicates the spectra through attenuation and so we simulated an optically thin regime by generating synthetic HI observations with the gas densities, most importantly $n_{\mathrm{HI}}$, reduced by a factor of 1000.

We then tested the ability of Zeeman analysis to recover the ground truth estimators of the field in these modified cubes. In the optically thin case, all HI gas is visible and the complete LOS magnetic field information is in principle able to influence the $V$ spectra from which $B_{zm}$ is derived. 


\begin{figure}[hbt!]
\centering
\includegraphics[width=0.68\linewidth]{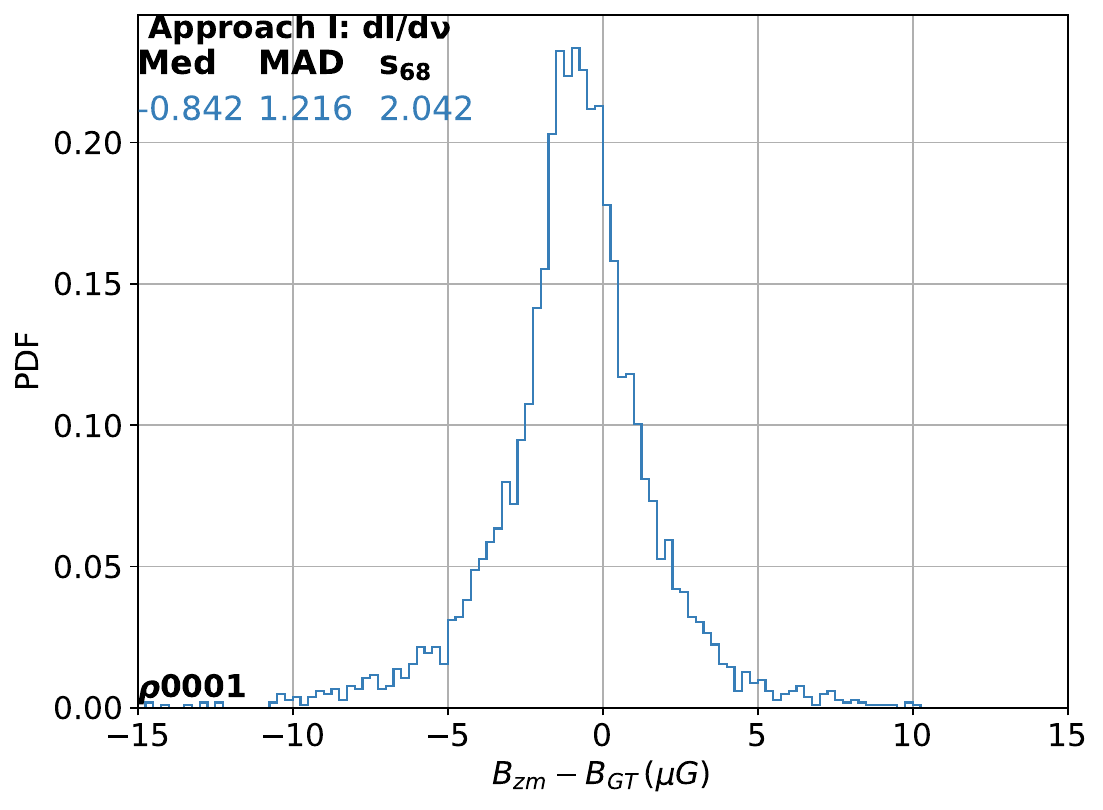}
\caption{Like Figure \ref{fig.hist_plot_Bmag_error_dI_dnu}, but after reducing the densities by a factor of 1000 (no noise case).
}
\label{fig.hist_plot_Bmag_error_dI_dnu_rho0001}
\end{figure}


As in the statistical study in Section \ref{Assessing Uncertainties in Approach I: Derivative-Based Stokes $I$ Fitting}, we derived $B_{zm}$ for spectra across the entire cube.
The ground truth values $B_{GT}$ used in Approach I, whatever the optical depth, are simply the mass-weighted LOS values (Equation \ref{eq.BGT_approch2}), which do not account for optical depth effects at all, and thus should be more appropriate here. Figure~\ref{fig.hist_plot_Bmag_error_dI_dnu_rho0001} presents the PDF of the difference between the Zeeman-estimated and ground truth LOS magnetic field strengths, for which the value of
$s_{\textrm{68}}$ 
is about 2.0~$\mu$G. This is marginally better than the results from the original (unscaled) density case that uses the same ground truth values, where $s_{\textrm{68}}$ is about 2.3~$\mu$G (Figure \ref{fig.hist_plot_Bmag_error_dI_dnu}). Optical depth does not appear to be the dominant source of uncertainty in the magnetic field estimation via Approach I.  The main issue is trying to describe the complex field by a single number.

\section{Zeeman Magnetic Field Estimates for Modified Cubes with Reduced Optical Depth: Approach II}
\label{app:approachII0001}

For the simulated optically thin regime with $n_{\mathrm{HI}}$ reduced by a factor of 1000 there are no clues from attenuation in the synthetic spectra, and so the order of components (Approach II) will not be meaningful. Gaussian decomposition has previously been applied to optically thin HI spectra at high Galactic latitude without considering magnetic fields (e.g., \citealt{2018A&A...619A..58K}; \citealt{2022ApJ...937...81T} using the spatially regularized ROHSA algorithm). In our case, we performed the decomposition using the full radiative-transfer model, rather than the simplified optically thin form where $T_{ex}$ is no longer a fitting parameter.

\subsection{Approach II: Spotlight on SR3\_$\rho$0001}
\label{Approach II: Component-Based Gaussian Decomposition Reduced Density}

As in Section \ref{Individual Region Analysis Spotlight on SR3}, we focus component-based analysis on the SR3 line of sight, now with the density reduced by 1000 (SR3\_$\rho$0001).

\begin{figure}[hbt!]
\centering
\includegraphics[width=0.48\linewidth]{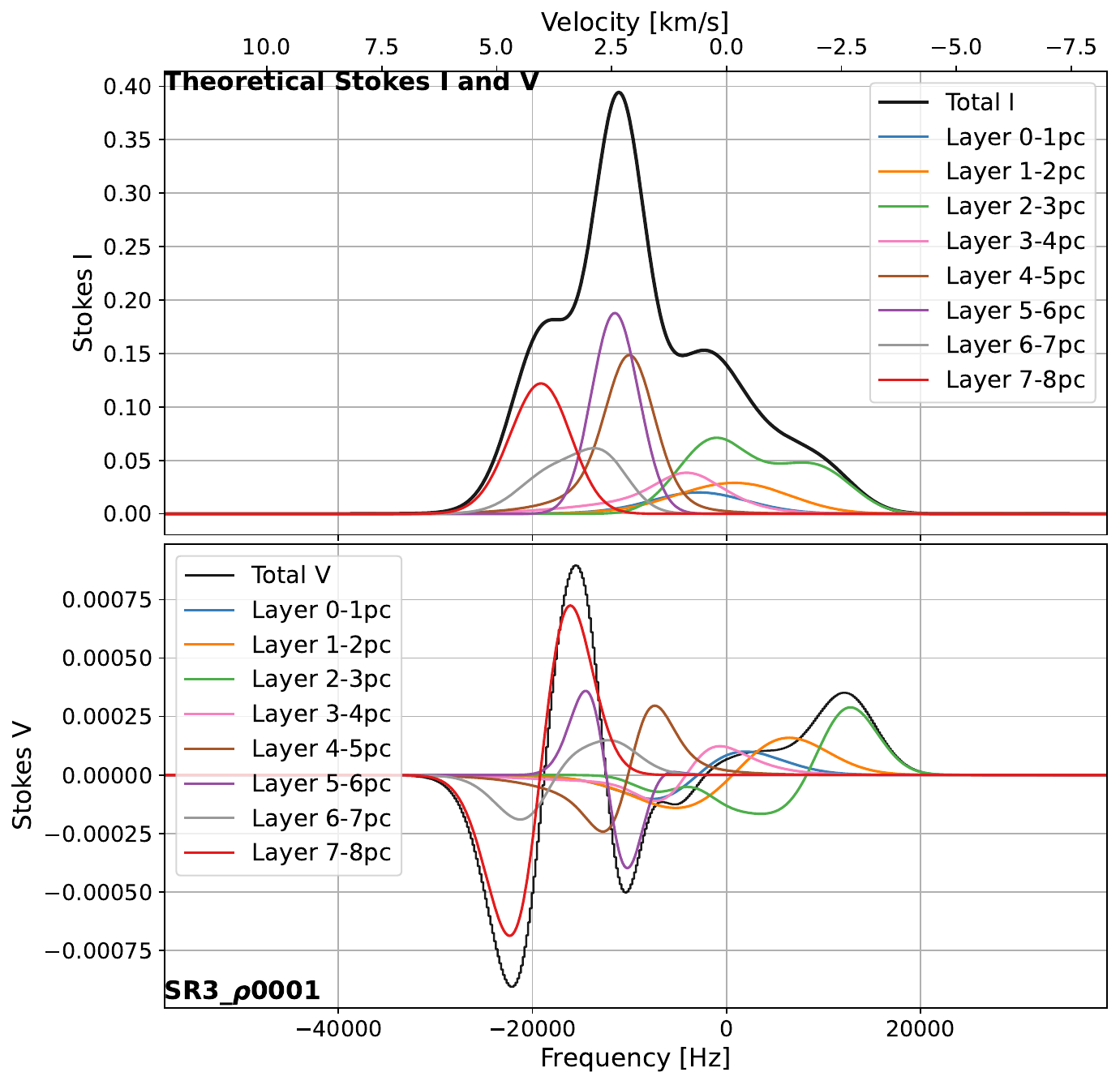}
\caption{Same as Figure~\ref{fig.SR3_GPB},  but for layer contributions in the SR3 subregion
with the gas density decreased by a factor of 1000 (optically thin emission unaffected by attenuation).}
\label{fig.SR3_rho0001_GPB}
\end{figure} 

\subsubsection{Ground truth layer contributions for SR3\_$\rho$0001}
\label{sec:gtlayerSR30001}

Figure \ref{fig.SR3_rho0001_GPB} shows the layer contributions to the spectra.  It is to be compared to Figure \ref{fig.SR3_GPB} Section \ref{sec:gtlsr3} where the effects of attenuation are pronounced.  One example is that here L45 and L56 both make strong contributions to the central peak in Stokes $I$.  But because the gas in these layers is at similar velocities, L56 (purple) in Figure \ref{fig.SR3_GPB} is very weak, being heavily attenuated by L45.

In the optically thin case all of the gas can be seen and so the properties of the layer contributions to $I$ and $V$, including the inferred magnetic field, can be interpreted readily in terms of the physical properties in Figure \ref{fig.physical_Region_All}: intrinsic emission $T_{ex}\cdot \tau_{HI,\, Center}$, velocity $v_z$ and LOS field $B_z$. 
The four layers (L23, L45, L56, and L78) where the column density is highest ($T_{ex}\cdot \tau_{HI,\, Center}$) make the strongest contributions.

The only place along the LOS that $v_z < 0$ is in L23, and so that layer (green) produces the negative velocity wing of the spectra. Likewise, gas in L78 has the highest $v_z$ and so dominates the high velocity wing in $I$ and the characteristic S shape there in $V$.

Because $B_z$ is mostly positive along the LOS, most layer contributions to $V$ have a familiar profile like L78, beginning with a negative excursion on the left and then cross over to a positive excursion at lower velocities. The $V$ profile for L56 is exceptionally reversed, because only in that layer is $B_z$ negative.

\begin{figure*}[hbt!]
\centering
\includegraphics[width=0.98\linewidth]{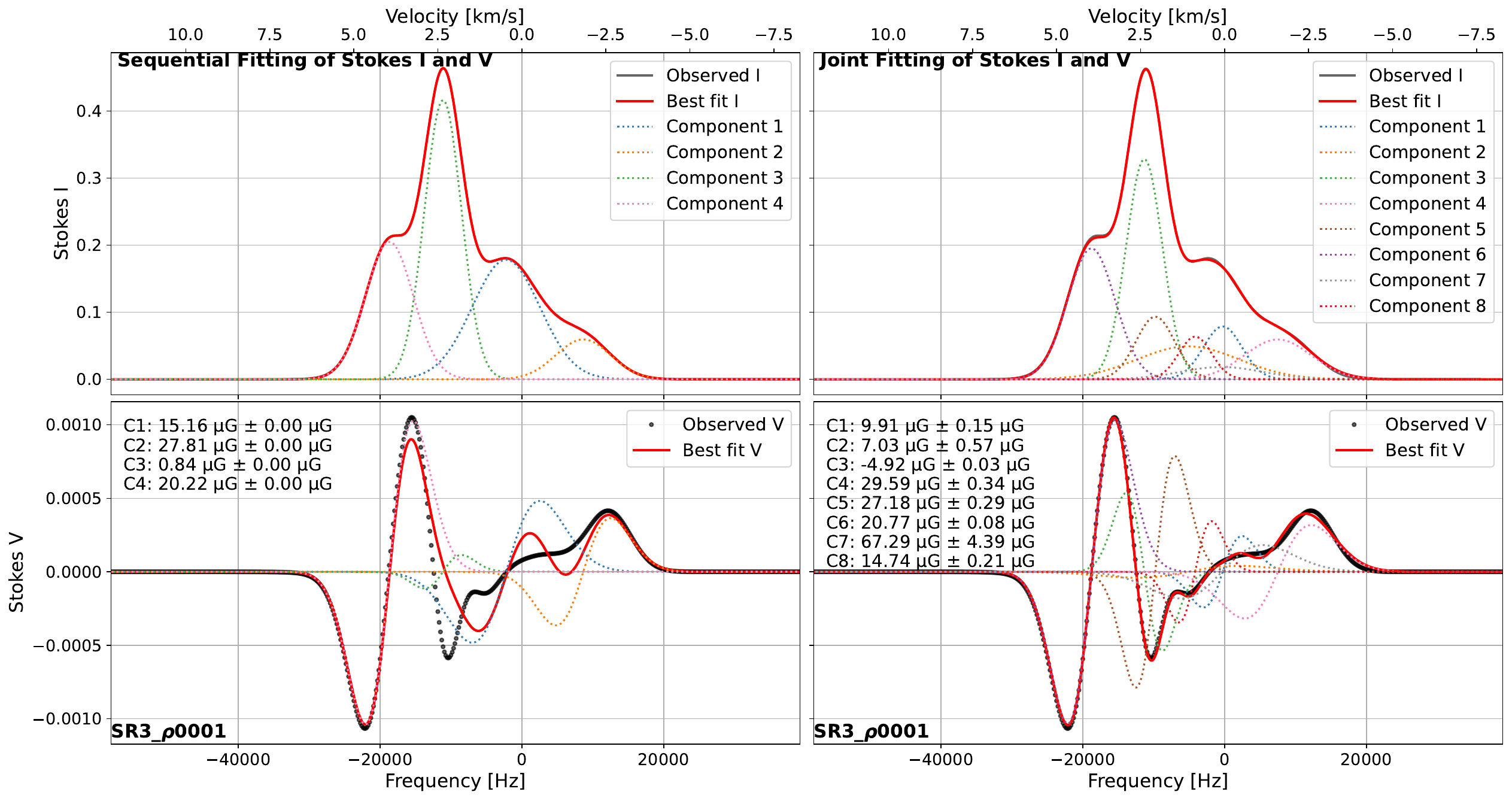}
\caption{Same as Figure~\ref{fig.ZeemanFitting_SR3_Breal_noise0}, \xd{but showing Approach II fitting results for the Stokes $I$ and Stokes $V$ spectra for} the SR3 subregion with gas densities decreased by a factor of 1000. The joint strategy yields a reasonably good fit for both Stokes $I$ and $V$, whereas the sequential strategy has difficulty accurately reproducing the Stokes $V$ spectrum.}
\label{fig.SR3_I_V_fitting_rho0001_}
\end{figure*} 

\subsubsection{Approach II Zeeman analysis results for SR3\_$\rho$0001}
\label{sec:resultsSR3reduced}

Paralleling the analysis in Figure \ref{fig.ZeemanFitting_SR3_Breal_noise0} Section \ref{sec:sr30}, Figure~\ref{fig.SR3_I_V_fitting_rho0001_} presents the results for the SR3 subregion at this reduced density. For the sequential strategy, the Stokes $I$ profile is well reproduced and plausibly interpreted in terms of Figure \ref{fig.physical_Region_All} lower left. But the Stokes $V$ profile shows substantial discrepancies.  As can be appreciated from the layer contributions, four components is not sufficient.

The joint strategy uses eight components, the same number as the number of layers.  Each component produces an S profile in $V$. 
Only C3 is reversed, with $B_z \simeq -5 \mu$G, quantitatively consistent with association with L56.  Like L56, it contributes to the second (from left) negative excursion in $V$.
The layer contributions are of course not Gaussian.  Nor are the components in any depth order.  Nevertheless, we can note some other correspondences.
C6 (purple) is reminiscent of L78 in both $I$ and $V$ and has $B_z \simeq 20 \mu$G as expected from Figure \ref{fig.physical_Region_All}.
C4 (pink) corresponds to the only negative excursion in $v_z$, thus part of L23; the expected $B_z$ would be about $30 \mu$G, close to that fitted.
C1 (blue) might correspond to the second peak of L23, from the deeper end with a low velocity and rising field.
Not all is well.  C7 (grey) is a weak broad component in $I$ that gets an unphysically high $B_z$ to contribute to a confused overlap region in $V$.

\begin{figure*}[hbt!]
\centering
\includegraphics[width=0.98\linewidth]{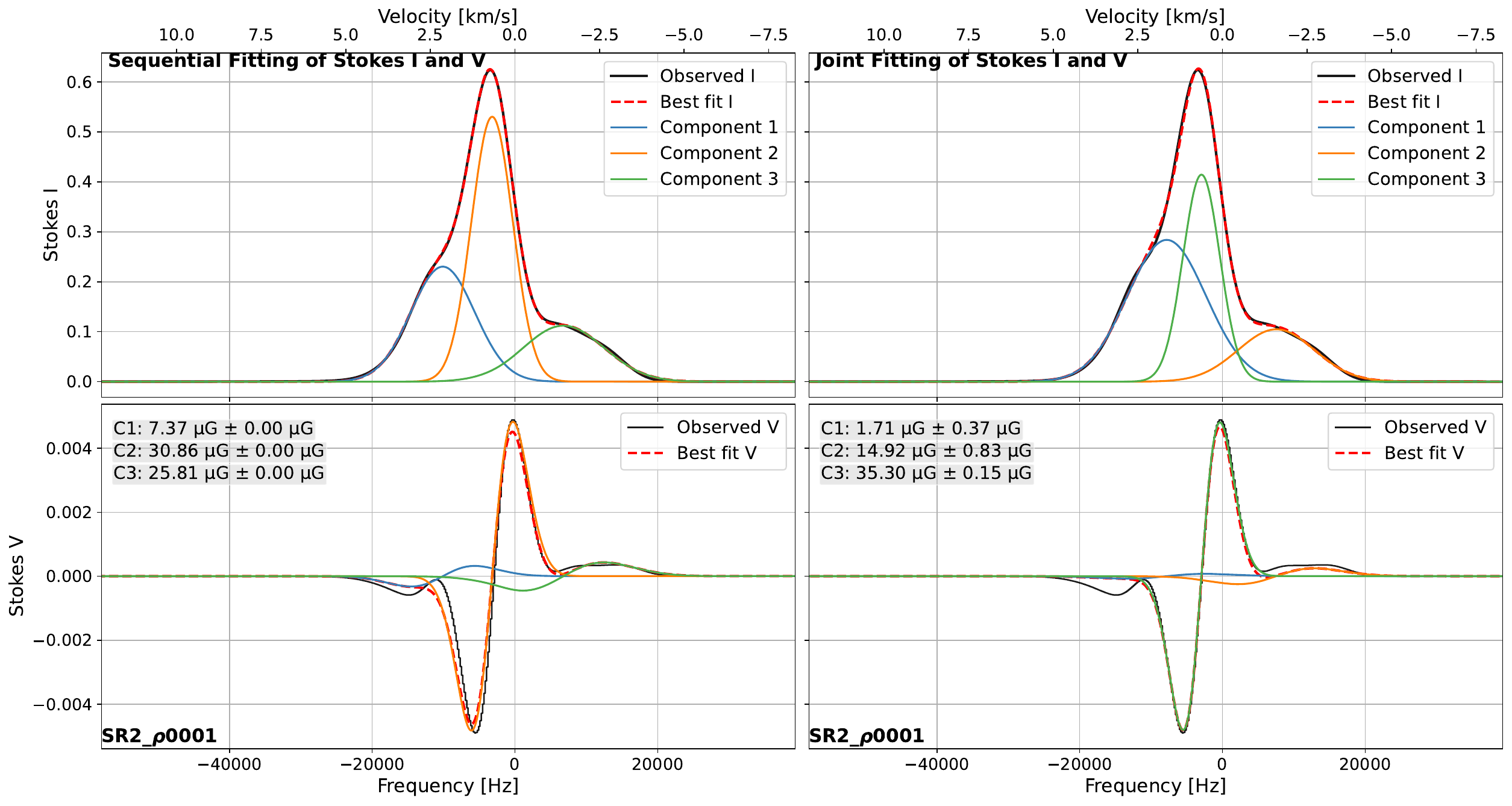}
\caption{Same as Figure~\ref{fig.ZeemanFitting_5_Layer_IV_noise0}, \xd{but showing Approach II fitting results for the Stokes $I$ and Stokes $V$ spectra for} the SR2 subregion with the gas densities decreased by a factor of 1000. Both the sequential and joint strategies provide reasonably good fits to both the Stokes $I$ and $V$ spectra with only a few components.}
\label{fig.SR2_I_V_fitting_rho0001_}
\end{figure*} 

\subsection{Approach II Zeeman analysis results for SR2\_$\rho$0001}
\label{sec:resultsSR2reduced}

Figure~\ref{fig.SR2_I_V_fitting_rho0001_} shows another example of the fitting results for the Stokes $I$ and $V$ spectra in the SR2 subregion under these reduced-density conditions. The original spectra for comparison are in Figure \ref{fig.ZeemanFitting_SR2_I_V_fitting_noise_All} (upper panels).  Note the relatively extended wings in the original beyond $\pm 20000$ Hz. These are actually optically thin and when reduced by a factor 1000 match the profile wings that are invisible at the scale in Figure~\ref{fig.SR2_I_V_fitting_rho0001_}.  But interior to the wings, the profiles are dramatically different because of the effects of optical depth.

A reduced number of Gaussian components is used in the spectral fit in the optically thin regime.  
There is no meaning to the order of the components, because there are no clues in the spectra produced by optical depth.  Note that C1, C2, and C3 from the sequential strategy correspond roughly to C2, C1, and C4 in the joint strategy.
Because all of the gas can be seen, the properties of the spectral components, including the inferred magnetic field, can be interpreted plausibly in terms of the properties in Figure \ref{fig.physical_Region_All}: intrinsic emission $T_{ex}\cdot \tau_{HI,\, Center}$, velocity $v_z$ and LOS field $B_z$.

While the spectral complexity is significantly reduced in the optically thin regime, Zeeman fitting fails to reproduce the Stokes $V$ profile with precision using either fitting strategy. 

\begin{figure}[hbt!]
\centering
\includegraphics[width=0.48\linewidth]{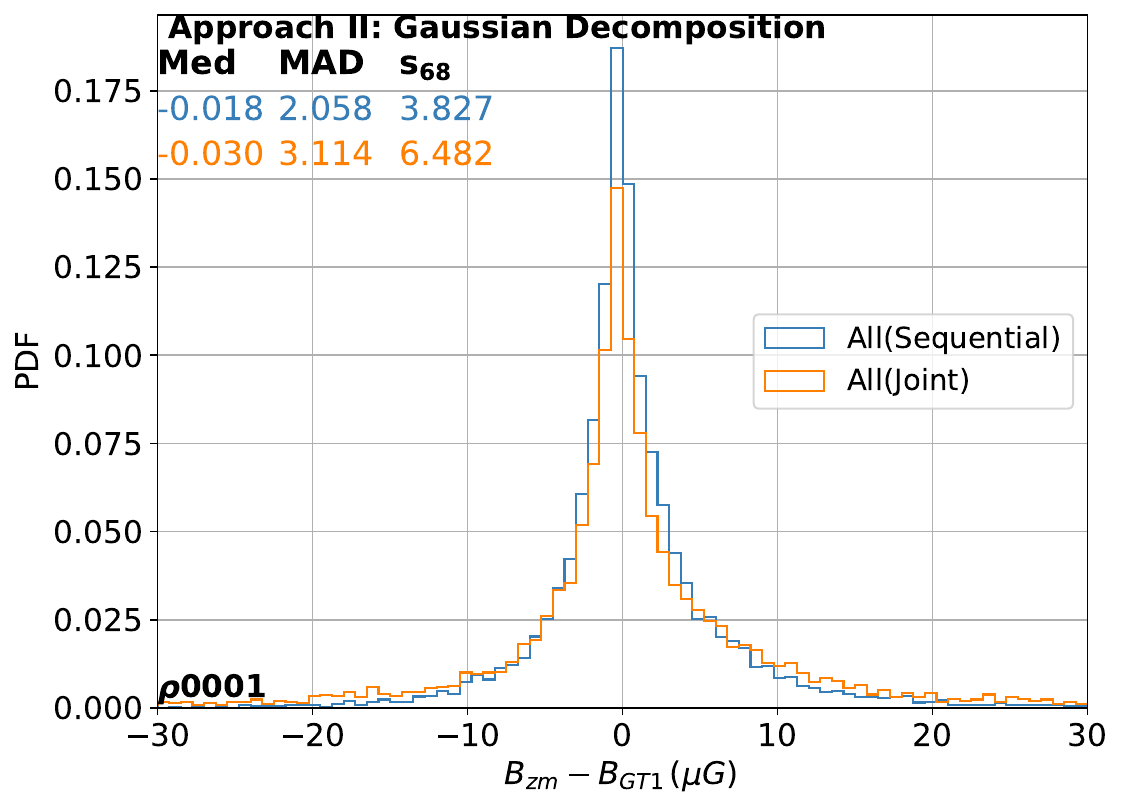}
\includegraphics[width=0.48\linewidth]{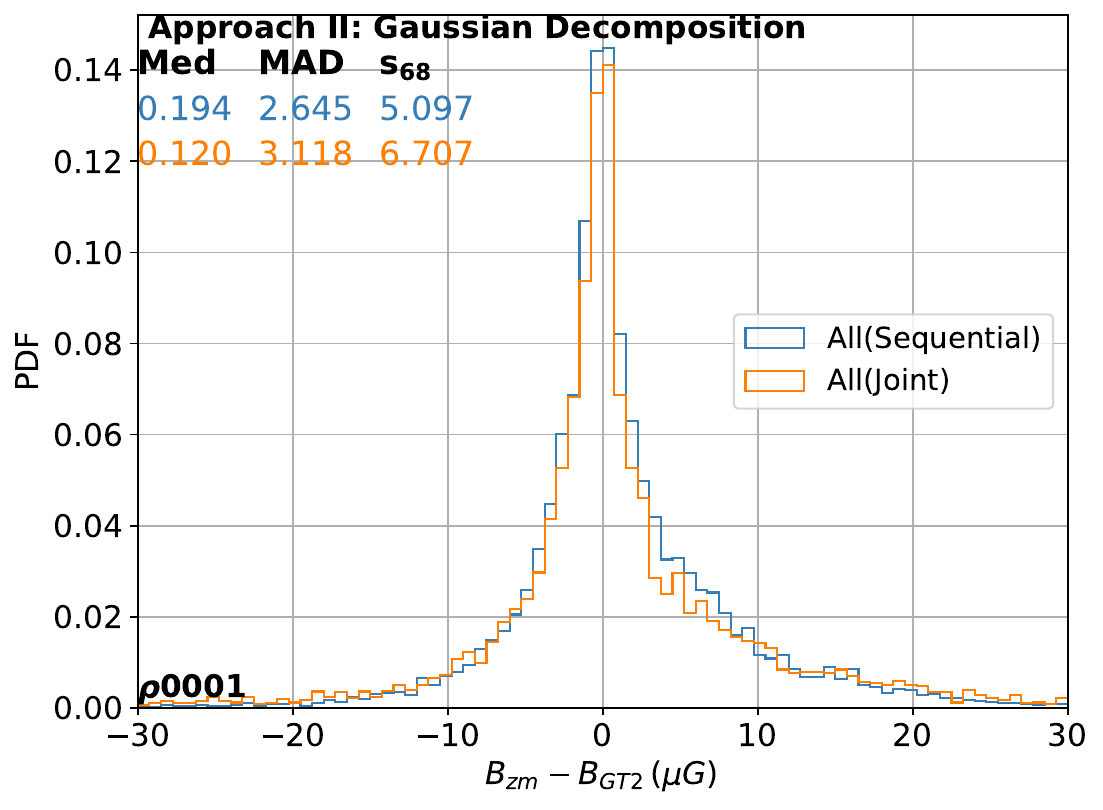}
\caption{Like PDFs in Figure \ref{fig.hist_plot_Bmag_error_GT1_GT2}, but after reducing the densities by a factor of 1000 and for the no noise case).
}
\label{fig.hist_plot_Bmag_error_GT1_GT2_rho0001}
\end{figure}

\subsection{Approach II: Statistics}
\label{app:appIIstats001}

Figure~\ref{fig.hist_plot_Bmag_error_GT1_GT2_rho0001} shows the distribution of the difference between Zeeman-estimated magnetic field strengths (via Approach II) and the ground truth values defined by Equations~\ref{eq.BGT1} and \ref{eq.BGT2}, using low-density, optically thin synthetic data. Despite the optically thin conditions, the heavy-tailed error distribution still exhibits a substantial width. Notably, in this test the sequential strategy achieves better accuracy than the joint strategy, with 
a value of $s_{\textrm{68}}$ about 3.8 $\mu$G and 5.1 $\mu$G 
compared to 6.5 $\mu$G and 6.7 $\mu$G 
for the two ground-truth estimators, respectively. 
For comparison, the corresponding 
values of $s_{\textrm{68}}$ 
for the original spectra (Figure \ref{fig.hist_plot_Bmag_error_GT1_GT2}) are 
about 5.2 $\mu$G and 6.1 $\mu$G (sequential) and about 4.7 $\mu$G and 4.8 $\mu$G (joint).
Note that the ground truth estimators in the two figures are now different. 
These results also support the conclusion that in Approach II optical depth is also not the dominant source of uncertainty in magnetic field estimation.

\bigskip\noindent
Overall, the findings here and in Appendix \ref{app:approachIstatistics0001} reinforce that optical depth, in and of itself, has limited impact on the uncertainty of Zeeman-based magnetic field estimates for both approaches and strategies, with uncertainty levels similar to those observed in the original (unscaled) density case discussed above. When the physical properties are changing in complicated ways as seen in Figure \ref{fig.physical_Region_All} it is challenging to quantify $B_z$ with just a few components.

\section{The SR3 LOS from the Reverse Perspective}
\label{app:flip}

\begin{figure}[hbt!]
\centering
\includegraphics[width=0.48\linewidth]{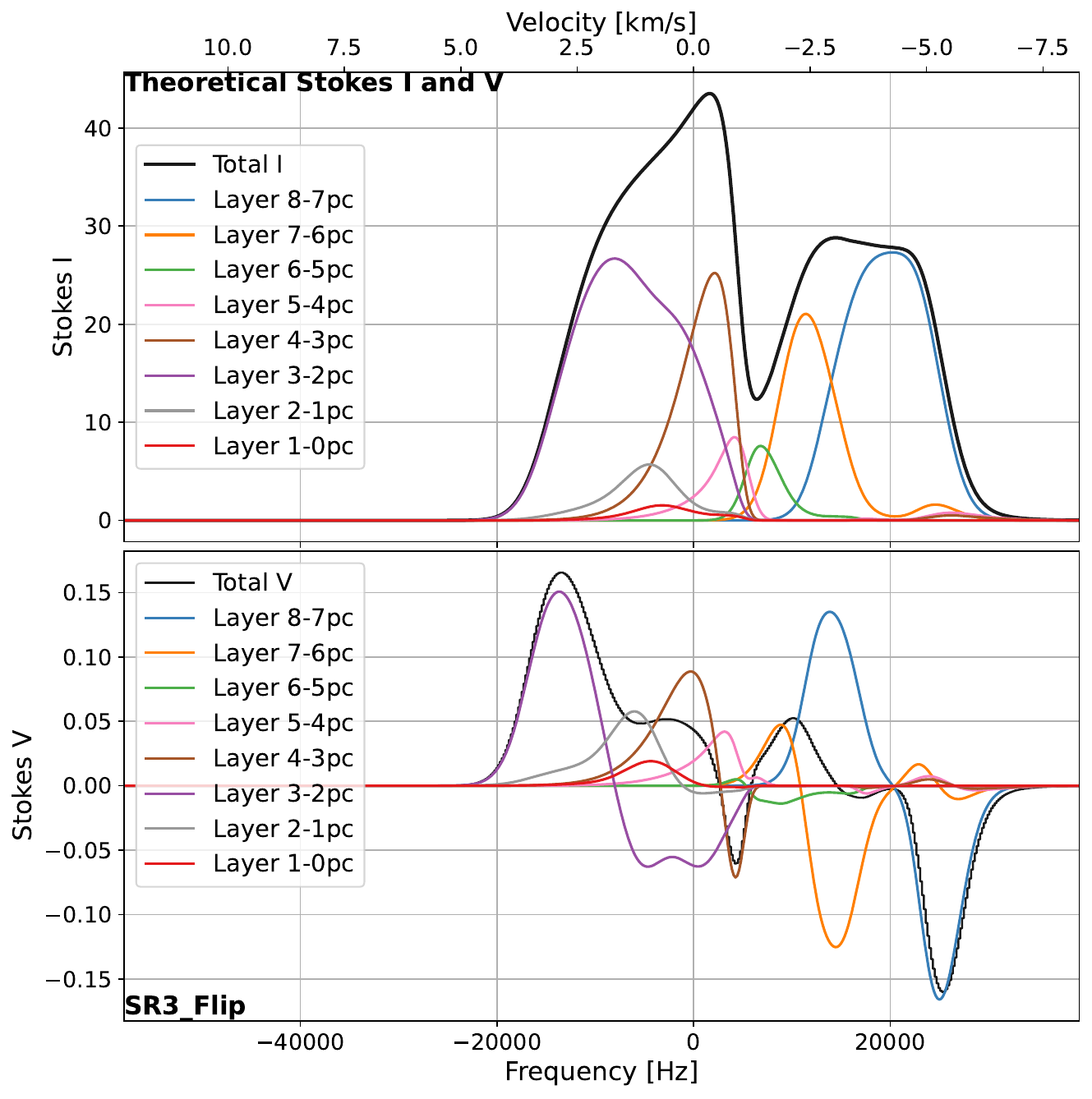}
\caption{Same as Figure~\ref{fig.SR3_GPB}, but \xd{showing the layer contributions} for the view along the $-z$ direction.  
}
\label{fig.SR3_flip_GPB}
\end{figure}

\begin{figure*}[hbt!]
\centering
\includegraphics[width=0.98\linewidth]{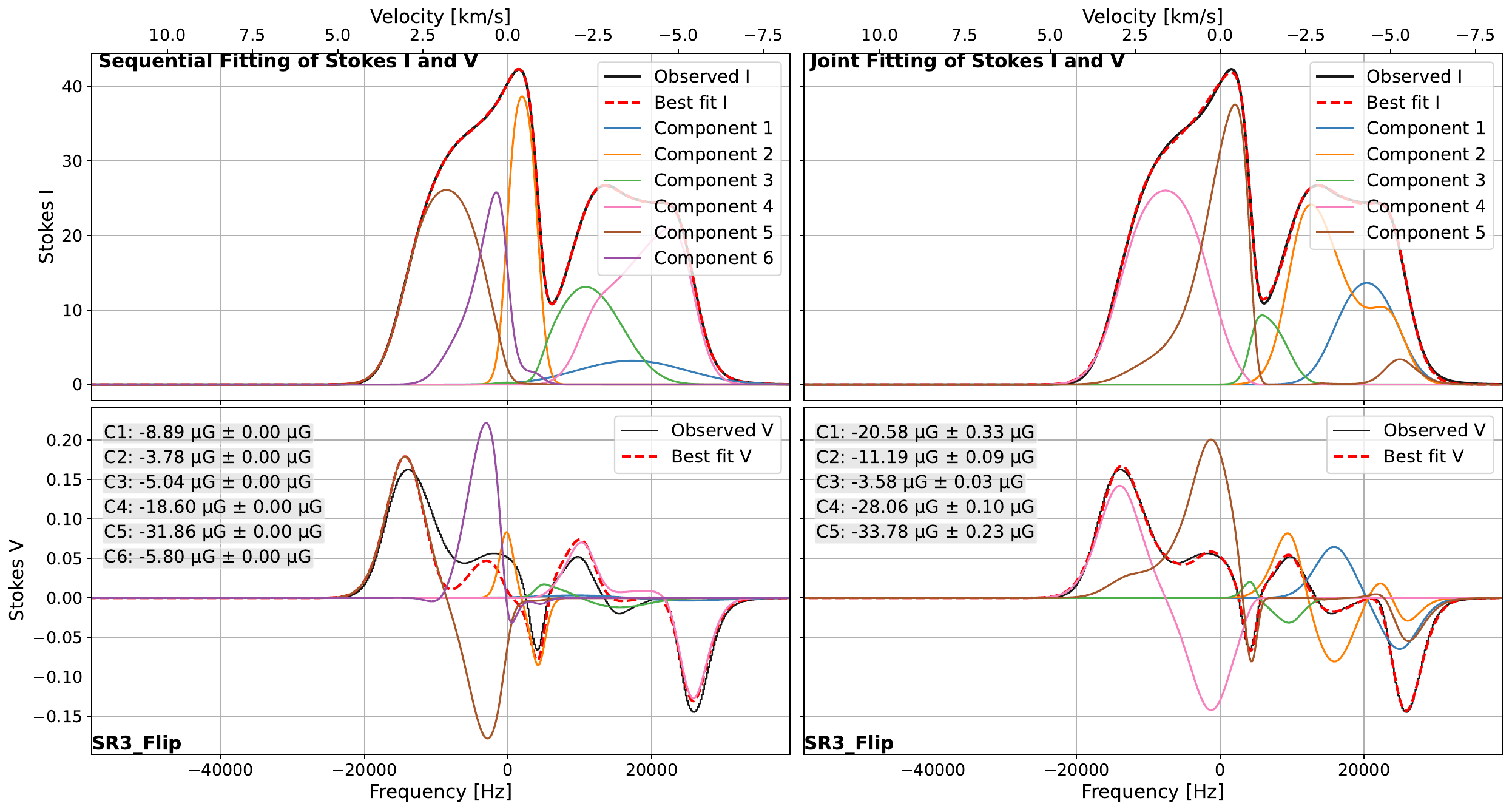}
\caption{Same as Figure~\ref{fig.ZeemanFitting_SR3_Breal_noise0}, \xd{but showing Approach II fitting results for the Stokes $I$ and Stokes $V$ spectra for} the view along the $-z$ direction. The Stokes $I$ and $V$ spectra differ markedly between the $+z$ and $-z$ viewing directions, accentuated by the effects of foreground attenuation. While both strategies provide reasonable fits to the Stokes $I$ spectrum, the joint strategy yields a more precise fit to the Stokes $V$ spectrum.}
\label{fig.ZeemanFitting_SR3_I_V_fitting_flip_}
\end{figure*} 

\begin{figure}[hbt!]
\centering
\includegraphics[width=0.48\linewidth]{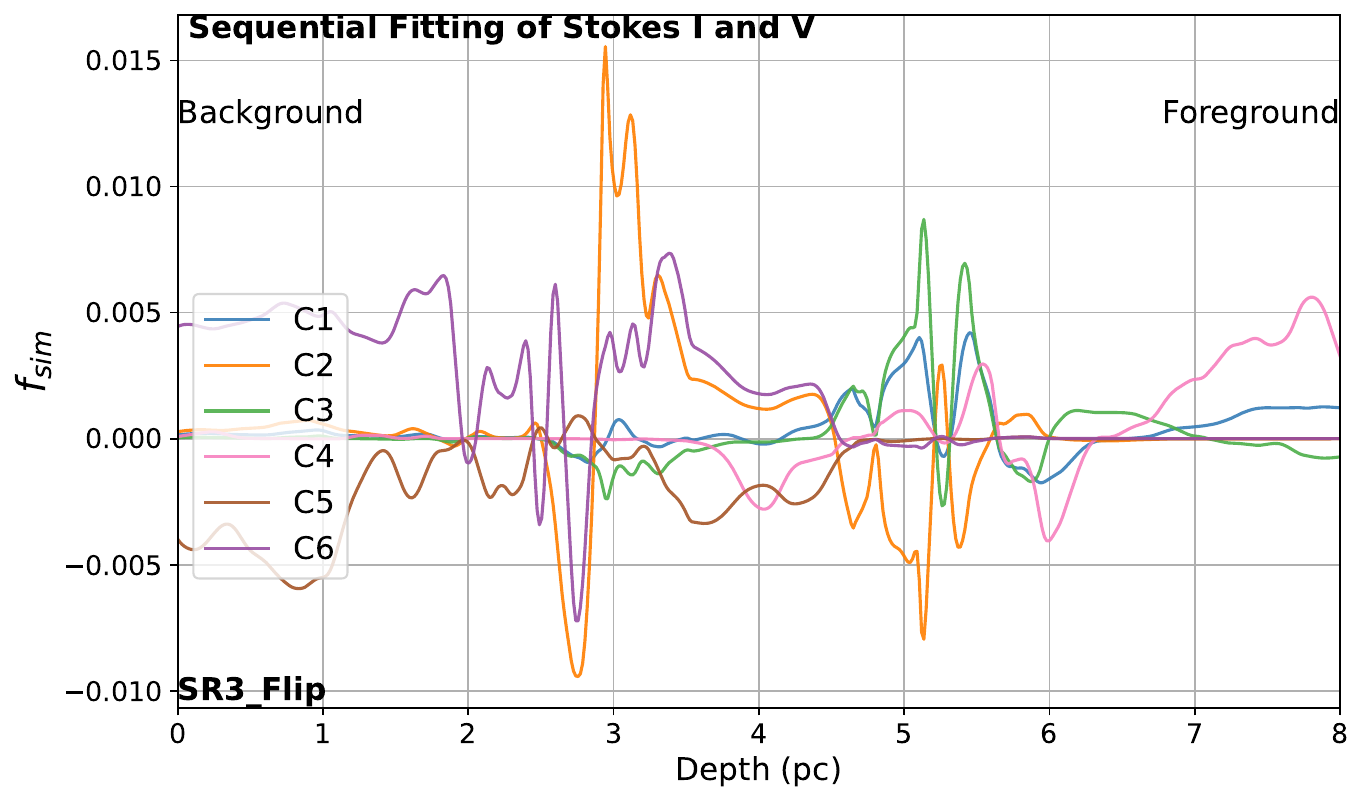}
\includegraphics[width=0.48\linewidth]{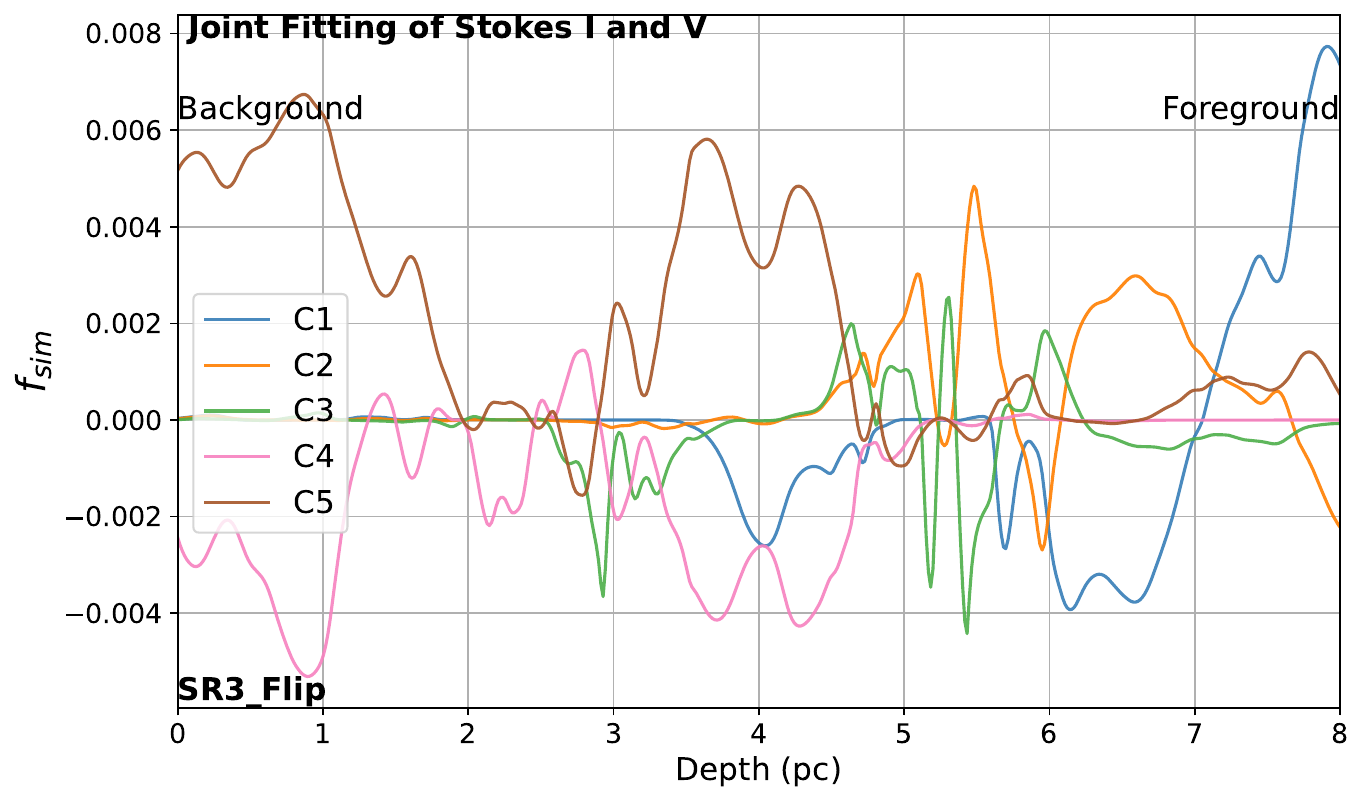}
\caption{Same as Figure~\ref{fig.f_sim_SR3}, but \xd{showing the profile of the similarity function $f_{\text{sim}}$ (Equation~\ref{eq.fsim})} for the view along the $-z$ direction. 
}
\label{fig.f_sim_SR3_flip}
\end{figure}

To test the physical reality of components and their magnetic fields we analyzed the same LOS (SR3) through the simulation cube from the opposite direction, where the foreground attenuation strongly influences the observed Stokes $I$ and $V$ spectra differently,
as shown in the layer contributions in Figure~\ref{fig.SR3_flip_GPB}.
The component contributions from the Approach II fitting shown in Figure \ref{fig.ZeemanFitting_SR3_I_V_fitting_flip_} are correspondingly different. 
Comparison with the depth-dependent physical properties, layer contributions, and $f_{\mathrm{sim}}$ (Figures \ref{fig.physical_Region_All} lower left,
\ref{fig.SR3_flip_GPB}, and \ref{fig.f_sim_SR3_flip}, respectively) again shows that the joint strategy tends to yield a more physically consistent association between fitted components and cloud layers than the sequential strategy.
%
Substantial spectral overlap and attenuation hinder finding a unique depth ordering or a clear one-to-one correspondence between fitted components from opposite viewing directions.

\section{Zeeman Fitting Gallery: Approach II Joint Strategy across Regions and Noise Levels}
\label{Zeeman Fitting Gallery: Approach II Across Noise and Regions}

\begin{figure}[hbt!]
\centering
\includegraphics[width=0.48\linewidth]{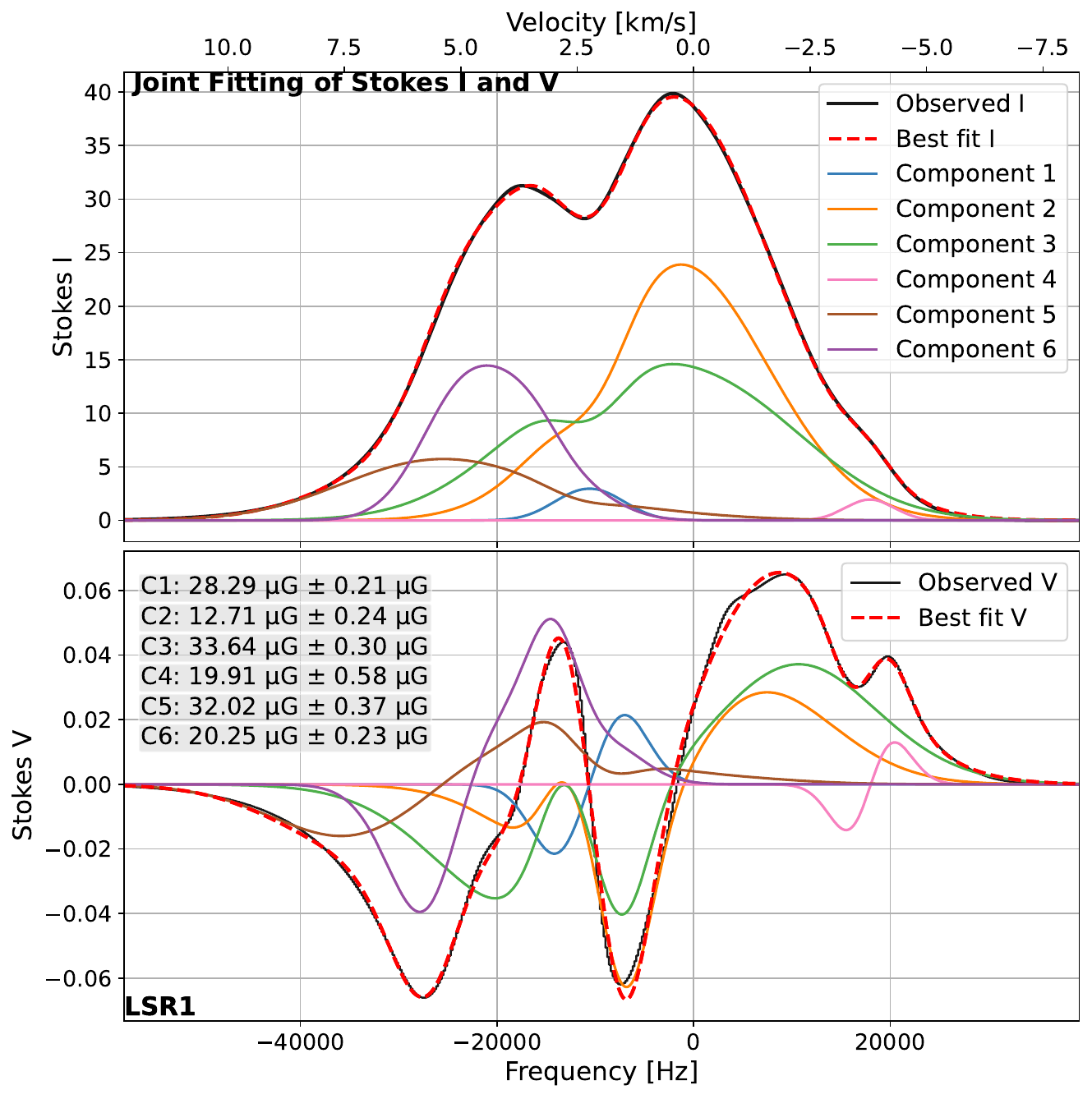}
\includegraphics[width=0.48\linewidth]{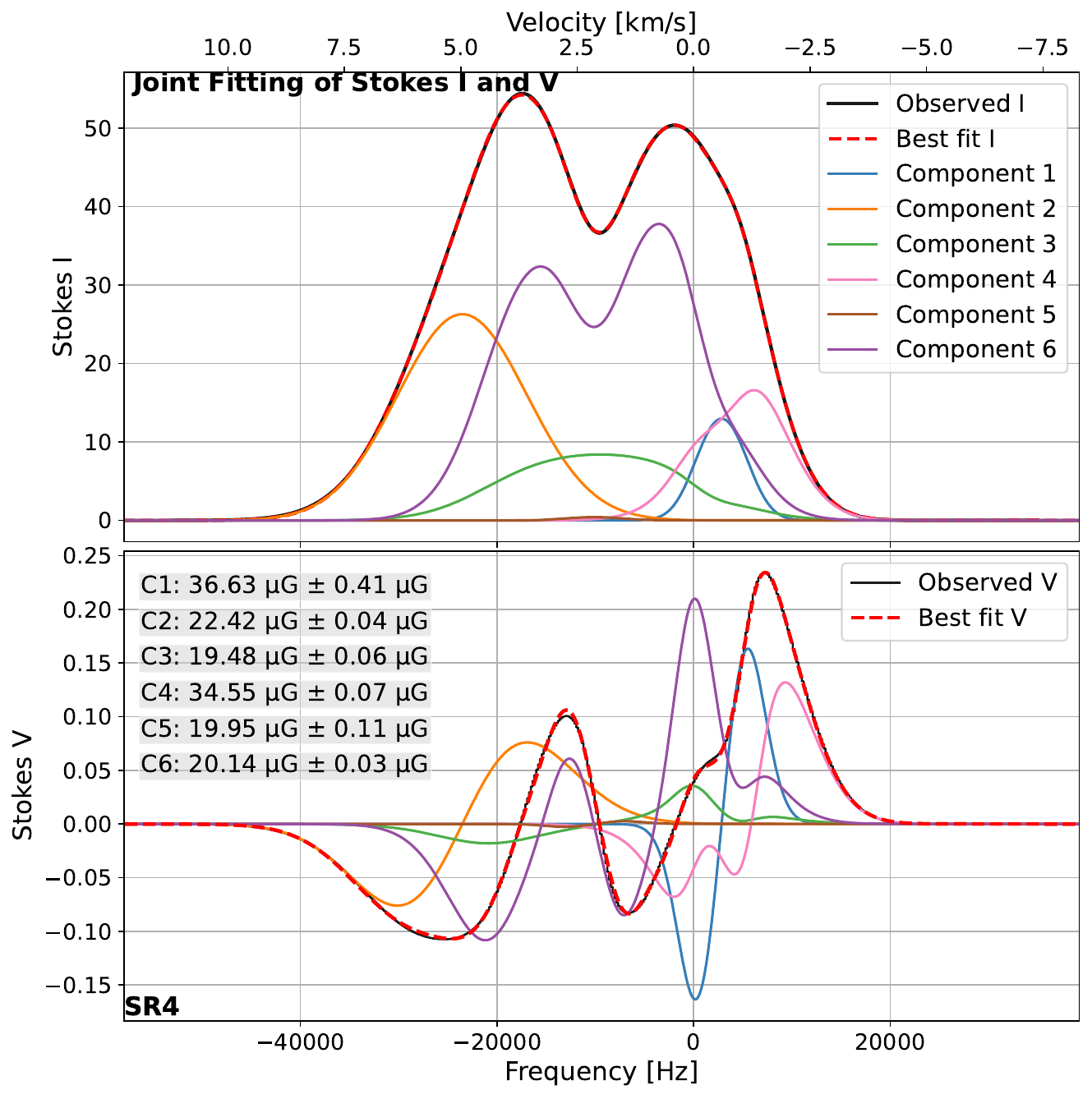}
\caption{
\xd{Approach II joint strategy fitting results for the Stokes $I$ and Stokes $V$ noise-free spectra of LSR1 and SR4, as for the noise-free results for SR3 in the right panel of Figure \ref{fig.ZeemanFitting_SR3_Breal_noise0} and SR2 in the upper left of Figure \ref{fig.ZeemanFitting_SR2_I_V_fitting_noise_All}.}}
\label{fig.ZeemanFitting_SR_ALL_I_V_fitting_nonoise}
\end{figure}

\begin{figure}[hbt!]
\centering
\includegraphics[width=0.32\linewidth]{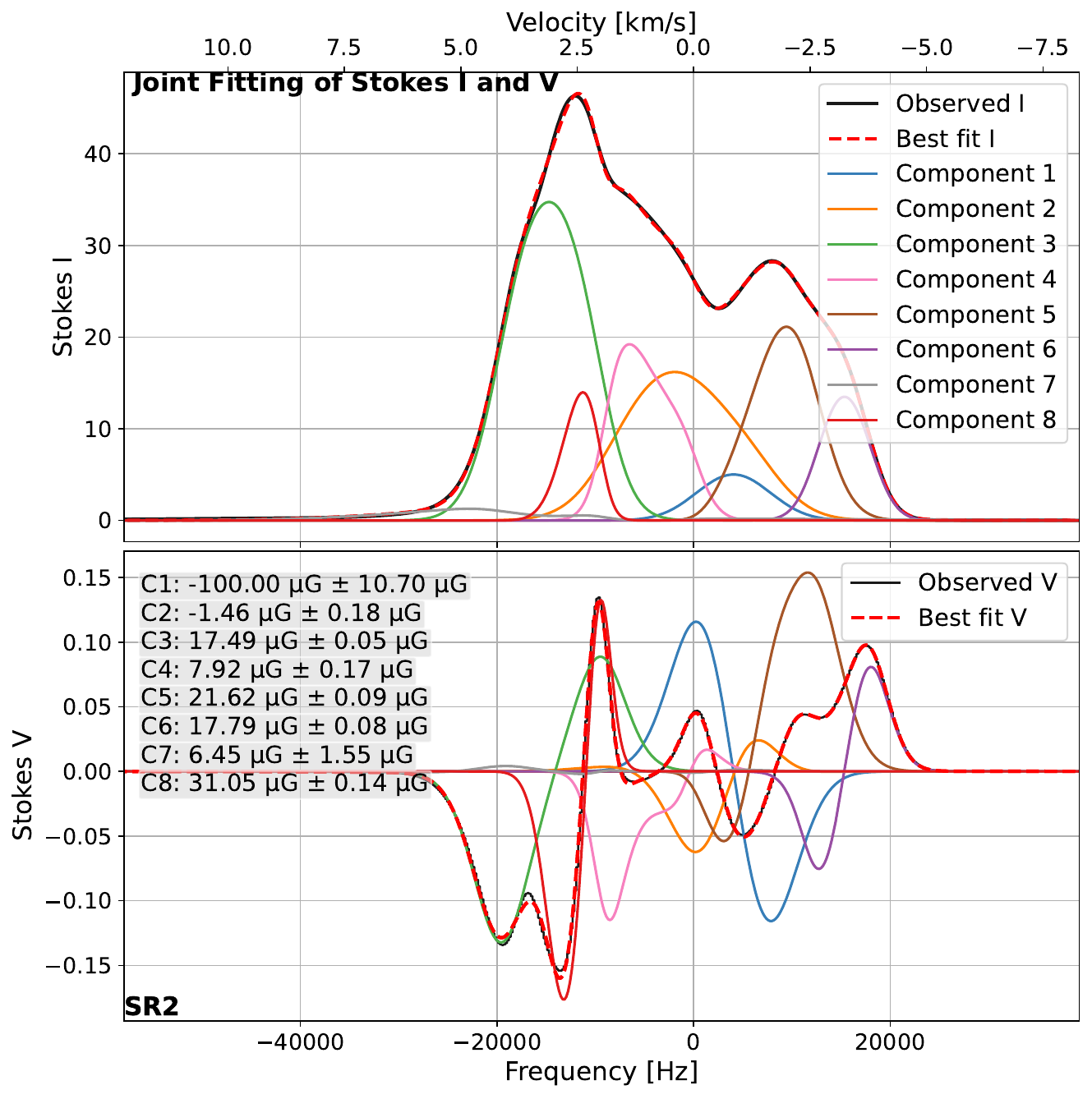}
\includegraphics[width=0.32\linewidth]{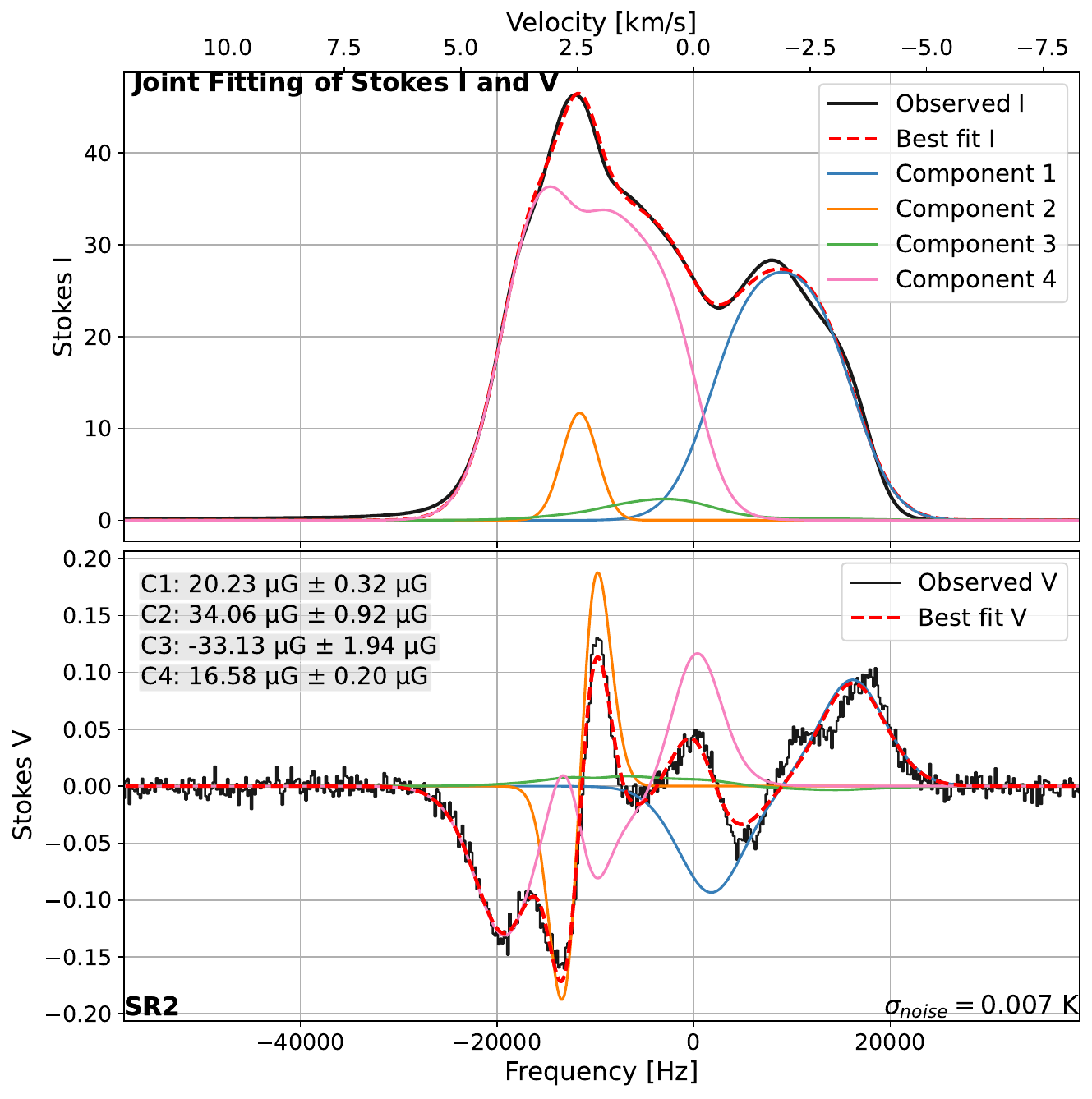}
\includegraphics[width=0.32\linewidth]{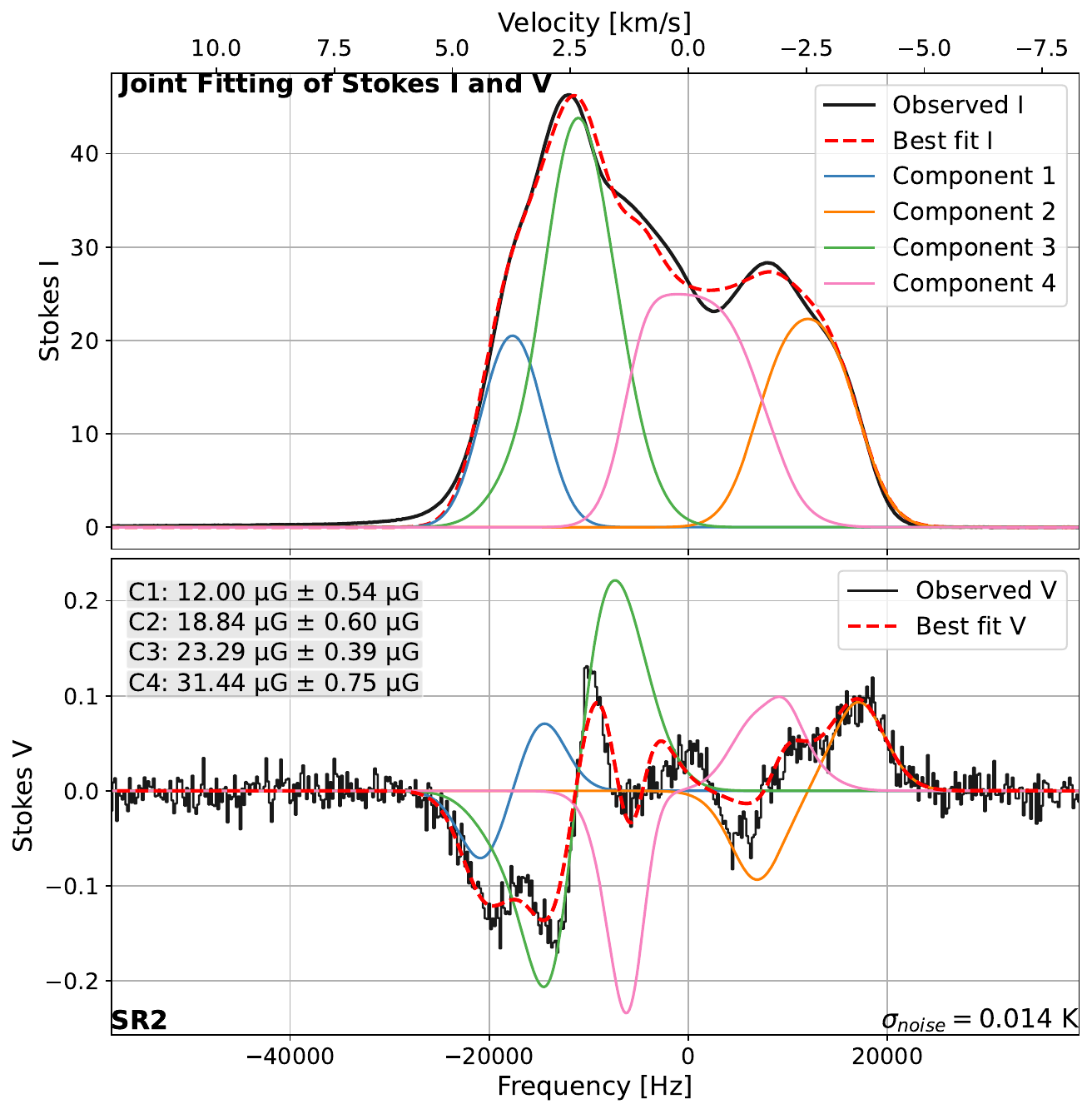}
\includegraphics[width=0.32\linewidth]{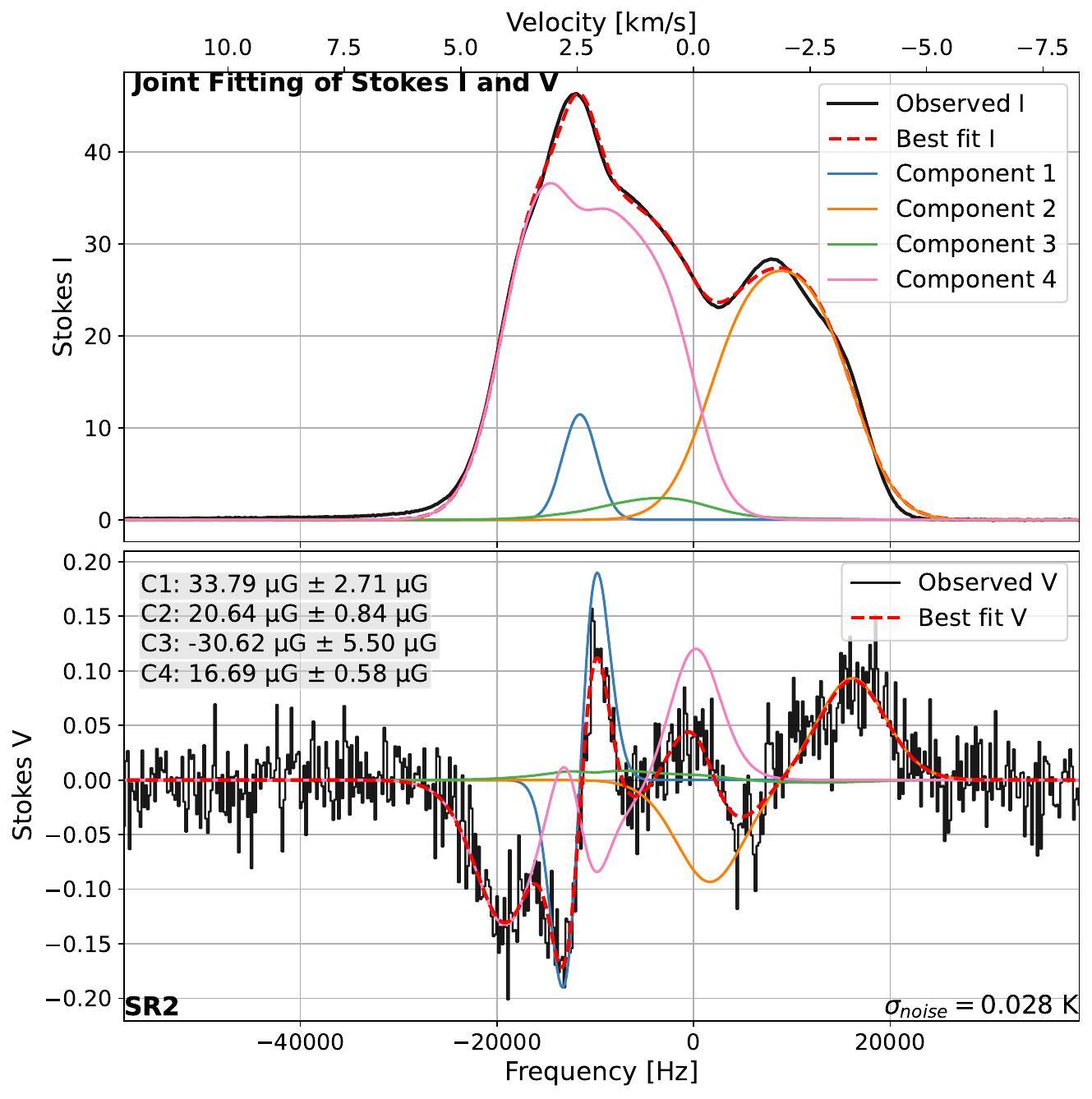}
\includegraphics[width=0.32\linewidth]{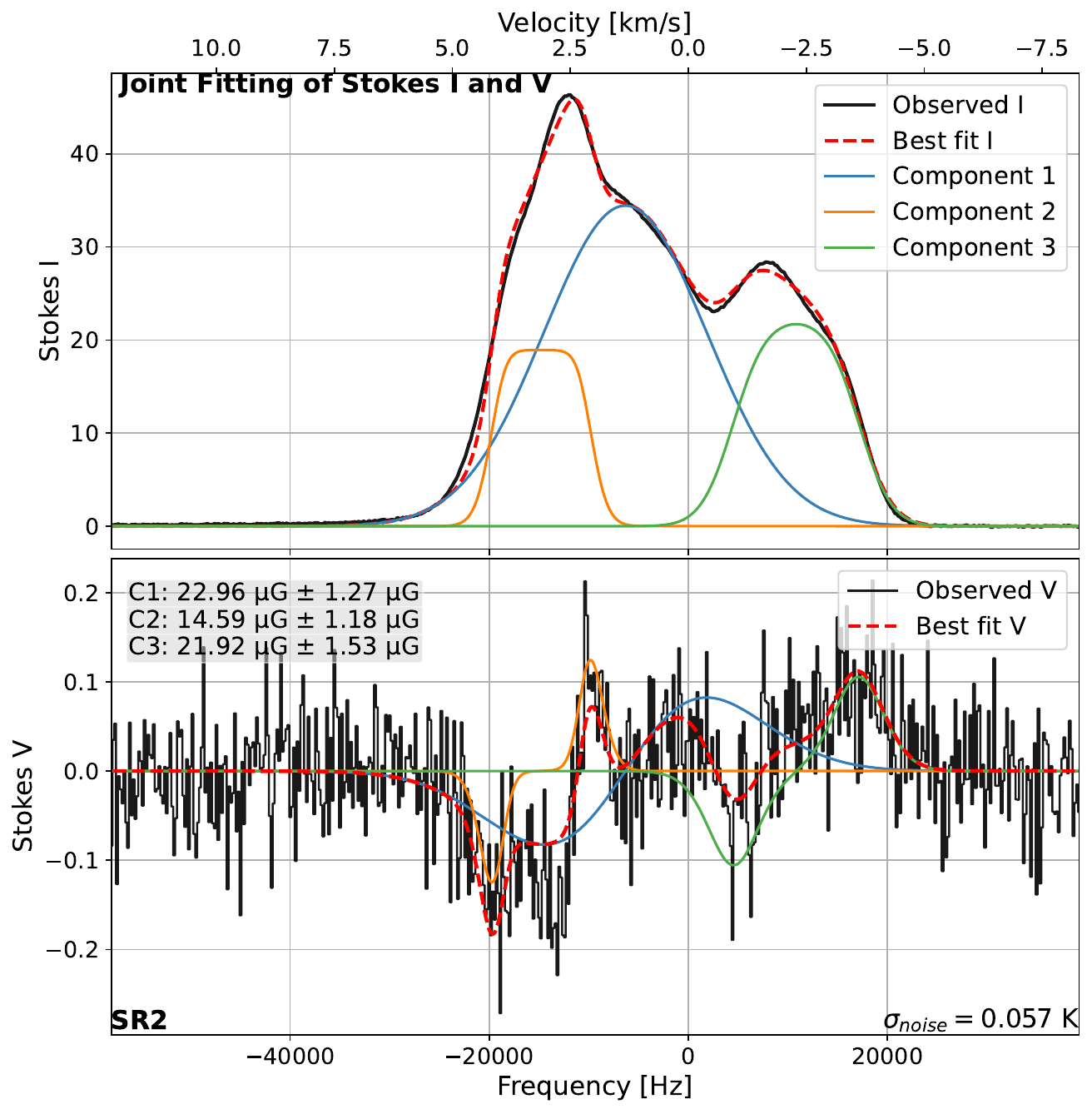}
\includegraphics[width=0.32\linewidth]{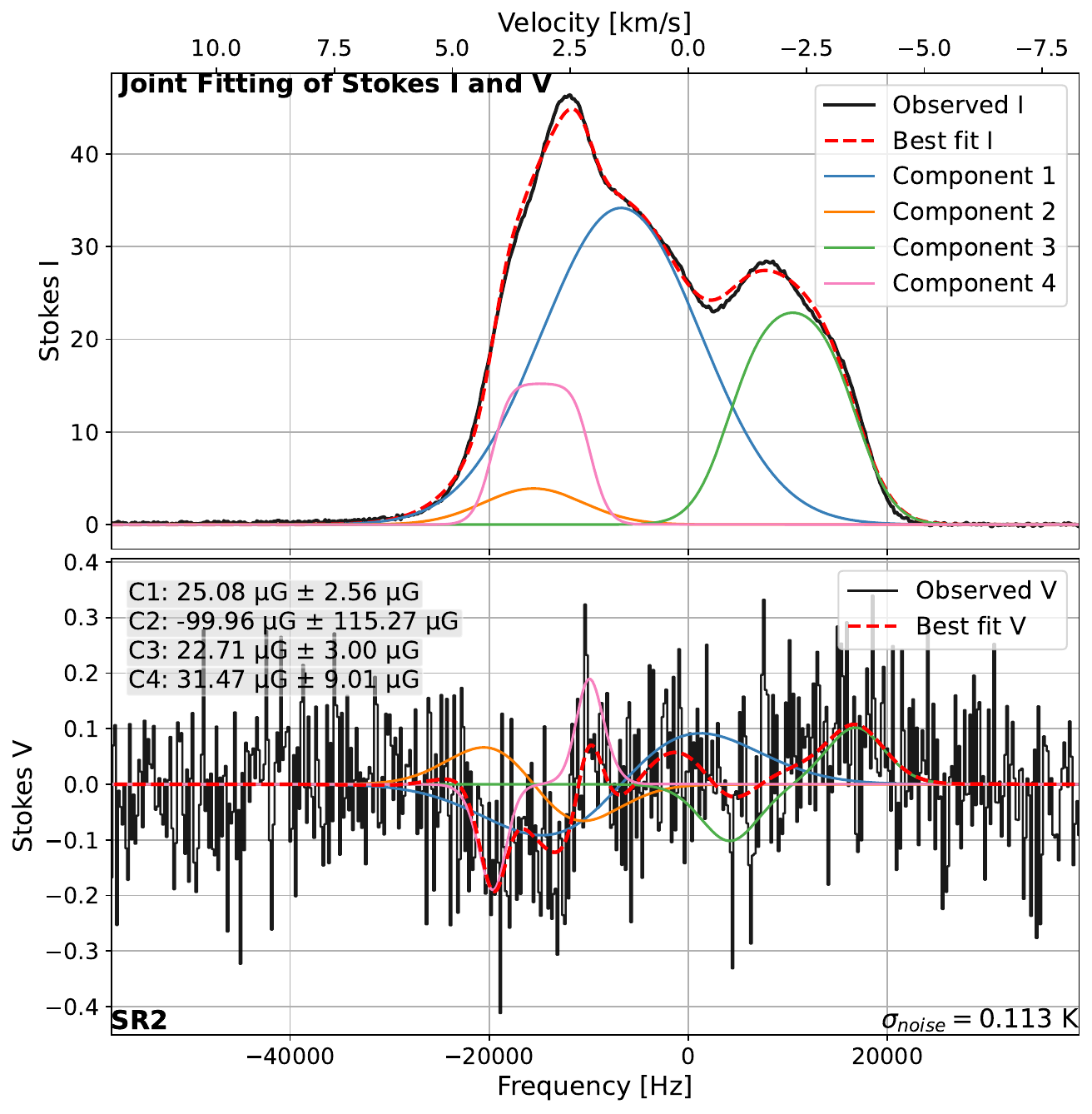}
\caption{Approach II joint strategy results for the SR2 subregion at different noise levels (Section \ref{sec:addnoise}) as indicated in lower right of the $V$ spectra.}
\label{fig.ZeemanFitting_SR2_I_V_fitting_noise_All}
\end{figure}

\begin{figure}[hbt!]
\centering
\includegraphics[width=0.32\linewidth]{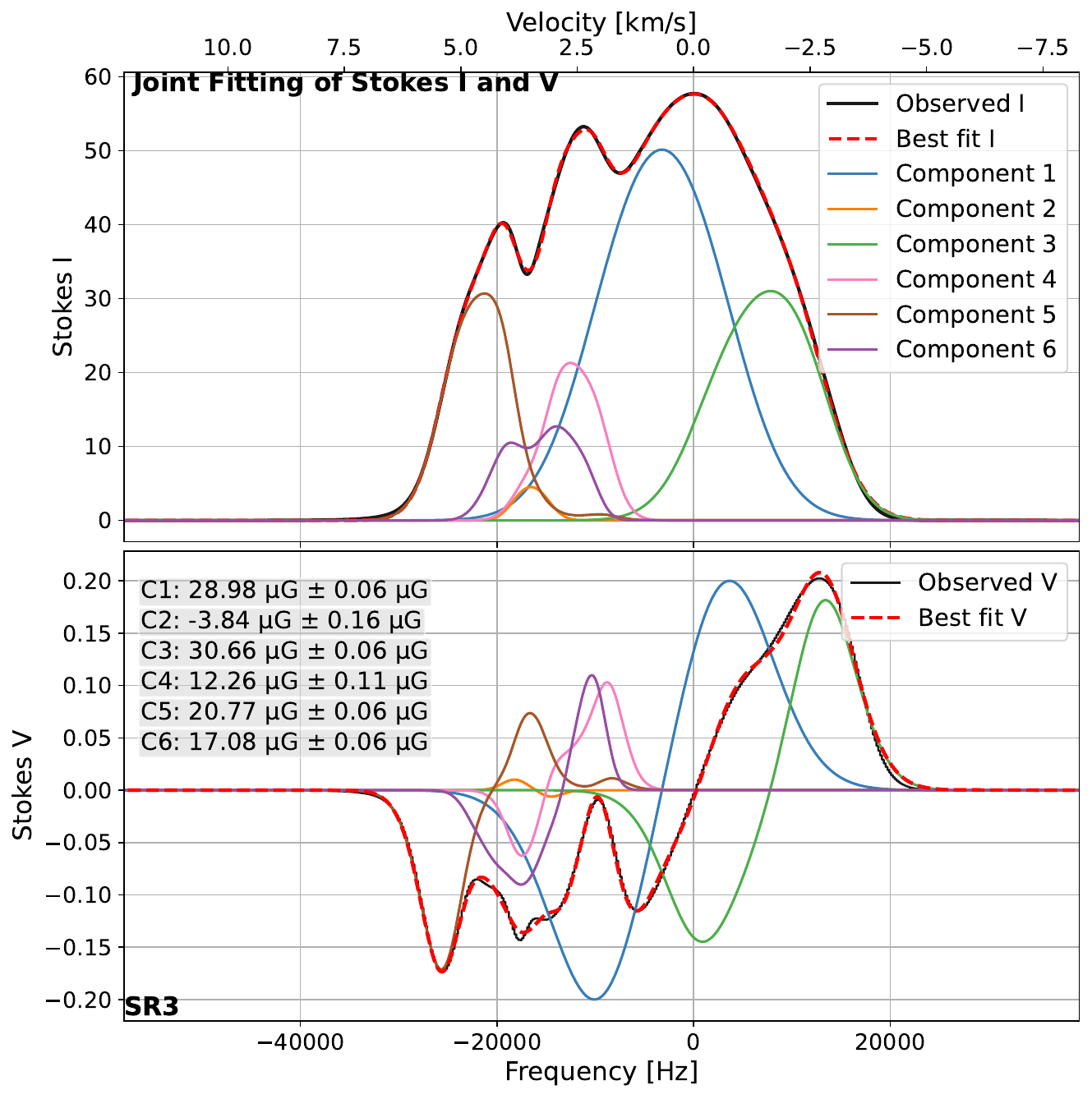}
\includegraphics[width=0.32\linewidth]{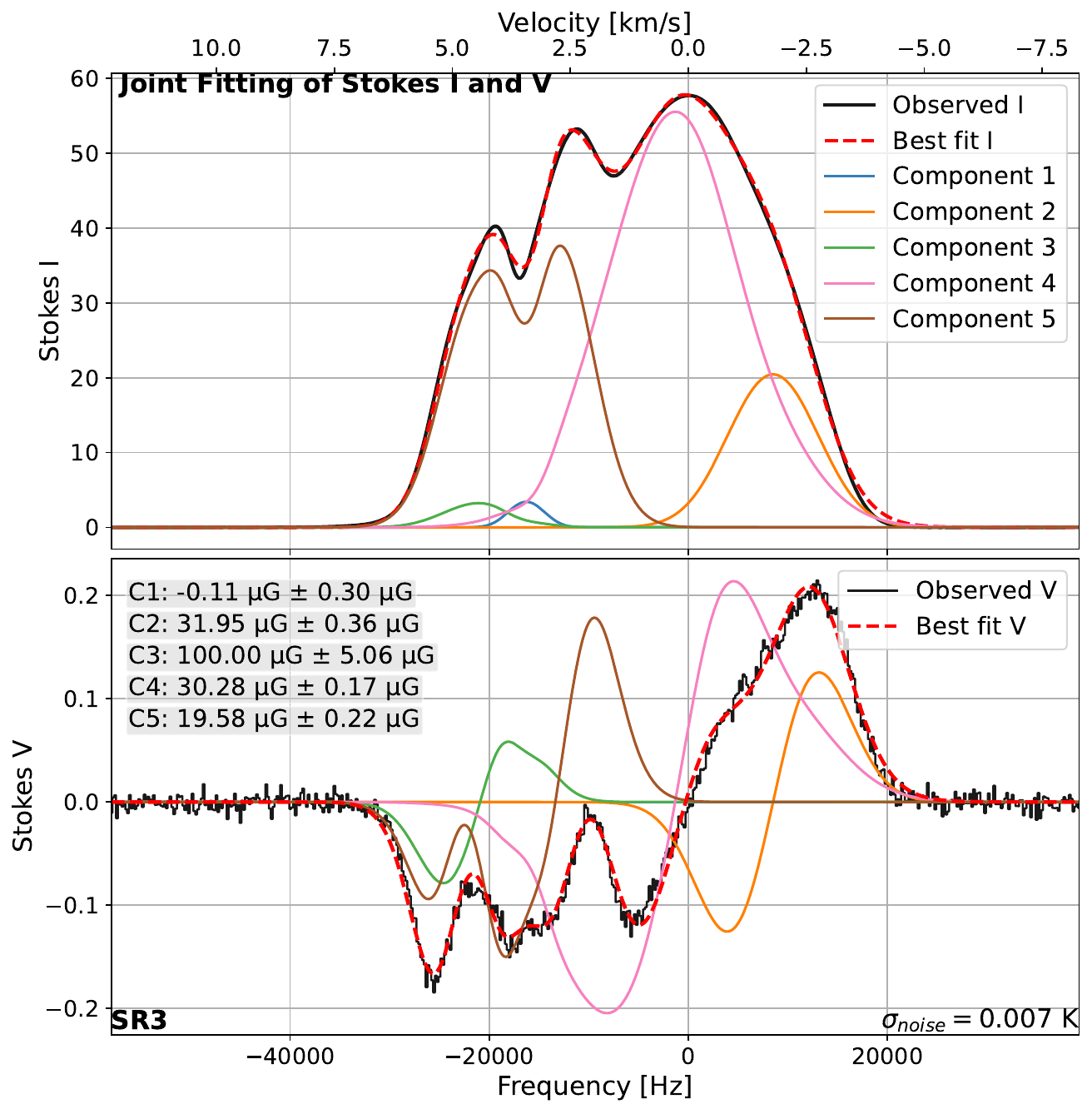}
\includegraphics[width=0.32\linewidth]{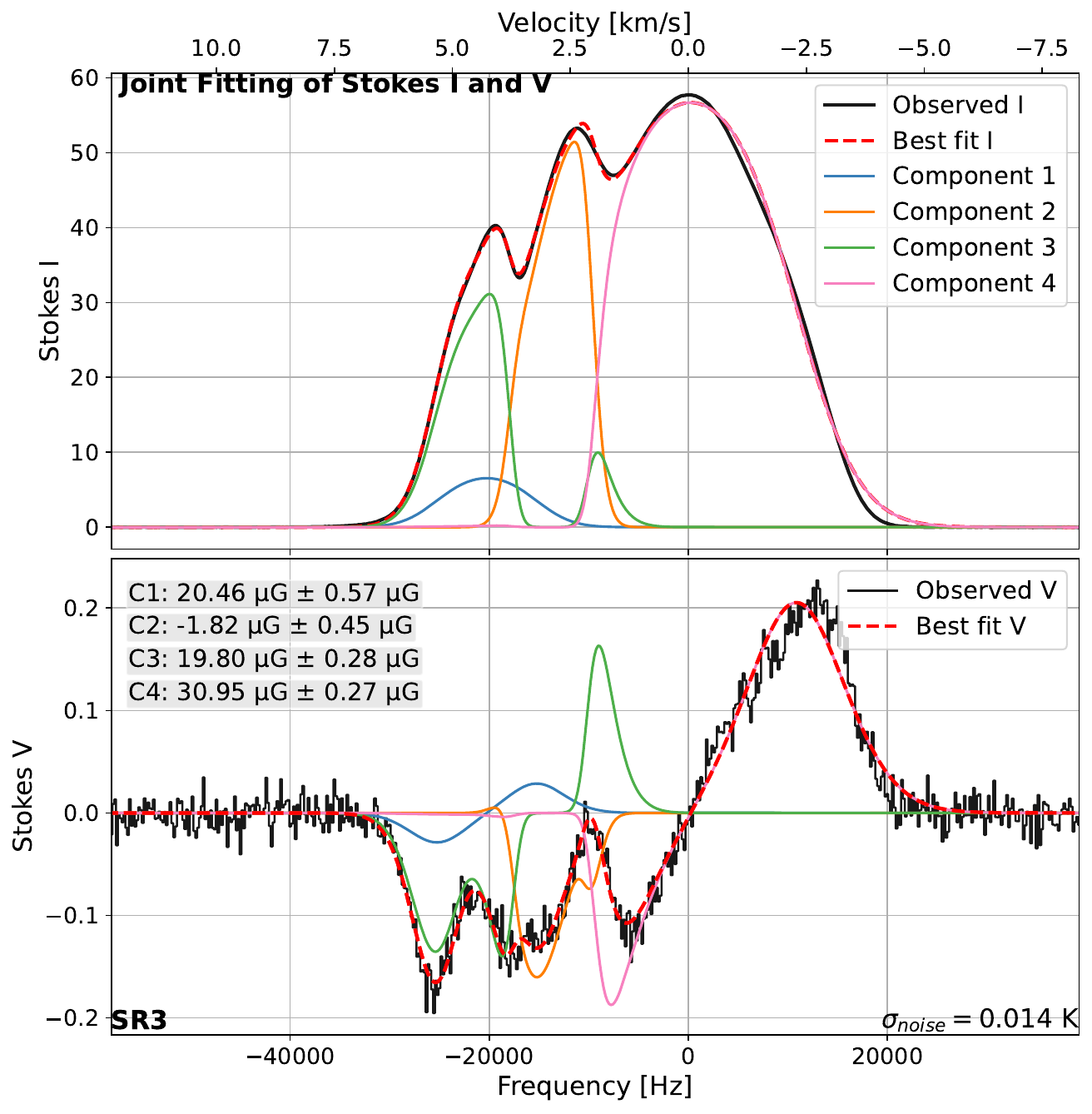}
\includegraphics[width=0.32\linewidth]{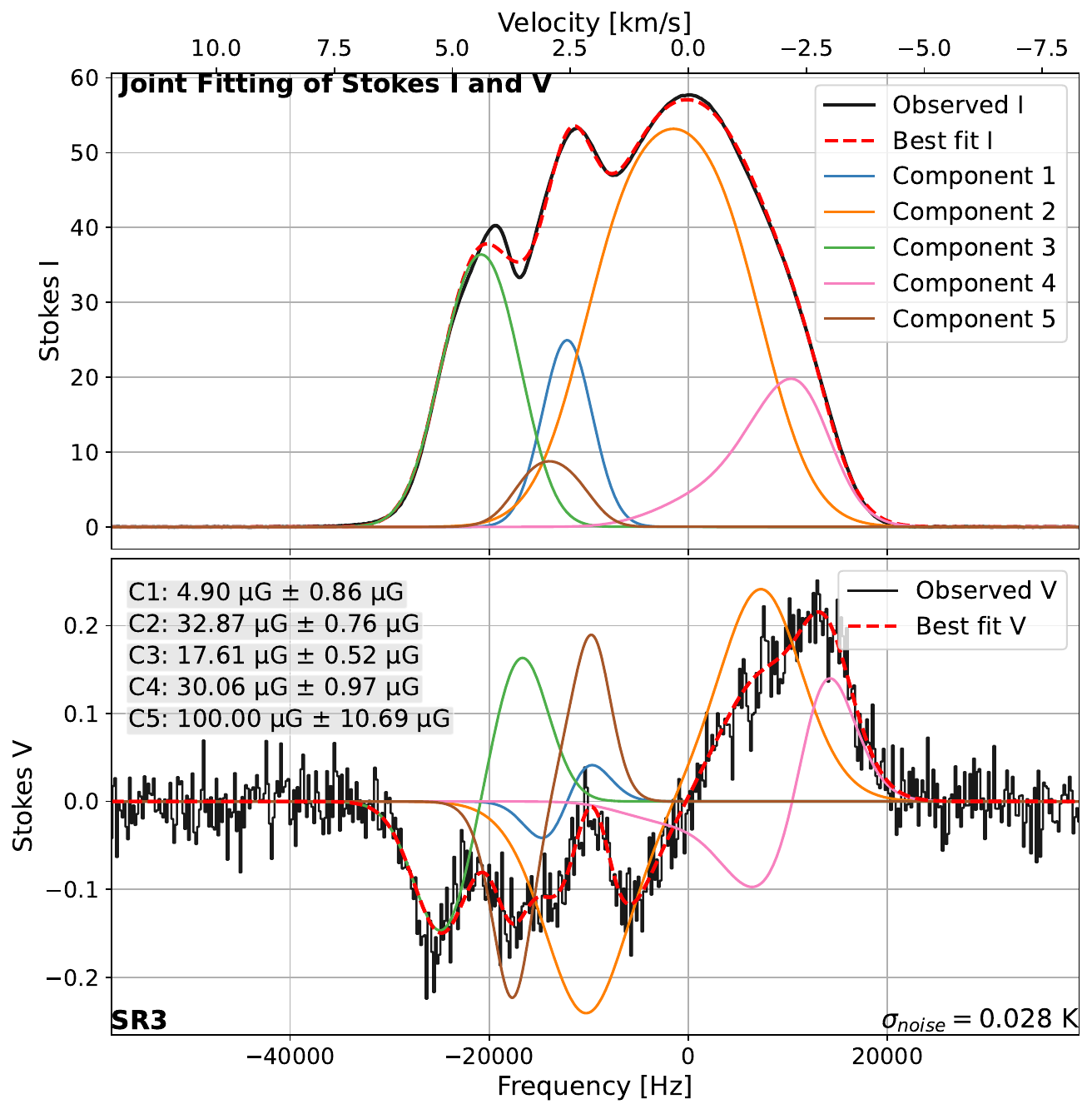}
\includegraphics[width=0.32\linewidth]{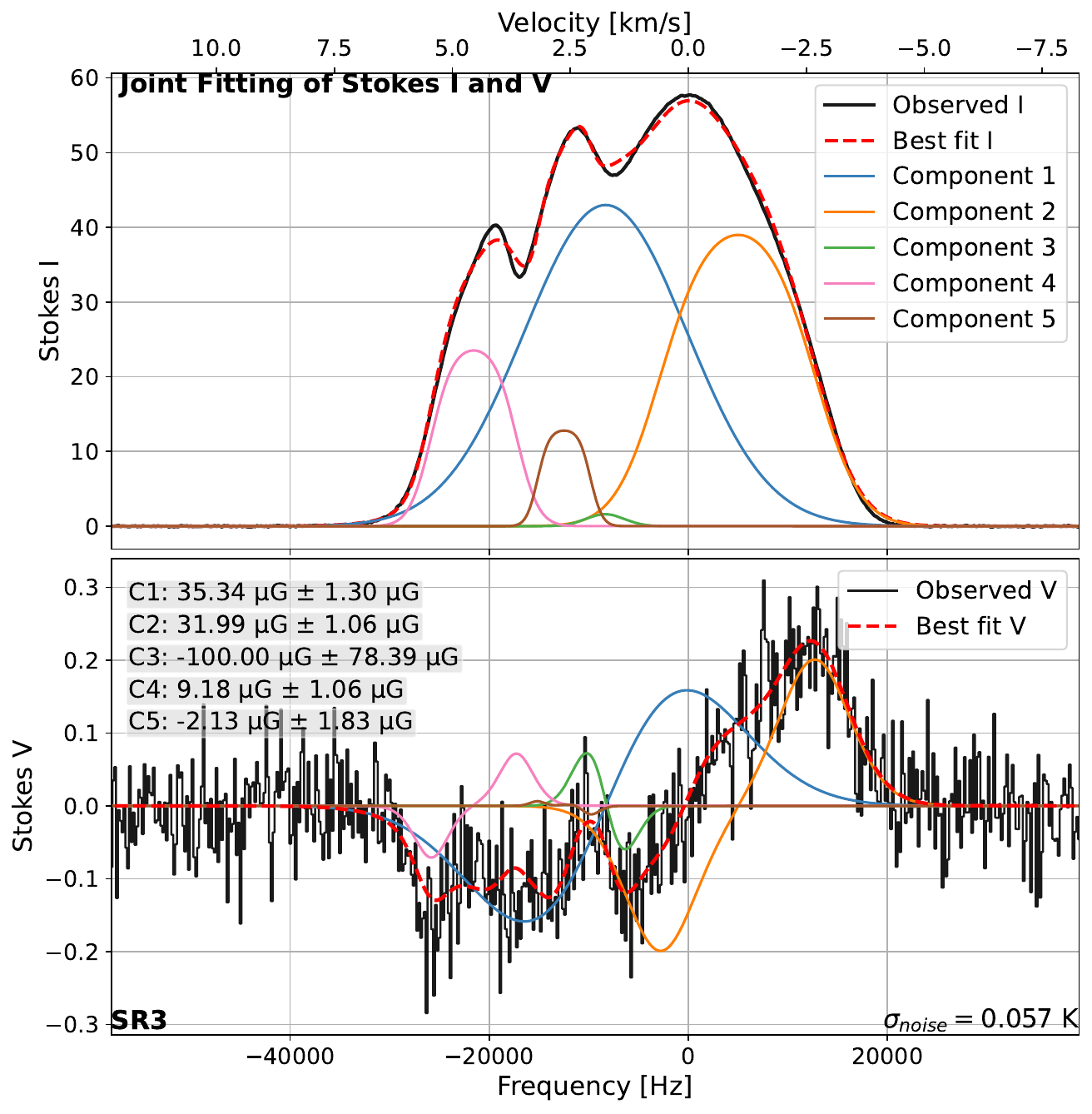}
\includegraphics[width=0.32\linewidth]{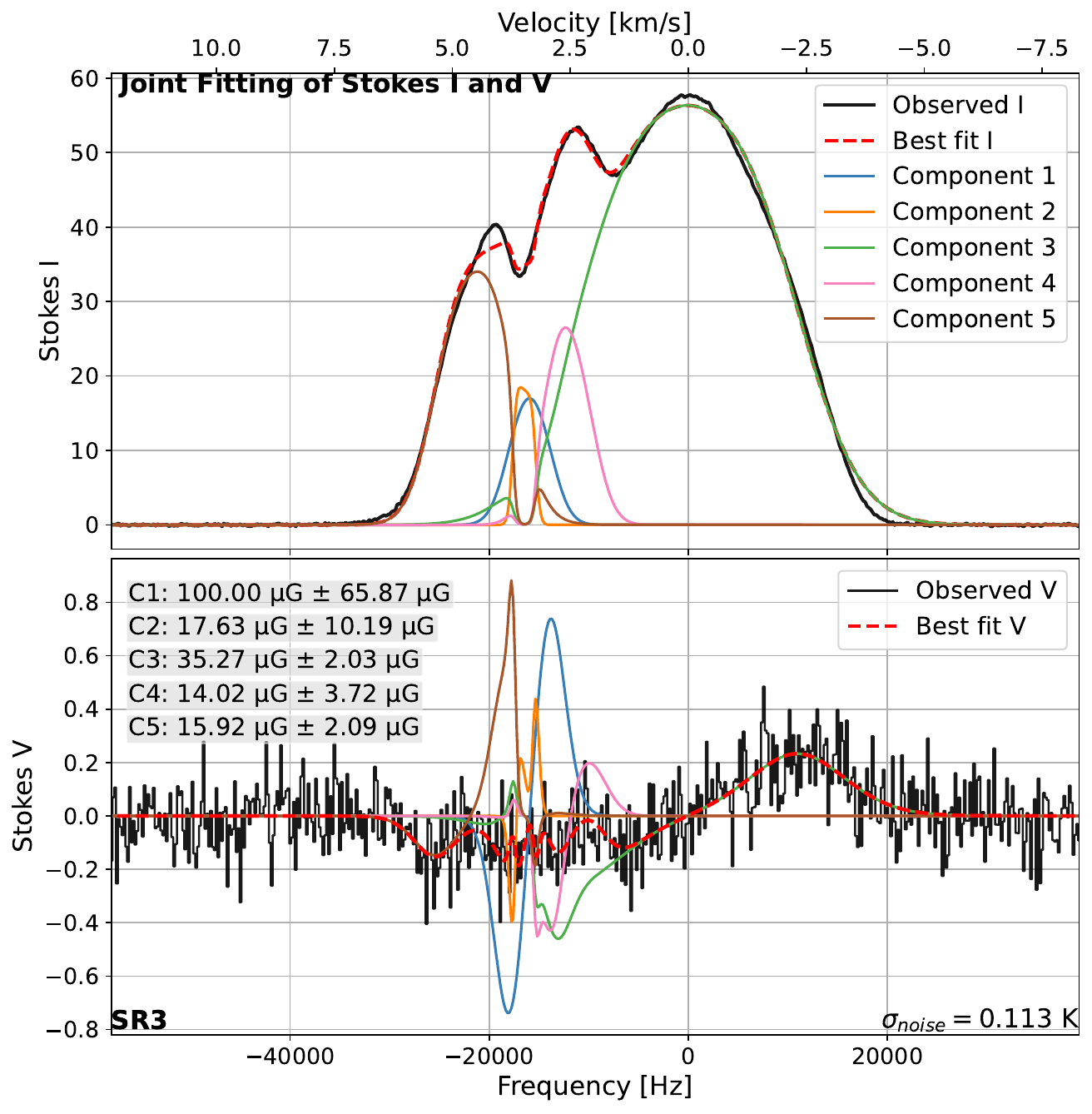}
\caption{Same as Figure~\ref{fig.ZeemanFitting_SR2_I_V_fitting_noise_All}, but for the SR3 subregion, the upper left and right panels being the right panels in Figures \ref{fig.ZeemanFitting_SR3_Breal_noise0} and \ref{fig.ZeemanFitting_SR3_Breal_noise01}, respectively.}
\label{fig.ZeemanFitting_SR3_I_V_fitting_noise_All}
\end{figure}

In this appendix, we present supplementary Approach II joint strategy results fitting Stokes $I$ and $V$.

Figure~\ref{fig.ZeemanFitting_SR_ALL_I_V_fitting_nonoise} shows the noise-free results obtained for subregions LSR1 and SR4, complementing the results for SR3 already examined in Figure \ref{fig.ZeemanFitting_SR3_Breal_noise0} and those for SR2 in the upper left of Figure \ref{fig.ZeemanFitting_SR2_I_V_fitting_noise_All}.  All fits of $I$ and $V$ are good.

Figure~\ref{fig.ZeemanFitting_SR2_I_V_fitting_noise_All} presents the results for the SR2 subregion for no noise and at noise levels of 0.007, 0.014, 0.028, 0.057, and 0.113 K (Section \ref{sec:addnoise}).  With the noise, solutions have just three or four components.  At 0.057 K and beyond, the fit of Stokes $I$ starts being compromised and some finer features in Stokes $V$ are lost.  The results for 0.007 K and 0.028 K are very similar, except for a swapping of C1 and C2. Generally, the ordering of components is not stable, lowering confidence in being able to extract the line of sight profile of the field.  However, the gas responsible for the negative velocity line wing has a fairly robust field near $20\, \mu$G.  The positive line wing is more challenging when multiple components are contributing, but when there is more clearly a narrower and more dominant component the field is more robust, nearer $15\, \mu$G. 

Figure \ref{fig.ZeemanFitting_SR3_I_V_fitting_noise_All} presents the results for the SR3 subregion for the same range of noise levels, including the no-noise and 0.014 K cases that are examined in detail in Section \ref{sec:sr30} (Figure \ref{fig.ZeemanFitting_SR3_Breal_noise0} right) and Section \ref{sec:sr301} (Figure \ref{fig.ZeemanFitting_SR3_Breal_noise01} right), respectively.  With the noise, solutions have four or five components.  At 0.113 K, the fit becomes contrived and unconvincing in recovering finer features in Stokes $V$.  The solutions for different noise levels do not have a stable ordering of the components.  However, the gas responsible for the negative velocity line wing has a fairly robust field near $30\, \mu$G.  The positive line wing is robust despite multiple components contributing, and when there is more clearly a narrower and more dominant component the field is confirmed to be near $20\, \mu$G. 

\section{Tables of Zeeman-estimated Fields and Corresponding Ground Truth Estimators}
\label{app:tables}

For the no noise case, Table~\ref{tab.Bz_B3D_sum_subregion} summarizes the estimated magnetic field strengths and corresponding ground truth estimators of the fitted spectral components obtained using both sequential and joint fitting strategies of Approach II across the four subregions. It is evident that Zeeman fitting does not always recover the ground truth estimators of magnetic field for these components, and in some cases, the fit can even infer an opposite LOS magnetic field direction compared to the estimators of the ground truth. 

Table~\ref{tab.Bz_B3D_sum_subregion_noise01} summarizes the results for the low-noise case (0.014 K). With the addition of noise, the number of components recovered by the sequential strategy remains largely unchanged relative to the noise-free case, whereas the joint strategy converges to fewer components, yielding four components for all subregions at this noise level. As in the noise-free case, the Zeeman fits at 0.014 K do not consistently recover the ground-truth magnetic field estimators for individual components, and in some cases even return magnetic field directions opposite to those of the corresponding ground truth.


\begin{deluxetable}{c|cccc|cccc}[p]
\label{tab.Bz_B3D_sum_subregion}
    \tablecaption{Summary of the ground truth LOS magnetic field and Zeeman fitting results for different subregions$^a$ }
\tablehead{ \multirow{3}{*}{{Subregion}} & \multicolumn{4}{c|}{Sequential Strategy} & \multicolumn{4}{c}{Joint Strategy} \\ \cline{2-9}
 & \multirow{2}{*}{Component}  & $B_{zm}$ & $B_{GT1}$ & $B_{GT2}$ & \multirow{2}{*}{Component}  & $B_{zm}$ & $B_{GT1}$ & $B_{GT2}$ \\ 
    &    & $(\mu G)$ & $(\mu G)$ & $(\mu G)$ &    & $(\mu G)$ & $(\mu G)$ & $(\mu G)$    }
  \startdata
        \multirow{7}{*} {LSR1} & C1 & 14.50 & 13.86 & 13.92 & C1 & 28.29 & 12.75 & 18.44 \\ 
        ~ & C2 & 24.41 & 20.01 & 17.82 & C2 & 12.71 & 14.67 & 11.70 \\ 
        ~ & C3 & 20.27 & 16.23 & 15.40 & C3 & 33.64 & 14.46 & 25.01  \\ 
        ~ & C4 & 21.23 & 13.24 & 18.60 & C4 & 19.91 & 20.76 & 19.06  \\ 
        ~ & C5 & 25.95 & 13.37 & 23.78 & C5 & 32.02 & 13.78 & 26.08 \\ 
        ~ & ~ & ~ & ~ & ~ & C6 & 20.25 & 13.52 & 20.23 \\ 
        ~ & ~ & ~ & ~ & ~ & ~ & ~ & ~ & ~ \\ \hline
        \multirow{8}{*}{SR2} & C1 & -4.98 & 20.84 & 3.11 & C1 & -100.00 & 26.37 & 28.40 \\ 
        ~ & C2 & 16.26 & 15.30 & 19.51 & C2 & -1.46 & 24.25 & -1.68 \\ 
        ~ & C3 & 9.02 & 25.12 & 3.62 & C3 & 17.49 & 13.80 & 17.98 \\ 
        ~ & C4 & 14.74 & 25.42 & 19.17 & C4 & 7.92 & 22.13 & 1.23 \\ 
        ~ & C5 & 19.25 & 21.16 & 23.02 & C5 & 21.62 & 23.03 & 24.41 \\ 
        ~ & C6 & 27.09 & 12.76 & 19.12 & C6 & 17.79 & 18.94 & 2.41 \\ 
        ~ & ~ & ~ & ~ & ~ & C7 & 6.45 & 17.06 & 4.93 \\ 
        ~ & ~ & ~ & ~ & ~ & C8 & 31.05 & 12.59 & 24.75 \\ 
        ~ & ~ & ~ & ~ & ~ & ~ & ~ & ~ & ~ \\ \hline
        \multirow{8}{*}{SR3} & C1 & 18.79 & 19.30 & 28.19 & C1 & 28.98 & 12.33 & 22.05 \\ 
        ~ & C2 & 23.55 & 8.34 & 23.69 & C2 & -3.84 & 13.66 & 10.07 \\ 
        ~ & C3 & 30.87 & 22.20 & 16.04 & C3 & 30.66 & 24.49 & 15.32 \\ 
        ~ & C4 & 12.44 & 4.27 & 16.54 & C4 & 12.26 & 4.63 & 17.20 \\ 
        ~ & C5 & 3.03 & 19.49 & 24.40 & C5 & 20.77 & 19.44 & 21.61 \\ 
        ~ & C6 & 28.99 & 28.94 & 1.07 & C6 & 17.08 & 9.92 & 19.65 \\ 
        ~ & C7 & 27.12 & 5.83 & 26.03 & ~ & ~ & ~ & ~ \\ 
        ~ & C8 & 28.94 & 13.62 & 21.84 & ~ & ~ & ~ & ~ \\ 
        ~ & ~ & ~ & ~ & ~ & ~ & ~ & ~ & ~ \\ \hline
        \multirow{6}{*}{SR4} & C1 & 23.00 & 21.92 & 23.45 & C1 & 36.63 & 30.26 & 23.35 \\ 
        ~ & C2 & 26.84 & 28.91 & 23.52 & C2 & 22.42 & 21.45 & 24.20 \\ 
        ~ & C3 & 46.56 & 30.73 & 21.41 & C3 & 19.48 & 25.18 & 19.39 \\ 
        ~ & C4 & 19.53 & 21.17 & 20.92 & C4 & 34.55 & 29.84 & 23.52 \\ 
        ~ & C5 & 17.45 & 26.98 & 19.68 & C5 & 19.95 & 22.56 & 21.24 \\ 
        ~ & ~ & ~ & ~ & ~ & C6 & 20.14 & 26.10 & 20.61 
\enddata
\tablenotetext{}{Note:}
\tablenotetext{}{$^a$ Summary of fitted components and magnetic field strengths derived from the sequential and joint fitting strategies for the four subregions shown in Figure~\ref{fig.column_density_full_4panel}. For each strategy, the table lists the fitted magnetic field values alongside the corresponding ground truth values defined by Equations~\ref{eq.BGT1} and~\ref{eq.BGT2}.}
\end{deluxetable}

\begin{deluxetable}{c|cccc|cccc}[p]
\label{tab.Bz_B3D_sum_subregion_noise01}
    \tablecaption{\xd{Similar to Table~\ref{tab.Bz_B3D_sum_subregion}, but for spectra with a noise level of 0.014 K}}
\tablehead{ \multirow{3}{*}{{Subregion}} & \multicolumn{4}{c|}{Sequential Strategy} & \multicolumn{4}{c}{Joint Strategy} \\ \cline{2-9}
 & \multirow{2}{*}{Component}  & $B_{zm}$ & $B_{GT1}$ & $B_{GT2}$ & \multirow{2}{*}{Component}  & $B_{zm}$ & $B_{GT1}$ & $B_{GT2}$ \\ 
    &    & $(\mu G)$ & $(\mu G)$ & $(\mu G)$ &    & $(\mu G)$ & $(\mu G)$ & $(\mu G)$    }
  \startdata
        \multirow{7}{*}{LSR1} & C1 & 27.43 & 13.23 & 24.14 & C1 & -27.70 & 13.81 & 25.21 \\ 
        ~ & C2 & 18.90 & 13.21 & 19.02 & C2 & 23.02 & 15.07 & 19.13 \\ 
        ~ & C3 & 21.12 & 18.46 & 17.81 & C3 & 22.42 & 14.87 & 19.83 \\ 
        ~ & C4 & 21.14 & 14.17 & 20.32 & C4 & 22.40 & 13.32 & 20.36 \\ 
        ~ & C5 & 21.31 & 16.79 & 19.92 & ~ & ~ & ~ & ~ \\ 
        ~ & C6 & 20.42 & 13.08 & 19.59 & ~ & ~ & ~ & ~ \\ 
        ~ & ~ & ~ & ~ & ~ & ~ & ~ & ~ & ~ \\ \hline
        \multirow{8}{*}{SR2} & C1 & 25.89 & 12.82 & 16.76 & C1 & 12.00 & 14.76 & 17.08 \\ 
        ~ & C2 & 8.49 & 25.60 & 20.95 & C2 & 18.84 & 21.40 & 22.90 \\ 
        ~ & C3 & 19.87 & 23.44 & 23.58 & C3 & 23.29 & 15.48 & 21.41 \\ 
        ~ & C4 & 16.11 & 21.60 & 13.64 & C4 & 31.44 & 26.71 & -6.84 \\ 
        ~ & C5 & 18.30 & 19.35 & 12.98 & ~ & ~ & ~ & ~ \\ 
        ~ & C6 & 16.93 & 15.14 & 20.82 & ~ & ~ & ~ & ~ \\ 
        ~ & C7 & 8.81 & 24.47 & 9.51 & ~ & ~ & ~ & ~ \\ 
        ~ & ~ & ~ & ~ & ~ & ~ & ~ & ~ & ~ \\ \hline
        \multirow{9}{*}{SR3} & C1 & 12.78 & 9.37 & 17.60 & C1 & 20.46 & 16.51 & 24.53 \\ 
        ~ & C2 & 31.44 & 9.95 & 34.54 & C2 & -1.82 & 5.13 & 13.73 \\ 
        ~ & C3 & -5.58 & 20.32 & 30.23 & C3 & 19.80 & 17.52 & 20.47 \\ 
        ~ & C4 & 24.95 & 29.64 & -6.11 & C4 & 30.95 & 17.56 & 21.60 \\ 
        ~ & C5 & 30.41 & 16.32 & 21.91 & ~ & ~ & ~ & ~ \\ 
        ~ & C6 & 31.34 & 25.04 & 14.06 & ~ & ~ & ~ & ~ \\ 
        ~ & C7 & 16.19 & 20.10 & 30.07 & ~ & ~ & ~ & ~ \\ 
        ~ & C8 & 8.70 & 4.10 & 15.19 & ~ & ~ & ~ & ~ \\ 
        ~ & ~ & ~ & ~ & ~ & ~ & ~ & ~ & ~ \\ \hline
        \multirow{7}{*}{SR4} & C1 & -0.33 & 20.66 & 21.81 & C1 & 21.33 & 22.18 & 22.27 \\ 
        ~ & C2 & 28.29 & 27.58 & 22.20 & C2 & 16.12 & 21.20 & 20.80 \\ 
        ~ & C3 & 21.28 & 21.67 & 22.08 & C3 & 30.08 & 29.81 & 23.01 \\ 
        ~ & C4 & 24.68 & 27.86 & 20.67 & C4 & 28.72 & 27.36 & 21.72 \\ 
        ~ & C5 & 16.91 & 20.95 & 20.65 & ~ & ~ & ~ & ~ \\ 
        ~ & C6 & 29.73 & 28.96 & 23.38 & ~ & ~ & ~ & ~ \\ 
\enddata
\end{deluxetable}

Table~\ref{tab.Bz_B3D_sum_subregion_flip} presents the inferred field strengths and corresponding ground truth estimators for fitted components for the four subregions when viewed from the $-z$ axis, using both the sequential and joint fitting strategies. As with Table~\ref{tab.Bz_B3D_sum_subregion}, the Zeeman fitting results do not consistently recover the ground truth magnetic field estimators.  The trends for the joint strategy appear to be the more interesting.

\begin{deluxetable}{c|cccc|cccc}[p]
\label{tab.Bz_B3D_sum_subregion_flip}
    \tablecaption{Similar to Table~\ref{tab.Bz_B3D_sum_subregion} but viewing from the $-z$ direction  }
\tablehead{ \multirow{3}{*}{{Subregion}} & \multicolumn{4}{c|}{Sequential Strategy} & \multicolumn{4}{c}{Joint Strategy} \\ \cline{2-9}
 & \multirow{2}{*}{Component}  & $B_{zm}$ & $B_{GT1}$ & $B_{GT2}$ & \multirow{2}{*}{Component}  & $B_{zm}$ & $B_{GT1}$ & $B_{GT2}$ \\ 
    &    & $(\mu G)$ & $(\mu G)$ & $(\mu G)$ &    & $(\mu G)$ & $(\mu G)$ & $(\mu G)$    }
  \startdata
        \multirow{7}{*}{LSR1}      & C1        & -18.38   & -13.40 & -19.21 & C1 & -34.49 & -13.94 & -25.66 \\
       & C2        & -16.09   & -16.76 & -14.17 & C2 & -0.63 & -13.24 & -0.84 \\
       & C3        & -32.09   & -13.34 & -26.25 & C3 & -18.16 & -18.10 & -16.20 \\
       & C4        & -26.24   & -21.04 & -19.04 & C4 & -20.43 & -20.64 & -12.00 \\
       & C5        & -22.50   & -20.04 & -17.14 & C5 & -27.94 & -15.19 & -22.74 \\
       & C6        & -8.32    & -12.92 & -7.26  & C6 & 100.00 & -14.75 & -28.20  \\
       & C7        & -18.78   & -13.62 & -21.84 &  C7 & -2.96 & -12.96 & -4.24   \\
       &           &          &        &        &           &          &        &        \\\hline
\multirow{8}{*}{SR2}      & C1 & -5.78 & -21.31 & -13.86 & C1        & -16.20   & -14.31 & -17.03 \\
       & C2 & -11.72 & -14.04 & -14.69 & C2        & 2.18     & -12.75 & -4.15  \\
       & C3 & -10.39 & -13.08 & -12.41 & C3        & -5.31    & -23.94 & -12.17 \\
       & C4 & -5.36 & -14.14 & -7.43   & C4        & -19.27   & -22.03 & -20.34 \\
       & C5 & -4.11 & -26.22 & -17.96  & C5        & -9.09    & -14.92 & -6.01  \\
       & C6 & -14.50 & -24.62 & -13.95   & C6        & -19.11   & -21.32 & -23.16 \\
       &  C7 & -23.49 & -22.25 & -23.21 & C7        & -30.41   & -24.13 & -24.01 \\
       & C8 & -10.91 & -25.64 & -18.05  &           &          &        &        \\
       & C9 & -194.96 & -19.10 & -31.29  &           &          &        &        \\
       &           &          &        &        &           &          &        &        \\\hline
\multirow{6}{*}{SR3}      & C1        & -8.89    & -10.67 & -13.70 & C1        & -20.58   & -17.25 & -29.54 \\
       & C2        & -3.78    & -16.51 & -5.77  & C2        & -11.19   & -9.36  & -19.08 \\
       & C3        & -5.04    & -6.75  & -10.82 & C3        & -3.58    & -7.45  & -12.03 \\
       & C4        & -18.60   & -14.59 & -29.16 & C4        & -28.06   & -25.53 & -13.64 \\
       & C5        & -31.86   & -27.05 & -8.83  & C5        & -33.78   & -17.11 & -22.32 \\
       & C6        & -5.80    & -21.72 & -5.26  &           &          &        &        \\
       &           &          &        &        &           &          &        &        \\\hline
\multirow{8}{*}{SR4}      & C1        & -34.41   & -30.72 & -23.23 & C1 & -25.82 & -21.85 & -20.20 \\
       & C2        & -28.43   & -24.12 & -21.62 & C2 & -28.13 & -22.68 & -20.10 \\
       & C3        & -27.21   & -29.95 & -21.18 & C3 & -10.82 & -20.82 & -21.16 \\
       & C4        & -23.75   & -26.84 & -22.69 & C4 & -19.79 & -28.37 & -22.25 \\
       & C5        & -39.43   & -20.77 & -21.71 & C5 & -29.79 & -30.70 & -23.20  \\
       & C6        & -22.26   & -22.10 & -22.01 & C6 & -21.97 & -22.99 & -22.08 \\
       &           &          &        &        & C7 & -37.79 & -28.97 & -21.51\\
\enddata
\end{deluxetable}

\bibliographystyle{aasjournalv7}
\bibliography{references}

\end{CJK*}

\end{document}